\documentclass[graybox]{svmult}

\usepackage{type1cm}        
\usepackage{makeidx}         
\usepackage{graphicx}        
\usepackage{siunitx}         
\usepackage{multicol}        
\usepackage[bottom]{footmisc}

\usepackage{newtxtext}       %
\usepackage[varvw]{newtxmath}       

\usepackage{aas_macros}
\usepackage[T1]{fontenc}
\usepackage{wrapfig}
\usepackage{mhchem}

\usepackage[round,colon,authoryear, book]{natbib} 
\usepackage[colorlinks,urlcolor=cyan,citecolor=blue,linkcolor=blue]{hyperref}
\DeclareSIUnit\year{yr}

\makeindex             
    
\begin{document}

\title*{Active moons in our Solar System and beyond - Io, Europa, Enceladus, Triton, and exomoons}
\titlerunning{Active moons in our Solar System and beyond}
\author{Caroline Haslebacher\orcidID{0000-0003-0674-1216} and\\ 
Emeline Bolmont \orcidID{0000-0001-5657-4503} and\\ 
Marco Cilibrasi \orcidID{0000-0002-4228-9936} and\\ 
Jonathan Grone \orcidID{0000-0001-5074-265X} and \\ 
Nico Haslebacher\orcidID{0000-0003-1802-9859} and \\ 
Ravit Helled  \orcidID{0000-0001-5555-2652} and\\ 
Mathilde Kervazo \orcidID{0000-0003-1196-6233} and\\ 
Niels F.W. Ligterink \orcidID{0000-0002-8385-9149} and\\ 
Christophe Lovis\orcidID{0000-0001-7120-5837} and\\ 
Lucio Mayer \orcidID{0000-0002-7078-2074} and\\ 
Lorenzo Obersnel\orcidID{0009-0003-3070-9002} and\\
Rafael Ottersberg\orcidID{0009-0002-0254-9411} and\\ 
Apurva V. Oza\orcidID{0000-0002-1655-0715} and\\ 
C.H. Lucas Patty \orcidID{0000-0002-0073-8879} and\\ 
Antoine Pommerol\orcidID{0000-0002-9165-9243} and\\ Ganna Portyankina\orcidID{0000-0002-1323-8195} and\\ Alyssa R. Rhoden\orcidID{0000-0003-2805-4994} and\\ Leander Schlarmann \orcidID{0000-0001-5800-132X} and \\ Yuhito Shibaike\orcidID{0000-0003-2993-5312} and\\ Vishaal Singh\orcidID{0000-0002-8009-8700} and\\ Audrey H. Vorburger \orcidID{0000-0002-7400-9142} and\\ Peter Wurz \orcidID{0000-0002-2603-1169} and\\ } 

\institute{Caroline Haslebacher (corresponding author) \at Division of Space Research and Planetary Sciences, Physics Institute, University of Bern, Sidlerstrasse 5, 3012 Bern, Switzerland, and 
\at Southwest Research Institute, 1301 Walnut St. Suite 400, 80302 Boulder, CO, USA \email{caroline.haslebacher@contractor.swri.org}
\and Emeline Bolmont \at Department of Astronomy, University of Geneva, Chemin Pegasi 51, CH-1290 Versoix, Switzerland \email{emeline.Bolmont@unige.ch}
\and Marco Cilibrasi \at Department of Astrophysics, University of Zurich,
Winterthurerstrasse 190, 805, Zurich, Switzerland, \email{marco.cilibrasi@uzh.ch} 
\and Jonathan Grone \at Center for Space and Habitability, University of Bern, Gesellschaftsstrasse 6, 3012 Bern, Switzerland \email{jonathan.grone@unibe.ch}
\and Nico Haslebacher \at Division of Space Research and Planetary Sciences, Physics Institute, University of Bern, Sidlerstrasse 5, 3012 Bern, Switzerland \email{nico.haslebacher@unibe.ch}
\and Ravit Helled \at Department of Astrophysics, University of Zurich,
Winterthurerstrasse 190, 805, Zurich, Switzerland \email{ravit.helled@uzh.ch}
\and Mathilde Kervazo \at Nantes Universite, Univ Angers, Le Mans Universite, CNRS, Laboratoire de Planetologie et Geosciences, LPG UMR 6112, 44000 Nantes, France \email{mathilde.kervazo@univ-nantes.fr}
\and Niels F.W. Ligterink \at Faculty of Aerospace Engineering, Delft University of Technology, Delft, The Netherlands \at Center for Space and Habitability, University of Bern, Bern, Switzerland,  \email{niels.ligterink@tudelft.nl}
\and Christophe Lovis \at Department of Astronomy, University of Geneva, Chemin Pegasi 51, CH-1290 Versoix, Switzerland \email{christophe.lovis@unige.ch}
\and Lucio Mayer \at Department of Astrophysics, University of Zurich,
Winterthurerstrasse 190, 805, Zurich, Switzerland, \email{lmayer@physik.uzh.ch} 
\and Lorenzo Obersnel \at Division of Space Research and Planetary Sciences, Physics Institute, University of Bern, Sidlerstrasse 5, 3012 Bern, Switzerland \email{lorenzo.obersnel@unibe.ch}
\and Rafael Ottersberg \at Division of Space Research and Planetary Sciences, Physics Institute, University of Bern, Sidlerstrasse 5, 3012 Bern, Switzerland \email{rafael.ottersberg@unibe.ch}
\and Apurva V. Oza \at Division of Geological and Planetary Sciences, California Institute of Technology, 1200 E. California Boulevard, MC 150-21, Pasadena CA 91125, USA
\at Jet Propulsion Laboratory, California Institute of Technology, 4800 Oak Grove Drive, Pasadena, CA 91109, USA
\email{oza@caltech.edu}
\and C.H. Lucas Patty \at ARTORG Center for Biomedical Engineering Research, University of Bern, Switzerland \at Center for Space and Habitability, University of Bern, Bern, Switzerland \email{lucas.patty@unibe.ch} 
\and Antoine Pommerol \at Division of Space Research and Planetary Sciences, Physics Institute, University of Bern, Sidlerstrasse 5, 3012 Bern, Switzerland \email{antoine.pommerol@unibe.ch}
\and Ganna Portyankina \at German Aerospace Center (DLR), Berlin \email{ganna.portyankina@dlr.de}
\and Alyssa R. Rhoden \at Southwest Research Institute, 1301 Walnut St Suite 400, 80302 Boulder, CO, USA \email{alyssa@boulder.swri.edu}
\and Leander Schlarmann \at Division of Space Research and Planetary Sciences, Physics Institute, University of Bern, Sidlerstrasse 5, 3012 Bern,  Switzerland \email{leander.schlarmann@unibe.ch} 
\and Yuhito Shibaike \at National Astronomical Observatory of Japan, 2-21-1 Osawa, Mitaka, Tokyo 181-8588, Japan, and \at Graduate School of Science and Engineering, Kagoshima University, 1-21-35 Korimoto, Kagoshima, Kagoshima 890-0065, Japan \email{yuhito.shibaike@sci.kagoshima-u.ac.jp}
\and Vishaal Singh \at Southwest Research Institute, 1301 Walnut St Suite 400, 80302 Boulder, CO, USA \email{vishaal.singh@swri.org}
\and Audrey H. Vorburger \at Division of Space Research and Planetary Sciences, Physics Institute, University of Bern, Sidlerstrasse 5, 3012 Bern,  Switzerland \email{audrey.vorburger@unibe.ch}
\and Peter Wurz \at Division of Space Research and Planetary Sciences, Physics Institute, University of Bern, Sidlerstrasse 5, 3012 Bern,  Switzerland \email{peter.wurz@unibe.ch}}
\authorrunning{Haslebacher et al.}
%
%

\maketitle

\abstract{The outgassing signatures of Io, Europa, Enceladus, Triton, and Io-like exomoons are the focus of this review chapter. The rocky volcanic world of Io is unique in our Solar System, with plumes reaching to hundreds of kilometres in altitude. Io-like exomoons could leave signatures strong enough to be detected with ground-based telescopes. The icy moons Europa and Enceladus, with their subsurface oceans, are currently the best candidates for life. Triton is different in many ways and raises unexplored questions.
Our knowledge of these active moons is derived from space- and ground-based observations. To understand their origin, we discuss moon formation in general, before examining evidence and signatures of plumes on these moons. Given the accessibility of subsurface oceanic material through the occurrence of plumes, we expand on possibilities to investigate biosignatures.
}

\section{Introduction}
\label{sec_Introduction}
The study of active processes on a satellite can inform our understanding of interior, atmosphere, and habitability, and constrain possible formation scenarios. In particular, cryovolcanic or volcanic eruptions into the atmosphere and onto the surface of a satellite provide access to subsurface material  \citep[e.g.][]{Fagents2003, Pappalardo2024}. However, accurate interpretation of such endogenic activity demands a deeper understanding of chemical alterations during transport, extrusion, and ejection, as well as exogenic activity. This is especially true for the study of habitability and the detection of biosignatures.
Here, we first provide context for active processes by describing Solar System satellite formation scenarios (Sect. \ref{sect_formation}). Next, we present the current evidence of activity for the Jovian moons Europa and Io, the Saturnian moon Enceladus, and the Neptunian moon Triton (Sect. \ref{sec_solar_system}), and review their specific activity signatures (Sect. \ref{sect_signatures_solar_system}). These sections provide a basis for exploring signatures of outgassing events relevant to biosignature detection (Sect. \ref{sect_biosignatures}) and the detection of exomoons (Sect. \ref{exomoon_detect_activity}).

\section{Formation of Moons in our Solar System}\label{sect_formation} 
Based on their similar mass and coplanar orbits, the Galilean moons of Jupiter (Io, Europa, Ganymede, and Callisto) likely formed in a circumplanetary disk (CPD). CPDs are small gas disks forming around gas accreting planets after the gas cannot keep their circumplanetary envelopes due to cooling \citep[e.g.][]{Szulagyi17gap}.
This scenario is illustrated in Figure 1A, which is applicable to the Galilean moons and perhaps also to Saturn's large moon, Titan \citep[e.g.][]{Canup06}. The original CPD models estimate the mass of the gas disk as 100 times the mass of the moons and considering situations isolated from the parental protoplanetary disk, called ``Minimum Mass Sub-nebular Model'', similar to the  ``Minimum Mass Solar Nebular'' in planet formation theory \citep[e.g.][]{Lunine1982}.
However, the icy moons must have formed in relatively cold environments to avoid sublimation of icy materials. To address this issue, \cite{Canup02} proposed the ``gas-starved disk model'' with smaller gas surface density and continuous supply of gas and solids to the CPD. Several (magneto-)~hydrodynamical (MHD) simulations 
support the scenario \citep[e.g.][]{Kley99,Tanigawa12,Szulagyi14,Cilibrasi23}.
Simulations also show that gas and solid accretion onto a CPD continues even after gaps form around the planet and CPD system \citep[e.g.][]{Szulagyi2022,Maeda2024DeliveryPlanets}. 
Ground-based observations also suggest that the gas and dust accrete into the vicinity of a planet beyond the gas gap of the extra-solar system PDS~70 protoplanetary system \citep[e.g.][]{Isella2019,Christiaens2024,Shibaike2024ConstraintsEvolution} and possibly in the AB~Aurigae system \citep[e.g.][]{Currie2022,Shibaike2025}.

The proposed formation scenarios of large moons in CPDs can be classified into two groups according to the dominant size of the accreting materials: satellitesimals (kilometre-sized) and pebbles (centimetre- to metre-sized). In the former scenario, large satellites form by accreting satellitesimals, and repeated cycles occur in which these satellites subsequently migrate inward and fall onto the central planet via the migration caused by gravitational interactions with the gas disk. The currently observed satellites are thought to represent the final generation that survived the dissipation of the gas disc. The satellitesimal scenario has roughly reproduced the observed mass, orbits, and icy/rocky compositions of the large moons of Jupiter and Saturn \citep[e.g.][]{Canup06,Sasaki10,Arakawa2019}.
However, the formation of satellitesimals is challenging due to the significant radial drift of small grains in the CPD \citep[e.g.][]{Shibaike2017}.
The following solutions have been proposed for the inward drift problem: 1) the accumulation of dust
in specific locations generated
by the outward gas flow in the CPD due to the viscous diffusion \citep{Drazkowska2018,Cilibrasi18,Cilibrasi21} and by the decretion disks outside the centrifugal radii \citep{Batygin2020FormationSatellites}, 2) an efficient collisional growth occurs on the mid-plane of the laminar CPD formed by their magnetic wind driven accretion \citep{Shibaike2023}, or 3) that planetesimals formed in protoplanetary disks are captured by the gas drag of the CPD and alternatively used as satellitesimals \citep[e.g.][]{Fujita2013}. On the other hand, recent pebble accretion scenarios directly use the radially drifting pebbles in the CPD as building material for the moons \citep[e.g.][]{Shibaike2019,Ronnet20}. The pebble accretion scenario can also reproduce the observed characteristics of the moons -- such as their masses, orbits, and ice-to-rock mass fractions -- under appropriate choices of model parameters.

Given that both the satellitesimal and pebble accretion scenarios are still viable, one way to distinguish the true history is by considering the internal structures of Europa and Callisto; satellitesimals could provide more heat during their formation resulting in full differentiation \citep[e.g.][]{Barr2008}. NASA’s Galileo spacecraft measured the gravitational field of the moons and estimated that their interiors could be only partially differentiated \citep[e.g.][]{Schubert2004,Petricca2025}. \citet{Shibaike2025Callisto} found that the partial differentiation of Callisto can only be maintained by the pebble accretion scenario. The gravity measurements by NASA's Europa Clipper mission \citep{Pappalardo2024} and ESA's JUICE mission \citep{Fletcher2023} will measure the degree of the moon's differentiation \citep{Cappuccio2022, Mazarico2023, VanHoolst2024GeophysicalExplorer}, shedding more light on the most likely process of CPD formation. 

\begin{figure}
    \centering
    \includegraphics[width=\linewidth]{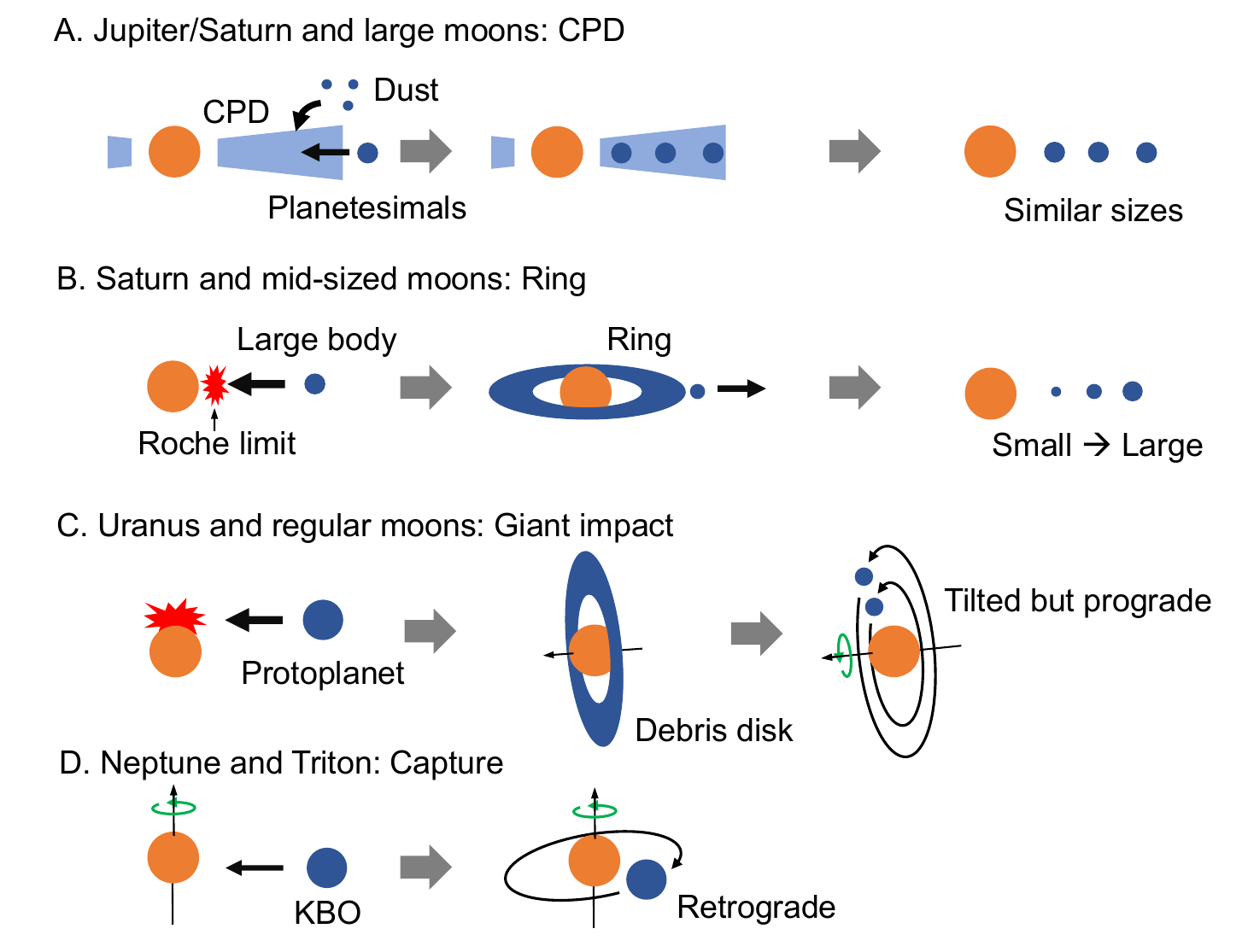}
    \caption{Schematic illustrations of leading hypotheses for formation scenarios of moons of the outer planets (not to scale). A. Formation in a circumplanetary disc is the likely formation mechanism for the large moons of Jupiter as well as large and/or initial moons around other giant planets. B. The mid-sized moons of Saturn likely formed out of a ring, which may have been the debris of a tidally-disrupted CPD moon. C. A giant impact might be responsible for Uranus' tilt, which would have destabilized any existing CPD moons. Its current moons could have formed out of a debris disk following the impact. D. Triton might be a captured KBO.}
    \label{fig:formation}
\end{figure}

The mid-sized moons of Saturn, Mimas, Enceladus, Thethys, Dione, Rhea, and Iapetus, may originate from a tidal disk (i.e., ring) around Saturn \citep[e.g.][]{Charnoz2010TheRings}, as illustrated in Figure 1B. In this model, ring materials that spread beyond the Roche radius of Saturn can grow larger by their mutual collisions to form a small moon. The moon continuously accretes material crossing the Roche radius, while migrating outward due to the exchange of angular momentum with the massive ring. Whether rocky material is the initial seed of the moons or is brought in by later impacts is currently debated \citep{Crida2012,Salmon2017}. In either case, as the ring evolves, the mass-distance distribution of the mid-sized moons can be reproduced.

A smooth particle hydrodynamics (SPH) simulation shows that the icy materials are provided to a ring by the scattered fragments of a Titan-mass differentiated object crossing the Roche radius \citep{Canup2010}. In that case, the ring had to form early in Saturn's history, when there was still gas to drive the inward migration of that satellite. Such an early start has implications for the age of the rings and the likelihood of the mid-sized moons surviving intact to the present day. It has also been suggested that a passing trans-neptunian object (TNO) was disrupted by Saturn to form the ring, which could happen somewhat later in solar system history, with the subsequent ring-moon evolution remaining the same.

The formation path of the present Uranian and Neptunian moons is expected to be different from that of Jupiter and Saturn \citep{Helled2020,Mousis2020,Eriksson2023},  although some works have explored the possibility of in-situ formation \citep[e.g.][]{Szulagyi18}. 
In addition, the satellite systems of Uranus and Neptune are rather different from each other. 

Uranus has five regular moons, which are similar in size to Saturn's mid-sized moons and display a similar mass gradient. The moons orbit in Uranus' equatorial plane, which is tilted by 98 degrees to the ecliptic, leading to the hypothesis that the current moons formed from a disk of debris after a giant impact tilted Uranus, similar to the hypothesized origin of Earth's Moon \citep{Morby2012,Ida2020}. Various simulations \citep[e.g.][]{Wayne1992, Woo2022} confirmed that a massive collision between Uranus and a large planetary body ($\thicksim$ 2--3~Earth masses) early in the solar system's history could have ejected material into orbit, which later coalesced into the planet’s regular satellites. However, the moons may also include debris from an earlier generation of moons that destabilized during tilting. These scenarios are supported by the moons' near-coplanar, prograde orbits, the planet's extreme axial tilt and its rotation rate, all of which are consistent with a violent impact event \citep{Kegerreis2018,Reinhardt2020}. The disk's evolution, governed by accretion dynamics and tidal interactions, likely led to the formation of the current Uranian satellite system. 

Triton, Neptune’s largest moon, is believed to be a captured Kuiper Belt Object (KBO) rather than a moon that formed in situ, although other formation scenarios are still possible \citep{Gomes2024}. Triton's retrograde orbit - opposite to Neptune’s rotation - suggests that it was initially an independent body before being gravitationally captured by Neptune. Triton’s geologically young surface, characterized by cryovolcanism and minimal cratering, indicates continued internal activity, possibly driven by tidal heating (Sect. \ref{subsec:tidal}).

Although Triton receives most of the attention of Neptune's moons, Neptune also has a small, close-in satellite, Proteus ($r= \SI{210}{km}$ orbiting prograde within 5~R$_{\text{Neptune}}$), which is similar in size to Mimas, Enceladus, and Uranus' moon Miranda. Proteus' surface is quite distinct from these other moons, with an extremely low albedo \citep{Nobis2025}, large craters, and a non-spherical shape. Proteus currently has negligible orbital eccentricity, suggesting it reassembled from debris that had settled into the equatorial plane of Neptune, perhaps after Triton's capture destabilized the existing satellite system \citep{Zhang2007, Zhang2008}. However, investigations into other possible origins for Proteus and their implications for its evolution have been limited. Hence, 
Proteus' origin and history are not well-understood, and its lack of similar activity to comparable moons is peculiar.

The formation conditions (as discussed above) affect later moon activity through processes like tidal heating (due to orbital resonances), internal differentiation (core formation, mantle/ice layering), and potential subsurface oceans. While many processes can complicate the link between formation and present-state, these formation scenarios do have implications that can aid in our interpretation of active processes on moons. In particular, the formation scenario can affect the contribution of different heat sources over time \citep[e.g., radiogenic decay compared to tides; as in ][]{Neveu2019}, which exert controls on the evolution of an ocean and ice shell. In addition, whether a moon formed from accretion in a CPD or emerged from a disk of debris can affect whether the rocky materials were ever exposed to liquid water. For example, the composition of Enceladus' plumes suggest on-going hydrothermal alteration within the ocean \citep[e.g.][]{Hsu2015OngoingEnceladus, Postberg2018}, which places limits on the historic mixing between rock and liquid water in Enceladus' past.

\section{Active Processes: Observations, Interpretations, and Environment} 
\label{sec_solar_system}

\subsection{Observations of Active Processes}\label{sec:obs}
\begin{figure}
    \centering
    \includegraphics[width=\linewidth]{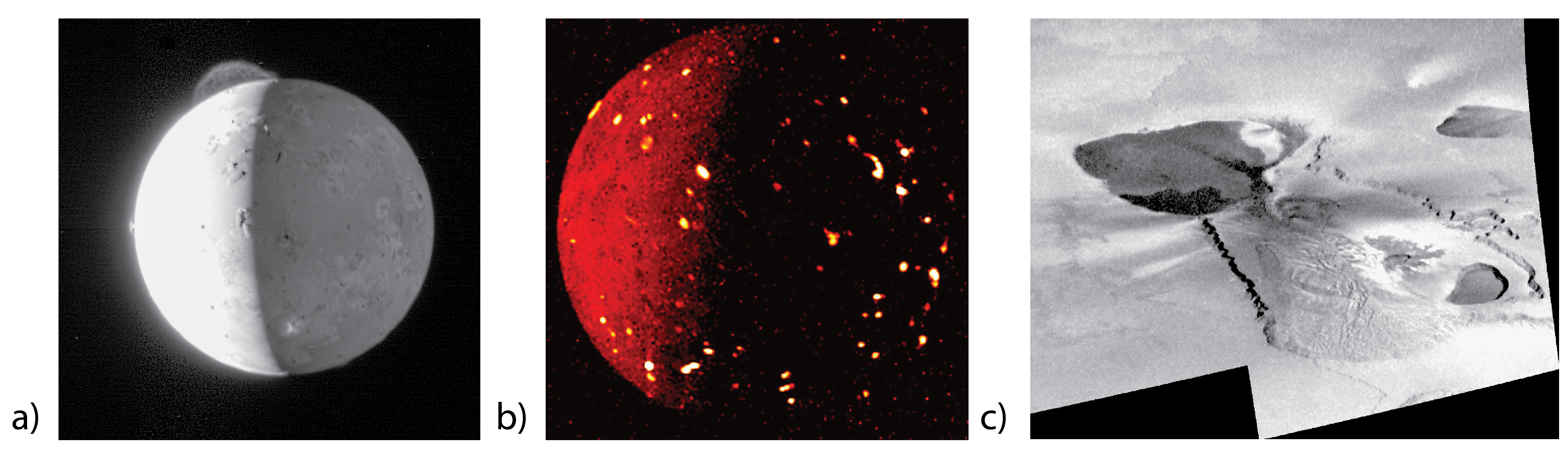}
    \caption{Io. a) Volcanic activity on Io near the terminator seen by the Long Range Reconnaissance Imager (LORRI) on New Horizons. The Tvashtar volcanic plume reaches to about 290~km altitude. (NASA photojournal \textit{PIA09248}) b) Hotspots as seen by the Jovian Infrared Auroral Mapper (JIRAM) instrument on Juno. (NASA photojournal \textit{PIA25698}) c) Tvashtar catena at 200~m/px imaged by Galileo. (NASA photojournal \textit{PIA03529})}
    \label{fig:Io}
\end{figure}

\begin{figure}
    \centering
    \includegraphics[width=\linewidth]{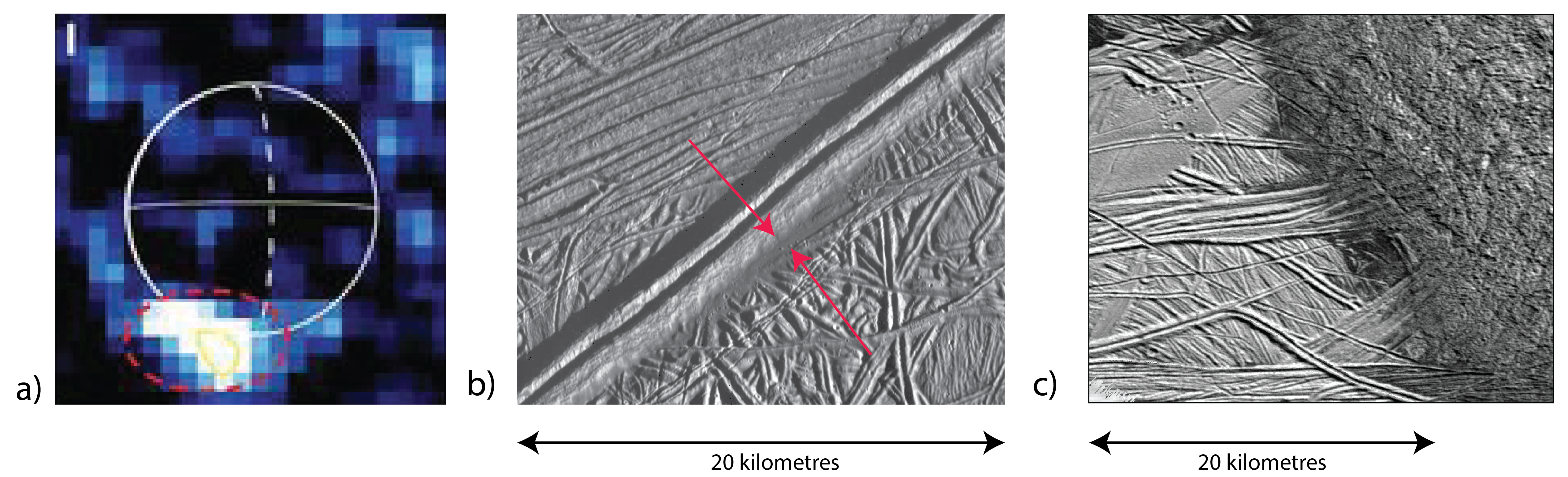}
    \caption{Europa. a) Possible plume detection with the HST telescope. Adapted from \citet{Roth2014}. b) Possible plume deposit adjacent to a double ridge (red arrows). (NASA photojournal \textit{PIA00589}) c) Ridged plains (left) are overprinted by chaos terrain (right), a possible cryolava source. Data from PDS \citep{PDS3_official}.}
    \label{fig:Europa}
\end{figure}

\begin{figure}
    \centering
    \includegraphics[width=\linewidth]{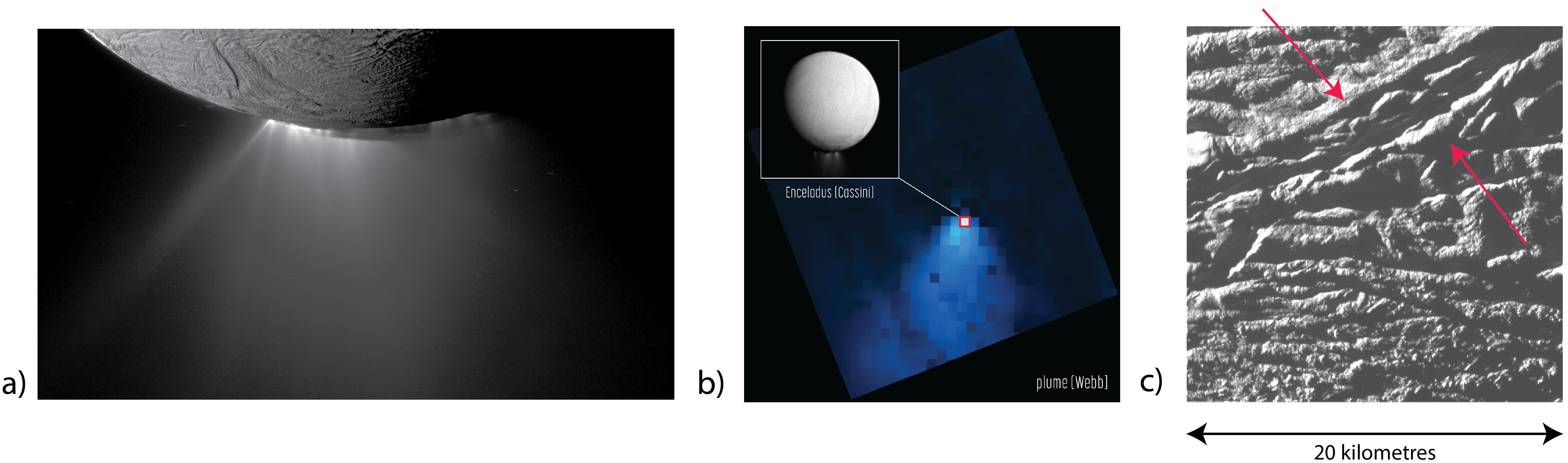}
    \caption{Enceladus. a) Plumes erupting from Enceladus SPT imaged by the Cassini narrow angle camera. (NASA photojournal \textit{PIA17184)}. b) Plumes seen with JWST (Image Credit: NASA, ESA, CSA, STScI, G. Villanueva (NASA’s Goddard Space Flight Center), A. Pagan (STScI)). c) A surface feature (Baghdad sulcus) where plumes are emitted from (red arrows). (NASA photojournal \textit{PIA11108})}
    \label{fig:Enceladus}
\end{figure}

This section focuses on evidence of activity of the selected Solar System moons Io (Fig. \ref{fig:Io}, Sect. \ref{subsub_Io}), Europa (Fig. \ref{fig:Europa}, Sect. \ref{subsub_Europa}), Enceladus (Fig. \ref{fig:Enceladus}, Sect. \ref{subsub_Enceladus}), and Triton (Fig. \ref{fig:Triton}, Sect. \ref{subsub_Triton}).

\subsubsection{Volcanic plumes on Io}\label{subsub_Io}
Io is the most volcanically active body in the Solar System (Fig. \ref{fig:Io}). \cite{peale1979} predicted correctly that the orbital eccentricity and tidal heating sustained by the Laplace resonance with Europa and Ganymede leads to a partially molten interior for Io and active surface volcanism of SO$_2$ \citep{lellouch1990}, with a mass flux of roughly $\SIrange{e6}{e7}{\kilo\gram\per\second}$ to the atmosphere estimated by \citep{Oza2019} based on the relative contribution of volcanic pressure \citep{Ingersoll1989} to the observed atmospheric column densities \citep{lellouch2015}. Io’s thermal activity has been studied with a variety of observation techniques, spanning different wavelengths, and include ground- and space-based telescopes \citep[e.g.][]{Rathbun2010, DeKleer2019, Tate2023}, and the NASA \textit{Voyager 1} \citep[e.g.][]{hanel1979,smith1979, morabito1979}, \textit{Galileo}, \citep[e.g.][]{Keszthelyi2001, Daviesb2012, Davies2023}, and \textit{Juno} missions \citep[e.g.][]{Mura2020, Zambon2023, Pettine2024, Park2025}. Recently, a global map of 343 volcanic hotspots was composed form these data \citep{Davies2024}.

The presence of a Na cloud in the vicinity of Io was postulated based on a detection of a sodium D-line emission doublet 
\citep[e.g.][]{Brown1974, Brown1975, Bouchez2000, thomas2004}. 
Na atoms are a trace element in Io's \ce{SO2} and sulphur-dominated atmosphere. Charged particle sputtering of Na from Io's surface has been proposed as a source of the Na cloud \citep{Matson1974, Haff1981}. However, it is still uncertain at what time and locations Io's atmosphere is thin enough to allow fast plasma particles of Jupiter’s magnetosphere to penetrate the surface \citep{lellouch2007,depater2023}. Moreover, NaCl has also been inferred as the main constituent of the dust of the Jovian stream particles that are considered to be emitted by Io into the Jovian system \citep{Postberg2006}.
 
Plume source observations suggest that the prominent Tvashtar plume (Fig. \ref{fig:Io}a) erupts from a fissure \citep{milazzo2005}, Prometheus forms by lava flow sublimation \citep{davies2006}, and Pele is fed by a lava lake \citep{mcewen2000, davies2001, radebaugh2004}. The combined effects of surface sublimation and volcanic plume activity contribute to increasing Io's atmospheric density with the exobase estimated to be at 
 a height of approximately 220 km \citep{Klaiber2024}. \citet{Wong2000} found that the exobase altitude ranges from 30 to 465 km while its
temperature ranges from 220 to 2800 K depending on the solar zenith angle and Io’s orbital location. Recent changes on Io can also be monitored with ground-based observatories, for example with the Large Binocular Telescope at a spatial resolution of 80~km \citep{Conrad2024}. Furthermore, the dust detected in the Jovian system correlates with the activity of Tvashtar-type volcanoes on Io \citep[e.g.][]{Graps2000, Postberg2006}.

\subsubsection{Evidence for Plumes on Europa}\label{subsub_Europa}
Whether plumes are active on Europa is still debated. 
\cite{Roth2014} reported on local surpluses in hydrogen and oxygen line emissions in limb observations of Europa at southern high-latitude regions, consistent with two 200-km-high water vapour plumes (Fig. \ref{fig:Europa}a). Follow-up observations with the Space Telescope Imaging Spectrograph (STIS) on the \textit{HST} showed tentative evidence consistent with plume activity \citep{Sparks2016, Sparks2017}. However, these signals could alternatively be explained by noise and misalignment \citep{Giono2020}. Indirect evidence of a plume on Europa was inferred from \textit{Galileo} magnetic field and plasma wave observations by \cite{jia2018EvidenceSignatures} and its influence on the
interaction between the corotating magnetospheric plasma and Europa’s atmosphere was modelled by \cite{bloecker2016}. Note that there were more unsuccessful than successful attempts at detecting plumes on Europa, most recently by the \textit{JWST} \citep{Villanueva2023}, the Atacama Large Millimeter/submillimeter Array \citep{Cordiner2024}, and \textit{Juno} \citep{kurth2023, Hansen2024}.
If plumes exist, they might be more difficult to detect by Earth-based telescopes, owing to a higher spatial and temporal variability \citep{Roth2014b, Rhoden2015} and may produce only half the water mass flux of Enceladus’ plumes \citep{Hansen2019}. Plumes at Europa might be short-lived and might also be much smaller than previously hypothesised, e.g.\ 26~km \citep{Quick2013} and $5-100$~km \citep{Southworth2015ModelingPlumes}, compared to $200\pm100$~km \citep{Roth2014}. However, with particle instrumentation even much smaller plumes can be detected during the JUICE flybys \citep{Huybrighs2017}. 

Cryovolcanic flows could still be present even if there are no plumes on Europa \citep{Lesage2025}. Flows may originate from pressure release after water sills in the upper ice layer freeze out \citep{Quick2013, Lesage2021}. Cryovolcanic flows are suspected along linear surface features \citep[][Fig. \ref{fig:Europa}b]{Craft2016, Kadel1998, Fagents2000} and in chaos terrain \citep[][Fig. \ref{fig:Europa}c]{Lesage2021, Collins2009}.
Plumes could offer a valuable way of accessing material from Europa's subsurface, if active \citep[either from a brine reservoir in the ice shell or the deep ocean][]{Lesage2025}). Europa is a key astrobiological target (Sect. \ref{sect_biosignatures}), and the sampling of plumes would facilitate a detection of biosignatures (Sect. \ref{sect_biosignatures}).

A search for plume deposits or cryovolcanic outflow can identify sites of recent activity. 
Potential sites can be investigated with future missions (Europa Clipper \citep{Daubar2024}, Juice \citep{Stephan2017, Tosi2024}). A search in existing observations, mostly in Galileo SSI images \citep{Belton2000, PDS3_official}, has revealed a young, smooth, quasi-circular patch \citep{Greeley1998, Hand2009}
and smooth deposits along lineae \citep[][Fig. \ref{fig:Europa}b]{Quick2020CharacterizingEuropa} that are all under debate for being plume deposits. The limb observation with HST by \citet{Sparks2017} corresponds to a region right above Pwyll crater, which may have a link to plume deposits.
However, other surface processes need to be considered as the source of smooth patches such as mass wasting on lineament slopes \citep{Jabaud2024}.

\subsubsection{The confirmed case of Enceladus}\label{subsub_Enceladus}
On Enceladus, Cassini data confirmed the eruption of plumes out of four large cracks in the south polar terrain (SPT), informally called Tiger Stripes \citep{Porco2006, Spencer2006, Porco2014HOWRELATED}. Detailed descriptions of the data, interpretations, and hypotheses as to the formation and on-going processes of the plumes can be found in reviews by \citep{Spencer2018} and \citep{Goldstein2018}, while the plume composition was reviewed by \citep{Postberg2018}.
The south polar eruptions were observed throughout the entirety of the Cassini mission, and their association with Saturn’s E-ring supports an even longer lifetime.  Detailed analyses of source regions along the Tiger Stripes suggest that eruptions occur in a combination of styles – including continuous “curtains” and individual jets \citep[e.g.][]{Spitale2007, Spitale2015, Helfenstein2015, Spitale2025}. The circuitous nature of the fractures, along with the viewing geometry of the Cassini images, can create so-called phantom jets, which must be accounted for when interpreting images of the plumes \citep{Spitale2015, Spitale2025, Goldstein2018}. 
The source material for the plumes is of key interest; an oceanic source would mean that the chemistry of plume particles is indicative of ocean chemistry and the habitability of Enceladus’ interior. A shallow liquid reservoir or a dry fracture system \citep[e.g.][]{Nimmo2007} would provide less astrobiological insight. The composition of plume materials is more consistent with liquid water in contact with rock \citep[e.g.][]{Glein2008, Hsu2015OngoingEnceladus, Glein2018, Postberg2018}, but there are no direct measurements that can confirm that the Tiger Stripes reach the ocean. 
Different models for the formation of the Tiger Stripe fractures have implications for the depth of the fractures, and thus, their plausible source. If the ocean is in a state of decline today, the overlying ice shell could be thickening, a process that creates large extensional stresses in the upper, elastic portion of the ice shell and pressurizes the underlying ocean, also suggested for Europa \citep{Nimmo2004,  Manga2007}. Radial fractures would form within the ice shell, propagating both upward and downward. Due to Enceladus’ low gravity, these fractures ought to be able to transit the entire ice shell for the range of thicknesses inferred from Cassini data \citep{Rudolph2022}. In addition, bending stresses induced by an initial fracture may create cascading parallel fractures akin to the present-day Tiger Stripes \citep{Hemingway2020}. Such models provide pathways to an oceanic source for Enceladus’ plumes. However, stresses from a freezing ice shell are isotropic, so there should not be a preferred orientation to the fractures if this were the only relevant stress mechanism affecting the formation of the Tiger Stripes (see Sect. \ref{subsec:tidal}).
Regardless of how they formed, the Tiger Stripes appear to have consistently provided source material for the plumes. Processes such as slipping, sloshing, boiling, and condensation along the walls of the fractures have likely affected their long-term eruptive stability \citep{Spencer2018, Soucek2024}. Over the span of the Cassini mission, measurements of eruptive activity suggest that sources do not turn on or off; eruptions persist along the Tiger Stripes throughout time. However, the total eruptive output from the region does appear to vary with time and in relation to Enceladus’ orbital cycle \citep[e.g.][]{Hedman2013, ingersoll2017, Spitale2025}. Variations in the eruptive flux of the plumes appear to correlate with the tidal cycle. However, no model of tidal-tectonic processes has fully matched the patterns of eruptive behaviour \citep[e.g.][]{Hurford2007Enc, Nimmo2007, Nimmo2014, Behounkova2015, Kite2016, Behounkova2017}.

\subsubsection{Active surface processes on Triton}\label{subsub_Triton}

\begin{figure}
    \centering
    \includegraphics[width=\linewidth]{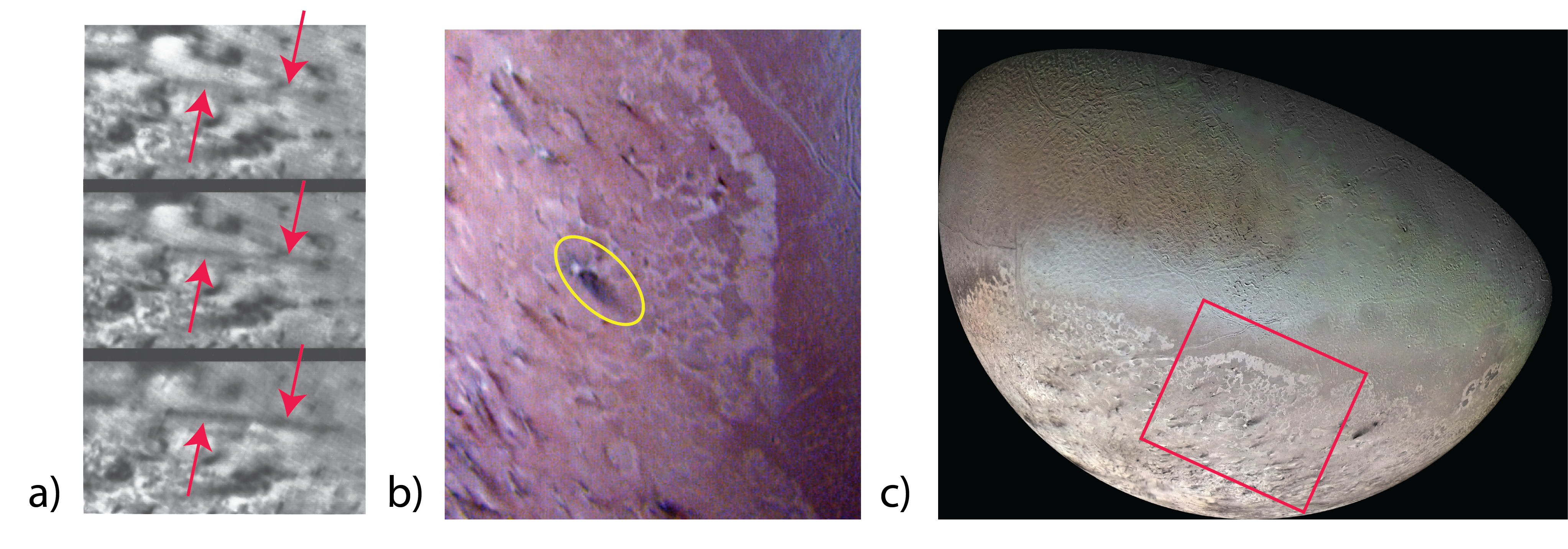}
    \caption{Triton. a) Triton's 8~km high and 150~km long plume appears after 45~min in the bottom image and indicates the tropopause (red arrows). (NASA photojournal \textit{PIA14449}) b) Surface streak from plume eruption (yellow ellipse) in a false colour image. (NASA photojournal\textit{PIA02214}) c) Dark surface streaks on a global view (red rectangle in b). (NASA photojournal \textit{PIA01538})}
    \label{fig:Triton}
\end{figure}

Triton’s relatively smooth surface, with few large impact craters, suggest that cryovolcanism and surface renewal processes have been occurring on geological timescales \citep[$\thicksim 10$~Myr;][]{Schenk_Zahnle_2007}.
The most striking evidence of active surface processes came in the form of geyser-like plumes, leaving behind dark streaks on the surface (Fig. \ref{fig:Triton}a) appearing to be composed primarily of nitrogen gas, with potential contributions from dust and methane particles \citep{Sagan1990}. These plumes were seen reaching heights of up to \SI{8}{km} before being carried horizontally by winds for distances up to \SI{150}{km} \citep{Hansen1990}, opposing the surface streak direction due to differing wind directions with altitude \citep{Ingersoll1990} (Fig. \ref{fig_linear_main}). 

A solid-state greenhouse effect was proposed as a mechanism to explain the plumes observed on Triton \citep{Kirk_1990, Brown_1990}. 
The effect occurs within a transparent or semi-transparent solid material, such as ice.
It works as follows: 
Sunlight penetrates Triton’s surface, which is covered by a layer of   nitrogen ice sufficiently transparent to allow solar energy to reach and heat darker material trapped beneath this ice layer.
When the built-up pressure exceeds the strength of the ice layer, it ruptures, releasing the trapped nitrogen gas explosively into the atmosphere. 
Along with the gas, small particles of ice and dust may be carried upward, forming the observed plumes. 
The released material gets carried by Triton’s winds, creating dark streaks on the surface (Fig. \ref{fig:Triton}bc).
Therefore, Triton's plumes were thought to be solar-driven; this model was later applied to cold \ce{CO2} jets eruption in the polar regions of Mars \citep[e.g.][]{Piqueux_2003, Kieffer_2007, Piqueux_2008}. Furthermore, the process of insolation-driven outgassing has been studied well on cometary surfaces \citep{Marschall2020, Herny2021}. 

Two alternative explanations for Triton's geyser-like plumes have gained attention in the years since the original plume discovery, due mainly to the discovery of Enceladus plumes: a) cryovolcanic activity and b) pressurized subsurface gas reservoirs. A recent overview of these hypotheses can be found in \cite{Hofgartner_2022}. It is worth noting that all hypotheses for Triton's plumes suggest (near-)surface mechanisms. 
The first alternative, the cryovolcanic hypothesis, posits that Triton’s interior may have retained enough heat—either from its tidal evolution from obliquity tides or after capture by Neptune, or from radiogenic decay—to sustain subsurface liquid reservoirs or slush. These cryovolcanic sources could periodically erupt through weaknesses in the icy crust, releasing volatile-rich material into the atmosphere.

The second alternative, the pressurized gas reservoir hypothesis suggests that volatile gases, particularly nitrogen, could accumulate in subsurface voids or fractures over time, independent of direct solar heating. These gases, originating from slow sublimation processes or residual geothermal gradients, may build up sufficient pressure to eventually breach the surface, producing eruptions. This mechanism is consistent with observations of localized and potentially episodic plume activity, especially in regions lacking the solar insolation required by the greenhouse model.  Though this hypothesis lacks direct evidence from Voyager 2, it is supported by Triton's youthful, resurfaced terrain and geological features indicative of internal mobility. 

The presence of plumes also suggests that Triton’s atmosphere may at least partially be maintained by surface activity rather than solely by sublimation from nitrogen ice deposits with relative proportions between these mechanisms vary through Triton’s seasons. The plumes eject primarily nitrogen gas, along with trace amounts of methane and entrained dust. They rise to altitudes of about 8 km before being deflected horizontally by atmospheric winds \citep{Smith1989}. This circulation distributes fresh material into the lower atmosphere and deposits dark streaks on the surface. The particles that constitute haze in Triton’s atmosphere \citep[i.e. ][]{Krasnopolsky1992} may also be the same dark material lofted into the atmosphere by plumes or, alternatively, invoking vigorous wind-driven transport and lofting of surface material is required.

\subsection{Processes Affecting Activity and Interpretations of Erupted Material}\label{sect_signatures_solar_system} 

We explore how activity signatures of moons in our Solar System (Europa, Enceladus, Triton, Io) compare to each other. 

\subsubsection{The Role of Tidal Heating}\label{subsec:tidal}
Tidal heating is a key driver of planetary evolution, shaping interior structure and geological activity \citep[e.g.][]{de2019}. As tidal forcing varies with orbital characteristics (semimajor axis, eccentricity, obliquity), its contribution to the internal heat budget can fluctuate significantly over geological timescales. In addition, the efficiency of tidal heating depends on the internal structure and rheology of the planetary body. The amount of mechanical energy converted into heat through internal friction is controlled by the viscoelastic properties of each internal layer \citep[e.g.][]{tobie2005, behounkova2010, kervazo2021, JaraOrue2011, Tobie2025}. These properties vary with temperature and melt fraction, leading in some cases to internal melting and surface activity. 

On Io, tidal dissipation explains its currently observed volcanic activity and exceptional heat flux \citep[e.g.][]{veeder1994, spencer2000, lainey2009, Pettine2024}. 
Most models attribute tidal heating primarily to solid-body dissipation in a viscoelastic mantle, with heat generated in a partially molten layer beneath the lithosphere as well as in the deep mantle \citep[e.g.][]{segatz1988, hamilton2013, steinke2020, kervazo2022}. 
An alternative model, proposing fluid-body tidal dissipation in a magma ocean \citep{tyler2015, aygun2024}, suggests that tidal heating is concentrated in a subsurface liquid layer (magma ocean), explaining Io’s extreme heat flux under broader geophysical conditions. However, recent gravity measurements from Juno spacecraft flybys provide new constraints on Io’s tidal deformation, with the measured $k_2$ value inconsistent with a shallow magma ocean \citep{Park2025}. These findings reinforce the solid-body dissipation model, implying that Io’s interior is predominantly solid, though likely containing a partially molten layer (see Fig. \ref{fig:interiors}). This refined understanding has critical implications for Io’s thermal evolution, volcanic activity, and our broader models of tidal heating in rocky bodies across the Solar System and beyond (Sect. \ref{exomoon_detect_activity}).

On Europa and Enceladus, tidal dissipation helps maintain a liquid water ocean and might feed hydrothermal vents at the ocean-rock interface \citep[e.g.][and references therein]{choblet2017, Behounkova2021TidallyEuropa, Tobie2025}. The models by \citet{choblet2017} and \citet{Tobie2025} only work if tidal dissipation occurs almost exclusively in a porous rocky mantle.
The Laplace resonance linking the orbit evolution of Io, Europa and Ganymede is known to play a key role in tidal dissipation by maintaining a non-zero orbital eccentricity of the moons \citep[e.g.][]{Griffin1920}. In contrast to the Laplace resonance, the 2:1 resonance of Enceladus with Dione is not stable \citep[e.g.][]{Zhang2009}, but currently has the same effect on Enceladus' eccentricity \citep{meyer2008, Zhang2009}. 

The orientations of the Tiger Stripes, while somewhat circuitous, largely follow patterns of eccentricity-driven tidal stress, suggesting they formed perpendicular to the maximum tensile stress direction caused by tidal deformation \citep[e.g.][]{Hurford2007Enc, Rhoden2020}, a process also suggested for Europa \citep{McEwen1986, Leith1996, Greenberg1998, Lee2005, Rhoden2013, Hurford2014}. Even if tidal forces regulate the activity, there are many unknowns, such as whether the volume of plume eruption is proportional to the fraction of open faults or to the normalised average stress \citep{Behounkova2015}. However, if tidal stresses alone were responsible for forming the Tiger Stripes, it would imply that Enceladus’ and Europa's ice shell have a failure strength of order 100 kPa rather than laboratory-derived values of 1 – 3 MPa for intact, pure water ice samples \citep{Schulson2006}. While it is possible that fatigue, scale, compositional, or environmental effects would allow for a low failure strength, it would suggest that the fractures are quite shallow because 100 kPa is exceeded by the compressive overburden pressure at roughly a kilometre depth. Hence, to link the ocean and the surface using tidal fractures would require an additional source of stress, such as shell thickening, to achieve failure at a higher stress and promote deeper propagation.
New work on the process of ocean growth and ice shell thinning within low-gravity bodies has shown that Enceladus’ ice shell could fail from the bottom up during the phase of ice shell melting \citep{Rudolph2025}, providing an additional potential mechanism for the initiation of the Tiger Stripe fractures. In this model, the volume difference between the solid ice shell and the meltwater creates a small gap, decoupling the ice shell from the growing ocean  \citep{Mckinnon2025, Rudolph2025}. 

Lineaments on Europa have long been thought to begin as tensile fractures in Europa's ice layer impacted by tidal stresses \citep{Figueredo2004, Bradak2023b, Greenberg1998, ProckterPatterson2009}. However, in contrast to Enceladus  \citep[although this was previously debated, see e.g.][]{Postberg2011}, Europa's thicker ice shell, larger radius and hence, larger surface gravity, impedes a direct radial fracture to the subsurface ocean \citep[e.g.][see Fig. \ref{fig_linear_main}]{CWalker2021}. While active, these cracks could have a connection to subsurface water reservoirs \citep[e.g.][]{Steinbrugge2025ShallowRidges} and let subsurface material emerge, e.g. as cryomagmatic outflow \citep{Fagents2000}. Diurnal cycling, the process controlling plume activity on Enceladus, was originally proposed for the cyclic build-up of Europa's double ridges \citep{Pappalardo1996, Greenberg1998}. 
Unique to Europa is the appearance of chaotic terrain: broken up pre-existing terrain \citep[e.g.][]{HaslebacherPSJ2025} and a background matrix material that suggests melting \citep[e.g.][; see Fig. \ref{fig_linear_main}]{Leonard2022}. Regardless of the underlying formation mechanism, chaos formation is also expected to bring subsurface material to the surface. There is evidence that young lineaments preferentially survive chaos formation \citep{HaslebacherPSJ2025}.
Because diurnal stress is presumed to govern the orientation of a fracture \citep{McEwen1986, Hoppa1999a}, unravelling the stress history is in principle possible using age relationships \citep{Rhoden2013}.
The nonparametric smoothing developed for axial data in \citet{Duembgen2025NonparametricData} shows that preferred lineament directions change locally.

On Triton, tidal dissipation through orbit circularization after capture (Sect. \ref{sect_formation}) would have contributed to sustaining a global water ocean in the past, and possibly up to billions of years with direct impact on resurfacing and habitability \citep{Hammond2024Triton}. Currently, radiogenic heating is the main heat source \citep[e.g.][]{Gaeman2012SustainabilityInterior}. 
The lack of impact craters and the presence of plumes (if they are considered as cryovolcanic features) suggest ongoing internal activity, possibly fuelled by an internal liquid layer. Hydrothermal vents might have existed \citep{Mandt2023}.

\subsubsection{Surface Composition and Processing}
\label{subsub_surfaceprocesses}
In this section, we first provide a high-level overview of Io's unique hot and cold surface, then dive into great detail about surface processes on the H$_2$O-dominated Europa and Enceladus, and finish the section with a high-level overview of Triton's surface processes. All of the mentioned processes alter the surface must be considered when interpreting evidence of purely endogenic activity.

Io's surface is of particular interest because of its volcanic activity constantly resurfacing the surface  \citep[$\leq 10$~cm/yr;][]{Johnson1982Io, Kirchoff2009}, which explains a lack of visible impact craters. The active outburst of volcanic material brings a unique dynamic into surface processes through thermal hotspots and plume and dust deposition. For example, surface age is 10-80 days around the most active Loki Patera \citep{Davies2003, deKleer2017}. 
The uppermost surface layer is dominated by sulphuric allotropes and SO$_2$ (see Sect. \ref{atmosphere}). Underneath is a silicate-rich surface layer \citep[e.g.][]{Carlson2007}, which shows that, like on Earth, volcanism on Io is silicate based \citep{Johnson1988, mcewen2000}. If the silicate volcanism, which is also inferred from volcano temperature measurements of 900~K \citep[or higher;][]{mcewen2000}, which is 400~K higher than the boiling point of sulphur \citep{Johnson1988}. 
Salts, namely NaCl and KCl, are found in some volcanic plumes \citep{Carlson2007}, feeding the neutral Na and K torus through photodissociation \citep{feaga2004, thomas2004, lellouch2003} and indicating a global variation in emitted subsurface material \citep[e.g.][]{Redwing2022}. Sputtering by energetic ions erodes the surface, in particular the frozen SO$_2$, which leads to a lower surface albedo \citep{Melcher1982}. Chemical alteration of salts and sulphur species by irradiation leads to hemispherical albedo variations \citep[e.g.][]{mcewen1988}. Heating of the surface by sunlight and endogenic volcanic activity results in diffusion of SO$_2$ into the subsurface, leaving behind a porous surface veneer and resulting in a volatile-rich subsurface layer \citep{Meade1991}. If the volatile-rich layer is exposed on mountain slopes facing the sun, heavy sublimation can lead to mass wasting events \citep[e.g.][]{McEwen2004}. 
During a 3-year observation period, the Jovian InfraRed Auroral Mapper (JIRAM) instrument on Juno found evidence for H$_2$S and suggests the presence of nitrile compounds as well as tholins \citep{Tosi2020}. 

The surface compositions of the icy moons observed today result from diverse endogenic and exogenic processes that compete over broad spatial and temporal scales and energy ranges. Global tectonic processes affect the entire thickness of the ice crust over geological timescales and potentially bring crystallised material from subsurface oceans toward the surface. Locally, cracks and associated jets can also release oceanic material to the surface at faster rates. There, continuous bombardment of the surface by radiation and particles in the magnetospheres of Jupiter, Saturn, or Uranus affect the surface through sputtering and chemical alteration. 
For Europa and Enceladus, this is especially interesting due to the salinity of the water oceans, inferred for Europa from magnetic induction \citep{Khurana1998} and observed in situ on Enceladus \citep{Postberg2009}.  
The salts (e.g. chlorides, carbonates, phosphates) are the result of ocean-rock interactions at the seafloor \citep{Hsu2015OngoingEnceladus, Postberg2023, Spiers2023}.

Telescopic observations of Europa have shown significant spectral differences between the two hemispheres \citep{1978Icar...36..271P}. While the spectra of the leading hemisphere indicate a high abundance of water ice, the trailing hemisphere shows distorted water absorption bands, suggesting the presence of other hydrates \citep{Carlson2009}. From Galileo data, hydrated sulfuric acid (H$_2$SO$_4$), magnesium sulfate and chloride (MgSO$_4$, MgCl$_2$), and sodium chloride (NaCl) have been proposed to explain the distinctive water bands and the reddish-brown colouration of the surface \citep{Mccord1998, Carlson2005, Hand2015, Ligier2016, King2022, King2025}. Recent telescopic observations further help constrain upper limits for some Cl-bearing salts \citep{Tan_2022}, with NaCl on Europa's surface inferred from observing distinctive radiation-induced crystalline defects known as colour centres with the Hubble Space Telescope \citep{Trumbo2019, 2022PSJ.....3...27T}. The quantitative interpretation of the data is supported by experimental work \citep{2017JGRE..122.2644P,Hibbitts2019ColorEuropa, Cerubini2022b}.

\begin{figure}
    \centering
    \includegraphics[width=\textwidth]{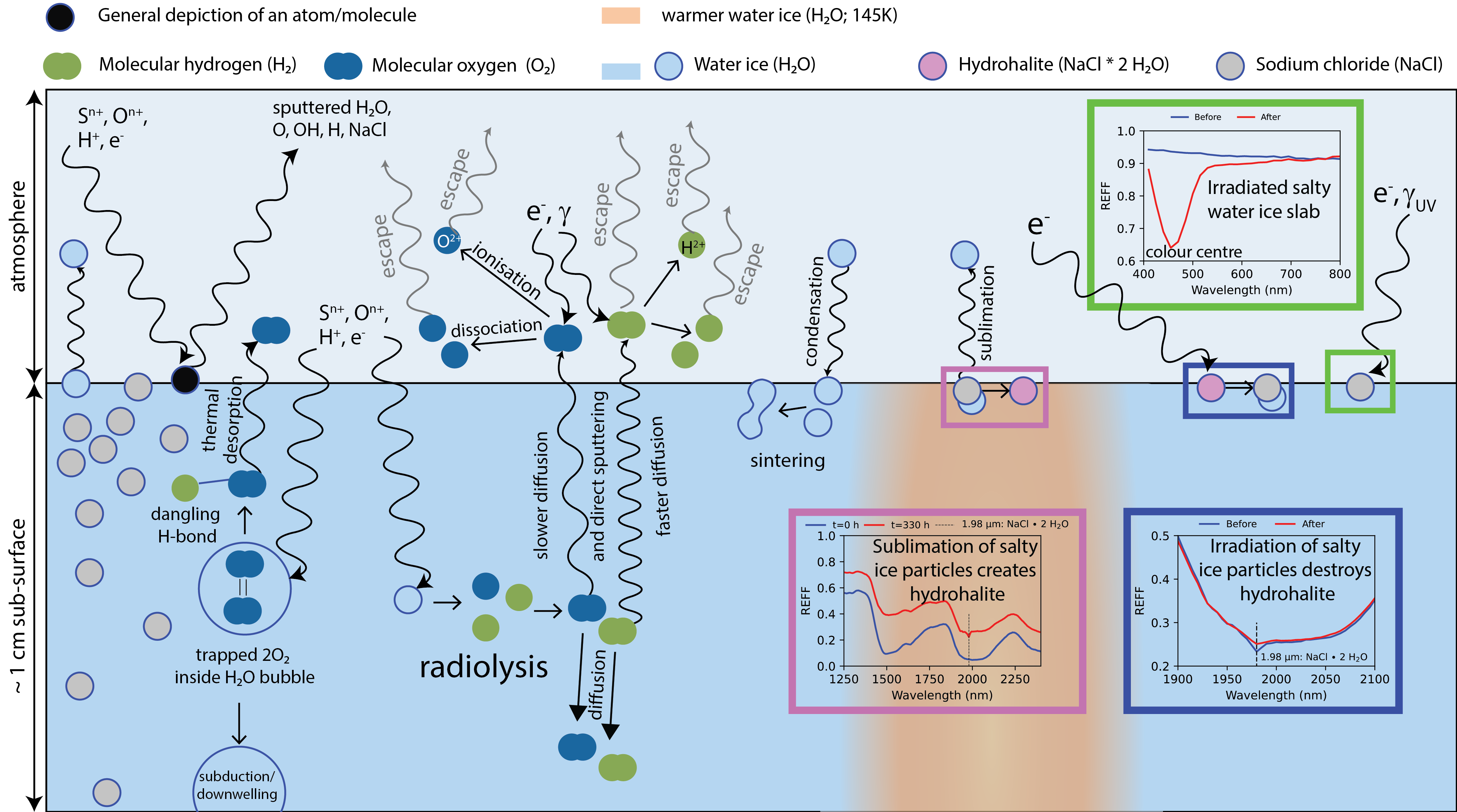}
    \caption{Visualisation of microscale processes in the atmosphere and the top~1~cm layer of Europa's surface (not to scale): (from left to right) the decreasing NaCl gradient due to sublimation, sputtering, trapping of \ce{O2}, radiolysis, ionization, dissociation, thermal escape, sintering, formation of hydrohalite in a hotspot (with data inlet from \citet{ottersberg2024sublimation}), destruction of hydrohalite in colder ice (with data inlet from \citet{ottersberg2024irradiation}) and irradiation and formation of colour centres in NaCl (with data inlet from \citet{cerubini2020vis}).}
    \label{fig_microscale}
\end{figure}

Laboratory experiments are indeed crucial to tie the effects of UV radiation, energetic electrons, and energetic ions to observations \citep{Pommerol2019ExperimentingMaterial,Galli2018,Gudipati2020,Poston2017,Obersnel2025}. Sputtering by hydrogen, oxygen, and sulphur ions is of particular relevance for Europa (Fig. \ref{fig_microscale}). Estimated particle sputtering rates are \SI{0.02}{\micro\meter/}yr \citep{2001Icar..149..133C}, while micrometeorite gardening reprocesses the top \SI{30}{\centi\meter} over tens of millions of years \citep{Costello2021}. \citet{galli2018first} measured the sputter yield of sulphur ions as two to three times higher than predicted by \citet{fama2008sputtering}. Applied to a simulation of Europa's exosphere, this caused a $\SI{30}{\percent}$ enhancement of the density of sputtered $\ce{O2}$ at the surface \citep{galli2018first}. Note, however, the large range in the measured yield among different studies on electron irradiation \citep{teolis2017water, Galli2017SputteringElectrons, Galli2018, Davis2021contribution}. 
In experiments conducted on thin films, the sputter yield decreases with electron energy after a maximum of a few hundred $\si{\eV}$, but this is not the case in experiments conducted with ice regolith \citep{Galli2018, Tinner2024Electron-InducedOxygen}. In pure ice irradiation experiments, only a minor part of the sputtered material is in the form of $\ce{H2O}$, while most of it leaves the ice as the radiolytically produced $\ce{O2}$ and $\ce{H2}$. Other radiolytic species, such as $\ce{HO}$, $\ce{H2O2}$, and $\ce{O3}$, may also be formed in trace quantities \citep{Obersnel2025}. The molecules produced by radiolysis may remain trapped in the ice, altering the surface composition. For example, \citet[][]{meier2020sputtering} noticed that $\ce{O_2}$ is radiolytically produced and trapped below the surface during electron irradiation of water ice films at $\SI{60}{\kelvin}$. $\ce{O_2}$ and other oxidants of astrobiological relevance were also detected on the surface of Europa and other Jovian icy moons \citep{Carlson1999HydrogenEuropa, spencer2002condensed}. The comae of comets 67P/C-G and 1P/Halley show an $\ce{O_2}$ to $\ce{H2O}$ ratio similar to the surfaces of the icy moons, suggesting a common radiolytic origin of the oxygen \citep{Oza2024CommonComets}. \citet{Tinner2024Electron-InducedOxygen} quantify residence time and saturation level for $\ce{O_2}$. In their experiments, the $\ce{O_2}$ rose faster when irradiated a second time, because of the saturation of $\ce{O_2}$ in the ice from the first irradiation. They report an $\ce{O_2}$ to $\ce{H2O}$ ratio of $\sim \num{0.01}$ in the irradiated ice, stable for at least $\SI{19}{\hour}$ at temperatures below $\SI{120}{\kelvin}$.

A mean surface temperature of \SI{106}{\kelvin} and peak brightness temperature of \SI{132}{\kelvin} were measured by the Galileo Photopolarimeter-Radiometer (PPR). Therefore, thermal segregation through sublimation is a significant process for the redistribution of ice at sub-kilometre scales \citep{Spencer1987ThermalSatellites}. \citet{Cerubini2022a} investigated how salty ice analogues for icy moons evolve when the water sublimates and reported the presence of highly hydrated compounds at the end of sublimation experiments. 
\citet{Ottersberg2025} constrained the sublimation kinetics and scaled the spectral changes to conditions present on the surface of Europa. The NaCl-bearing sample developed a narrow absorption band attributed to hydrohalite (\ce{NaCl * 2 H2O}), a hydrated form of NaCl also found on Ceres \citep{Zolotov2017Ceres, DeSanctis2020}. The formation of this feature is expected within thousands of years at Europa's surface temperatures \citep{Rathbun2010,Ashkenazy2023}, inconsistent with remote sensing data, which show no sign of this narrow feature even though it should be easily detectable. Only a weak shift of the minimum of the \SI{2.0}{\micro\meter} band to a lower wavelength has been observed at Tara Regio \citep{Ligier2016, Cartwrigth2025JWST}.
A radiation dose equivalent to a few years of surface exposure on Europa efficiently dehydrates the NaCl in addition to the formation of colour centres \citep{Ottersberg2025}. Future detection of hydrohalite on Europa would suggest that the sublimation rate is higher than the dehydration rate through irradiation, requiring a hotspot with mean surface temperatures $\geq$~\SI{145}{\kelvin}. Even at an area of \SI{100}{\km^2}, such hotspots cannot be ruled out by Galileo photopolarimeter–radiometer (PPR) data \citep{2010Icar..210..763R}. On Europa, hotspots like this could coincide with small chaos features \citep{Abramov2013, Singer2021}, while on Enceladus, such hotspots are directly linked to active processes. The Europa Thermal Emission Imaging System (E-THEMIS) onboard Europa Clipper \citep{christensen_europa_2024} will be able to detect such thermal hotspots if present on Europa. The Mapping Imaging Spectrometer Experiment (MISE) will be able to detect local compositional variations, such as the local occurrence of hydrohalite \citep{blaney_mapping_2024}.
While $\ce{H2O}$ dominates the composition of icy areas of the surface, $\ce{CO2}$ is also detected in its solid state. Recent studies by \cite{Trumbo2023}, \cite{Villanueva2023}, and \citet{Cartwrigth2025JWST} mapped the $\ce{CO2}$ distribution on Europa's surface using the \textit{James Webb Space Telescope} (\textit{JWST}), analysis of which indicates an endogenic carbon source. Besides composition, spectroscopy is also useful for retrieving the size distribution of the icy regolith. A range of 40--\SI{400}{\micro\meter} derived from spectral modelling of observational data \citep{Ligier2016, King2022} is consistent with studies based on the comparison of polarimetric phase curves of laboratory sample analogues with observational data \citep{Poch2018}.

In contrast to Europa's salty surface, Enceladus's surface is characterized by spectrometry with the presence of nearly pure water ice along with \ce{CO2}, \ce{NH3}, and aliphatic organics   \citep{Brown2006, Postberg2018}. Spectroscopic observations with the VIMS instrument suggest particle sizes up to \SI{0.2}{\micro\meter}, which can be correlated with the relative age of geological features (larger grains closer to plume sources) but are potentially overestimated due to the presence of micrometer and submicrometer ice grains which enhance interpreted band depths
\citep{Jaumann2008, Scipioni2017}. Grain size is predicted to decrease towards the north pole \citep{Southworth2019, Jabaud2024}. \citet{Poch2018} showed that the polarimetric phase curve of an analogue sample covered with micrometer-sized frost was the best match for the surface of Enceladus. Plume deposition rates range from \SI{0.5}{\milli\meter\per yr} close to the vents and in the SPT, and \SI{10}{\nano\meter\per yr} at regions north of the equator, with broad plume deposits below 45S and two split patterns centered at 45W and 225W \citep{Kempf2010, Southworth2019}. 
The ice plume deposits are expected to produce mostly crystalline ice on the surface, with  \ce{H2O} plume vapor deposits generating small-scale amorphous ice, as in the tigers stripes region and potential amorphisation in the south pole \citep{Newman2008} However, detections of amorphous ice are complicated again due to the presence of submicrometer ice particles. 
While the peak subsolar brightness temperature is around \SI{76}{\kelvin}, the Cassini Composite Infrared Spectrometer detected thermal emission from a hotspot with temperatures of up to \SI{195}{\kelvin}, located in the tiger stripes region \citep{2006Sci...311.1401S, Gougen2013}. This shows a strong link between endogenically sourced thermal hotspots and active sources.
Enceladus' low gravity \citep[\SI{0.113}{m/s^2},][]{McKinnon2013} leads to a loose deposit of plume material on the surface, facilitating sampling \citep{Choukroun2021}. This deposit can be studied through plume deposit accumulation into open fractures and the resulting formation of circular or elliptic pit chains. Pit chain development suggests that Enceladus' tiger stripes eruptions were either more active in the past, or more sites of active jets exist or existed \citep{Martin2023}.

Triton’s surface composition suggests an active volatile cycle involving sublimation and deposition, as well as cryovolcanism reshaping its landscape.  
Its similarities to Pluto and other Kuiper Belt objects support the hypothesis that Triton was captured from the Kuiper Belt and has since undergone significant resurfacing \citep{Hammond2024Triton, McKinnon_1995}.
Triton’s surface is primarily covered by frozen nitrogen (N$_2$), similar to Pluto, making it one of the brightest bodies in the solar system due to its high albedo \citep{Cruikshank1993,McKinnon_2014, Holler_2016}. The presence of methane (CH$_4$) was confirmed through infrared spectroscopy, and its interactions with nitrogen ice influence the moon’s surface temperature and phase transitions \citep{Tryka_1993}.
Spectral analysis has revealed carbon monoxide (CO) and carbon dioxide (CO$_2$) ices mixed with nitrogen, suggesting atmospheric cycling of these volatiles \citep{Quirico_1999}. CO is particularly interesting because it can serve as a tracer for sublimation and seasonal processes affecting Triton’s thin atmosphere.
Although nitrogen and methane dominate, water ice (H$_2$O) has been detected, possibly forming the bedrock beneath volatile ices \citep{Cruikshank_2000}. Irradiation of such ice mixtures with energetic particles has been shown to result in the formation of volatile molecules, such as hydrocarbons and refractory organic residues \citep{Kipfer2024ComplexMethane}.  Additionally, some darker material, likely tholins or organic compounds, is present on Triton’s surface, possibly resulting from radiation-driven chemistry involving methane \citep{McDonald_1994}. These organic materials could provide insights into prebiotic chemistry in the outer solar system. Prebiotic chemistry here refers to the formation of complex organics (e.g., hydrocarbons, nitriles, polycyclic aromatics) from simple starting molecules like CH$_4$ and N$_2$. Studying these materials on Triton could therefore provide important clues to the kinds of organic molecules that may have been widespread in the early solar system and potentially delivered to the young Earth by comets and icy bodies. In this way, Triton serves as a natural laboratory for examining pathways of prebiotic organic synthesis in cold, irradiated environments far from the Sun.

\subsubsection{Atmospheric Composition}\label{atmosphere}

\begin{figure}
    \centering
    \includegraphics[width=\linewidth]{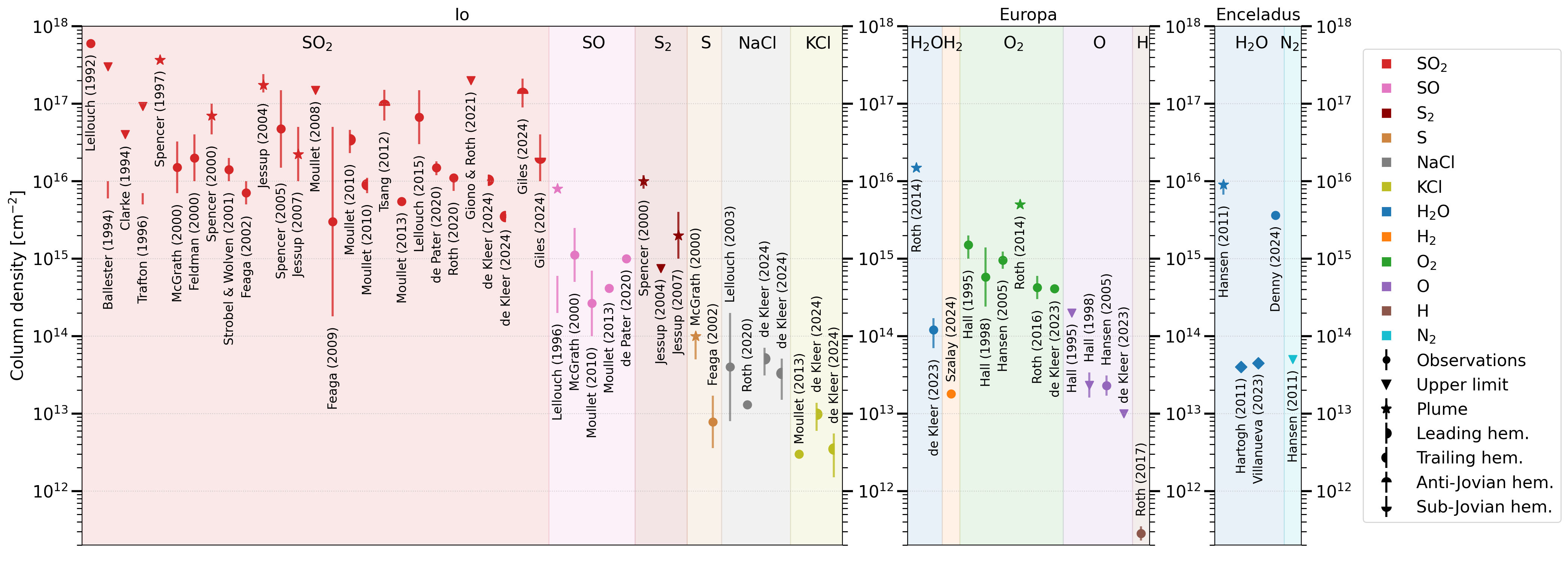}
    \caption{Comparison of the column density of SO$_2$, SO, S$_2$, S, NaCl, KCl, H$_2$O, H$_2$, O$_2$, O, H, and N$_2$ derived from observations for Io (left), Europa (middle), and Enceladus (right). The data and references used for this plot can be found in Table~\ref{tab:data} in the appendix. This plot is intended as an overview of the column densities from remote observations to compare the atmospheres and does not claim to be complete. Especially for Enceladus, mass spectrometry measurements gave a more detailed picture of the plume gas composition.} 
    \label{fig:columndensities}
\end{figure}

Figure \ref{fig:columndensities} shows an overview of the different species that have been observed in the atmospheres of Io, Europa, and Enceladus. 
In the following paragraphs, the atmospheric composition of the four active moons is described in detail. 

Io's tenuous atmosphere consists primarily of SO$_2$ \citep[$\sim$90--97\%,][]{lellouch2007, depater2023,dekleer2024} and atoms and molecules associated with SO$_2$. NaCl and KCl were detected in Io's atmosphere at the level of permile with respect to SO$_2$ \cite{depater2023, roth2020}. The main source of Io's SO$_2$ atmosphere is still being debated, as it is unclear if it is driven by sublimation of SO$_2$ surface frost \citep[e.g.,][]{tsang2012, tsang2016} or by volcanic activity \citep[e.g.,][]{mcgrath2000, spencer2000, jessup2007}. Most of the SO$_2$ frost may have been ultimately provided by volcanoes \citep{lellouch2007, depater2020, depater2023}. 
The SO$_2$ atmosphere is expected to collapse on the nightside and when Io moves through Jupiter's shadow \citep{clarke1994, retherford2007,jessup2015}, as SO$_2$ would likely condense on the surface at the lower nightside temperatures. However, the details and magnitude of this process are not well understood. Non-condensible gases, such as O$_2$ and SO, could survive and buffer atmospheric collapse \citep{davies2022}. Furthermore, SO$_2$ dissociates into SO, S, O, and O$_2$ by interactions with electrons and photons. Most of these species have been observed in Io's atmosphere, with the exception of O$_2$. 
Direct escape of volcanic plume gases and dust from Io is marginal, because the plume ejection velocities are generally below Io's escape velocity of 2.6 km/s \citep{roth2020}. Nevertheless, in the analysis of dust particles from the Jovian system  NaCl was identified, accompanied by sulphurous and potassium-bearing components, which were interpreted to have an origin on Io \citep{Postberg2006}.

Based on high-resolution line-of-sight SO$_2$ observations of Io by \citet{lellouch2015} and an approximation of its near-surface atmosphere \citep{Ingersoll1989}, \citet{Oza2019} inferred that roughly 10$^{6}$--10$^{7}$ kg/s of SO$_2$ is provided by volcanic activity to the surface. The majority of atmospheric SO$_2$ appears to result from sublimation, additionally variable due to Jupiter's eccentricity \citep{Giles2024}. 
Furthermore, volcanic outgassing can strongly impact the atmosphere and dominate locally as was concluded  from simulations \citep{Walker2012, McDoniel2015, Klaiber2024}. 

SO$_2$ has been observed at millimetre \citep[e.g.,][]{lellouch1992, dekleer2024}, infrared \citep[e.g.,][]{tsang2012, lellouch2015} and ultraviolet \citep[e.g.,][]{ballester1994, giono2021}
wavelengths with ground-based 
and space-based 
telescopes. Other observed atmospheric species include SO \citep[e.g.][]{moullet2010, depater2020}, S$_2$ \citep{spencer2000}, S \citep{feaga2002}, O \citep[e.g.][]{roth2014a}, NaCl \citep[e.g.][]{roth2020} and KCl \citep[e.g.][]{moullet2013}. NaCl and KCl were observed in confined locations connected with volcanic outgassing \citep{dekleer2024, Thelen2024}. \\

The tenuous atmosphere of Europa consists mainly of O$_2$, with a uniform spatial distribution. O$_2$ was not observed directly, its presence was inferred from the observation of the 
emission lines of atomic oxygen 
\citep[e.g.][]{Hall1995, Hansen2005, deKleer2023TheCallisto}. 
While O$_2$ likely dominates the atmosphere close to the surface (up to a few 100~km), an extended corona of lighter H$_2$ molecules is expected \citep[e.g.][]{Vorburger2018}. 
\cite{Szalay2024} observed H$_2^+$ and O$_2^+$ pickup ions with the Jovian Auroral Distributions Experiment (JADE) of the Juno mission. Furthermore, an atomic H corona was observed at Europa with HST/STIS \citep{Roth2017}. 
The non-condensing H$_2$ and O$_2$ molecules do not freeze out when they return to the icy surface, not even on the night side. Instead, they are thermalised to the local surface temperature and re-emitted into the atmosphere. Therefore, their atmospheric enrichment is mainly limited by different loss processes, such as Jeans escape or the ionisation and fragmentation by photons and electrons \citep[e.g.][]{Vorburger2018}. The main source of atmospheric H$_2$O on the illuminated hemisphere is expected to be sublimation of the surface ice, with minor global contributions from sputtering \citep{Vorburger2018, plainaki2018}. Furthermore, interactions with photons and electrons can ionise and dissociate molecules in the atmosphere. \\ 

Measurements with the Cassini Ion Neutral Mass Spectrometer \citep[INMS,][]{waite2004} showed that Enceladus' plume gas consists mainly of H$_2$O (96--99\%), with CO$_2$ (0.3--0.8\%), CH$_4$ (0.1--0.3\%), NH$_2$ (0.4--1.3\%), and H$_2$ (0.4--1.4\%) \citep{waite2017cassini}. 
The Cosmic Dust Analyzer \citep[CDA;][]{Srama2004} and the Cassini Plasma Spectrometer \citep[CAPS;][]{Young2004}  assessed the icy solid component of Enceladus' plume, with CDA detecting the presence of three major distinct families of ice grains based on compositional differences: 60--70\% Type I (with <0.0001\% Na, K), 20--30\% Type II (with 0.00001--10\% organics), 10\% Type III (with 0.5--2\% Na and K salts), and a high concentration of SiO$_2$ nanograins. CAPS measurements in the plumes reveal both positively and negatively charged nanophase dust grains, with positive ion composition dominated by water group ions (O+, OH+, H$_2$O+, and H$_3$O+) and negative water cluster ions (OH-, O-, H-). \citet{Hansen2011} observed the water vapour plumes in the Extreme Ultraviolet (EUV) with the Cassini Ultraviolet Imaging Spectrograph (UVIS), presenting an upper limit for the undetected N$_2$. The most recent compilation of the composition of Enceladus' plume gas is given by \cite{peter2024detection}, who also find significant evidence for HCN ($0.11{\pm}0.02$\%), C$_2$H$_2$ ($0.023{\pm}0.005$\%), and C$_3$H$_6$ ($0.004{\pm}0.002$\%) from INMS data. 
\cite{Denny2024ConstrainingSpectrometer} showed that the \SI{2.7}{\micro\meter} plume water-vapour emission feature is also detected in Cassini’s Visual and Infrared Mapping Spectrometer (VIMS) spectra.
A water vapour torus was also detected for Enceladus in the far- \citep{hartogh2011} and near-infrared  \citep{Villanueva2023Enc}. Estimated plume production rates from Enceladus are about 200 kg~s$^{-1}$ for the gas \citep{Hansen2011} and up to 50 kg~s$^{-1}$ for the dust \citep{Ingersoll2011TotalImages}. Given the low gravity of Enceladus, most of the plume gas escapes Enceladus, whereas only about 10\% of the dust particles escape, and most of them fall back to the surface \citep{spencer2013}. Thus, Saturn's E ring is mostly populated by the gas released from the plumes.  \\

Triton’s atmosphere is primarily composed of nitrogen (N$_2$), with trace amounts of methane (CH$_4$) and carbon monoxide (CO) \citep{Broadfoot1989, herbert1991, lellouch2010}. These gases are in thermal equilibrium with volatile ice deposits on Triton’s surface, particularly with the nitrogen ice, which undergoes sublimation and condensation depending on seasonal changes. Atmospheric temperatures range from $\sim$~\SI{40}{\kelvin} near the surface to $\sim$~\SI{95}{\kelvin} at higher altitudes (Yelle et al., 1991). The atmosphere has a troposphere with a surface pressure of \SI{1.4}{Pa} \citep{Broadfoot1989}, stratosphere, and thermosphere, which transitions into an exosphere at $\sim$~\SI{850}{\kilo\meter} \citep{Strobel2017}.
As Triton follows its inclined orbit around Neptune, different hemispheres receive varying amounts of solar radiation over its 165-year orbital period. This likely causes periodic expansion and contraction of the atmosphere \citep{Elliot1998, Bertrand_2022}. Some models suggest that during colder periods, a significant portion of the atmosphere may collapse onto the surface as frost \citep{Olkin1997}.

\begin{figure}
    \centering
    \includegraphics[width=\textheight,angle=90]{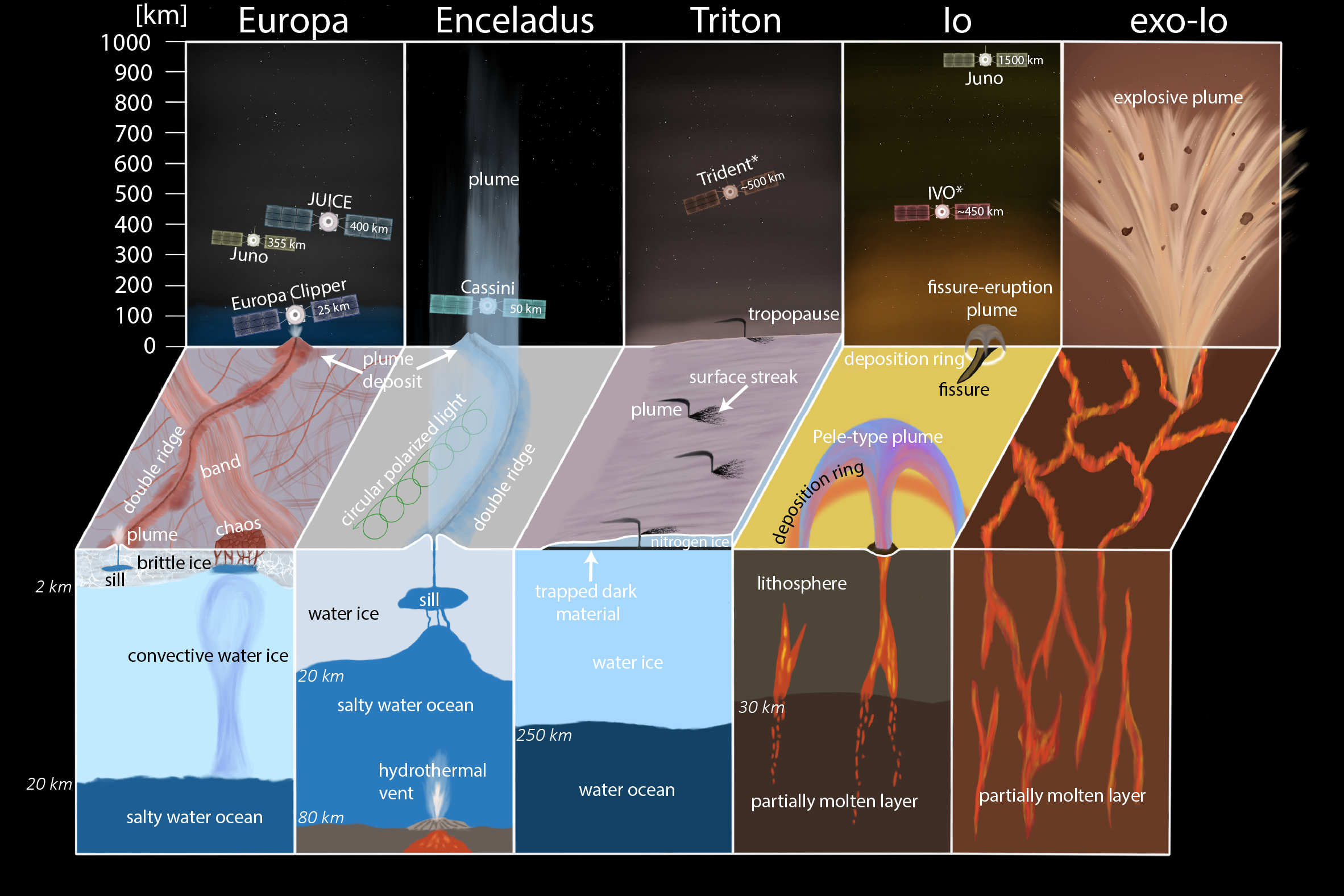}
    \caption{Interior, surface and atmosphere visualisation of Europa, Enceladus, Triton, Io, and exo-Io. Best-estimate plume heights are displayed to scale. Individual scales are visible for interior structures. For interior structures to scale, see Fig. \ref{fig:interiors}.}
    \label{fig_linear_main}
\end{figure}

\subsubsection{Impact of the Magnetosphere of the Host Planet}
The influence of the magnetospheres of Jupiter, Saturn, and Neptune impact the environments of their moons in different ways. We do not discuss the magnetosphere of Uranus here. 

Jupiter hosts the largest and most powerful magnetosphere in the Solar System. Plasma particles bring additional energy to Io's atmosphere and volcanic plumes exciting molecules in the atmosphere as seen in HST observations \citep[][]{Geissler1999Io} and heating the atmosphere \citep[e.g.][]{Strobel1994, Klaiber2024}.
Io’s intense volcanic activity releases large amounts of sulphur- and oxygen-rich gases into its atmosphere. A fraction of the atmospheric species are ionised through interactions with solar radiation and collisions with electrons and ions in Jupiter’s magnetosphere \citep{thomas2004, saur2004}. The newly formed ions are mostly O$^+$ and S$^+$ that escape Io’s atmosphere and are forced into co-rotation with Jupiter's plasma disk, forming the Io Plasma Torus (IPT) \citep{bagenal1980, intriligator1981}. 
Galileo flybys confirmed that Io lacks an intrinsic magnetic field, making these plasma interactions the dominant factor in its electromagnetic environment \citep{kivelson1996, kivelson2001}.

Enceladus orbits Saturn in a far less intense radiation environment than Europa in Jupiter's magnetosphere. Therefore, the energetic particles produce far more O$_2$ in Europa's $\thicksim \SI{1}{cm}$ sub-surface ice, which becomes trapped \citep[][Fig. \ref{fig_microscale}]{Johnson1981, Johnson1982, Bagenal2004}. The current downward transport rate, or downwelling rate, of O$_2$ at Europa and Enceladus is estimated to be
 $\leq$~10~kg/s \citep{Hesse2022DownwardPercolation}. 

Neptune’s magnetic axis is tilted by \SI{47}{\degree} relative to its rotation axis and is offset from the planet’s centre, creating a complex and varying magnetospheric environment.
Considering that Triton does not have its own magnetic field, interactions with Neptune's magnetosphere are expected to be significant for Triton’s thin atmosphere \citep{Delitsky1989}.
Sputtering processes may modify Triton’s tenuous atmosphere,
might alter the atmosphere and charged particles may reach the surface, knocking off nitrogen molecules and contributing to atmospheric loss over time \citep{Lammer1995}.
Furthermore, if the assumption of solar-powered plumes is incorrect, and they are instead tidally-driven, the indirect and mutual influence between the magnetic field and subsurface ocean that connects to the outgassing activity should be revisited.

\subsection{Biosignatures of Active Moons}\label{sect_biosignatures}

The search for life beyond Earth has increasingly focused on the active moons of our Solar System, where subsurface oceans, proposed hydrothermal activity, and dynamic surface processes create conditions that may be conducive to biological activity \citep[e.g.][]{Greenberg1998, Xu2025}. We discuss a selection of possible biosignatures: the homochirality of life (Sect. \ref{sect_detecting_biosign_homochir}) and the detection of biogenic trace molecules, and organic macromolecules through mass spectrometry (Sect. \ref{sect_detecting_biosign_mass_spec}).

The term ``habitability'' refers to the potential of a planetary body or a specific environment to host life. 
Our understanding of planetary habitability is rudimentary and closely linked to life as we know it. 
Terrestrial life primarily requires liquid water, the chemical elements C, H, N, O, P, S,
and a source of free energy \citep[electromagnetic, thermal, or chemical, e.g.][]{longo2020factoring,remick2023elements}. 
Surface habitability is primarily constrained by the availability of liquid water, which led to the definition of the so-called ``habitable zone''. In the Solar System the habitable zone extends approximately from 0.95 to 1.37~AU
, or even up to 2.4 AU \citep{zsom2013toward,ramirez2017volcanic}. 
Yet, the habitable zone does not exclusively define all potentially habitable environments within a system. Areas below the surface can harbour habitats with liquid water, thermal and/or chemical energy, complex chemistry, and shielding from strong UV and particle radiation of interplanetary space. Such subsurface environments are suspected to exist on bodies like Enceladus, Europa, and potentially other ocean moons of the Solar System \citep{nimmo2016ocean, PetriccaTitans2025}. Chemical composition of these environments has been investigated by sampling the atmospheric (including ions and solid dust particles; Sect. \ref{atmosphere}) and surface composition (Sect. \ref{subsub_surfaceprocesses}). On Enceladus for example, the determined plume composition not only indicates a chemically diverse and dynamic environment with many building blocks of life, but it suggests the presence of hydrothermal sites \citep{Hsu2015OngoingEnceladus,waite2017cassini}, unless Enceladus' plumes are not representing ocean composition \citep{Meyer2025}. On Earth hydrothermal vents are considered the birthplace of life \citep[e.g.][]{baross1981hypothesis, Martin2008}.

\subsubsection{Detection of Biosignatures: Homochirality}\label{sect_detecting_biosign_homochir}
A very promising biosignature of life on icy moons and beyond is homochirality, the presence of only one of two enantiomers of a chiral molecule (left- and right-handedness). Homochirality is generally considered to be a universal and distinctive feature of life \citep[e.g.][]{cahn1956specification,blackmond2010origin}. Whether life emerged from an environment with an inherent enantiomeric excess \citep[like that found on several meteorites,][]{engel1982distribution,pizzarello2000non,glavin2009enrichment,pizzarello2012large} or whether biological processes themselves catalysed an increasing enantiomeric excess in their surroundings remains an open question \citep[e.g.][]{bailey2001astronomical,blackmond2010origin,chen2020origin}.
Regardless of its origins, homochirality is essential for life as we know it \citep{blackmond2010origin}. 

Every organism depends on a vast array of functional macromolecules with highly organized and repetitive structures, where even a single deviation can lead to cell death. For example, only left-hand amino acids are used by life to construct proteins
\citep[e.g.][]{cahn1956specification,blackmond2010origin}.
The molecular asymmetry of life has a specific response to electromagnetic radiation. Biomolecules, particularly when organized within cellular structures, exhibit optical activity \citep{pasteur1848relations, wald1957origin} and circular dichroism \citep{velluz1965optical}, and can even induce a slight circular polarization, up to the permil level, in scattered, initially unpolarised light \citep{pospergelis1969spectroscopic,Wolstencroft1974}, e.g.\ starlight, and as such can be sensed remotely.

The ability to detect induced circular polarization in originally unpolarised light has significantly expanded the potential applications of polarisation-based remote sensing in biology, planetary sciences, and eventually astrobiology \citep{kemp1971circular,swedlund1972circular,sparks2005search,patty2019circular}. Different instrument designs that measure chirality as a biosignature are underway.
One instrument is the Mid-InfraRed AdvanCed Life-detection Explorer (MIRACLE) \citep{Phal2021ConcurrentImaging, Phal2023QuantumCapabilities}, which measures circular polarisation in transmittance. 
Another instrument is FlyPol, a highly sensitive and accurate spectropolarimeter for transmittance or reflectance \citep[][see Patty et al., this collection]{patty2017circular,kuhn2020monitoring,Patty2021BiosignaturesLife}. 
FlyPol was designed for deployment on airborne platforms such as helicopters and balloons and has demonstrated the ability to differentiate between abiotic and biotic surfaces from the air \citep{Patty2021BiosignaturesLife,mulder2022spectropolarimetry}. 
Current developments of FlyPol aim to deepen the understanding of the measured circular polarisation features and prepare the instrument for deployment in space. 
In the long term, FlyPol may contribute to missions targeting the icy moons Enceladus or Europa \citep{grone2024modeling, grone2024full}.  
Although FlyPol's current wavelength range of 420–850 nm prevents the detection of universally conserved biomolecules such as DNA, RNA, and proteins, which absorb in the mid-UV, it is highly effective in detecting and characterizing biopigments. Most biopigments are chiral macromolecules embedded within protein structures, playing essential roles in light harvesting and photoprotection. Light harvesting, the partial conversion of photons into chemical energy, is widely proposed as a fundamental metabolic strategy for potential life across the universe, given the ubiquity of light energy \citep{schwieterman2018exoplanet}. Even in environments far away from direct sunlight (or star light), organisms could sustain themselves through phototrophic metabolic pathways, utilising ambient far-red light from chemoluminescence or thermal radiation near hydrothermal vent systems \citep{beatty2005obligately}.

Currently, Enceladus is the most likely ocean world to possess an active hydrothermal vent system at its seafloor, especially below the plume sites at the southern pole \citep[][visualised in Fig. \ref{fig_linear_main}]{Hsu2015OngoingEnceladus,waite2017cassini}. These vents could supply the necessary thermal and chemical energy to sustain a localised ecosystem \citep{orcutt2011microbial}, where light harvesting and light sensing may represent viable strategies for occupying ecological niches \citep{beatty2005obligately, perez2013potential}. Thus, biopigments would be a likely phenotypic adaptation to efficiently absorb light, which could be detectable if sufficient amounts of cellular material get ejected at plume sites \citep{porco2017could, cable2021science}. 
However, this process could be hampered by ocean stratification \citep{Ames2025}.
\citet{porco2017could} hypothesize that bubbles from hydrothermal vents could carry a significant number of cells from depth to the surface, where they would be ejected and be present in plume material and/or deposited on the surface of Enceladus. Bubble bursting might further amplify the concentration of organic compounds and cells \citep{cable2021science, postberg2018macromolecular}. Such cells could be potentially detectable through spectropolarimetry, either in reflectance from the surface or in transmittance through the plume (visualised in Fig. \ref{fig_linear_main}).

\begin{figure}
    \centering
    \includegraphics[width=0.95\linewidth]{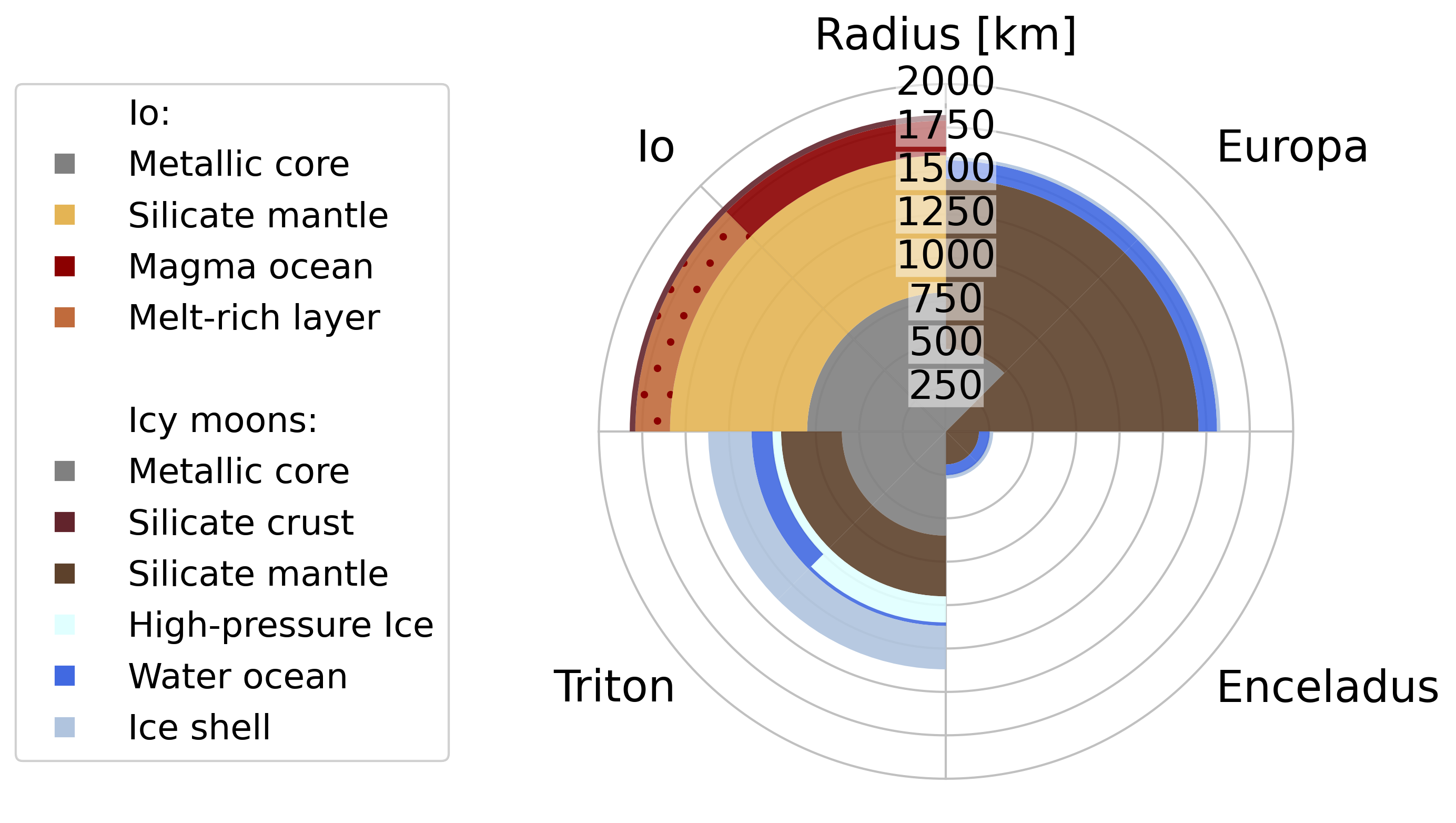}
    \caption{A visualisation of the assumed interior structures of Io, Europa, Enceladus, and Triton (to scale). Two possible interior structures are shown for Io and Triton. \citep[Sources: ][]{Schubert2009, Hemingway2018, McKinnon_2014, Park2025}}
    \label{fig:interiors}
\end{figure}

\subsubsection{Detection of Biosignatures: Mass Spectrometry}\label{sect_detecting_biosign_mass_spec}

Mass spectrometry has been integral to space research \citep[see reviews by][]{Ren2018,Vorburger2020} and is crucial for detecting bio-relevant molecules on planetary bodies. 
While gas-chromatography mass spectrometry (GC-MS) is very common in laboratories, and was used on the Viking landers \citep{Biemann1977}, the selection of the chromatographic column sets the range of different compounds that can be separated, and thus introduces bias in compounds hat can be identfied. In contrast, Laser-based Mass Spectrometry (LIMS) is an unbiased, agnostic analytical technique for the detection of complex chemical compounds. 
Recent technological advancements made LIMS competitive both for laboratory research \citep{Meneghin2022} and space missions, although separation techniques remain essential for secure molecule detection. \citet{Wurz2021} developed compact laser-based mass spectrometry instruments for planetary exploration (see also Riedo et al., this collection). Their LIMS instrument provides sensitive chemical (elements, isotopes, and molecules) analysis to identify life signatures on planetary surfaces \citep{Wurz2022}. This compact system features simple, robust operation and can be integrated into rover payloads or landed spacecraft with common sample delivery systems \citep{Wurz2021, Wurz2022, Riedo2025, Riedo2025SSR}.
\citet{Riedo2013} and \citet{Tulej2014} demonstrated quantitative measurements of almost all elements in laser ablation mode. A significant advancement came from replacing nanosecond laser pulses with femtosecond pulses for ablation and ionisation, enhancing the capabilities for element and isotope analysis \citep{Riedo2013, Grimaudo2015}. Further improvements have been made by employing high mass resolution Orbitrap \citep{briois2016orbitrap,arevalo2018orbitrap,ray2024characterization} or two-laser systems for separate desorption and ionisation steps \citep{getty2012compact}. LIMS instruments are flying on various upcoming missions, such as the MOMA instrument on ESA's Rosalind Franklin Mars rover \citep{goesmann2017mars,li2017mars} and the DraMS instrument on the Dragonfly Titan rotorcraft of NASA \citep{grubisic2021laser}. 

The ORganics Information Gathering INstrument (ORIGIN) represents an adaptation of the LIMS technology specifically designed for biosignature detection on planetary surfaces \citep{ligterink2020origin}. ORIGIN employs nanosecond pulsed lasers for gentle desorption and ionisation of molecules from sample surfaces, with subsequent mass analysis performed by a miniature reflectron time-of-flight mass spectrometer \citep{Rohner2003}. This approach enables sensitive detection of intact biomolecules while minimising molecule fragmentation.
ORIGIN has successfully demonstrated detection capabilities for biosignature molecules including amino acids with detection limits in the femtomole~mm$^{-2}$ range \citep{ligterink2020origin}, polycyclic aromatic hydrocarbons (PAHs) \citep{kipfer2022toward}, lipids \citep{boeren2022detecting}, nucleobases \citep{Boeren2025}, and various salts \citep{ligterink2022origin}. The instrument can analyse complex natural samples and employs correlation network analysis to separate different molecular components \citep{schwander2022correlation}. Given its compact size, sensitivity, and ability to analyse samples without extensive preparation, Laser desorption and ablation-based instruments like ORIGIN, CORALS \citep{willhite2021corals}, and CRATER \citep{ray2024characterization} are well-suited for inclusion on future planetary lander missions to potentially habitable ocean worlds like Europa and Enceladus, where biosignatures may be present in surface ice deposits \citep{mackenzie2021enceladus,mousis2022moonraker}.

The biosignature might be present also in the grains ejected with the plume gas. Indeed, Cassini spacecraft measurements of dust grains in Enceladus' plume with the Cosmic Dust Analyzer (CDA) were interpreted to contain concentrated and complex macromolecular organic material with molecular masses above 200 atomic mass units \citep{postberg2018macromolecular}. In dust instruments, like the CDA, the dust grains impact a target surface inside the instrument at high velocity, about 8 km/s for the Cassini flyby of Enceladus,  
which causes the formation of an impact plume at several thousend Kelvin. Ionised species in the plume are then mass analysed and recorded by the instrument. The high impact speed (referred to as hyper velocity impacts, HVI)  and the plasma plume may cause fragmentation, change of the chemical nature, or even formation of new molecules. Thus, the interpretation of dust mass spectra has to be done with caution. The initial kinetic energy of a single charged species at m/z 1 (H$^+$) is 0.5 eV at 10 km/s and 2.1 eV at 20 km/s, and scales linearly with mass. These energies contrast the bond-dissociation energy of a carbon–carbon bond of 3.6 eV and about 11 eV for the triple bonds \citep{Fausch2023}. Whereas some studies report the onset velocities of HVI induced bond-dissociation already starting at 900 m/s \citep{Furukawa2015}, it is generally accepted that the vast majority of species will be subject to fragmentation at speeds above 5 to 6 km/s \citep{Jaramillo-Botero2021}, depending on the impact conditions. \cite{Ulibarri2023} found that the fragmentation rates of amino acids rise significantly beyond 6.1 km/s for bare amino acids and beyond 8.5 km/s for water ice-shielded amino acids, as exemplarily studied on histidine-monohydrochloride. Also the formation of biomolecules from abiological material has been observed for hyperveloctity impacts of dust particles on solid surfaces \citep[e.g.][]{Martins2013, Managadze2016, Mieno2025}.

\section{Detection of Exomoons}\label{exomoon_detect_activity}
Given how common it is for a planet in our Solar System to host one or more satellites, we might expect an abundance of extrasolar satellites. 
However, the first extrasolar satellite, or exomoon, has not yet been unambiguously identified. 
We discuss here in depth the detection of an exomoon with the mass of Io. 
The very low mass of an exomoon compared to the exoplanet fundamentally biases transit-timing variation (TTV) detections towards massive exomoons ($\sim R_{\text{Uranus}}$) as claimed at Kepler-1625  \citep{Teachey2018} and Kepler-1708 \citep{Kipping2022}, although not confirmed by independent analyses \citep{Heller2019, Kreidberg2019, HellerHipke2024}. For more details on TTV see the chapter by Ulmer-Moll et al., this collection. 
As demonstrated by \citet{Sucerquia2025} for the small candidate Io-sized exomoon, WASP-49 A b I, the TTV signal is negligible. While microlensing can be used to detect exomoons, especially those around free-floating planets \citep[FFPs: e.g.][]{Bachelet2022}, this technique does not yield 
characterisation beyond their mass.
This raises the necessity for an \textit{indirect} detection technique, for example due to the presence of a neutral sodium (Na I) signature of an active moon \citep{Charbonneau2002DetectionAtmosphere, johnsonhuggins06, Oza2019, Oza2024RedshiftedTransit}. Indeed, \citet{Oza2019} show that an exo-Io with a planetary tidal $Q_p \lesssim 10^{10}$, corresponding to gas giant host planets with an orbital distance $\gtrsim 0.2$~au, would have a 10$^{3}$ -- 10$^{7}$ higher mass-loss rate than Io nurturing the Na exosphere and easing detection. 

\begin{figure}
    \centering
    \includegraphics[width=\linewidth]{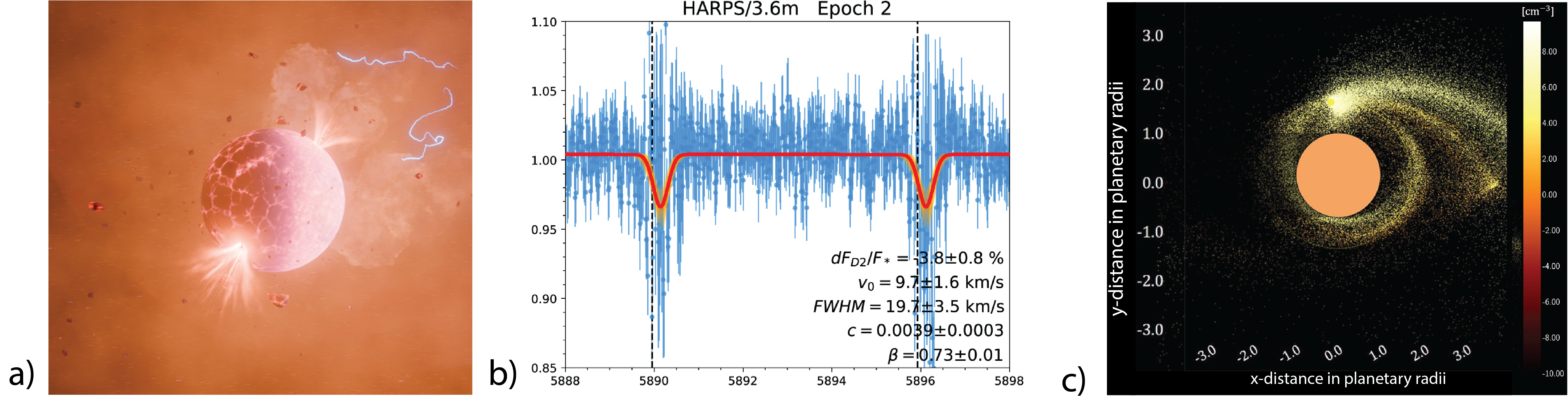}
    \caption{An exo-Io as a possible explanation for a neutral sodium (Na I) signature. a) An artist's impression of an exo-Io (with the host star and host planet next to it). Credits: NASA/JPL-Caltech/K. Miller (Caltech/IPAC). b) Redshifted \ce{Na} transmission spectrum observed with HARPS instrument on the ESO La Silla 3.6~m telescope interpreted by \citet{Oza2024RedshiftedTransit} as an indication of exo-Io WASP-49 b I. Credits: \citet{Oza2024RedshiftedTransit}. c) Simulated bird's eye view of the toroidal \ce{Na} density by an active WASP-49 b I. Credits: From \citet{MeyerZuWestram2024}, licensed under CC-BY 4.0. }
    \label{fig_exoIo}
\end{figure}

By modelling Na and K outgassing and escape, \citet{Oza2019} have shown that Na I and K I clouds, such as the ones we observe around Io (Sect. \ref{subsub_Io}),
and contributions from volcanic plumes \citep{Mendillo2007}, can serve as a unique detection method for a geologically active exo-Io signature. The alkali exospheres of hot Jupiters
may be uniquely supplied by a hidden natural satellite \citep{Oza2019}. First, the alkali metals Na I and K I have been attributed to the H/He envelopes of gas giant atmospheres assumed to have metallicities of solar abundance  \citep{Charbonneau2002}. It was soon shown that these alkali metals, due to photoionisation so close to the star \citep{johnsonhuggins06}, rather originate from a natural satellite \citep{Gebek2020AlkalineSpectra, Oza2026}. 

\citet{Oza2024RedshiftedTransit} observed a transient \ce{Na} signal close to a close-in hot Saturn around WASP-49 A with the HARPS instrument on the ESO La Silla 3.6~m telescope \citep{Wyttenbach2017HotWASP-49b} and additional observations showed the surprising disappearance of the \ce{Na} signal with the larger aperture ESPRESSO instrument of the Very Large Telescope (VLT). The transient absorption signal indicates a Doppler velocity of 9.7 $\pm$ 1.6 km/s, which corrected for radiation pressure, corresponds to a net radial velocity of ~\SI{+15.4}{km/s}, very close to the orbit of a close-in natural satellite. In this way, they infer the presence of a natural satellite, fuelling the Doppler shift. This is further backed up by the cloud travelling against the line-of-sight and against the vector of radiation pressure due to its measured redshift.  The fact that the transient \ce{Na} signature is \textit{red}shifted (against radiation pressure, which blueshifts atomic lines) has been suggested to be an unambiguous signature of a natural satellite interacting with a magnetosphere 
\citep{Schmidt2022Doppler-ShiftedExomoons}.
The observed signal (Fig.~\ref{fig_exoIo}b) varies from night to night by 1--4~\% in Na~D, and rarely coincides with planetary transit. Furthermore, the Doppler shift occurs at the precise moment the transient \ce{Na} absorption occurs, lasting for roughly $\sim$ 40 minutes. It is unclear whether the \ce{Na} gas is a cloud or jet, and further toroidal atmosphere simulations are needed to better constrain the behaviour in this system \citep[][Fig.~\ref{fig_exoIo}c]{MeyerZuWestram2024}.  
If the roughly 8 hour orbit is confirmed, this would be the first detection of an exomoon, depicted as an exo-Io in Fig. \ref{fig_linear_main} and \ref{fig_exoIo}a. Recently, KECK/HIRES observations have confirmed the variable Doppler shifted behaviour and transient Na I behaviour of $\sim$ 40 minutes \citep{Unni2025}.

Every claim of an exomoon detection around close-in gas giants raises the question about the system's formation and long-term survival \citep[e.g.][]{2020ApJ...902L..20Q,2021PASP..133i4401D,2022MNRAS.tmp.2636H,2023A&A...672A..78M}.
\citet{Bolmont2025A&A...704A...9B} investigated whether such satellites could endure planetary migration within a protoplanetary disk. 
Using an extended version of the N-body code \textit{Posidonius} \citep{2020A&A...635A.117B,2024A&A...691L...3R}, which includes tidal interactions and planetary migration, they explored key factors such as initial satellite-planet distance, disk lifetime, satellite mass, and tidal dissipation. 
They also examined how different tidal prescriptions for the star and planet affect satellite survival. 
In particular, they compared scenarios where only the equilibrium tide was considered in both bodies to those where the dynamical tide was also included \citep[e.g.,][]{2013MNRAS.429..613O}.
Their results indicate that the satellite survival probability is low if its tidal dissipation is above zero, with only those starting at least 0.6 times the Jupiter-Io separation and orbiting a planet beyond ~0.1 AU having a chance to persist. 
Most satellites are either tidally disrupted following a slow inward migration toward the planet until reaching the Roche limit or dynamically disrupted following an eccentricity excitation due to the close-in massive star. 
The study suggests that if satellites are found around close-in exoplanets, they may have been captured rather than formed in situ.

In light of their results, they find that the survival of such a satellite as proposed by \citet{Oza2024RedshiftedTransit} is extremely challenging.
If this satellite is confirmed, then one might have to invoke new avenues of planet-satellite formation, such as 
 a late in-situ formation or a late migration process of the planet-satellite pair, except for high-eccentricity migration \citep{2020MNRAS.499.4195T}. Assuming the satellite survived formation and evolutionary stages, recent N-body simulations considering stellar perturbations and general relativistic effects do indeed find a stability region near the Roche limit $\gtrsim $ 1.2 R$_p$ at WASP-49 A b, noting a chaos region at $\sim $ 1.4 R$_p$ \citep{Sucerquia2025}. Resonant orbits as studied by \citet{Cassidy2009MassiveExoplanets} might be especially interesting to consider in future works.

\section{PlanetS Legacy}\label{sec_planetslegacy}
The NCCR \textit{PlanetS} has contributed uniquely to the study of active moons and their relevance for exoplanetary systems. 
In particular, \textit{PlanetS} has advanced the study of moon formation through HD and MHD simulations \citep{Cilibrasi23, Maeda2024DeliveryPlanets} and impact simulations \citep{Reinhardt2020, Woo2022}, the presentation of a possible solution to the inward drift problem of dust for a satellitesimal scenario \citep{Shibaike2017, Cilibrasi18, Cilibrasi21, Shibaike2023}, a new pebble accretion formation scenario \citep{Shibaike2019} and the effects of photophoresis \citep{Arakawa2019}, and ground-based observations of the  PDS~70 system \citep{Shibaike2024ConstraintsEvolution} and the AB~Aurigae system \citep{Shibaike2025}.

On icy moons, \textit{PlanetS} contributions include the characterisation of the surface through laboratory studies of icy analogues \citep{Poch2018, Pommerol2019ExperimentingMaterial, Cerubini2022a, Cerubini2022b, kipfer2022toward, Kipfer2024ComplexMethane, Ottersberg2025}, the investigation of the composition of their tenuous atmospheres \citep{Schlarmann2024}, and the acceleration of lineament mapping with deep learning \citep{Haslebacher2024a, Duembgen2025NonparametricData, HaslebacherPSJ2025, Haslebacher2025MapsBook}, which in the future can be used for a fast and data-driven prediction and identification of the location of freshest surface areas, and possibly, plumes.
Furthermore, \textit{PlanetS} is advancing the detection of biosignatures, which are best accessible on active moons, through the development of homochirality-detecting instruments \citep{grone2024modeling, grone2024full} and laser-based mass spectrometry instruments \citep{ligterink2022origin, schwander2022correlation, boeren2022detecting, Boeren2025, Riedo2025}.   
On Io, the nature of volcanic plumes and interactions with the atmosphere was studied within \textit{PlanetS} through numerical modelling \citep{Klaiber2024}. The study of Io’s atmospheric signature (\ce{Na} I and \ce{K} I clouds) were crucial for the possible detection of an Io-like exo-moon around WASP-49 A b \citep{Oza2019}. Theoretical work is in progress to investigate the survivability of exomoons around close-in planets \citep{2020A&A...635A.117B,2024A&A...691L...3R, Bolmont2025A&A...704A...9B}.

For future ground-based observations, \citet{Lovis2024} are developing the \textit{PlanetS}-affiliated RISTRETTO instrument, an innovative visible (620--840 nm), high-resolution (R=140,000) spectrograph. It will be fed by an extreme adaptive optics system and installed at the VLT as a visitor instrument by 2030.

\section{Topics of Future Interest}
Broad connections are waiting to be pursued. How can outgassing events be linked to sub-surface, interior, and formation processes? 
How can outgassing events be used to constrain habitability? Missions on cruise (Europa Clipper, JUICE) or in operation (JWST) at the time of writing as well as missions and instruments in development (ELT) and already proposed (IVO, Trident) and advertised future flagship missions (Uranus Orbiter and Probe, ESA's L4 mission) will deliver observations of unprecedented quality. Interpreting these observations requires preparation in the form of modelling, laboratory work, and data analysis based on pre-existing data. In the following, we will introduce some specific questions.

The presence of salty water oceans on icy moons raises the grand questions of habitability and life beyond Earth, and even beyond the Solar System. Because this is a non-falsifiable problem, the understanding of the icy environments, for example through modelling and laboratory work, is critical. Since claims of a detection of life should be verified with complementary experimental methods, the development of different life-detecting instruments is necessary.  

Due to the low abundance of craters, a fundamental unknown of icy moons Europa and Enceladus is their surface age. Studies on icy analogues can support the determination of relative and absolute surface age. Such studies link the remotely probed top \SI{1}{m} surface composition to surface processes in the top \SI{1}{cm}. Especially for Europa, the prediction of hotspot areas based on currently available data would be outstanding. Furthermore, a possible link of hotspots to chaos formation warrants further investigation, for example through modelling.

A better estimation of the downwelling rate of \ce{O2} into Europa's ocean would be helpful to constrain the oxidation state of its ocean. For this, work on diffusion constants for different states of ice, from the surface regolith down to ice near the melting point at the ice-ocean interface, is needed. 
The cracking of Europa's ice shell is not yet well understood. What governs the lineament azimuths, and why are some lineaments straight, while others are in a cycloidal shape? Did the ice shell undergo periodic thinning and thickening, allowing for different cracking mechanisms \citep{Hussmann2004, Rudolph2022}? What is the influence of periodically changing conditions on a possible development of life?

On Io, the interactions between plasma and plumes are not yet fully understood. More detailed modelling can guide future investigations.

Some Uranian moons (Miranda, Ariel, Titania) show intriguing geologic surface features, which raise the question of past or current subsurface oceans. Because imaging data are limited, modelling possible long-term evolutions of orbital dynamics and tidal dissipation of the Uranian moons is needed. 
The study of Uranian moons is advertised by NASA's decadal survey with the Uranus Orbiter and Probe \citep{OriginsWorldsandLife:ADecadalStrategy}.

RISTRETTO (Sect. \ref{sec_planetslegacy}) will provide a spatial resolution of about 120 km at the Jupiter system and 240 km at the Saturn system, which is remarkable for such a sensitive spectrograph. Specific science questions for observations of active moons with this instrument need yet to be worked out.
ESA’s JUICE and NASA's Europa Clipper spacecraft each carry a comprehensive suite of instruments to investigate the Galilean moons, with the potential to investigate plume activity on Europa and Io. 
With the recent discovery that the Saturnian moon Mimas is likely a young, evolving ocean world \citep{Rhoden2024, Lainey2024}, further studies that investigate active surface processes on Mimas are needed. The new generation of planetary missions, ongoing advancements in instrumentation, and the development of increasingly sophisticated modelling frameworks, the next decades will be exciting for the investigation of active moons in our Solar System and beyond.

The Voyage 2050 Senior Committee report recommends that ESA's next large-class mission L4 
\citep{tacconi2021voyage} should focus on the moons of the giant planets. This mission would aim to explore the habitability and evidence of life on ocean worlds, search for biosignatures, and study the connection between moon interiors, near-surface environments, and the exchange of mass and energy within the broader moon-planet system.
The L4 expert committee identified Europa, Ganymede, Titan as possible targets and Enceladus as the recommended target for this mission. 
A lander near the south pole was deemed both feasible and well-aligned with the mission's scientific goals.
The expert committee strongly suggests that flybys of both moons be conducted. Any planned mission to the Saturnian system is recommended to include Enceladus plume flybys and an additional Moon Tour focusing on other moons with potential subsurface oceans. With the recent discovery that Mimas is likely a young, evolving ocean world \citep{Rhoden2024, Lainey2024}, the committee also advises prioritizing a larger number of close Mimas flybys.

At the frontier of exomoon detection, there are two directions to develop. On the one hand, each exomoon candidate needs to be verified with independent methods.
On the other hand, exomoon detections can inform formation theories of Solar System moon formation.  
Another large unknown for formation theories is the structure of the gas flow in CPDs. Satellite formation strongly depends on the gas and its flow structures. The structures can be analysed with HD and MHD simulations, but depend on viscosity, opacity (i.e., dust amount), and magnetic field, which are also unknown.

\begin{acknowledgement}
This work has been carried out within the framework of the National Centre of Competence in Research PlanetS supported by the Swiss National Science Foundation under grants 51NF40\_205606. The authors acknowledge the financial support of the SNSF.
C. Haslebacher furthermore acknowledges the financial support of the SNSF through grant P500PT\_225447. Y.S. is supported by JSPS KAKENHI Grant Number JP24K22907.
We thank Jason Hofgartner and John Spencer for their helpful insights, and the four anonymous reviewers for their constructive feedback that improved this manuscript.
Inspiration for Fig. \ref{fig_linear_main} came from \citet{DePater2021AIo, Hofgartner2022, depater2023, Park2025}. 
\end{acknowledgement}
\ethics{Competing Interests}{The authors have no conflicts of interest to declare that are relevant to the content of this chapter.}

\bibliography{references} 

\begin{thebibliography}{483}
\providecommand{\natexlab}[1]{#1}
\providecommand{\url}[1]{\texttt{#1}}
\expandafter\ifx\csname urlstyle\endcsname\relax
  \providecommand{\doi}[1]{doi: #1}\else
  \providecommand{\doi}{doi: \begingroup \urlstyle{rm}\Url}\fi

\bibitem[Abramov et~al.(2013)Abramov, Rathbun, Schmidt, and Spencer]{Abramov2013}
O.~Abramov, J.~A. Rathbun, B.~E. Schmidt, and J.~R. Spencer.
\newblock {Detectability of thermal signatures associated with active formation of ‘chaos terrain’ on Europa}.
\newblock \emph{Earth and Planetary Science Letters}, 384:\penalty0 37--41, 12 2013.
\newblock ISSN 0012-821X.
\newblock \doi{10.1016/J.EPSL.2013.09.027}.

\bibitem[Ames et~al.(2025)Ames, Ferreira, Czaja, and Masters]{Ames2025}
F.~Ames, D.~Ferreira, A.~Czaja, and A.~Masters.
\newblock {Ocean stratification impedes particulate transport to the plumes of Enceladus}.
\newblock \emph{Communications Earth {\&} Environment 2025 6:1}, 6\penalty0 (1):\penalty0 1--11, 2 2025.
\newblock ISSN 2662-4435.
\newblock \doi{10.1038/s43247-025-02036-3}.
\newblock URL \url{https://www.nature.com/articles/s43247-025-02036-3}.

\bibitem[Arakawa and Shibaike(2019)]{Arakawa2019}
S.~Arakawa and Y.~Shibaike.
\newblock {Photophoresis in the circumjovian disk and its impact on the orbital configuration of the Galilean satellites}.
\newblock \emph{Astronomy {\&} Astrophysics}, 629:\penalty0 A106, 9 2019.
\newblock ISSN 0004-6361.
\newblock \doi{10.1051/0004-6361/201936202}.
\newblock URL \url{https://www.aanda.org/articles/aa/full_html/2019/09/aa36202-19/aa36202-19.html https://www.aanda.org/articles/aa/abs/2019/09/aa36202-19/aa36202-19.html}.

\bibitem[Arevalo~Jr et~al.(2018)Arevalo~Jr, Selliez, Briois, Carrasco, Thirkell, Cherville, Colin, Gaubicher, Farcy, Li, et~al.]{arevalo2018orbitrap}
R.~Arevalo~Jr, L.~Selliez, C.~Briois, N.~Carrasco, L.~Thirkell, B.~Cherville, F.~Colin, B.~Gaubicher, B.~Farcy, X.~Li, et~al.
\newblock An orbitrap-based laser desorption/ablation mass spectrometer designed for spaceflight.
\newblock \emph{Rapid Communications in Mass Spectrometry}, 32\penalty0 (21):\penalty0 1875--1886, 2018.

\bibitem[Ashkenazy et~al.(2023)Ashkenazy, Tziperman, and Nimmo]{Ashkenazy2023}
Y.~Ashkenazy, E.~Tziperman, and F.~Nimmo.
\newblock {Non-Synchronous Rotation on Europa Driven by Ocean Currents}.
\newblock \emph{AGU Advances}, 4\penalty0 (3):\penalty0 e2022AV000849, 6 2023.
\newblock ISSN 2576-604X.
\newblock \doi{10.1029/2022AV000849}.
\newblock URL \url{https://onlinelibrary.wiley.com/doi/full/10.1029/2022AV000849 https://onlinelibrary.wiley.com/doi/abs/10.1029/2022AV000849 https://agupubs.onlinelibrary.wiley.com/doi/10.1029/2022AV000849}.

\bibitem[Ayg{\"u}n and {\v{C}}adek(2024)]{aygun2024}
B.~Ayg{\"u}n and O.~{\v{C}}adek.
\newblock Tidal heating in a subsurface magma ocean on io revisited.
\newblock \emph{Geophysical Research Letters}, 51\penalty0 (10):\penalty0 e2023GL107869, 2024.

\bibitem[Bachelet et~al.(2022)Bachelet, Specht, Penny, Hundertmark, Awiphan, Beaulieu, Dominik, Kerins, Maoz, Meade, Nucita, Poleski, Ranc, Rhodes, and Robin]{Bachelet2022}
E.~Bachelet, D.~Specht, M.~Penny, M.~Hundertmark, S.~Awiphan, J.~P. Beaulieu, M.~Dominik, E.~Kerins, D.~Maoz, E.~Meade, A.~A. Nucita, R.~Poleski, C.~Ranc, J.~Rhodes, and A.~C. Robin.
\newblock {Euclid-Roman joint microlensing survey: Early mass measurement, free floating planets, and exomoons}.
\newblock \emph{Astronomy and Astrophysics}, 664:\penalty0 A136, 8 2022.
\newblock ISSN 14320746.
\newblock \doi{10.1051/0004-6361/202140351}.
\newblock URL \url{https://ui.adsabs.harvard.edu/abs/2022A&A...664A.136B/abstract}.

\bibitem[Bagenal et~al.(1980)Bagenal, Sullivan, and Siscoe]{bagenal1980}
F.~Bagenal, J.~D. Sullivan, and G.~L. Siscoe.
\newblock Spatial distribution of plasma in the io torus.
\newblock \emph{Geophysical Research Letters}, 7\penalty0 (1):\penalty0 41--44, 1980.

\bibitem[Bagenal et~al.(2004)Bagenal, Dowling, and McKinnon]{Bagenal2004}
F.~Bagenal, T.~E. Dowling, and W.~B. McKinnon.
\newblock \emph{{Jupiter : the planet, satellites and magnetosphere}}.
\newblock Cambridge University Press, Cambridge, UK, 2004.

\bibitem[Bailey(2001)]{bailey2001astronomical}
J.~Bailey.
\newblock Astronomical sources of circularly polarized light and the origin of homochirality.
\newblock \emph{Origins of Life and Evolution of the Biosphere}, 31:\penalty0 167--183, 2001.

\bibitem[Ballester et~al.(1994)Ballester, McGrath, Strobel, Zhu, Feldman, and Moos]{ballester1994}
G.~E. Ballester, M.~McGrath, D.~Strobel, X.~Zhu, P.~D. Feldman, and H.~Moos.
\newblock Detection of the so2 atmosphere on io with the hubble space telescope.
\newblock \emph{Icarus}, 111\penalty0 (1):\penalty0 2--17, 1994.

\bibitem[Baross et~al.(1981)Baross, Hoffman, et~al.]{baross1981hypothesis}
J.~Baross, S.~Hoffman, et~al.
\newblock An hypothesis concerning the relationships between submarine hot springs and the origin of life on earth.
\newblock In \emph{Oceanologica Acta, Special issue}. Gauthier-Villars, 1981.

\bibitem[Barr and Canup(2008)]{Barr2008}
A.~C. Barr and R.~M. Canup.
\newblock Constraints on gas giant satellite formation from the interior states of partially differentiated satellites.
\newblock \emph{Icarus}, 198\penalty0 (1):\penalty0 163--177, 2008.

\bibitem[Batygin and Morbidelli(2020)]{Batygin2020FormationSatellites}
K.~Batygin and A.~Morbidelli.
\newblock {Formation of giant planet satellites}.
\newblock \emph{Astrophys J}, 894\penalty0 (2):\penalty0 143, 5 2020.
\newblock ISSN 15384357.
\newblock \doi{10.3847/1538-4357/ab8937}.

\bibitem[Beatty et~al.(2005)Beatty, Overmann, Lince, Manske, Lang, Blankenship, Van~Dover, Martinson, and Plumley]{beatty2005obligately}
J.~T. Beatty, J.~Overmann, M.~T. Lince, A.~K. Manske, A.~S. Lang, R.~E. Blankenship, C.~L. Van~Dover, T.~A. Martinson, and F.~G. Plumley.
\newblock An obligately photosynthetic bacterial anaerobe from a deep-sea hydrothermal vent.
\newblock \emph{Proceedings of the National Academy of Sciences}, 102\penalty0 (26):\penalty0 9306--9310, 2005.

\bibitem[B{\v{e}}hounkov{\'a} et~al.(2010)B{\v{e}}hounkov{\'a}, Tobie, Choblet, and {\v{C}}adek]{behounkova2010}
M.~B{\v{e}}hounkov{\'a}, G.~Tobie, G.~Choblet, and O.~{\v{C}}adek.
\newblock Coupling mantle convection and tidal dissipation: Applications to enceladus and earth-like planets.
\newblock \emph{Journal of Geophysical Research: Planets}, 115\penalty0 (E9), 2010.

\bibitem[Behounkov{\'{a}} et~al.(2015)Behounkov{\'{a}}, Tobie, Cadek, Choblet, Porco, and Nimmo]{Behounkova2015}
M.~Behounkov{\'{a}}, G.~Tobie, O.~Cadek, G.~Choblet, C.~Porco, and F.~Nimmo.
\newblock {Timing of water plume eruptions on Enceladus explained by interior viscosity structure}.
\newblock \emph{Nature Geoscience}, 8\penalty0 (8):\penalty0 601--604, 8 2015.
\newblock ISSN 17520908.
\newblock \doi{10.1038/NGEO2475}.

\bibitem[Behounkova et~al.(2017)Behounkova, Soucek, Hron, and Cadec]{Behounkova2017}
M.~Behounkova, O.~Soucek, J.~Hron, and O.~Cadec.
\newblock {Plume Activity and Tidal Deformation on Enceladus Influenced by Faults and Variable Ice Shell Thickness}.
\newblock \emph{Astrobiology}, 17\penalty0 (9):\penalty0 941, 9 2017.
\newblock ISSN 15311074.
\newblock \doi{10.1089/AST.2016.1629}.
\newblock URL \url{https://pmc.ncbi.nlm.nih.gov/articles/PMC5610426/}.

\bibitem[B{\v{e}}hounkov{\'{a}} et~al.(2021)B{\v{e}}hounkov{\'{a}}, Tobie, Choblet, Kervazo, Melwani~Daswani, Dumoulin, and Vance]{Behounkova2021TidallyEuropa}
M.~B{\v{e}}hounkov{\'{a}}, G.~Tobie, G.~Choblet, M.~Kervazo, M.~Melwani~Daswani, C.~Dumoulin, and S.~D. Vance.
\newblock {Tidally Induced Magmatic Pulses on the Oceanic Floor of Jupiter's Moon Europa}.
\newblock \emph{Geophysical Research Letters}, 48\penalty0 (3), 2 2021.
\newblock ISSN 19448007.
\newblock \doi{10.1029/2020GL090077}.

\bibitem[Belton et~al.(2000)Belton, Anger, Carr, Chapman, Davies, Greeley, Greenberg, Head~III, Klaasen, Neukum, Pilcher, Thomas, Veverka, Gierasch, Ingersoll, Fanale, McEwen, Morrison, Schubert, Beebe, Burns, Johnson, West, Ip, McElroy, and Orton]{Belton2000}
M.~J. Belton, C.~Anger, M.~Carr, C.~Chapman, M.~Davies, R.~Greeley, R.~Greenberg, J.~Head~III, K.~Klaasen, G.~Neukum, C.~Pilcher, P.~Thomas, J.~Veverka, P.~Gierasch, A.~Ingersoll, F.~Fanale, A.~McEwen, D.~Morrison, G.~Schubert, R.~Beebe, J.~Burns, T.~Johnson, R.~West, W.-H. Ip, M.~McElroy, and G.~Orton.
\newblock {Resutls of the Galileo Solid State Imaging (SSI) Experiment}.
\newblock \emph{Advances in Space Research}, 26\penalty0 (10):\penalty0 1641--1647, 2000.
\newblock URL \url{www.elsevier.nl/locate/asr}.

\bibitem[Bertrand et~al.(2022)Bertrand, Lellouch, Holler, Young, Schmitt, {Marques Oliveira}, Sicardy, Forget, Grundy, Merlin, Vangvichith, Millour, Schenk, Hansen, White, Moore, Stansberry, Oza, Dubois, Quirico, and Cruikshank]{Bertrand_2022}
T.~Bertrand, E.~Lellouch, B.~Holler, L.~Young, B.~Schmitt, J.~{Marques Oliveira}, B.~Sicardy, F.~Forget, W.~Grundy, F.~Merlin, M.~Vangvichith, E.~Millour, P.~Schenk, C.~Hansen, O.~White, J.~Moore, J.~Stansberry, A.~Oza, D.~Dubois, E.~Quirico, and D.~Cruikshank.
\newblock Volatile transport modeling on triton with new observational constraints.
\newblock \emph{Icarus}, 373:\penalty0 114764, 2022.
\newblock ISSN 0019-1035.
\newblock \doi{https://doi.org/10.1016/j.icarus.2021.114764}.
\newblock URL \url{https://www.sciencedirect.com/science/article/pii/S0019103521004164}.

\bibitem[Biemann et~al.(1977)Biemann, Oro, Toulmin~III, Orgel, Nier, Anderson, Simmonds, Flory, Diaz, Rushneck, Biller, and Lafleur]{Biemann1977}
K.~Biemann, J.~Oro, P.~Toulmin~III, L.~E. Orgel, A.~O. Nier, D.~M. Anderson, P.~G. Simmonds, D.~Flory, A.~V. Diaz, D.~R. Rushneck, J.~E. Biller, and A.~L. Lafleur.
\newblock The search for organic substances and inorganic volatile compounds in the surface of mars.
\newblock \emph{Journal of Geophysical Research (1896-1977)}, 82\penalty0 (28):\penalty0 4641--4658, 1977.
\newblock \doi{https://doi.org/10.1029/JS082i028p04641}.
\newblock URL \url{https://agupubs.onlinelibrary.wiley.com/doi/abs/10.1029/JS082i028p04641}.

\bibitem[Blackmond(2010)]{blackmond2010origin}
D.~G. Blackmond.
\newblock The origin of biological homochirality.
\newblock \emph{Cold Spring Harbor perspectives in biology}, 2\penalty0 (5):\penalty0 a002147, 2010.

\bibitem[Blaney et~al.(2024)Blaney, Hibbitts, Diniega, Davies, Clark, Green, Hedman, Langevin, Lunine, McCord, Murchie, Paranicas, Seelos, Soderblom, Cable, Eckert, Thompson, Trumbo, Bruce, Lundeen, Bender, Helmlinger, Moore, Mouroulis, Small, Tang, Van~Gorp, Sullivan, Zareh, Rodriquez, McKinley, Hahn, Bowers, Hourani, Bryce, Nuding, Bailey, Rettura, and Zarate]{blaney_mapping_2024}
D.~L. Blaney, K.~Hibbitts, S.~Diniega, A.~G. Davies, R.~N. Clark, R.~O. Green, M.~Hedman, Y.~Langevin, J.~Lunine, T.~B. McCord, S.~Murchie, C.~Paranicas, F.~Seelos, J.~M. Soderblom, M.~L. Cable, R.~Eckert, D.~R. Thompson, S.~K. Trumbo, C.~Bruce, S.~R. Lundeen, H.~A. Bender, M.~C. Helmlinger, L.~B. Moore, P.~Mouroulis, Z.~Small, H.~Tang, B.~Van~Gorp, P.~W. Sullivan, S.~Zareh, J.~I. Rodriquez, I.~McKinley, D.~V. Hahn, M.~Bowers, R.~Hourani, B.~A. Bryce, D.~Nuding, Z.~Bailey, A.~Rettura, and E.~D. Zarate.
\newblock The {Mapping} {Imaging} {Spectrometer} for {Europa} ({MISE}).
\newblock \emph{Space Science Reviews}, 220\penalty0 (7):\penalty0 80, Oct. 2024.
\newblock ISSN 0038-6308, 1572-9672.
\newblock \doi{10.1007/s11214-024-01097-8}.
\newblock URL \url{https://link.springer.com/10.1007/s11214-024-01097-8}.

\bibitem[Bl\"ocker et~al.(2016)Bl\"ocker, Saur, and Roth]{bloecker2016}
A.~Bl\"ocker, J.~Saur, and L.~Roth.
\newblock {Europa's plasma interaction with an inhomogeneous atmosphere: Development of Alfv\'en winglets within the Alfv\'en wings}.
\newblock \emph{Journal of Geophysical Research: Space Physics}, 121\penalty0 (10):\penalty0 9794--9828, 2016.
\newblock \doi{10.1002/2016JA022479}.

\bibitem[Boeren et~al.(2022)Boeren, Gruchola, de~Koning, Schmidt, Kipfer, Ligterink, Tulej, Wurz, and Riedo]{boeren2022detecting}
N.~J. Boeren, S.~Gruchola, C.~P. de~Koning, P.~K. Schmidt, K.~A. Kipfer, N.~F. Ligterink, M.~Tulej, P.~Wurz, and A.~Riedo.
\newblock Detecting lipids on planetary surfaces with laser desorption ionization mass spectrometry.
\newblock \emph{The Planetary Science Journal}, 3\penalty0 (10):\penalty0 241, 2022.

\bibitem[Boeren et~al.(2025)Boeren, Keresztes~Schmidt, Tulej, Wurz, and Riedo]{Boeren2025}
N.~J. Boeren, P.~Keresztes~Schmidt, M.~Tulej, P.~Wurz, and A.~Riedo.
\newblock Origin: Laser desorption ionization mass spectrometry of nucleobases for in situ space exploration.
\newblock \emph{The Planetary Science Journal}, 6\penalty0 (2):\penalty0 28, jan 2025.
\newblock \doi{10.3847/PSJ/ad9de9}.
\newblock URL \url{https://dx.doi.org/10.3847/PSJ/ad9de9}.

\bibitem[{Bolmont} et~al.(2020){Bolmont}, {Demory}, {Blanco-Cuaresma}, {Agol}, {Grimm}, {Auclair-Desrotour}, {Selsis}, and {Leleu}]{2020A&A...635A.117B}
E.~{Bolmont}, B.~O. {Demory}, S.~{Blanco-Cuaresma}, E.~{Agol}, S.~L. {Grimm}, P.~{Auclair-Desrotour}, F.~{Selsis}, and A.~{Leleu}.
\newblock {Impact of tides on the transit-timing fits to the TRAPPIST-1 system}.
\newblock \emph{\aap}, 635:\penalty0 A117, Mar. 2020.

\bibitem[{Bolmont} et~al.(2025){Bolmont}, {Galantay}, {Blanco-Cuaresma}, {Oza}, and {Mordasini}]{Bolmont2025A&A...704A...9B}
E.~{Bolmont}, E.~{Galantay}, S.~{Blanco-Cuaresma}, A.~V. {Oza}, and C.~{Mordasini}.
\newblock {Survival of satellites during the migration of a hot Jupiter}.
\newblock \emph{\aap}, 704:\penalty0 A9, Nov. 2025.
\newblock \doi{10.1051/0004-6361/202554625}.

\bibitem[Bouchez et~al.(2000)Bouchez, Brown, and Schneider]{Bouchez2000}
A.~H. Bouchez, M.~E. Brown, and N.~M. Schneider.
\newblock Eclipse spectroscopy of io's atmosphere.
\newblock \emph{Icarus}, 148\penalty0 (1):\penalty0 316--319, 2000.
\newblock ISSN 0019-1035.
\newblock \doi{https://doi.org/10.1006/icar.2000.6518}.
\newblock URL \url{https://www.sciencedirect.com/science/article/pii/S0019103500965187}.

\bibitem[Bradák et~al.(2023)Bradák, Ákos Kereszturi, and Gomez]{Bradak2023b}
B.~Bradák, Ákos Kereszturi, and C.~Gomez.
\newblock Tectonic analysis of a newly identified putative cryovolcanic field on europa.
\newblock \emph{Advances in Space Research}, 72\penalty0 (9):\penalty0 4064--4073, 2023.
\newblock ISSN 0273-1177.
\newblock \doi{https://doi.org/10.1016/j.asr.2023.07.062}.
\newblock URL \url{https://www.sciencedirect.com/science/article/pii/S0273117723006117}.

\bibitem[Briois et~al.(2016)Briois, Thissen, Thirkell, Aradj, Bouabdellah, Boukrara, Carrasco, Chalumeau, Chapelon, Colin, et~al.]{briois2016orbitrap}
C.~Briois, R.~Thissen, L.~Thirkell, K.~Aradj, A.~Bouabdellah, A.~Boukrara, N.~Carrasco, G.~Chalumeau, O.~Chapelon, F.~Colin, et~al.
\newblock Orbitrap mass analyser for in situ characterisation of planetary environments: Performance evaluation of a laboratory prototype.
\newblock \emph{Planetary and Space Science}, 131:\penalty0 33--45, 2016.

\bibitem[Broadfoot et~al.(1989)Broadfoot, Atreya, Bertaux, Blamont, Dessler, Donahue, Forrester, Hall, Herbert, Holberg, et~al.]{Broadfoot1989}
A.~Broadfoot, S.~Atreya, J.~Bertaux, J.~Blamont, A.~Dessler, T.~Donahue, W.~Forrester, D.~Hall, F.~Herbert, J.~Holberg, et~al.
\newblock Ultraviolet spectrometer observations of neptune and triton.
\newblock \emph{Science}, 246\penalty0 (4936):\penalty0 1459--1466, 1989.

\bibitem[Brown(1974)]{Brown1974}
R.~A. Brown.
\newblock \emph{{Optical line emission from Io}}, pages 527--531.
\newblock Reidel, 1974.

\bibitem[Brown et~al.(1975)Brown, Goody, Murcray, and Chaffee]{Brown1975}
R.~A. Brown, R.~M. Goody, F.~J. Murcray, and J.~Chaffee, F.~H.
\newblock {Further studies of line emission from Io}.
\newblock \emph{Astrophysical Journal}, 200\penalty0 (2):\penalty0 L49--L53, 1975.
\newblock \doi{10.1086/181894}.

\bibitem[Brown et~al.(1990)Brown, Kirk, Johnson, and Soderblom]{Brown_1990}
R.~H. Brown, R.~L. Kirk, T.~V. Johnson, and L.~A. Soderblom.
\newblock Energy sources for triton's geyser-like plumes.
\newblock \emph{Science}, 250\penalty0 (4979):\penalty0 431--435, 1990.
\newblock \doi{10.1126/science.250.4979.431}.
\newblock URL \url{https://doi.org/10.1126/science.250.4979.431}.

\bibitem[Brown et~al.(2006)Brown, Clark, Buratti, Cruikshank, Barnes, Mastrapa, Bauer, Newman, Momary, Baines, Bellucci, Capaccioni, Cerroni, Combes, Coradini, Drossart, Formisano, Jaumann, Langavin, Matson, McCord, Nelson, Nicholson, Sicardy, and Sotin]{Brown2006}
R.~H. Brown, R.~N. Clark, B.~J. Buratti, D.~P. Cruikshank, J.~W. Barnes, R.~M. Mastrapa, J.~Bauer, S.~Newman, T.~Momary, K.~H. Baines, G.~Bellucci, F.~Capaccioni, P.~Cerroni, M.~Combes, A.~Coradini, P.~Drossart, V.~Formisano, R.~Jaumann, Y.~Langavin, D.~L. Matson, T.~B. McCord, R.~M. Nelson, P.~D. Nicholson, B.~Sicardy, and C.~Sotin.
\newblock {Composition and physical properties of Enceladus' surface}.
\newblock \emph{Science}, 311\penalty0 (5766):\penalty0 1425--1428, 3 2006.
\newblock ISSN 00368075.
\newblock \doi{10.1126/SCIENCE.1121031;PAGE:STRING:ARTICLE/CHAPTER}.
\newblock URL \url{/doi/pdf/10.1126/science.1121031}.

\bibitem[Cable et~al.(2021)Cable, Porco, Glein, German, MacKenzie, Neveu, Hoehler, Hofmann, Hendrix, Eigenbrode, et~al.]{cable2021science}
M.~L. Cable, C.~Porco, C.~R. Glein, C.~R. German, S.~M. MacKenzie, M.~Neveu, T.~M. Hoehler, A.~E. Hofmann, A.~R. Hendrix, J.~Eigenbrode, et~al.
\newblock The science case for a return to enceladus.
\newblock \emph{The planetary science journal}, 2\penalty0 (4):\penalty0 132, 2021.

\bibitem[Cahn et~al.(1956)Cahn, Ingold, and Prelog]{cahn1956specification}
R.~S. Cahn, C.~K. Ingold, and V.~Prelog.
\newblock The specification of asymmetric configuration in organic chemistry.
\newblock \emph{Experientia}, 12:\penalty0 81--94, 1956.

\bibitem[Canup(2010)]{Canup2010}
R.~M. Canup.
\newblock Origin of saturn's rings and inner moons by mass removal from a lost titan-sized satellite.
\newblock \emph{Nature}, 468\penalty0 (7326):\penalty0 943--946, 2010.

\bibitem[{Canup} and {Ward}(2002)]{Canup02}
R.~M. {Canup} and W.~R. {Ward}.
\newblock {Formation of the Galilean Satellites: Conditions of Accretion}.
\newblock \emph{\aj}, 124\penalty0 (6):\penalty0 3404--3423, Dec. 2002.
\newblock \doi{10.1086/344684}.

\bibitem[{Canup} and {Ward}(2006)]{Canup06}
R.~M. {Canup} and W.~R. {Ward}.
\newblock {A common mass scaling for satellite systems of gaseous planets}.
\newblock \emph{\nat}, 441\penalty0 (7095):\penalty0 834--839, June 2006.
\newblock \doi{10.1038/nature04860}.

\bibitem[Cappuccio et~al.(2022)Cappuccio, Di~Benedetto, Durante, and Iess]{Cappuccio2022}
P.~Cappuccio, M.~Di~Benedetto, D.~Durante, and L.~Iess.
\newblock {Callisto and Europa Gravity Measurements from JUICE 3GM Experiment Simulation}.
\newblock \emph{The Planetary Science Journal}, 3\penalty0 (8):\penalty0 199, 8 2022.
\newblock ISSN 2632-3338.
\newblock \doi{10.3847/PSJ/AC83C4}.
\newblock URL \url{https://iopscience.iop.org/article/10.3847/PSJ/ac83c4 https://iopscience.iop.org/article/10.3847/PSJ/ac83c4/meta}.

\bibitem[Carlson et~al.(1999)Carlson, Anderson, Johnson, Smythe, Hendrix, Barth, Soderblom, Hansen, McCord, Dalton, Clark, Shirley, Ocampo, and Matson]{Carlson1999HydrogenEuropa}
R.~W. Carlson, M.~S. Anderson, R.~E. Johnson, W.~D. Smythe, A.~R. Hendrix, C.~A. Barth, L.~A. Soderblom, G.~B. Hansen, T.~B. McCord, J.~B. Dalton, R.~N. Clark, J.~H. Shirley, A.~C. Ocampo, and D.~L. Matson.
\newblock {Hydrogen peroxide on the surface of Europa}.
\newblock \emph{Science}, 283\penalty0 (5410):\penalty0 2062--2064, 3 1999.
\newblock ISSN 00368075.
\newblock \doi{10.1126/science.283.5410.2062}.

\bibitem[Carlson et~al.(2005)Carlson, Anderson, Mehlman, and Johnson]{Carlson2005}
R.~W. Carlson, M.~S. Anderson, R.~Mehlman, and R.~E. Johnson.
\newblock {Distribution of hydrate on Europa: Further evidence for sulfuric acid hydrate}.
\newblock \emph{Icarus}, 177\penalty0 (2):\penalty0 461--471, 10 2005.
\newblock ISSN 0019-1035.
\newblock \doi{10.1016/J.ICARUS.2005.03.026}.

\bibitem[Carlson et~al.(2007)Carlson, Kargel, Dout{\'{e}}, Soderblom, and Dalton]{Carlson2007}
R.~W. Carlson, J.~S. Kargel, S.~Dout{\'{e}}, L.~A. Soderblom, and J.~B. Dalton.
\newblock {Io's surface composition}.
\newblock In R.~M.~C. Lopes and J.~R. Spencer, editors, \emph{Io After Galileo: A New View of Jupiter's Volcanic Moon}, page 193. 2007.
\newblock \doi{10.1007/978-3-540-48841-5{\_}9}.

\bibitem[Carlson et~al.(2009)Carlson, Calvin, Dalton, Hansen, Hudson, Johnson, McCord, and Moore]{Carlson2009}
R.~W. Carlson, W.~M. Calvin, J.~B. Dalton, G.~B. Hansen, R.~L. Hudson, R.~E. Johnson, T.~B. McCord, and M.~H. Moore.
\newblock {Europa's Surface Composition}.
\newblock In R.~T. Pappalardo, W.~B. McKinnon, and K.~Khurana, editors, \emph{Europa}, pages 283--327. University of Arizona Press, 2009.
\newblock URL \url{https://muse.jhu.edu/book/57223}.

\bibitem[Cartwright et~al.(2025)Cartwright, Hibbitts, Holler, Raut, Nordheim, Neveu, Protopapa, Glein, Leonard, Roth, Beddingfield, and Villanueva]{Cartwrigth2025JWST}
R.~J. Cartwright, C.~A. Hibbitts, B.~J. Holler, U.~Raut, T.~A. Nordheim, M.~Neveu, S.~Protopapa, C.~R. Glein, E.~J. Leonard, L.~Roth, C.~B. Beddingfield, and G.~L. Villanueva.
\newblock {JWST Reveals Spectral Tracers of Recent Surface Modification on Europa}.
\newblock \emph{The Planetary Science Journal}, 6\penalty0 (5):\penalty0 125, 5 2025.
\newblock ISSN 2632-3338.
\newblock \doi{10.3847/PSJ/ADCAB9}.
\newblock URL \url{https://iopscience.iop.org/article/10.3847/PSJ/adcab9 https://iopscience.iop.org/article/10.3847/PSJ/adcab9/meta}.

\bibitem[Cassidy et~al.(2009)Cassidy, Mendez, Arras, Johnson, and Skrutskie]{Cassidy2009MassiveExoplanets}
T.~A. Cassidy, R.~Mendez, P.~Arras, R.~E. Johnson, and M.~F. Skrutskie.
\newblock {Massive satellites of close-in gas giant exoplanets}.
\newblock \emph{ApJ}, 704\penalty0 (2):\penalty0 1341, 2009.
\newblock ISSN 15384357.
\newblock \doi{10.1088/0004-637x/704/2/1341}.

\bibitem[Cerubini(2020)]{cerubini2020vis}
R.~Cerubini.
\newblock Vis reflectance spectra collected during electron irradiation experiments of salty ice particles (spherical, 67\,\textmu{}m average diameter) prepared by freezing solutions of nacl with different concentrations, 2020.

\bibitem[Cerubini et~al.(2022{\natexlab{a}})Cerubini, Pommerol, Galli, Jost, Wurz, and Thomas]{Cerubini2022b}
R.~Cerubini, A.~Pommerol, A.~Galli, B.~Jost, P.~Wurz, and N.~Thomas.
\newblock {VIS spectroscopy of NaCl – water ice mixtures irradiated with 1 and 5 keV electrons under Europa's conditions: Formation of colour centres and Na colloids}.
\newblock \emph{Icarus}, 379:\penalty0 114977, 6 2022{\natexlab{a}}.
\newblock ISSN 0019-1035.
\newblock \doi{10.1016/J.ICARUS.2022.114977}.

\bibitem[Cerubini et~al.(2022{\natexlab{b}})Cerubini, Pommerol, Yoldi, and Thomas]{Cerubini2022a}
R.~Cerubini, A.~Pommerol, Z.~Yoldi, and N.~Thomas.
\newblock {Near-infrared reflectance spectroscopy of sublimating salty ice analogues. Implications for icy moons}.
\newblock \emph{Planetary and Space Science}, 211:\penalty0 105391, 2 2022{\natexlab{b}}.
\newblock ISSN 0032-0633.
\newblock \doi{10.1016/J.PSS.2021.105391}.

\bibitem[{Charbonneau} et~al.(2002){Charbonneau}, {Brown}, {Noyes}, and {Gilliland}]{Charbonneau2002}
D.~{Charbonneau}, T.~M. {Brown}, R.~W. {Noyes}, and R.~L. {Gilliland}.
\newblock {Detection of an Extrasolar Planet Atmosphere}.
\newblock \emph{\apj}, 568\penalty0 (1):\penalty0 377--384, Mar. 2002.
\newblock \doi{10.1086/338770}.

\bibitem[Charbonneau et~al.(2002)Charbonneau, Brown, Noyes, and Gilliland]{Charbonneau2002DetectionAtmosphere}
D.~Charbonneau, T.~M. Brown, R.~W. Noyes, and R.~L. Gilliland.
\newblock {Detection of an Extrasolar Planet Atmosphere}.
\newblock \emph{ApJ}, 568\penalty0 (1):\penalty0 377, 3 2002.
\newblock ISSN 0004-637X.
\newblock \doi{10.1086/338770}.

\bibitem[Charnoz et~al.(2010)Charnoz, Salmon, and Crida]{Charnoz2010TheRings}
S.~Charnoz, J.~Salmon, and A.~Crida.
\newblock {The recent formation of Saturn's moonlets from viscous spreading of the main rings}.
\newblock \emph{Nature}, 465\penalty0 (7299):\penalty0 752--754, 6 2010.
\newblock ISSN 00280836.
\newblock \doi{10.1038/nature09096}.

\bibitem[Chen and Ma(2020)]{chen2020origin}
Y.~Chen and W.~Ma.
\newblock The origin of biological homochirality along with the origin of life.
\newblock \emph{PLOS Computational Biology}, 16\penalty0 (1):\penalty0 e1007592, 2020.

\bibitem[Choblet et~al.(2017)Choblet, Tobie, Sotin, B{\v{e}}hounkov{\'a}, {\v{C}}adek, Postberg, and Sou{\v{c}}ek]{choblet2017}
G.~Choblet, G.~Tobie, C.~Sotin, M.~B{\v{e}}hounkov{\'a}, O.~{\v{C}}adek, F.~Postberg, and O.~Sou{\v{c}}ek.
\newblock Powering prolonged hydrothermal activity inside enceladus.
\newblock \emph{Nature Astronomy}, 1\penalty0 (12):\penalty0 841--847, 2017.

\bibitem[Choukroun et~al.(2021)Choukroun, Backes, Cable, Fayolle, Hodyss, Murdza, chulson, Badescu, Malaska, Marteau, Molaro, Moreland, Noell, Nordheim, Okamoto, Riccobono, and Zacny]{Choukroun2021}
M.~Choukroun, P.~Backes, M.~L. Cable, E.~C. Fayolle, R.~Hodyss, A.~Murdza, E.~M. chulson, M.~Badescu, M.~J. Malaska, E.~Marteau, J.~L. Molaro, S.~J. Moreland, A.~C. Noell, T.~A. Nordheim, T.~Okamoto, D.~Riccobono, and K.~Zacny.
\newblock {Sampling Plume Deposits on Enceladus’ Surface to Explore Ocean Materials and Search for Traces of Life or Biosignatures}.
\newblock \emph{The Planetary Science Journal}, 2\penalty0 (3):\penalty0 100, 5 2021.
\newblock ISSN 2632-3338.
\newblock \doi{10.3847/PSJ/ABF2C5}.
\newblock URL \url{https://iopscience.iop.org/article/10.3847/PSJ/abf2c5 https://iopscience.iop.org/article/10.3847/PSJ/abf2c5/meta}.

\bibitem[Christensen et~al.(2024)Christensen, Spencer, Mehall, Patel, Anwar, Brick, Bowles, Farkas, Fisher, Gjellum, Holmes, Kubik, Larson, Levy, Madril, Masini, McEwen, Miner, Nickles, O’Donnell, Ortiz, Osterman, Pelham, Rudeen, Saunders, Woodward, Abramov, Hayne, Howett, Mellon, Nimmo, Piqueux, and Rathbun]{christensen_europa_2024}
P.~R. Christensen, J.~R. Spencer, G.~L. Mehall, M.~Patel, S.~Anwar, M.~Brick, H.~Bowles, Z.~Farkas, T.~Fisher, D.~Gjellum, A.~Holmes, I.~Kubik, M.~Larson, A.~Levy, E.~Madril, P.~Masini, T.~McEwen, M.~Miner, N.~Nickles, W.~O’Donnell, C.~Ortiz, D.~Osterman, D.~Pelham, A.~Rudeen, T.~Saunders, R.~Woodward, O.~Abramov, P.~O. Hayne, C.~J.~A. Howett, M.~T. Mellon, F.~Nimmo, S.~Piqueux, and J.~A. Rathbun.
\newblock The {Europa} {Thermal} {Emission} {Imaging} {System} ({E}-{THEMIS}) {Investigation} for the {Europa} {Clipper} {Mission}.
\newblock \emph{Space Science Reviews}, 220\penalty0 (4):\penalty0 38, June 2024.
\newblock ISSN 0038-6308, 1572-9672.
\newblock \doi{10.1007/s11214-024-01074-1}.
\newblock URL \url{https://link.springer.com/10.1007/s11214-024-01074-1}.

\bibitem[Christiaens et~al.(2024)Christiaens, Samland, Henning, Portilla-Revelo, Perotti, Matthews, Absil, Decin, Kamp, Boccaletti, et~al.]{Christiaens2024}
V.~Christiaens, M.~Samland, T.~Henning, B.~Portilla-Revelo, G.~Perotti, E.~Matthews, O.~Absil, L.~Decin, I.~Kamp, A.~Boccaletti, et~al.
\newblock Minds: Jwst/nircam imaging of the protoplanetary disk pds 70.
\newblock \emph{arXiv preprint arXiv:2403.04855}, 2024.

\bibitem[{Cilibrasi} et~al.(2018){Cilibrasi}, {Szul{\'a}gyi}, {Mayer}, {Dra{\.z}kowska}, {Miguel}, and {Inderbitzi}]{Cilibrasi18}
M.~{Cilibrasi}, J.~{Szul{\'a}gyi}, L.~{Mayer}, J.~{Dra{\.z}kowska}, Y.~{Miguel}, and P.~{Inderbitzi}.
\newblock {Satellites form fast \&amp; late: a population synthesis for the Galilean moons}.
\newblock \emph{\mnras}, 480\penalty0 (4):\penalty0 4355--4368, Nov. 2018.
\newblock \doi{10.1093/mnras/sty2163}.

\bibitem[{Cilibrasi} et~al.(2021){Cilibrasi}, {Szul{\'a}gyi}, {Grimm}, and {Mayer}]{Cilibrasi21}
M.~{Cilibrasi}, J.~{Szul{\'a}gyi}, S.~L. {Grimm}, and L.~{Mayer}.
\newblock {An N-body population synthesis framework for the formation of moons around Jupiter-like planets}.
\newblock \emph{\mnras}, 504\penalty0 (4):\penalty0 5455--5474, July 2021.
\newblock \doi{10.1093/mnras/stab1179}.

\bibitem[{Cilibrasi} et~al.(2023){Cilibrasi}, {Flock}, and {Szul{\'a}gyi}]{Cilibrasi23}
M.~{Cilibrasi}, M.~{Flock}, and J.~{Szul{\'a}gyi}.
\newblock {Meridional circulation driven by planetary spiral wakes in radiative and magnetized protoplanetary discs}.
\newblock \emph{\mnras}, 523\penalty0 (2):\penalty0 2039--2058, Aug. 2023.
\newblock \doi{10.1093/mnras/stad1477}.

\bibitem[Clarke et~al.(1994)Clarke, Ajello, Luhmann, Schneider, and Kanik]{clarke1994}
J.~T. Clarke, J.~Ajello, J.~Luhmann, N.~Schneider, and I.~Kanik.
\newblock Hubble space telescope uv spectral observations of io passing into eclipse.
\newblock \emph{Journal of Geophysical Research: Planets}, 99\penalty0 (E4):\penalty0 8387--8402, 1994.

\bibitem[Collins and Nimmo(2009)]{Collins2009}
G.~Collins and F.~Nimmo.
\newblock {Chaotic Terrain on Europa}.
\newblock \emph{Europa}, pages 259--281, 2009.
\newblock \doi{10.2307/j.ctt1xp3wdw.17}.

\bibitem[Conrad et~al.(2024)Conrad, Pedichini, Li~Causi, Antoniucci, de~Pater, Davies, de~Kleer, Piazzesi, Testa, Vaccari, Vicinanza, Power, Ertel, Shields, Ragland, Giorgi, Jefferies, Hope, Perry, Williams, and Nelson]{Conrad2024}
A.~Conrad, F.~Pedichini, G.~Li~Causi, S.~Antoniucci, I.~de~Pater, A.~G. Davies, K.~de~Kleer, R.~Piazzesi, V.~Testa, P.~Vaccari, M.~Vicinanza, J.~Power, S.~Ertel, J.~C. Shields, S.~Ragland, F.~Giorgi, S.~M. Jefferies, D.~Hope, J.~Perry, D.~A. Williams, and D.~M. Nelson.
\newblock {Observation of Io's Resurfacing via Plume Deposition Using Ground-Based Adaptive Optics at Visible Wavelengths With LBT SHARK-VIS}.
\newblock \emph{Geophysical Research Letters}, 51, 6 2024.
\newblock ISSN 19448007.
\newblock \doi{10.1029/2024GL108609}.

\bibitem[{Cooper} et~al.(2001){Cooper}, {Johnson}, {Mauk}, {Garrett}, and {Gehrels}]{2001Icar..149..133C}
J.~F. {Cooper}, R.~E. {Johnson}, B.~H. {Mauk}, H.~B. {Garrett}, and N.~{Gehrels}.
\newblock {Energetic Ion and Electron Irradiation of the Icy Galilean Satellites}.
\newblock \emph{\icarus}, 149\penalty0 (1):\penalty0 133--159, Jan. 2001.
\newblock \doi{10.1006/icar.2000.6498}.

\bibitem[Cordiner et~al.(2024)Cordiner, Thelen, Lai, Tseng, Nixon, Kuan, Villanueva, Paganini, Charnley, and Retherford]{Cordiner2024}
M.~A. Cordiner, A.~E. Thelen, I.~L. Lai, W.~L. Tseng, C.~A. Nixon, Y.~J. Kuan, G.~L. Villanueva, L.~Paganini, S.~B. Charnley, and K.~D. Retherford.
\newblock {ALMA Spectroscopy of Europa: A Search for Active Plumes}.
\newblock \emph{arXiv}, 4 2024.
\newblock URL \url{https://arxiv.org/abs/2404.05525v1}.

\bibitem[Costello et~al.(2021)Costello, Phillips, Lucey, and Ghent]{Costello2021}
E.~S. Costello, C.~B. Phillips, P.~G. Lucey, and R.~R. Ghent.
\newblock {Impact gardening on Europa and repercussions for possible biosignatures}.
\newblock \emph{Nature Astronomy}, 2021.
\newblock ISSN 2397-3366.
\newblock \doi{10.1038/s41550-021-01393-1}.
\newblock URL \url{http://dx.doi.org/10.1038/s41550-021-01393-1}.

\bibitem[Craft et~al.(2016)Craft, Patterson, Lowell, and Germanovich]{Craft2016}
K.~L. Craft, G.~W. Patterson, R.~P. Lowell, and L.~Germanovich.
\newblock {Fracturing and flow: Investigations on the formation of shallow water sills on Europa}.
\newblock \emph{Icarus}, 274:\penalty0 297--313, 8 2016.
\newblock ISSN 0019-1035.
\newblock \doi{10.1016/J.ICARUS.2016.01.023}.

\bibitem[Crida and Charnoz(2012)]{Crida2012}
A.~Crida and S.~Charnoz.
\newblock Formation of regular satellites from ancient massive rings in the solar system.
\newblock \emph{Science}, 338\penalty0 (6111):\penalty0 1196--1199, 2012.

\bibitem[Cruikshank et~al.(1993)Cruikshank, Roush, Owen, Geballe, De~Bergh, Schmitt, Brown, and Bartholomew]{Cruikshank1993}
D.~P. Cruikshank, T.~L. Roush, T.~C. Owen, T.~R. Geballe, C.~De~Bergh, B.~Schmitt, R.~H. Brown, and M.~J. Bartholomew.
\newblock {Ices on the Surface of Triton}.
\newblock \emph{Science}, 261\penalty0 (5122):\penalty0 742--745, 8 1993.
\newblock ISSN 00368075.
\newblock \doi{10.1126/SCIENCE.261.5122.742}.
\newblock URL \url{https://www.science.org/doi/10.1126/science.261.5122.742}.

\bibitem[Cruikshank et~al.(2000)Cruikshank, Schmitt, Roush, Owen, Quirico, Geballe, {de Bergh}, Bartholomew, {Dalle Ore}, Douté, and Meier]{Cruikshank_2000}
D.~P. Cruikshank, B.~Schmitt, T.~L. Roush, T.~C. Owen, E.~Quirico, T.~R. Geballe, C.~{de Bergh}, M.~J. Bartholomew, C.~M. {Dalle Ore}, S.~Douté, and R.~Meier.
\newblock Water ice on triton.
\newblock \emph{Icarus}, 147\penalty0 (1):\penalty0 309--316, 2000.
\newblock ISSN 0019-1035.
\newblock \doi{https://doi.org/10.1006/icar.2000.6451}.
\newblock URL \url{https://www.sciencedirect.com/science/article/pii/S0019103500964510}.

\bibitem[Currie et~al.(2022)Currie, Lawson, Schneider, Lyra, Wisniewski, Grady, Guyon, Tamura, Kotani, Kawahara, Brandt, Uyama, Muto, Dong, Kudo, Hashimoto, Fukagawa, Wagner, Lozi, Chilcote, Tobin, Groff, Ward-Duong, Januszewski, Norris, Tuthill, van~der Marel, Sitko, Deo, Vievard, Jovanovic, Martinache, and Skaf]{Currie2022}
T.~Currie, K.~Lawson, G.~Schneider, W.~Lyra, J.~Wisniewski, C.~Grady, O.~Guyon, M.~Tamura, T.~Kotani, H.~Kawahara, T.~Brandt, T.~Uyama, T.~Muto, R.~Dong, T.~Kudo, J.~Hashimoto, M.~Fukagawa, K.~Wagner, J.~Lozi, J.~Chilcote, T.~Tobin, T.~Groff, K.~Ward-Duong, W.~Januszewski, B.~Norris, P.~Tuthill, N.~van~der Marel, M.~Sitko, V.~Deo, S.~Vievard, N.~Jovanovic, F.~Martinache, and N.~Skaf.
\newblock {Images of embedded Jovian planet formation at a wide separation around AB Aurigae}.
\newblock \emph{Nature Astronomy}, 6\penalty0 (6):\penalty0 751--759, 6 2022.
\newblock ISSN 23973366.
\newblock \doi{10.1038/S41550-022-01634-X;SUBJMETA=33,34,4122,639,766,862;KWRD=ASTROPHYSICAL+DISKS,EXOPLANETS}.
\newblock URL \url{https://www.nature.com/articles/s41550-022-01634-x}.

\bibitem[Daubar et~al.(2024)Daubar, Hayes, Collins, Craft, Rathbun, Spencer, Wyrick, Bland, Davies, Ernst, Howell, Leonard, McEwen, Moore, Phillips, Prockter, Quick, Scully, Soderblom, Brooks, Cable, Cameron, Chan, Chivers, Choukroun, Cochrane, Diniega, Dombard, Elder, Gerekos, Glein, Greathouse, Grima, Gudipati, Hand, Hansen, Hayne, Hedman, Hughson, Jia, Lawrence, Meyer, Miller, Parekh, Patterson, Persaud, Piqueux, Retherford, Scanlan, Schenk, Schmidt, Schroeder, Steinbr{\"{u}}gge, Stern, Tobie, Withers, Young, Buratti, Korth, Senske, and Pappalardo]{Daubar2024}
I.~J. Daubar, A.~G. Hayes, G.~C. Collins, K.~L. Craft, J.~A. Rathbun, J.~R. Spencer, D.~Y. Wyrick, M.~T. Bland, A.~G. Davies, C.~M. Ernst, S.~M. Howell, E.~J. Leonard, A.~S. McEwen, J.~M. Moore, C.~B. Phillips, L.~M. Prockter, L.~C. Quick, J.~E. Scully, J.~M. Soderblom, S.~M. Brooks, M.~Cable, M.~E. Cameron, K.~Chan, C.~J. Chivers, M.~Choukroun, C.~J. Cochrane, S.~Diniega, A.~J. Dombard, C.~M. Elder, C.~Gerekos, C.~Glein, T.~K. Greathouse, C.~Grima, M.~S. Gudipati, K.~P. Hand, C.~Hansen, P.~Hayne, M.~Hedman, K.~Hughson, X.~Jia, J.~Lawrence, H.~M. Meyer, K.~Miller, R.~Parekh, G.~W. Patterson, D.~M. Persaud, S.~Piqueux, K.~D. Retherford, K.~M. Scanlan, P.~Schenk, B.~Schmidt, D.~Schroeder, G.~Steinbr{\"{u}}gge, A.~Stern, G.~Tobie, P.~Withers, D.~A. Young, B.~Buratti, H.~Korth, D.~Senske, and R.~Pappalardo.
\newblock {Planned Geological Investigations of the Europa Clipper Mission}.
\newblock \emph{SSR}, 220\penalty0 (1):\penalty0 1--55, 2 2024.
\newblock ISSN 15729672.
\newblock \doi{10.1007/S11214-023-01036-Z/METRICS}.
\newblock URL \url{https://link.springer.com/article/10.1007/s11214-023-01036-z}.

\bibitem[Davies et~al.(2012)Davies, Veeder, Matson, and Johnson]{Daviesb2012}
A.~Davies, G.~Veeder, D.~Matson, and T.~Johnson.
\newblock Io: Charting thermal emission variability with the galileo nims io thermal emission database (nited): Loki patera.
\newblock \emph{Geophysical research letters}, 39\penalty0 (1), 2012.

\bibitem[Davies(2003)]{Davies2003}
A.~G. Davies.
\newblock {Temperature, age and crust thickness distributions of Loki Patera on Io from Galileo NIMS data: Implications for resurfacing mechanism}.
\newblock \emph{Geophysical Research Letters}, 30\penalty0 (21):\penalty0 2133, 11 2003.
\newblock ISSN 1944-8007.
\newblock \doi{10.1029/2003GL018371}.
\newblock URL \url{/doi/pdf/10.1029/2003GL018371 https://onlinelibrary.wiley.com/doi/abs/10.1029/2003GL018371 https://agupubs.onlinelibrary.wiley.com/doi/10.1029/2003GL018371}.

\bibitem[Davies and Veeder(2023)]{Davies2023}
A.~G. Davies and G.~J. Veeder.
\newblock Near infrared spectral radiance at multiple wavelengths from io's volcanoes 1: The low spatial resolution night-time galileo nims data set.
\newblock \emph{Journal of Geophysical Research: Planets}, 128\penalty0 (8):\penalty0 e2023JE007839, 2023.

\bibitem[Davies and Vorburger(2022)]{davies2022}
A.~G. Davies and A.~H. Vorburger.
\newblock Io’s volcanic activity and atmosphere.
\newblock \emph{Elements: An International Magazine of Mineralogy, Geochemistry, and Petrology}, 18\penalty0 (6):\penalty0 379--384, 2022.

\bibitem[Davies et~al.(2001)Davies, Keszthelyi, Williams, Phillips, McEwen, Lopes, Smythe, Kamp, Soderblom, and Carlson]{davies2001}
A.~G. Davies, L.~P. Keszthelyi, D.~A. Williams, C.~B. Phillips, A.~S. McEwen, R.~M. Lopes, W.~D. Smythe, L.~W. Kamp, L.~A. Soderblom, and R.~W. Carlson.
\newblock Thermal signature, eruption style, and eruption evolution at pele and pillan on io.
\newblock \emph{Journal of Geophysical Research: Planets}, 106\penalty0 (E12):\penalty0 33079--33103, 2001.

\bibitem[Davies et~al.(2006)Davies, Wilson, Matson, Leone, Keszthelyi, and Jaeger]{davies2006}
A.~G. Davies, L.~Wilson, D.~Matson, G.~Leone, L.~Keszthelyi, and W.~Jaeger.
\newblock The heartbeat of the volcano: The discovery of episodic activity at prometheus on io.
\newblock \emph{Icarus}, 184\penalty0 (2):\penalty0 460--477, 2006.

\bibitem[Davies et~al.(2024)Davies, Perry, Williams, Veeder, and Nelson]{Davies2024}
A.~G. Davies, J.~E. Perry, D.~A. Williams, G.~J. Veeder, and D.~M. Nelson.
\newblock {New Global Map of Io’s Volcanic Thermal Emission and Discovery of Hemispherical Dichotomies}.
\newblock \emph{The Planetary Science Journal}, 5\penalty0 (5):\penalty0 121, 5 2024.
\newblock ISSN 2632-3338.
\newblock \doi{10.3847/PSJ/AD4346}.
\newblock URL \url{https://iopscience.iop.org/article/10.3847/PSJ/ad4346 https://iopscience.iop.org/article/10.3847/PSJ/ad4346/meta}.

\bibitem[Davis et~al.(2021)Davis, Meier, Cooper, and Loeffler]{Davis2021contribution}
M.~Davis, R.~Meier, J.~Cooper, and M.~Loeffler.
\newblock The contribution of electrons to the sputter-produced o2 exosphere on europa.
\newblock \emph{The Astrophysical Journal Letters}, 908\penalty0 (2):\penalty0 L53, 2021.

\bibitem[De~Kleer et~al.(2017)De~Kleer, Skrutskie, Leisenring, Davies, Conrad, De~Pater, Resnick, Bailey, Defr{\`{e}}re, Hinz, Skemer, Spalding, Vaz, Veillet, and Woodward]{deKleer2017}
K.~De~Kleer, M.~Skrutskie, J.~Leisenring, A.~G. Davies, A.~Conrad, I.~De~Pater, A.~Resnick, V.~Bailey, D.~Defr{\`{e}}re, P.~Hinz, A.~Skemer, E.~Spalding, A.~Vaz, C.~Veillet, and C.~E. Woodward.
\newblock {Multi-phase volcanic resurfacing at Loki Patera on Io}.
\newblock \emph{Nature}, 545\penalty0 (7653):\penalty0 199--202, 5 2017.
\newblock ISSN 14764687.
\newblock \doi{10.1038/NATURE22339;TECHMETA}.
\newblock URL \url{https://www.nature.com/articles/nature22339}.

\bibitem[de~Kleer et~al.(2019{\natexlab{a}})de~Kleer, de~Pater, Molter, Banks, Davies, Alvarez, Campbell, Aycock, Pelletier, Stickel, et~al.]{DeKleer2019}
K.~de~Kleer, I.~de~Pater, E.~M. Molter, E.~Banks, A.~G. Davies, C.~Alvarez, R.~Campbell, J.~Aycock, J.~Pelletier, T.~Stickel, et~al.
\newblock Io’s volcanic activity from time domain adaptive optics observations: 2013--2018.
\newblock \emph{The Astronomical Journal}, 158\penalty0 (1):\penalty0 29, 2019{\natexlab{a}}.

\bibitem[de~Kleer et~al.(2019{\natexlab{b}})de~Kleer, McEwen, Park, Bierson, Davies, DellaGustina, Ermakov, Fuller, Hamilton, Harris, et~al.]{de2019}
K.~de~Kleer, A.~S. McEwen, R.~S. Park, C.~J. Bierson, A.~G. Davies, D.~N. DellaGustina, A.~I. Ermakov, J.~Fuller, C.~W. Hamilton, C.~Harris, et~al.
\newblock Tidal heating: Lessons from io and the jovian system-final report.
\newblock \emph{Keck Institute for Space Studies}, 2019{\natexlab{b}}.

\bibitem[de~Kleer et~al.(2023)de~Kleer, Milby, Schmidt, Camarca, and Brown]{deKleer2023TheCallisto}
K.~de~Kleer, Z.~Milby, C.~Schmidt, M.~Camarca, and M.~E. Brown.
\newblock {The Optical Aurorae of Europa, Ganymede, and Callisto}.
\newblock \emph{PSJ}, 4\penalty0 (2):\penalty0 37, 2 2023.
\newblock ISSN 26323338.
\newblock \doi{10.3847/psj/acb53c}.

\bibitem[de~Kleer et~al.(2024)de~Kleer, Hughes, Nimmo, Eiler, Hofmann, Luszcz-Cook, and Mandt]{dekleer2024}
K.~de~Kleer, E.~C. Hughes, F.~Nimmo, J.~Eiler, A.~E. Hofmann, S.~Luszcz-Cook, and K.~Mandt.
\newblock Isotopic evidence of long-lived volcanism on io.
\newblock \emph{Science}, 384\penalty0 (6696):\penalty0 682--687, 2024.

\bibitem[de~Pater et~al.(2020)de~Pater, Luszcz-Cook, Rojo, Redwing, De~Kleer, and Moullet]{depater2020}
I.~de~Pater, S.~Luszcz-Cook, P.~Rojo, E.~Redwing, K.~De~Kleer, and A.~Moullet.
\newblock Alma observations of io going into and coming out of eclipse.
\newblock \emph{The Planetary Science Journal}, 1\penalty0 (3):\penalty0 60, 2020.

\bibitem[De~Pater et~al.(2021)De~Pater, Keane, De~Kleer, and Davies]{DePater2021AIo}
I.~De~Pater, J.~T. Keane, K.~De~Kleer, and A.~G. Davies.
\newblock {A 2020 observational perspective of io}.
\newblock \emph{Annual Review of Earth and Planetary Sciences}, 49\penalty0 (Volume 49, 2021):\penalty0 643--678, 5 2021.
\newblock ISSN 00846597.
\newblock \doi{10.1146/ANNUREV-EARTH-082420-095244/CITE/REFWORKS}.
\newblock URL \url{https://www.annualreviews.org/content/journals/10.1146/annurev-earth-082420-095244}.

\bibitem[de~Pater et~al.(2023)de~Pater, Goldstein, and Lellouch]{depater2023}
I.~de~Pater, D.~Goldstein, and E.~Lellouch.
\newblock \emph{The plumes and atmosphere of Io}, pages 233--290.
\newblock Springer, 2023.

\bibitem[De~Sanctis et~al.(2020)De~Sanctis, Ammannito, Raponi, Frigeri, Ferrari, Carrozzo, Ciarniello, Formisano, Rousseau, Tosi, Zambon, Raymond, and Russell]{DeSanctis2020}
M.~C. De~Sanctis, E.~Ammannito, A.~Raponi, A.~Frigeri, M.~Ferrari, F.~G. Carrozzo, M.~Ciarniello, M.~Formisano, B.~Rousseau, F.~Tosi, F.~Zambon, C.~A. Raymond, and C.~T. Russell.
\newblock {Fresh emplacement of hydrated sodium chloride on Ceres from ascending salty fluids}.
\newblock \emph{Nature Astronomy}, 4\penalty0 (8):\penalty0 786--793, 8 2020.
\newblock ISSN 2397-3366.
\newblock \doi{10.1038/s41550-020-1138-8}.
\newblock URL \url{https://www.nature.com/articles/s41550-020-1138-8}.

\bibitem[Delitsky et~al.(1989)Delitsky, Eviatar, and Richardson]{Delitsky1989}
M.~L. Delitsky, A.~Eviatar, and J.~D. Richardson.
\newblock {A predicted Triton plasma torus in Neptune's magnetosphere}.
\newblock \emph{GeoRL}, 16\penalty0 (2):\penalty0 215--218, 1989.
\newblock ISSN 19448007.
\newblock \doi{10.1029/GL016I002P00215}.
\newblock URL \url{https://scixplorer.org/abs/1989GeoRL..16..215D/abstract}.

\bibitem[Denny et~al.(2024)Denny, Hedman, Bockel{\'{e}}e-Morvan, Filacchione, and Capaccioni]{Denny2024ConstrainingSpectrometer}
K.~E. Denny, M.~M. Hedman, D.~Bockel{\'{e}}e-Morvan, G.~Filacchione, and F.~Capaccioni.
\newblock {Constraining Time Variations in Enceladus's Water-vapor Plume with Near-infrared Spectra from Cassini's Visual and Infrared Mapping Spectrometer}.
\newblock 2024.
\newblock \doi{10.3847/PSJ/ad4c69}.
\newblock URL \url{https://doi.org/10.3847/PSJ/ad4c69}.

\bibitem[{Dobos} et~al.(2021){Dobos}, {Charnoz}, {P{\'a}l}, {Roque-Bernard}, and {Szab{\'o}}]{2021PASP..133i4401D}
V.~{Dobos}, S.~{Charnoz}, A.~{P{\'a}l}, A.~{Roque-Bernard}, and G.~M. {Szab{\'o}}.
\newblock {Survival of Exomoons Around Exoplanets}.
\newblock \emph{\pasp}, 133\penalty0 (1027):\penalty0 094401, Sept. 2021.

\bibitem[Dra{\.z}kowska and Szul{\'a}gyi(2018)]{Drazkowska2018}
J.~Dra{\.z}kowska and J.~Szul{\'a}gyi.
\newblock Dust evolution and satellitesimal formation in circumplanetary disks.
\newblock \emph{The Astrophysical Journal}, 866\penalty0 (2):\penalty0 142, 2018.

\bibitem[Duembgen and Haslebacher(2025)]{Duembgen2025NonparametricData}
L.~Duembgen and C.~Haslebacher.
\newblock {Nonparametric Smoothing of Directional and Axial Data}.
\newblock \emph{Statistica Neerlandica}, (to appear), 2025.
\newblock URL \url{https://arxiv.org/abs/2501.17463v1}.

\bibitem[Elliot et~al.(1998)Elliot, Hammel, Wasserman, Franz, McDonald, Person, Olkin, Dunham, Spencer, Stansberry, Bule, Pasachoff, Babcock, and McConnochle]{Elliot1998}
J.~L. Elliot, H.~B. Hammel, L.~H. Wasserman, O.~G. Franz, S.~W. McDonald, M.~J. Person, C.~B. Olkin, E.~W. Dunham, J.~R. Spencer, J.~A. Stansberry, M.~W. Bule, J.~M. Pasachoff, B.~A. Babcock, and T.~H. McConnochle.
\newblock {Global warming on Triton}.
\newblock \emph{Nature}, 393\penalty0 (6687):\penalty0 765--767, 6 1998.
\newblock ISSN 00280836.
\newblock \doi{10.1038/31651;KWRD}.
\newblock URL \url{https://www.nature.com/articles/31651}.

\bibitem[Engel and Nagy(1982)]{engel1982distribution}
M.~H. Engel and B.~Nagy.
\newblock Distribution and enantiomeric composition of amino acids in the murchison meteorite.
\newblock \emph{Nature}, 296\penalty0 (5860):\penalty0 837--840, 1982.

\bibitem[{Eriksson} et~al.(2023){Eriksson}, {Mol Lous}, {Shibata}, and {Helled}]{Eriksson2023}
L.~E.~J. {Eriksson}, M.~A.~S. {Mol Lous}, S.~{Shibata}, and R.~{Helled}.
\newblock {Can Uranus and Neptune form concurrently via pebble, gas, and planetesimal accretion?}
\newblock \emph{\mnras}, 526\penalty0 (4):\penalty0 4860--4876, Dec. 2023.
\newblock \doi{10.1093/mnras/stad3007}.

\bibitem[Fagents(2003)]{Fagents2003}
S.~A. Fagents.
\newblock {Considerations for effusive cryovolcanism on Europa: The post-Galileo perspective}.
\newblock \emph{Journal of Geophysical Research E: Planets}, 108\penalty0 (12), 12 2003.
\newblock ISSN 01480227.
\newblock \doi{10.1029/2003JE002128}.

\bibitem[Fagents et~al.(2000)Fagents, Greeley, Sullivan, Pappalardo, and Prockter]{Fagents2000}
S.~A. Fagents, R.~Greeley, R.~J. Sullivan, R.~T. Pappalardo, and L.~M. Prockter.
\newblock {Cryomagmatic mechanisms for the formation of Rhadamanthys Linea, triple band margins, and other low albedo features on Europa}.
\newblock \emph{Icarus}, 144\penalty0 (1):\penalty0 54--88, 3 2000.
\newblock ISSN 00191035.
\newblock \doi{10.1006/icar.1999.6254}.

\bibitem[Fam{\'a} et~al.(2008)Fam{\'a}, Shi, and Baragiola]{fama2008sputtering}
M.~Fam{\'a}, J.~Shi, and R.~Baragiola.
\newblock Sputtering of ice by low-energy ions.
\newblock \emph{Surface Science}, 602\penalty0 (1):\penalty0 156--161, 2008.

\bibitem[Fausch et~al.(2023)Fausch, Schertenleib, and Wurz]{Fausch2023}
R.~Fausch, J.~Schertenleib, and P.~Wurz.
\newblock Reliably analyzing the chemical composition of plumes during flybys at velocities exceeding 5 km/s.
\newblock In \emph{2023 IEEE Aerospace Conference}, pages 1--8, 2023.
\newblock \doi{10.1109/AERO55745.2023.10115795}.

\bibitem[Feaga et~al.(2002)Feaga, McGrath, and Feldman]{feaga2002}
L.~M. Feaga, M.~A. McGrath, and P.~D. Feldman.
\newblock The abundance of atomic sulfur in the atmosphere of io.
\newblock \emph{The Astrophysical Journal}, 570\penalty0 (1):\penalty0 439, 2002.

\bibitem[Feaga et~al.(2004)Feaga, McGrath, Feldman, and Strobel]{feaga2004}
L.~M. Feaga, M.~A. McGrath, P.~D. Feldman, and D.~F. Strobel.
\newblock { Detection of Atomic Chlorine in Io’s Atmosphere with the Hubble Space Telescope GHRS }.
\newblock \emph{The Astrophysical Journal}, 610\penalty0 (2):\penalty0 1191--1198, 8 2004.
\newblock ISSN 0004-637X.
\newblock \doi{10.1086/421862/FULLTEXT/}.
\newblock URL \url{https://iopscience.iop.org/article/10.1086/421862 https://iopscience.iop.org/article/10.1086/421862/meta}.

\bibitem[Feaga et~al.(2009)Feaga, McGrath, and Feldman]{feaga2009}
L.~M. Feaga, M.~McGrath, and P.~D. Feldman.
\newblock Io's dayside so2 atmosphere.
\newblock \emph{Icarus}, 201\penalty0 (2):\penalty0 570--584, 2009.

\bibitem[Feldman et~al.(2000)Feldman, Strobel, Moos, Retherford, Wolven, McGrath, Roesler, Woodward, Oliversen, and Ballester]{feldman2000}
P.~D. Feldman, D.~F. Strobel, H.~W. Moos, K.~D. Retherford, B.~C. Wolven, M.~A. McGrath, F.~L. Roesler, R.~C. Woodward, R.~J. Oliversen, and G.~E. Ballester.
\newblock Lyman-$\alpha$ imaging of the so2 distribution on io.
\newblock \emph{Geophysical research letters}, 27\penalty0 (12):\penalty0 1787--1790, 2000.

\bibitem[Figueredo and Greeley(2004)]{Figueredo2004}
P.~H. Figueredo and R.~Greeley.
\newblock {Resurfacing history of Europa from pole-to-pole geological mapping}.
\newblock \emph{Icarus}, 167\penalty0 (2):\penalty0 287--312, 2004.
\newblock \doi{10.1016/J.ICARUS.2003.09.016}.

\bibitem[Fletcher et~al.(2023)Fletcher, Cavali{\'{e}}, Grassi, Hueso, Lara, Kaspi, Galanti, Greathouse, Molyneux, Galand, Vallat, Witasse, Lorente, Hartogh, Poulet, Langevin, Palumbo, Gladstone, Retherford, Dougherty, Wahlund, Barabash, Iess, Bruzzone, Hussmann, Gurvits, Santolik, Kolmasova, Fischer, M{\"{u}}ller-Wodarg, Piccioni, Fouchet, G{\'{e}}rard, S{\'{a}}nchez-Lavega, Irwin, Grodent, Altieri, Mura, Drossart, Kammer, Giles, Cazaux, Jones, Smirnova, Lellouch, Medvedev, Moreno, Rezac, Coustenis, and Costa]{Fletcher2023}
L.~N. Fletcher, T.~Cavali{\'{e}}, D.~Grassi, R.~Hueso, L.~M. Lara, Y.~Kaspi, E.~Galanti, T.~K. Greathouse, P.~M. Molyneux, M.~Galand, C.~Vallat, O.~Witasse, R.~Lorente, P.~Hartogh, F.~Poulet, Y.~Langevin, P.~Palumbo, G.~R. Gladstone, K.~D. Retherford, M.~K. Dougherty, J.~E. Wahlund, S.~Barabash, L.~Iess, L.~Bruzzone, H.~Hussmann, L.~I. Gurvits, O.~Santolik, I.~Kolmasova, G.~Fischer, I.~M{\"{u}}ller-Wodarg, G.~Piccioni, T.~Fouchet, J.~C. G{\'{e}}rard, A.~S{\'{a}}nchez-Lavega, P.~G. Irwin, D.~Grodent, F.~Altieri, A.~Mura, P.~Drossart, J.~Kammer, R.~Giles, S.~Cazaux, G.~Jones, M.~Smirnova, E.~Lellouch, A.~S. Medvedev, R.~Moreno, L.~Rezac, A.~Coustenis, and M.~Costa.
\newblock {Jupiter Science Enabled by ESA’s Jupiter Icy Moons Explorer}.
\newblock \emph{SSR}, 219\penalty0 (7):\penalty0 1--75, 9 2023.
\newblock ISSN 1572-9672.
\newblock \doi{10.1007/S11214-023-00996-6}.
\newblock URL \url{https://link.springer.com/article/10.1007/s11214-023-00996-6}.

\bibitem[Fujita et~al.(2013)Fujita, Ohtsuki, Tanigawa, and Suetsugu]{Fujita2013}
T.~Fujita, K.~Ohtsuki, T.~Tanigawa, and R.~Suetsugu.
\newblock Capture of planetesimals by gas drag from circumplanetary disks.
\newblock \emph{The Astronomical Journal}, 146\penalty0 (6):\penalty0 140, 2013.
\newblock URL \url{http://stacks.iop.org/1538-3881/146/i=6/a=140}.

\bibitem[Furukawa et~al.(2015)Furukawa, Nakazawa, Sekine, Kobayashi, and Kakegawa]{Furukawa2015}
Y.~Furukawa, H.~Nakazawa, T.~Sekine, T.~Kobayashi, and T.~Kakegawa.
\newblock Nucleobase and amino acid formation through impacts of meteorites on the early ocean.
\newblock \emph{Earth Planet. Sci. Lett.}, 429:\penalty0 216–--222, 2015.
\newblock \doi{10.1016/j.epsl.2015.07.049}.

\bibitem[Gaeman et~al.(2012)Gaeman, Hier-Majumder, and Roberts]{Gaeman2012SustainabilityInterior}
J.~Gaeman, S.~Hier-Majumder, and J.~H. Roberts.
\newblock {Sustainability of a subsurface ocean within Triton’s interior}.
\newblock \emph{Icarus}, 220\penalty0 (2):\penalty0 339--347, 8 2012.
\newblock ISSN 0019-1035.
\newblock \doi{10.1016/J.ICARUS.2012.05.006}.

\bibitem[Galli et~al.(2017)Galli, Vorburger, Wurz, and Tulej]{Galli2017SputteringElectrons}
A.~Galli, A.~Vorburger, P.~Wurz, and M.~Tulej.
\newblock {Sputtering of water ice films: A re-assessment with singly and doubly charged oxygen and argon ions, molecular oxygen, and electrons}.
\newblock \emph{Icarus}, 291:\penalty0 36--45, 7 2017.
\newblock ISSN 0019-1035.
\newblock \doi{10.1016/J.ICARUS.2017.03.018}.

\bibitem[Galli et~al.(2018{\natexlab{a}})Galli, Vorburger, Wurz, Cerubini, and Tulej]{galli2018first}
A.~Galli, A.~Vorburger, P.~Wurz, R.~Cerubini, and M.~Tulej.
\newblock First experimental data of sulphur ions sputtering water ice.
\newblock \emph{Icarus}, 312:\penalty0 1--6, 2018{\natexlab{a}}.

\bibitem[Galli et~al.(2018{\natexlab{b}})Galli, Vorburger, Wurz, Pommerol, Cerubini, Jost, Poch, Tulej, and Thomas]{Galli2018}
A.~Galli, A.~Vorburger, P.~Wurz, A.~Pommerol, R.~Cerubini, B.~Jost, O.~Poch, M.~Tulej, and N.~Thomas.
\newblock 0.2 to 10 kev electrons interacting with water ice: Radiolysis, sputtering, and sublimation.
\newblock \emph{Planetary and Space Science}, 155:\penalty0 91--98, 2018{\natexlab{b}}.
\newblock ISSN 0032-0633.
\newblock \doi{https://doi.org/10.1016/j.pss.2017.11.016}.
\newblock URL \url{https://www.sciencedirect.com/science/article/pii/S0032063317301290}.
\newblock Surfaces, atmospheres and magnetospheres of the outer planets, their satellites and ring systems: Part XII.

\bibitem[Gebek and Oza(2020)]{Gebek2020AlkalineSpectra}
A.~Gebek and A.~V. Oza.
\newblock {Alkaline exospheres of exoplanet systems: Evaporative transmission spectra}.
\newblock \emph{MNRAS}, 497\penalty0 (4):\penalty0 5271, 10 2020.
\newblock ISSN 13652966.
\newblock \doi{10.1093/mnras/staa2193}.

\bibitem[Geissler et~al.(1999)Geissler, McEwen, Ip, Belton, Johnson, Smyth, and Ingersoll]{Geissler1999Io}
P.~E. Geissler, A.~S. McEwen, W.~Ip, M.~J. Belton, T.~V. Johnson, W.~H. Smyth, and A.~P. Ingersoll.
\newblock {Galileo imaging of atmospheric emissions from Io}.
\newblock \emph{Science}, 285\penalty0 (5429):\penalty0 870--874, 8 1999.
\newblock ISSN 00368075.
\newblock \doi{10.1126/SCIENCE.285.5429.870/SUPPL{\_}FILE/1039991.XHTML}.
\newblock URL \url{https://www.science.org/doi/10.1126/science.285.5429.870}.

\bibitem[Getty et~al.(2012)Getty, Brinckerhoff, Cornish, Ecelberger, and Floyd]{getty2012compact}
S.~A. Getty, W.~B. Brinckerhoff, T.~Cornish, S.~Ecelberger, and M.~Floyd.
\newblock Compact two-step laser time-of-flight mass spectrometer for in situ analyses of aromatic organics on planetary missions.
\newblock \emph{Rapid Communications in Mass Spectrometry}, 26\penalty0 (23):\penalty0 2786--2790, 2012.

\bibitem[Giles et~al.(2024)Giles, Spencer, Tsang, Greathouse, Lellouch, and L{\'o}pez-Valverde]{Giles2024}
R.~S. Giles, J.~R. Spencer, C.~C. Tsang, T.~K. Greathouse, E.~Lellouch, and M.~A. L{\'o}pez-Valverde.
\newblock Seasonal and longitudinal variability in io’s so2 atmosphere from 22 years of irtf/texes observations.
\newblock \emph{Icarus}, 418:\penalty0 116151, 2024.

\bibitem[Giono and Roth(2021)]{giono2021}
G.~Giono and L.~Roth.
\newblock Io's so2 atmosphere from hst lyman-$\alpha$ images: 1997 to 2018.
\newblock \emph{Icarus}, 359:\penalty0 114212, 2021.

\bibitem[Giono et~al.(2020)Giono, Roth, Ivchenko, Saur, Retherford, Schlegel, Ackland, and Strobel]{Giono2020}
G.~Giono, L.~Roth, N.~Ivchenko, J.~Saur, K.~Retherford, S.~Schlegel, M.~Ackland, and D.~Strobel.
\newblock {An Analysis of the Statistics and Systematics of Limb Anomaly Detections in HST/STIS Transit Images of Europa}.
\newblock \emph{The Astronomical Journal}, 159\penalty0 (4):\penalty0 155, 3 2020.
\newblock ISSN 1538-3881.
\newblock \doi{10.3847/1538-3881/AB7454}.
\newblock URL \url{https://iopscience.iop.org/article/10.3847/1538-3881/ab7454 https://iopscience.iop.org/article/10.3847/1538-3881/ab7454/meta}.

\bibitem[Glavin and Dworkin(2009)]{glavin2009enrichment}
D.~P. Glavin and J.~P. Dworkin.
\newblock Enrichment of the amino acid l-isovaline by aqueous alteration on ci and cm meteorite parent bodies.
\newblock \emph{Proceedings of the National Academy of Sciences}, 106\penalty0 (14):\penalty0 5487--5492, 2009.

\bibitem[Glein et~al.(2008)Glein, Zolotov, and Shock]{Glein2008}
C.~R. Glein, M.~Y. Zolotov, and E.~L. Shock.
\newblock {The oxidation state of hydrothermal systems on early Enceladus}.
\newblock \emph{Icarus}, 197\penalty0 (1):\penalty0 157--163, 9 2008.
\newblock ISSN 00191035.
\newblock \doi{10.1016/j.icarus.2008.03.021}.
\newblock URL \url{https://ui.adsabs.harvard.edu/abs/2008Icar..197..157G/abstract}.

\bibitem[Glein et~al.(2018)Glein, Postberg, and Vance]{Glein2018}
C.~R. Glein, F.~Postberg, and S.~D. Vance.
\newblock {The Geochemistry of Enceladus: Composition and Controls}.
\newblock In P.~M. Schenk, R.~N. Clark, C.~J.~A. Howett, A.~J. Verbiscer, and J.~H. Waite, editors, \emph{Enceladus and the Icy Moons of Saturn}, pages 39--56. The University of Arizona Press, 2018.
\newblock \doi{10.2458/azu{\_}uapress{\_}9780816537075-ch003}.

\bibitem[Goesmann et~al.(2017)Goesmann, Brinckerhoff, Raulin, Goetz, Danell, Getty, Siljestr{\"o}m, Mi{\ss}bach, Steininger, Arevalo~Jr, et~al.]{goesmann2017mars}
F.~Goesmann, W.~B. Brinckerhoff, F.~Raulin, W.~Goetz, R.~M. Danell, S.~A. Getty, S.~Siljestr{\"o}m, H.~Mi{\ss}bach, H.~Steininger, R.~D. Arevalo~Jr, et~al.
\newblock The mars organic molecule analyzer (moma) instrument: characterization of organic material in martian sediments.
\newblock \emph{Astrobiology}, 17\penalty0 (6-7):\penalty0 655--685, 2017.

\bibitem[Goguen et~al.(2013)Goguen, Buratti, Brown, Clark, Nicholson, Hedman, Howell, Sotin, Cruikshank, Baines, Lawrence, Spencer, and Blackburn]{Gougen2013}
J.~D. Goguen, B.~J. Buratti, R.~H. Brown, R.~N. Clark, P.~D. Nicholson, M.~M. Hedman, R.~R. Howell, C.~Sotin, D.~P. Cruikshank, K.~H. Baines, K.~J. Lawrence, J.~R. Spencer, and D.~G. Blackburn.
\newblock {The temperature and width of an active fissure on Enceladus measured with Cassini VIMS during the 14 April 2012 South Pole flyover}.
\newblock \emph{Icarus}, 226\penalty0 (1):\penalty0 1128--1137, 9 2013.
\newblock ISSN 0019-1035.
\newblock \doi{10.1016/J.ICARUS.2013.07.012}.
\newblock URL \url{https://www.sciencedirect.com/science/article/pii/S0019103513003138}.

\bibitem[Goldstein et~al.(2018)Goldstein, Hedman, Manga, Perry, Spitale, Teolis, Goldstein, Hedman, Manga, Perry, Spitale, and Teolis]{Goldstein2018}
D.~B. Goldstein, M.~Hedman, M.~Manga, M.~Perry, J.~Spitale, B.~Teolis, D.~B. Goldstein, M.~Hedman, M.~Manga, M.~Perry, J.~Spitale, and B.~Teolis.
\newblock {Enceladus Plume Dynamics: From Surface to Space}.
\newblock In P.~Schenk, R.~Clark, C.~Howett, A.~Verbiscer, and J.~Waite, editors, \emph{Enceladus and the Icy Moons of Saturn}, pages 175--194. The University of Arizona Press, 2018.
\newblock \doi{10.2458/AZU{\_}UAPRESS{\_}9780816537075-CH009}.
\newblock URL \url{https://ui.adsabs.harvard.edu/abs/2018eims.book..175G/abstract}.

\bibitem[{Gomes} and {Morbidelli}(2024)]{Gomes2024}
R.~{Gomes} and A.~{Morbidelli}.
\newblock {Was Triton originally a regular satellite of Neptune?}
\newblock \emph{\icarus}, 420:\penalty0 116142, Sept. 2024.
\newblock \doi{10.1016/j.icarus.2024.116142}.

\bibitem[{Graps} et~al.(2000){Graps}, L., {Gr{\"{u}}n}, {E.}, {Svedhem}, {H.}, {Kr{\"{u}}ger}, {H.}, {Hor{\'{a}}nyi}, {M.}, {Heck}, {A.}, {Lammers}, and {S.}]{Graps2000}
{Graps}, A.~L., {Gr{\"{u}}n}, {E.}, {Svedhem}, {H.}, {Kr{\"{u}}ger}, {H.}, {Hor{\'{a}}nyi}, {M.}, {Heck}, {A.}, {Lammers}, and {S.}
\newblock {Io as a source of the jovian dust streams}.
\newblock \emph{Nature}, 405\penalty0 (6782):\penalty0 48--50, 5 2000.
\newblock ISSN 00280836.
\newblock \doi{10.1038/35011008}.
\newblock URL \url{https://scixplorer.org/abs/2000Natur.405...48G/abstract}.

\bibitem[Greeley et~al.(1998)Greeley, Sullivan, Klemaszewski, Homan, Head, Pappalardo, Veverka, Clark, Johnson, Klaasen, Belton, Moore, Asphaug, Carr, Neukum, Denk, Chapman, Pilcher, Geissler, Greenberg, and Tufts]{Greeley1998}
R.~Greeley, R.~Sullivan, J.~Klemaszewski, K.~Homan, J.~W. Head, R.~T. Pappalardo, J.~Veverka, B.~E. Clark, T.~V. Johnson, K.~P. Klaasen, M.~Belton, J.~Moore, E.~Asphaug, M.~H. Carr, G.~Neukum, T.~Denk, C.~R. Chapman, C.~B. Pilcher, P.~E. Geissler, R.~Greenberg, and R.~Tufts.
\newblock {Europa: Initial Galileo Geological Observations}.
\newblock \emph{Icarus}, 135\penalty0 (1):\penalty0 4--24, 9 1998.
\newblock ISSN 0019-1035.
\newblock \doi{10.1006/ICAR.1998.5969}.

\bibitem[Greenberg et~al.(1998)Greenberg, Geissler, Hoppa, Tufts, Durda, Pappalardo, Head, Greeley, Sullivan, and Carr]{Greenberg1998}
R.~Greenberg, P.~Geissler, G.~Hoppa, B.~R. Tufts, D.~D. Durda, R.~Pappalardo, J.~W. Head, R.~Greeley, R.~Sullivan, and M.~H. Carr.
\newblock {Tectonic Processes on Europa: Tidal Stresses, Mechanical Response, and Visible Features}.
\newblock \emph{Icarus}, 135\penalty0 (1):\penalty0 64--78, 9 1998.
\newblock ISSN 0019-1035.
\newblock \doi{10.1006/ICAR.1998.5986}.

\bibitem[Griffin(1920)]{Griffin1920}
F.~L. Griffin.
\newblock {Certain Periodic Orbits of k Finite Bodies revolving about a relatively Large Central Mass}.
\newblock In F.~R. Moulton, editor, \emph{Periodic Orbits}, pages 425--456. Carnegie Institution, Washington, 1920.

\bibitem[Grimaudo et~al.(2015)Grimaudo, Moreno-García, Riedo, Neuland, Tulej, Broekmann, and Wurz]{Grimaudo2015}
V.~Grimaudo, P.~Moreno-García, A.~Riedo, M.~B. Neuland, M.~Tulej, P.~Broekmann, and P.~Wurz.
\newblock High-resolution chemical depth profiling of solid material using a miniature laser ablation/ionization mass spectrometer.
\newblock \emph{Analytical Chemistry}, 87\penalty0 (4):\penalty0 2037--2041, 2015.
\newblock \doi{10.1021/ac504403j}.
\newblock URL \url{https://doi.org/10.1021/ac504403j}.
\newblock PMID: 25642789.

\bibitem[Grone et~al.(2024{\natexlab{a}})Grone, Patty, Pommerol, Kitzmann, Brandenburg, Fatton, Schroffenegger, and Demory]{grone2024modeling}
J.~Grone, C.~L. Patty, A.~Pommerol, D.~Kitzmann, L.~Brandenburg, M.~Fatton, U.~Schroffenegger, and B.-O. Demory.
\newblock Modeling circular polarization in enceladus' plumes: Implications for ground-based detection of solar system biosignatures.
\newblock In \emph{2024 EANA Conference}. EANA, 2024{\natexlab{a}}.

\bibitem[Grone et~al.(2024{\natexlab{b}})Grone, Patty, Brandenburg, Fatton, Schroffenegger, Meyer, Junier, and Demory]{grone2024full}
J.~Grone, L.~Patty, L.~Brandenburg, M.~Fatton, U.~Schroffenegger, E.~Meyer, P.~Junier, and B.-O. Demory.
\newblock Full-stokes spectropolarimetric modelling of potential ocean-world microbial biosignatures.
\newblock In \emph{2024 Astrobiology Science Conference}. AGU, 2024{\natexlab{b}}.

\bibitem[Grubisic et~al.(2021)Grubisic, Trainer, Li, Brinckerhoff, van Amerom, Danell, Costa, Castillo, Kaplan, and Zacny]{grubisic2021laser}
A.~Grubisic, M.~G. Trainer, X.~Li, W.~B. Brinckerhoff, F.~H. van Amerom, R.~M. Danell, J.~T. Costa, M.~Castillo, D.~Kaplan, and K.~Zacny.
\newblock Laser desorption mass spectrometry at saturn’s moon titan.
\newblock \emph{International Journal of Mass Spectrometry}, 470:\penalty0 116707, 2021.

\bibitem[Gudipati et~al.(2020)Gudipati, Henderson, and Bateman]{Gudipati2020}
M.~S. Gudipati, B.~L. Henderson, and F.~B. Bateman.
\newblock {Laboratory predictions for the night-side surface ice glow of Europa}.
\newblock \emph{Nature Astronomy 2020 5:3}, 5\penalty0 (3):\penalty0 276--282, 11 2020.
\newblock ISSN 2397-3366.
\newblock \doi{10.1038/s41550-020-01248-1}.
\newblock URL \url{https://www.nature.com/articles/s41550-020-01248-1}.

\bibitem[Haff et~al.(1981)Haff, Watson, and Yung]{Haff1981}
P.~K. Haff, C.~C. Watson, and Y.~L. Yung.
\newblock Sputter ejection of matter from io.
\newblock \emph{J. Geophys. Res.}, 86:\penalty0 6933--6983, 1981.

\bibitem[Hall et~al.(1995)Hall, Strobel, Feldman, Mc~Grath, and Weaver]{Hall1995}
D.~T. Hall, D.~F. Strobel, P.~D. Feldman, M.~A. Mc~Grath, and H.~A. Weaver.
\newblock {Detection of an oxygen atmosphere on Jupiter's moon Europa}.
\newblock \emph{Nature 1995 373:6516}, 373\penalty0 (6516):\penalty0 677--679, 2 1995.
\newblock ISSN 1476-4687.
\newblock \doi{10.1038/373677a0}.
\newblock URL \url{https://www.nature.com/articles/373677a0}.

\bibitem[Hall et~al.(1998)Hall, Feldman, McGrath, and Strobel]{hall1998}
D.~T. Hall, P.~Feldman, M.~A. McGrath, and D.~Strobel.
\newblock The far-ultraviolet oxygen airglow of europa and ganymede.
\newblock \emph{The Astrophysical Journal}, 499\penalty0 (1):\penalty0 475, 1998.

\bibitem[Hamilton et~al.(2013)Hamilton, Beggan, Still, Beuthe, Lopes, Williams, Radebaugh, and Wright]{hamilton2013}
C.~W. Hamilton, C.~D. Beggan, S.~Still, M.~Beuthe, R.~M. Lopes, D.~A. Williams, J.~Radebaugh, and W.~Wright.
\newblock Spatial distribution of volcanoes on io: Implications for tidal heating and magma ascent.
\newblock \emph{Earth and Planetary Science Letters}, 361:\penalty0 272--286, 2013.

\bibitem[Hammond and Collins(2024)]{Hammond2024Triton}
N.~P. Hammond and G.~C. Collins.
\newblock {Triton’s Captured Youth: Tidal Heating Kept Triton Warm and Active for Billions of Years}.
\newblock \emph{The Planetary Science Journal}, 5\penalty0 (9):\penalty0 200, 9 2024.
\newblock ISSN 2632-3338.
\newblock \doi{10.3847/PSJ/AD6744}.
\newblock URL \url{https://iopscience.iop.org/article/10.3847/PSJ/ad6744 https://iopscience.iop.org/article/10.3847/PSJ/ad6744/meta}.

\bibitem[Hand and Carlson(2015)]{Hand2015}
K.~P. Hand and R.~W. Carlson.
\newblock {Europa's surface color suggests an ocean rich with sodium chloride}.
\newblock \emph{Geophysical Research Letters}, 42\penalty0 (9):\penalty0 3174--3178, 5 2015.
\newblock ISSN 19448007.
\newblock \doi{10.1002/2015GL063559}.

\bibitem[Hand et~al.(2009)Hand, Chyba, Carlson, and Nealson]{Hand2009}
K.~P. Hand, C.~Chyba, R.~W. Carlson, and K.~H. Nealson.
\newblock {Astrobiology and the Potential for Life on Europa}.
\newblock In \emph{Europa}. University of Arizona Press, 2009.
\newblock URL \url{https://www.researchgate.net/publication/252433874}.

\bibitem[Hanel et~al.(1979)Hanel, Conrath, Flasar, Kunde, Lowman, Maguire, Pearl, Pirraglia, Samuelson, Gautier, et~al.]{hanel1979}
R.~Hanel, B.~Conrath, M.~Flasar, V.~Kunde, P.~Lowman, W.~Maguire, J.~Pearl, J.~Pirraglia, R.~Samuelson, D.~Gautier, et~al.
\newblock Infrared observations of the jovian system from voyager 1.
\newblock \emph{Science}, 204\penalty0 (4396):\penalty0 972--976, 1979.

\bibitem[{Hansen}(2022)]{2022MNRAS.tmp.2636H}
B.~M.~S. {Hansen}.
\newblock {Consequences of dynamically unstable moons in extrasolar systems}.
\newblock \emph{\mnras}, Oct. 2022.

\bibitem[Hansen et~al.(1990)Hansen, McEwen, Ingersoll, and Terrile]{Hansen1990}
C.~J. Hansen, A.~S. McEwen, A.~P. Ingersoll, and R.~J. Terrile.
\newblock {Surface and Airborne Evidence for Plumes and Winds on Triton}.
\newblock \emph{Science}, 250\penalty0 (4979):\penalty0 421--424, 10 1990.
\newblock ISSN 0036-8075.
\newblock \doi{10.1126/SCIENCE.250.4979.421}.
\newblock URL \url{https://www.science.org/doi/10.1126/science.250.4979.421}.

\bibitem[Hansen et~al.(2005)Hansen, Shemansky, and Hendrix]{Hansen2005}
C.~J. Hansen, D.~E. Shemansky, and A.~Hendrix.
\newblock Cassini uvis observations of europa's oxygen atmosphere and torus.
\newblock \emph{Icarus}, 176\penalty0 (2):\penalty0 305--315, 2005.

\bibitem[Hansen et~al.(2011)Hansen, Shemansky, Esposito, Stewart, Lewis, Colwell, Hendrix, West, Waite, Teolis, and Magee]{Hansen2011}
C.~J. Hansen, D.~E. Shemansky, L.~W. Esposito, A.~I. Stewart, B.~R. Lewis, J.~E. Colwell, A.~R. Hendrix, R.~A. West, J.~H. Waite, B.~Teolis, and B.~A. Magee.
\newblock {The composition and structure of the Enceladus plume}.
\newblock \emph{Geophysical Research Letters}, 38\penalty0 (11), 6 2011.
\newblock ISSN 1944-8007.
\newblock \doi{10.1029/2011GL047415}.
\newblock URL \url{https://onlinelibrary.wiley.com/doi/full/10.1029/2011GL047415 https://onlinelibrary.wiley.com/doi/abs/10.1029/2011GL047415 https://agupubs.onlinelibrary.wiley.com/doi/10.1029/2011GL047415}.

\bibitem[Hansen et~al.(2019)Hansen, Esposito, and Hendrix]{Hansen2019}
C.~J. Hansen, L.~W. Esposito, and A.~R. Hendrix.
\newblock {Ultraviolet observation of Enceladus' plume in transit across Saturn, compared to Europa}.
\newblock \emph{Icarus}, 330:\penalty0 256--260, 9 2019.
\newblock ISSN 0019-1035.
\newblock \doi{10.1016/J.ICARUS.2019.04.031}.

\bibitem[Hansen et~al.(2024)Hansen, Ravine, Schenk, Collins, Leonard, Phillips, Caplinger, Tosi, Bolton, and J{\'{o}}nsson]{Hansen2024}
C.~J. Hansen, M.~A. Ravine, P.~M. Schenk, G.~C. Collins, E.~J. Leonard, C.~B. Phillips, M.~A. Caplinger, F.~Tosi, S.~J. Bolton, and B.~J{\'{o}}nsson.
\newblock {Juno’s JunoCam Images of Europa}.
\newblock \emph{The Planetary Science Journal}, 5\penalty0 (3):\penalty0 76, 3 2024.
\newblock ISSN 2632-3338.
\newblock \doi{10.3847/PSJ/AD24F4}.
\newblock URL \url{https://iopscience.iop.org/article/10.3847/PSJ/ad24f4 https://iopscience.iop.org/article/10.3847/PSJ/ad24f4/meta}.

\bibitem[Hartogh et~al.(2011)Hartogh, Lellouch, Moreno, Bockel{\'e}e-Morvan, Biver, Cassidy, Rengel, Jarchow, Cavali{\'e}, Crovisier, et~al.]{hartogh2011}
P.~Hartogh, E.~Lellouch, R.~Moreno, D.~Bockel{\'e}e-Morvan, N.~Biver, T.~Cassidy, M.~Rengel, C.~Jarchow, T.~Cavali{\'e}, J.~Crovisier, et~al.
\newblock Direct detection of the enceladus water torus with herschel.
\newblock \emph{Astronomy \& Astrophysics}, 532:\penalty0 L2, 2011.

\bibitem[Haslebacher et~al.(2024)Haslebacher, Thomas, and Bickel]{Haslebacher2024a}
C.~Haslebacher, N.~Thomas, and V.~T. Bickel.
\newblock {LineaMapper: A deep learning-powered tool for mapping linear surface features on Europa}.
\newblock \emph{Icarus}, 410\penalty0 (115722), 3 2024.
\newblock ISSN 0019-1035.
\newblock \doi{10.1016/j.icarus.2023.115722}.
\newblock URL \url{https://www.sciencedirect.com/science/article/pii/S0019103523002993}.

\bibitem[Haslebacher et~al.(2025{\natexlab{a}})Haslebacher, Tejero, Prockter, Leonard, Rhoden, and Thomas]{HaslebacherPSJ2025}
C.~Haslebacher, J.~G. Tejero, L.~M. Prockter, E.~J. Leonard, A.~R. Rhoden, and N.~Thomas.
\newblock {Length, Width, and Relative Age Analysis of Lineaments in the Galileo Regional Maps with LineaMapper}.
\newblock \emph{The Planetary Science Journal}, 6\penalty0 (6):\penalty0 133, 6 2025{\natexlab{a}}.
\newblock ISSN 2632-3338.
\newblock \doi{10.3847/PSJ/ADD349}.
\newblock URL \url{https://iopscience.iop.org/article/10.3847/PSJ/add349 https://iopscience.iop.org/article/10.3847/PSJ/add349/meta}.

\bibitem[Haslebacher et~al.(2025{\natexlab{b}})Haslebacher, Thomas, Rossi, Leonard, Prockter, Sabbeth, Hasnain, and Tejero]{Haslebacher2025MapsBook}
C.~Haslebacher, N.~Thomas, C.~Rossi, E.~J. Leonard, L.~M. Prockter, L.~Sabbeth, Z.~Hasnain, and J.~G. Tejero.
\newblock {A deep-learning powered global lineament map of Europa and applications to other planetary bodies}.
\newblock In L.~Marinangeli and M.~Pantaloni, editors, \emph{Atlas of Geological Planetary Mapping}, chapter (under review). 2025{\natexlab{b}}.

\bibitem[Hedman et~al.(2013)Hedman, Gosmeyer, Nicholson, Sotin, Brown, Clark, Baines, Buratti, and Showalter]{Hedman2013}
M.~M. Hedman, C.~M. Gosmeyer, P.~D. Nicholson, C.~Sotin, R.~H. Brown, R.~N. Clark, K.~H. Baines, B.~J. Buratti, and M.~R. Showalter.
\newblock {An observed correlation between plume activity and tidal stresses on Enceladus}.
\newblock \emph{Nature 2013 500:7461}, 500\penalty0 (7461):\penalty0 182--184, 7 2013.
\newblock ISSN 1476-4687.
\newblock \doi{10.1038/nature12371}.
\newblock URL \url{https://www.nature.com/articles/nature12371}.

\bibitem[Helfenstein and Porco(2015)]{Helfenstein2015}
P.~Helfenstein and C.~C. Porco.
\newblock {ENCELADUS’ GEYSERS: RELATION TO GEOLOGICAL FEATURES}.
\newblock \emph{The Astronomical Journal}, 150\penalty0 (3):\penalty0 96, 8 2015.
\newblock ISSN 1538-3881.
\newblock \doi{10.1088/0004-6256/150/3/96}.
\newblock URL \url{https://iopscience.iop.org/article/10.1088/0004-6256/150/3/96 https://iopscience.iop.org/article/10.1088/0004-6256/150/3/96/meta}.

\bibitem[{Helled} et~al.(2020){Helled}, {Nettelmann}, and {Guillot}]{Helled2020}
R.~{Helled}, N.~{Nettelmann}, and T.~{Guillot}.
\newblock {Uranus and Neptune: Origin, Evolution and Internal Structure}.
\newblock \emph{\ssr}, 216\penalty0 (3):\penalty0 38, Mar. 2020.
\newblock \doi{10.1007/s11214-020-00660-3}.

\bibitem[{Heller} and {Hippke}(2024)]{HellerHipke2024}
R.~{Heller} and M.~{Hippke}.
\newblock {Large exomoons unlikely around Kepler-1625 b and Kepler-1708 b}.
\newblock \emph{Nature Astronomy}, 8:\penalty0 193--206, Feb. 2024.
\newblock \doi{10.1038/s41550-023-02148-w}.

\bibitem[Heller et~al.(2019)Heller, Rodenbeck, and Bruno]{Heller2019}
R.~Heller, K.~Rodenbeck, and G.~Bruno.
\newblock {An alternative interpretation of the exomoon candidate signal in the combined Kepler and Hubble data of Kepler-1625}.
\newblock \emph{Astronomy {\&} Astrophysics}, 624:\penalty0 A95, 4 2019.
\newblock ISSN 0004-6361.
\newblock \doi{10.1051/0004-6361/201834913}.
\newblock URL \url{https://www.aanda.org/articles/aa/full_html/2019/04/aa34913-18/aa34913-18.html https://www.aanda.org/articles/aa/abs/2019/04/aa34913-18/aa34913-18.html}.

\bibitem[Hemingway et~al.(2018)Hemingway, Iess, Tajeddine, and Tobie]{Hemingway2018}
D.~Hemingway, L.~Iess, R.~Tajeddine, and G.~Tobie.
\newblock {The Interior of Enceladus}.
\newblock In P.~M. Schenk, R.~N. Clark, C.~J.~A. Howett, A.~J. Verbiscer, and J.~H. Waite, editors, \emph{Enceladus and the Icy Moons of Saturn}, page~57. The University of Arizona Press, 2018.
\newblock \doi{10.2458/azu{\_}uapress{\_}9780816537075-ch004}.

\bibitem[Hemingway et~al.(2020)Hemingway, Rudolph, and Manga]{Hemingway2020}
D.~J. Hemingway, M.~L. Rudolph, and M.~Manga.
\newblock {Cascading parallel fractures on Enceladus}.
\newblock \emph{Nature Astronomy}, 4\penalty0 (3):\penalty0 234--239, 2020.
\newblock ISSN 2397-3366.
\newblock \doi{10.1038/s41550-019-0958-x}.
\newblock URL \url{https://www.nature.com/articles/s41550-019-0958-x}.

\bibitem[Herbert and Sandel(1991)]{herbert1991}
F.~Herbert and B.~Sandel.
\newblock Ch${_4}$ and haze in triton's lower atmosphere.
\newblock \emph{Journal of Geophysical Research: Space Physics}, 96\penalty0 (S01):\penalty0 19241--19252, 1991.

\bibitem[Herny et~al.(2021)Herny, Mousis, Marschall, Thomas, Rubin, Pinz{\'{o}}n-Rodr{\'{i}}guez, and Wright]{Herny2021}
C.~Herny, O.~Mousis, R.~Marschall, N.~Thomas, M.~Rubin, O.~Pinz{\'{o}}n-Rodr{\'{i}}guez, and I.~P. Wright.
\newblock {New constraints on the chemical composition and outgassing of 67P/Churyumov-Gerasimenko}.
\newblock \emph{Planetary and Space Science}, 200:\penalty0 105194, 6 2021.
\newblock ISSN 0032-0633.
\newblock \doi{10.1016/J.PSS.2021.105194}.

\bibitem[Hesse et~al.(2022)Hesse, Jordan, Vance, and Oza]{Hesse2022DownwardPercolation}
M.~A. Hesse, J.~S. Jordan, S.~D. Vance, and A.~V. Oza.
\newblock {Downward Oxidant Transport Through Europa's Ice Shell by Density-Driven Brine Percolation}.
\newblock \emph{Geophysical Research Letters}, 49\penalty0 (5):\penalty0 e2021GL095416, 3 2022.
\newblock ISSN 1944-8007.
\newblock \doi{10.1029/2021GL095416}.
\newblock URL \url{https://onlinelibrary.wiley.com/doi/full/10.1029/2021GL095416 https://onlinelibrary.wiley.com/doi/abs/10.1029/2021GL095416 https://agupubs.onlinelibrary.wiley.com/doi/10.1029/2021GL095416}.

\bibitem[Hibbitts et~al.(2019)Hibbitts, Stockstill-Cahill, Wing, and Paranicas]{Hibbitts2019ColorEuropa}
C.~A. Hibbitts, K.~Stockstill-Cahill, B.~Wing, and C.~Paranicas.
\newblock {Color centers in salts—evidence for the presence of sulfates on Europa}.
\newblock \emph{Icarus}, 326:\penalty0 37--47, 7 2019.
\newblock ISSN 10902643.
\newblock \doi{10.1016/j.icarus.2019.02.022}.

\bibitem[Hofgartner et~al.(2022{\natexlab{a}})Hofgartner, Birch, Castillo, Grundy, Hansen, Hayes, Howett, Hurford, Martin, Mitchell, Nordheim, Poston, Prockter, Quick, Schenk, Schindhelm, and Umurhan]{Hofgartner2022}
J.~D. Hofgartner, S.~P. Birch, J.~Castillo, W.~M. Grundy, C.~J. Hansen, A.~G. Hayes, C.~J. Howett, T.~A. Hurford, E.~S. Martin, K.~L. Mitchell, T.~A. Nordheim, M.~J. Poston, L.~M. Prockter, L.~C. Quick, P.~Schenk, R.~N. Schindhelm, and O.~M. Umurhan.
\newblock {Hypotheses for Triton's plumes: New analyses and future remote sensing tests}.
\newblock \emph{Icarus}, 375:\penalty0 114835, 3 2022{\natexlab{a}}.
\newblock ISSN 0019-1035.
\newblock \doi{10.1016/J.ICARUS.2021.114835}.

\bibitem[Hofgartner et~al.(2022{\natexlab{b}})Hofgartner, Birch, Castillo, Grundy, Hansen, Hayes, Howett, Hurford, Martin, Mitchell, Nordheim, Poston, Prockter, Quick, Schenk, Schindhelm, and Umurhan]{Hofgartner_2022}
J.~D. Hofgartner, S.~P.~D. Birch, J.~Castillo, W.~M. Grundy, C.~J. Hansen, A.~G. Hayes, C.~J.~A. Howett, T.~A. Hurford, E.~S. Martin, K.~L. Mitchell, T.~A. Nordheim, M.~J. Poston, L.~M. Prockter, L.~C. Quick, P.~Schenk, R.~N. Schindhelm, and O.~M. Umurhan.
\newblock Hypotheses for triton's plumes: New analyses and future remote sensing tests.
\newblock \emph{Icarus}, 376:\penalty0 114859, 2022{\natexlab{b}}.
\newblock \doi{10.1016/j.icarus.2021.114859}.
\newblock URL \url{https://doi.org/10.1016/j.icarus.2021.114859}.

\bibitem[Holler et~al.(2016)Holler, Young, Grundy, and Olkin]{Holler_2016}
B.~Holler, L.~Young, W.~Grundy, and C.~Olkin.
\newblock On the surface composition of triton’s southern latitudes.
\newblock \emph{Icarus}, 267:\penalty0 255--266, 2016.
\newblock ISSN 0019-1035.
\newblock \doi{https://doi.org/10.1016/j.icarus.2015.12.027}.
\newblock URL \url{https://www.sciencedirect.com/science/article/pii/S0019103515005849}.

\bibitem[Hoppa et~al.(1999)Hoppa, Tufts, Greenberg, and Geissler]{Hoppa1999a}
G.~V. Hoppa, B.~R. Tufts, R.~Greenberg, and P.~E. Geissler.
\newblock {Formation of cycloidal features on Europa}.
\newblock \emph{Science (New York, N.Y.)}, 285\penalty0 (5435):\penalty0 1899--1902, 9 1999.
\newblock ISSN 0036-8075.
\newblock \doi{10.1126/SCIENCE.285.5435.1899}.
\newblock URL \url{https://pubmed.ncbi.nlm.nih.gov/10489365/}.

\bibitem[Hsu et~al.(2015)Hsu, Postberg, Sekine, Shibuya, Kempf, Hor{\'{a}}nyi, Juh{\'{a}}sz, Altobelli, Suzuki, Masaki, Kuwatani, Tachibana, Sirono, Moragas-Klostermeyer, and Srama]{Hsu2015OngoingEnceladus}
H.~W. Hsu, F.~Postberg, Y.~Sekine, T.~Shibuya, S.~Kempf, M.~Hor{\'{a}}nyi, A.~Juh{\'{a}}sz, N.~Altobelli, K.~Suzuki, Y.~Masaki, T.~Kuwatani, S.~Tachibana, S.~I. Sirono, G.~Moragas-Klostermeyer, and R.~Srama.
\newblock {Ongoing hydrothermal activities within Enceladus}.
\newblock \emph{Nature}, 519\penalty0 (7542):\penalty0 207--210, 3 2015.
\newblock ISSN 14764687.
\newblock \doi{10.1038/nature14262}.

\bibitem[Hurford and Brunt(2014)]{Hurford2014}
T.~A. Hurford and K.~M. Brunt.
\newblock {Antarctic analog for dilational bands on Europa}.
\newblock \emph{Earth and Planetary Science Letters}, 401:\penalty0 275--283, 9 2014.
\newblock ISSN 0012-821X.
\newblock \doi{10.1016/J.EPSL.2014.05.015}.

\bibitem[Hurford et~al.(2007)Hurford, Helfenstein, Hoppa, Greenberg, and Bills]{Hurford2007Enc}
T.~A. Hurford, P.~Helfenstein, G.~V. Hoppa, R.~Greenberg, and B.~G. Bills.
\newblock {Eruptions arising from tidally controlled periodic openings of rifts on Enceladus}.
\newblock \emph{Nature}, 447\penalty0 (7142):\penalty0 292--294, 5 2007.
\newblock ISSN 14764687.
\newblock \doi{10.1038/nature05821}.

\bibitem[Hussmann and Spohn(2004)]{Hussmann2004}
H.~Hussmann and T.~Spohn.
\newblock {Thermal-orbital evolution of Io and Europa}.
\newblock \emph{Icarus}, 171\penalty0 (2):\penalty0 391--410, 10 2004.
\newblock ISSN 00191035.
\newblock \doi{10.1016/j.icarus.2004.05.020}.

\bibitem[Huybrighs et~al.(2017)Huybrighs, Futaana, Barabash, Wieser, Wurz, Krupp, Glassmeier, and Vermeersen]{Huybrighs2017}
H.~L. Huybrighs, Y.~Futaana, S.~Barabash, M.~Wieser, P.~Wurz, N.~Krupp, K.-H. Glassmeier, and B.~Vermeersen.
\newblock On the in-situ detectability of europa's water vapour plumes from a flyby mission.
\newblock \emph{Icarus}, 289:\penalty0 270--280, 2017.
\newblock ISSN 0019-1035.
\newblock \doi{https://doi.org/10.1016/j.icarus.2016.10.026}.
\newblock URL \url{https://www.sciencedirect.com/science/article/pii/S0019103516301968}.

\bibitem[{Ida} et~al.(2020){Ida}, {Ueta}, {Sasaki}, and {Ishizawa}]{Ida2020}
S.~{Ida}, S.~{Ueta}, T.~{Sasaki}, and Y.~{Ishizawa}.
\newblock {Uranian satellite formation by evolution of a water vapour disk generated by a giant impact}.
\newblock \emph{Nature Astronomy}, 4:\penalty0 880--885, Mar. 2020.
\newblock \doi{10.1038/s41550-020-1049-8}.

\bibitem[Ingersoll(1989)]{Ingersoll1989}
A.~P. Ingersoll.
\newblock Io meteorology: How atmospheric pressure is controlled locally by volcanos and surface frosts.
\newblock \emph{Icarus}, 81\penalty0 (2):\penalty0 298--313, 1989.

\bibitem[Ingersoll(1990)]{Ingersoll1990}
A.~P. Ingersoll.
\newblock {Dynamics of Triton's atmosphere}.
\newblock \emph{Nature}, 344\penalty0 (6264):\penalty0 315--317, 1990.
\newblock ISSN 1476-4687.
\newblock \doi{10.1038/344315a0}.
\newblock URL \url{https://www.nature.com/articles/344315a0}.

\bibitem[Ingersoll and Ewald(2011)]{Ingersoll2011TotalImages}
A.~P. Ingersoll and S.~P. Ewald.
\newblock {Total particulate mass in Enceladus plumes and mass of Saturn’s E ring inferred from Cassini ISS images}.
\newblock \emph{Icarus}, 216\penalty0 (2):\penalty0 492--506, 12 2011.
\newblock ISSN 0019-1035.
\newblock \doi{10.1016/J.ICARUS.2011.09.018}.

\bibitem[Ingersoll and Ewald(2017)]{ingersoll2017}
A.~P. Ingersoll and S.~P. Ewald.
\newblock {Decadal timescale variability of the Enceladus plumes inferred from Cassini images}.
\newblock \emph{Icarus}, 282:\penalty0 260--275, 1 2017.
\newblock ISSN 0019-1035.
\newblock \doi{10.1016/J.ICARUS.2016.09.018}.
\newblock URL \url{https://www.sciencedirect.com/science/article/pii/S0019103516305905}.

\bibitem[Intriligator and Miller(1981)]{intriligator1981}
D.~Intriligator and W.~Miller.
\newblock Detection of the io plasma torus by pioneer 10.
\newblock \emph{Geophysical Research Letters}, 8\penalty0 (4):\penalty0 409--412, 1981.

\bibitem[Isella et~al.(2019)Isella, Benisty, Teague, Bae, Keppler, Facchini, and P{\'e}rez]{Isella2019}
A.~Isella, M.~Benisty, R.~Teague, J.~Bae, M.~Keppler, S.~Facchini, and L.~P{\'e}rez.
\newblock Detection of continuum submillimeter emission associated with candidate protoplanets.
\newblock \emph{The Astrophysical Journal Letters}, 879\penalty0 (2):\penalty0 L25, 2019.

\bibitem[Jabaud et~al.(2024)Jabaud, Artoni, Tobie, Le~Menn, and Richard]{Jabaud2024}
B.~Jabaud, R.~Artoni, G.~Tobie, E.~Le~Menn, and P.~Richard.
\newblock {Cohesive properties of ice powders analogous to fresh plume deposits on Enceladus and Europa}.
\newblock \emph{Icarus}, 409:\penalty0 115859, 2 2024.
\newblock ISSN 0019-1035.
\newblock \doi{10.1016/J.ICARUS.2023.115859}.

\bibitem[Jara-Oru{\'{e}} and Vermeersen(2011)]{JaraOrue2011}
H.~M. Jara-Oru{\'{e}} and B.~L. Vermeersen.
\newblock {Effects of low-viscous layers and a non-zero obliquity on surface stresses induced by diurnal tides and non-synchronous rotation: The case of Europa}.
\newblock \emph{Icarus}, 215\penalty0 (1):\penalty0 417--438, 9 2011.
\newblock ISSN 00191035.
\newblock \doi{10.1016/j.icarus.2011.05.034}.

\bibitem[Jaramillo-Botero et~al.(2021)Jaramillo-Botero, Cable, Hofmann, Malaska, Hodyss, and Lunine]{Jaramillo-Botero2021}
A.~Jaramillo-Botero, M.~L. Cable, A.~E. Hofmann, M.~Malaska, R.~Hodyss, and J.~Lunine.
\newblock Understanding hypervelocity sampling of biosignatures in space missions.
\newblock \emph{Astrobiology}, 21\penalty0 (4):\penalty0 421–--442, 2021.
\newblock \doi{10.1089/ast.2020.2301}.

\bibitem[Jaumann et~al.(2008)Jaumann, Stephan, Hansen, Clark, Buratti, Brown, Baines, Newman, Bellucci, Filacchione, Coradini, Cruikshank, Griffith, Hibbitts, McCord, Nelson, Nicholson, Sotin, and Wagner]{Jaumann2008}
R.~Jaumann, K.~Stephan, G.~B. Hansen, R.~N. Clark, B.~J. Buratti, R.~H. Brown, K.~H. Baines, S.~F. Newman, G.~Bellucci, G.~Filacchione, A.~Coradini, D.~P. Cruikshank, C.~A. Griffith, C.~A. Hibbitts, T.~B. McCord, R.~M. Nelson, P.~D. Nicholson, C.~Sotin, and R.~Wagner.
\newblock {Distribution of icy particles across Enceladus' surface as derived from Cassini-VIMS measurements}.
\newblock \emph{Icarus}, 193\penalty0 (2):\penalty0 407--419, 2 2008.
\newblock ISSN 0019-1035.
\newblock \doi{10.1016/J.ICARUS.2007.09.013}.

\bibitem[Jessup and Spencer(2015)]{jessup2015}
K.~L. Jessup and J.~R. Spencer.
\newblock Spatially resolved hst/stis observations of io’s dayside equatorial atmosphere.
\newblock \emph{Icarus}, 248:\penalty0 165--189, 2015.

\bibitem[Jessup et~al.(2004)Jessup, Spencer, Ballester, Howell, Roesler, Vigel, and Yelle]{jessup2004}
K.~L. Jessup, J.~R. Spencer, G.~E. Ballester, R.~R. Howell, F.~Roesler, M.~Vigel, and R.~Yelle.
\newblock The atmospheric signature of io's prometheus plume and anti-jovian hemisphere: evidence for a sublimation atmosphere.
\newblock \emph{Icarus}, 169\penalty0 (1):\penalty0 197--215, 2004.

\bibitem[Jessup et~al.(2007)Jessup, Spencer, and Yelle]{jessup2007}
K.~L. Jessup, J.~Spencer, and R.~Yelle.
\newblock Sulfur volcanism on io.
\newblock \emph{Icarus}, 192\penalty0 (1):\penalty0 24--40, 2007.

\bibitem[Jia et~al.(2018)Jia, Kivelson, Khurana, and Kurth]{jia2018EvidenceSignatures}
X.~Jia, M.~G. Kivelson, K.~K. Khurana, and W.~S. Kurth.
\newblock {Evidence of a plume on Europa from Galileo magnetic and plasma wave signatures}.
\newblock \emph{Nature Astronomy}, 2\penalty0 (6):\penalty0 459--464, 6 2018.
\newblock ISSN 23973366.
\newblock \doi{10.1038/S41550-018-0450-Z}.

\bibitem[{Johnson} and {Huggins}(2006)]{johnsonhuggins06}
R.~E. {Johnson} and P.~J. {Huggins}.
\newblock {Toroidal Atmospheres around Extrasolar Planets}.
\newblock \emph{\pasp}, 118:\penalty0 1136--1143, Aug. 2006.
\newblock \doi{10.1086/506183}.

\bibitem[Johnson et~al.(1981)Johnson, Lanzerotti, Brown, and Armstrong]{Johnson1981}
R.~E. Johnson, L.~J. Lanzerotti, W.~L. Brown, and T.~P. Armstrong.
\newblock {Erosion of Galilean Satellite Surfaces by Jovian Magnetosphere Particles}.
\newblock \emph{Science}, 212\penalty0 (4498):\penalty0 1027--1030, 5 1981.
\newblock ISSN 00368075.
\newblock \doi{10.1126/SCIENCE.212.4498.1027}.
\newblock URL \url{https://www.science.org/doi/10.1126/science.212.4498.1027}.

\bibitem[Johnson et~al.(1982)Johnson, Lanzerotti, and Brown]{Johnson1982}
R.~E. Johnson, L.~J. Lanzerotti, and W.~L. Brown.
\newblock {Planetary applications of ion induced erosion of condensed-gas frosts}.
\newblock \emph{Nuclear Instruments and Methods in Physics Research}, 198\penalty0 (1):\penalty0 147--157, 7 1982.
\newblock ISSN 0167-5087.
\newblock \doi{10.1016/0167-5087(82)90066-7}.

\bibitem[Johnson and Soderblom(1982)]{Johnson1982Io}
T.~V. Johnson and L.~A. Soderblom.
\newblock {Volcanic eruptions on Io - Implications for surface evolution and mass loss}.
\newblock In D.~Morrison, editor, \emph{Satellites of Jupiter}, pages 634--646, 1 1982.

\bibitem[Johnson et~al.(1988)Johnson, Veeder, Matson, Brown, Nelson, and Morrison]{Johnson1988}
T.~V. Johnson, G.~J. Veeder, D.~L. Matson, R.~H. Brown, R.~M. Nelson, and D.~Morrison.
\newblock {Io: evidence for silicate volcanism in 1986}.
\newblock \emph{Science}, 242\penalty0 (4883):\penalty0 1280--1283, 1988.
\newblock ISSN 0036-8075.
\newblock \doi{10.1126/SCIENCE.242.4883.1280}.
\newblock URL \url{https://pubmed.ncbi.nlm.nih.gov/17817074/}.

\bibitem[Kadel et~al.(1998)Kadel, Fagents, Greeley, and {Galileo SSI Team}]{Kadel1998}
S.~D. Kadel, S.~A. Fagents, R.~Greeley, and {Galileo SSI Team}.
\newblock {Trough-Bounding Ridge Pairs on Europa -- Considerations for an Endogenic Model of Formation}.
\newblock \emph{29th Lunar Planetary Science Conference}, page 1078, 1998.

\bibitem[{Kegerreis} et~al.(2018){Kegerreis}, {Teodoro}, {Eke}, {Massey}, {Catling}, {Fryer}, {Korycansky}, {Warren}, and {Zahnle}]{Kegerreis2018}
J.~A. {Kegerreis}, L.~F.~A. {Teodoro}, V.~R. {Eke}, R.~J. {Massey}, D.~C. {Catling}, C.~L. {Fryer}, D.~G. {Korycansky}, M.~S. {Warren}, and K.~J. {Zahnle}.
\newblock {Consequences of Giant Impacts on Early Uranus for Rotation, Internal Structure, Debris, and Atmospheric Erosion}.
\newblock \emph{\apj}, 861\penalty0 (1):\penalty0 52, July 2018.
\newblock \doi{10.3847/1538-4357/aac725}.

\bibitem[Kemp et~al.(1971)Kemp, Wolstencroft, and Swedlund]{kemp1971circular}
J.~C. Kemp, R.~D. Wolstencroft, and J.~B. Swedlund.
\newblock Circular polarization: Jupiter and other planets.
\newblock \emph{Nature}, 232\penalty0 (5307):\penalty0 165--168, 1971.

\bibitem[Kempf et~al.(2010)Kempf, Beckmann, and Schmidt]{Kempf2010}
S.~Kempf, U.~Beckmann, and J.~Schmidt.
\newblock {How the Enceladus dust plume feeds Saturn’s E ring}.
\newblock \emph{Icarus}, 206\penalty0 (2):\penalty0 446--457, 4 2010.
\newblock ISSN 0019-1035.
\newblock \doi{10.1016/J.ICARUS.2009.09.016}.

\bibitem[Kervazo et~al.(2021)Kervazo, Tobie, Choblet, Dumoulin, and B{\v{e}}hounkov{\'a}]{kervazo2021}
M.~Kervazo, G.~Tobie, G.~Choblet, C.~Dumoulin, and M.~B{\v{e}}hounkov{\'a}.
\newblock Solid tides in io’s partially molten interior-contribution of bulk dissipation.
\newblock \emph{Astronomy \& Astrophysics}, 650:\penalty0 A72, 2021.

\bibitem[Kervazo et~al.(2022)Kervazo, Tobie, Choblet, Dumoulin, and B{\v{e}}hounkov{\'a}]{kervazo2022}
M.~Kervazo, G.~Tobie, G.~Choblet, C.~Dumoulin, and M.~B{\v{e}}hounkov{\'a}.
\newblock Inferring io’s interior from tidal monitoring.
\newblock \emph{Icarus}, 373:\penalty0 114737, 2022.

\bibitem[Keszthelyi et~al.(2001)Keszthelyi, McEwen, Phillips, Milazzo, Geissler, Turtle, Radebaugh, Williams, Simonelli, Breneman, et~al.]{Keszthelyi2001}
L.~Keszthelyi, A.~McEwen, C.~Phillips, M.~Milazzo, P.~Geissler, E.~Turtle, J.~Radebaugh, D.~Williams, D.~Simonelli, H.~Breneman, et~al.
\newblock Imaging of volcanic activity on jupiter's moon io by galileo during the galileo europa mission and the galileo millennium mission.
\newblock \emph{Journal of Geophysical Research: Planets}, 106\penalty0 (E12):\penalty0 33025--33052, 2001.

\bibitem[Khurana et~al.(1998)Khurana, Kivelson, Stevenson, Schubert, Russell, Walker, and Polanskey]{Khurana1998}
K.~K. Khurana, M.~G. Kivelson, D.~J. Stevenson, G.~Schubert, C.~T. Russell, R.~J. Walker, and C.~Polanskey.
\newblock {Induced magnetic fields as evidence for subsurface oceans in Europa and Callisto}.
\newblock \emph{Nature}, 395\penalty0 (6704):\penalty0 777, 10 1998.
\newblock ISSN 00280836.
\newblock \doi{10.1038/27394}.

\bibitem[Kieffer(2007)]{Kieffer_2007}
H.~H. Kieffer.
\newblock Cold jets in the martian polar caps.
\newblock \emph{Journal of Geophysical Research: Planets}, 112\penalty0 (E8):\penalty0 E08005, 2007.
\newblock \doi{10.1029/2006JE002816}.
\newblock URL \url{https://doi.org/10.1029/2006JE002816}.

\bibitem[King et~al.(2022)King, Fletcher, and Ligier]{King2022}
O.~King, L.~N. Fletcher, and N.~Ligier.
\newblock {Compositional Mapping of Europa Using MCMC Modeling of Near-IR VLT/SPHERE and Galileo/NIMS Observations}.
\newblock \emph{The Planetary Science Journal}, 3\penalty0 (3):\penalty0 72, 3 2022.
\newblock ISSN 2632-3338.
\newblock \doi{10.3847/PSJ/AC596D}.
\newblock URL \url{https://iopscience.iop.org/article/10.3847/PSJ/ac596d https://iopscience.iop.org/article/10.3847/PSJ/ac596d/meta}.

\bibitem[King et~al.(2025)King, Fletcher, Clarke, and Hidalgo]{King2025}
O.~R. King, L.~N. Fletcher, F.~Clarke, and A.~Hidalgo.
\newblock {Spatially Resolved Visible Wavelength Spectroscopy of the Galilean Moons With VLT/MUSE}.
\newblock \emph{Journal of Geophysical Research: Planets}, 130\penalty0 (3):\penalty0 e2024JE008511, 3 2025.
\newblock ISSN 2169-9100.
\newblock \doi{10.1029/2024JE008511}.
\newblock URL \url{https://onlinelibrary.wiley.com/doi/full/10.1029/2024JE008511 https://onlinelibrary.wiley.com/doi/abs/10.1029/2024JE008511 https://agupubs.onlinelibrary.wiley.com/doi/10.1029/2024JE008511}.

\bibitem[Kipfer et~al.(2022)Kipfer, Ligterink, Bouwman, Schwander, Grimaudo, de~Koning, Boeren, Schmidt, Lukmanov, Tulej, et~al.]{kipfer2022toward}
K.~A. Kipfer, N.~F.~W. Ligterink, J.~Bouwman, L.~Schwander, V.~Grimaudo, C.~P. de~Koning, N.~J. Boeren, P.~K. Schmidt, R.~Lukmanov, M.~Tulej, et~al.
\newblock Toward detecting polycyclic aromatic hydrocarbons on planetary objects with origin.
\newblock \emph{The planetary science journal}, 3\penalty0 (2):\penalty0 43, 2022.

\bibitem[Kipfer et~al.(2024)Kipfer, Galli, Riedo, Tulej, Wurz, and Ligterink]{Kipfer2024ComplexMethane}
K.~A. Kipfer, A.~Galli, A.~Riedo, M.~Tulej, P.~Wurz, and N.~F. Ligterink.
\newblock {Complex Ice Chemistry: A comparative study of electron irradiated planetary ice analogues containing methane}.
\newblock \emph{Icarus}, 410:\penalty0 115742, 3 2024.
\newblock ISSN 0019-1035.
\newblock \doi{10.1016/J.ICARUS.2023.115742}.

\bibitem[{Kipping} et~al.(2022){Kipping}, {Bryson}, {Burke}, {Christiansen}, {Hardegree-Ullman}, {Quarles}, {Hansen}, {Szul{\'a}gyi}, and {Teachey}]{Kipping2022}
D.~{Kipping}, S.~{Bryson}, C.~{Burke}, J.~{Christiansen}, K.~{Hardegree-Ullman}, B.~{Quarles}, B.~{Hansen}, J.~{Szul{\'a}gyi}, and A.~{Teachey}.
\newblock {An exomoon survey of 70 cool giant exoplanets and the new candidate Kepler-1708 b-i}.
\newblock \emph{Nature Astronomy}, 6:\penalty0 367--380, Jan. 2022.
\newblock \doi{10.1038/s41550-021-01539-1}.

\bibitem[Kirchoff and McKinnon(2009)]{Kirchoff2009}
M.~R. Kirchoff and W.~B. McKinnon.
\newblock {Formation of mountains on Io: Variable volcanism and thermal stresses}.
\newblock \emph{Icarus}, 201\penalty0 (2):\penalty0 598--614, 6 2009.
\newblock ISSN 0019-1035.
\newblock \doi{10.1016/J.ICARUS.2009.02.006}.
\newblock URL \url{https://www.sciencedirect.com/science/article/pii/S0019103509000633}.

\bibitem[Kirk et~al.(1990)Kirk, Brown, and Soderblom]{Kirk_1990}
R.~L. Kirk, R.~H. Brown, and L.~A. Soderblom.
\newblock Subsurface energy storage and transport for solar-powered geysers on triton.
\newblock \emph{Science}, 250\penalty0 (4979):\penalty0 424--429, 1990.
\newblock \doi{10.1126/science.250.4979.424}.
\newblock URL \url{https://doi.org/10.1126/science.250.4979.424}.

\bibitem[Kite and Rubin(2016)]{Kite2016}
E.~S. Kite and A.~M. Rubin.
\newblock {Sustained eruptions on Enceladus explained by turbulent dissipation in tiger stripes}.
\newblock \emph{Proceedings of the National Academy of Sciences}, 113\penalty0 (15):\penalty0 3972--3975, 4 2016.
\newblock ISSN 10916490.
\newblock \doi{10.1073/PNAS.1520507113}.
\newblock URL \url{/doi/pdf/10.1073/pnas.1520507113?download=true}.

\bibitem[Kivelson et~al.(1996)Kivelson, Khurana, Walker, Warnecke, Russell, Linker, Southwood, and Polanskey]{kivelson1996}
M.~Kivelson, K.~Khurana, R.~Walker, J.~Warnecke, C.~Russell, J.~Linker, D.~Southwood, and C.~Polanskey.
\newblock Io's interaction with the plasma torus: Galileo magnetometer report.
\newblock \emph{Science}, 274\penalty0 (5286):\penalty0 396--398, 1996.

\bibitem[Kivelson et~al.(2001)Kivelson, Khurana, Russell, Joy, Volwerk, Walker, Zimmer, and Linker]{kivelson2001}
M.~G. Kivelson, K.~K. Khurana, C.~T. Russell, S.~P. Joy, M.~Volwerk, R.~J. Walker, C.~Zimmer, and J.~A. Linker.
\newblock Magnetized or unmagnetized: Ambiguity persists following galileo's encounters with io in 1999 and 2000.
\newblock \emph{Journal of Geophysical Research: Space Physics}, 106\penalty0 (A11):\penalty0 26121--26135, 2001.

\bibitem[Klaiber(2024)]{Klaiber2024}
L.~Klaiber.
\newblock \emph{{Three Dimensional DSMC Modelling of the Dynamics of Io's Atmosphere}}.
\newblock PhD thesis, University of Bern, Bern, 2 2024.

\bibitem[{Kley}(1999)]{Kley99}
W.~{Kley}.
\newblock {Mass flow and accretion through gaps in accretion discs}.
\newblock \emph{\mnras}, 303\penalty0 (4):\penalty0 696--710, Mar. 1999.
\newblock \doi{10.1046/j.1365-8711.1999.02198.x}.

\bibitem[Krasnopolsky et~al.(1992)Krasnopolsky, Sandel, and Herbert]{Krasnopolsky1992}
V.~A. Krasnopolsky, B.~R. Sandel, and F.~Herbert.
\newblock {Properties of haze in the atmosphere of Triton}.
\newblock \emph{Journal of Geophysical Research, Volume 97, Issue E7, p. 11695-11700}, 97\penalty0 (E7):\penalty0 11695, 7 1992.
\newblock ISSN 0148-0227.
\newblock \doi{10.1029/92JE00945}.
\newblock URL \url{https://ui.adsabs.harvard.edu/abs/1992JGR....9711695K/abstract}.

\bibitem[Kreidberg et~al.(2019)Kreidberg, Luger, and Bedell]{Kreidberg2019}
L.~Kreidberg, R.~Luger, and M.~Bedell.
\newblock {No Evidence for Lunar Transit in New Analysis of Hubble Space Telescope Observations of the Kepler-1625 System}.
\newblock \emph{The Astrophysical Journal Letters}, 877\penalty0 (2):\penalty0 L15, 5 2019.
\newblock ISSN 2041-8205.
\newblock \doi{10.3847/2041-8213/AB20C8}.
\newblock URL \url{https://iopscience.iop.org/article/10.3847/2041-8213/ab20c8 https://iopscience.iop.org/article/10.3847/2041-8213/ab20c8/meta}.

\bibitem[K{\"u}hn et~al.(2020)K{\"u}hn, Patty, Demory, Pommerol, Snik, Keller, Hoeijmakers, Poch, Stam, Pallichadath, et~al.]{kuhn2020monitoring}
J.~K{\"u}hn, L.~Patty, B.-O. Demory, A.~Pommerol, F.~Snik, C.~Keller, J.~Hoeijmakers, O.~Poch, D.~Stam, V.~Pallichadath, et~al.
\newblock Monitoring the earth's diverse environments with full-stokes spectro-polarimetry: the mermoz project.
\newblock \emph{Ground-based and Airborne Instrumentation for Astronomy VIII}, 11447:\penalty0 114479L, 2020.

\bibitem[Kurth et~al.(2023)Kurth, Wilkinson, Hospodarsky, Santolik, Averkamp, Sulaiman, Menietti, Connerney, Allegrini, Mauk, et~al.]{kurth2023}
W.~S. Kurth, D.~R. Wilkinson, G.~B. Hospodarsky, O.~Santolik, T.~F. Averkamp, A.~H. Sulaiman, J.~D. Menietti, J.~E. Connerney, F.~Allegrini, B.~H. Mauk, et~al.
\newblock Juno plasma wave observations at europa.
\newblock \emph{Geophysical research letters}, 50\penalty0 (24):\penalty0 e2023GL105775, 2023.

\bibitem[Lainey et~al.(2009)Lainey, Arlot, Karatekin, and Van~Hoolst]{lainey2009}
V.~Lainey, J.-E. Arlot, {\"O}.~Karatekin, and T.~Van~Hoolst.
\newblock Strong tidal dissipation in io and jupiter from astrometric observations.
\newblock \emph{Nature}, 459\penalty0 (7249):\penalty0 957--959, 2009.

\bibitem[Lainey et~al.(2024)Lainey, Rambaux, Tobie, Cooper, Zhang, Noyelles, and Bailli{\'{e}}]{Lainey2024}
V.~Lainey, N.~Rambaux, G.~Tobie, N.~Cooper, Q.~Zhang, B.~Noyelles, and K.~Bailli{\'{e}}.
\newblock {A recently formed ocean inside Saturn’s moon Mimas}.
\newblock \emph{Nature 2024 626:7998}, 626\penalty0 (7998):\penalty0 280--282, 2 2024.
\newblock ISSN 1476-4687.
\newblock \doi{10.1038/s41586-023-06975-9}.
\newblock URL \url{https://www.nature.com/articles/s41586-023-06975-9}.

\bibitem[{Lammer} and {H.}(1995)]{Lammer1995}
{Lammer} and {H.}
\newblock {Mass loss of N2 molecules from Triton by magnetospheric plasma interaction}.
\newblock \emph{P{\&}SS}, 43\penalty0 (7):\penalty0 845--850, 1995.
\newblock ISSN 00320633.
\newblock \doi{10.1016/0032-0633(94)00214-C}.
\newblock URL \url{https://scixplorer.org/abs/1995P&SS...43..845L/abstract}.

\bibitem[Lee et~al.(2005)Lee, Pappalardo, and Makris]{Lee2005}
S.~Lee, R.~T. Pappalardo, and N.~C. Makris.
\newblock {Mechanics of tidally driven fractures in Europa's ice shell}.
\newblock \emph{Icarus}, 177\penalty0 (2):\penalty0 367--379, 10 2005.
\newblock ISSN 0019-1035.
\newblock \doi{10.1016/J.ICARUS.2005.07.003}.

\bibitem[Leith and Mckinnon(1996)]{Leith1996}
A.~C. Leith and W.~B. Mckinnon.
\newblock {Is There Evidence for Polar Wander on Europa?}
\newblock \emph{ICARUS}, 120:\penalty0 387--398, 1996.

\bibitem[Lellouch et~al.(1990)Lellouch, Belton, de~Pater, Gulkis, and Encrenaz]{lellouch1990}
E.~Lellouch, M.~Belton, I.~de~Pater, S.~Gulkis, and T.~Encrenaz.
\newblock Io's atmosphere from microwave detection of so2.
\newblock \emph{Nature}, 346\penalty0 (6285):\penalty0 639--641, 1990.

\bibitem[Lellouch et~al.(1992)Lellouch, Belton, De~Pater, Paubert, Gulkis, and Encrenaz]{lellouch1992}
E.~Lellouch, M.~Belton, I.~De~Pater, G.~Paubert, S.~Gulkis, and T.~Encrenaz.
\newblock The structure, stability, and global distribution of io's atmosphere.
\newblock \emph{Icarus}, 98\penalty0 (2):\penalty0 271--295, 1992.

\bibitem[Lellouch et~al.(1996)Lellouch, Strobel, Belton, Summers, Paubert, and Moreno]{lellouch1996}
E.~Lellouch, D.~Strobel, M.~Belton, M.~Summers, G.~Paubert, and R.~Moreno.
\newblock Detection of sulfur monoxide in io's atmosphere.
\newblock \emph{The Astrophysical Journal}, 459\penalty0 (2):\penalty0 L107, 1996.

\bibitem[Lellouch et~al.(2003)Lellouch, Paubert, Moses, Schneider, and Strobel]{lellouch2003}
E.~Lellouch, G.~Paubert, J.~I. Moses, N.~M. Schneider, and D.~Strobel.
\newblock Volcanically emitted sodium chloride as a source for io's neutral clouds and plasma torus.
\newblock \emph{Nature}, 421\penalty0 (6918):\penalty0 45--47, 2003.

\bibitem[Lellouch et~al.(2007)Lellouch, McGrath, and Jessup]{lellouch2007}
E.~Lellouch, M.~A. McGrath, and K.~L. Jessup.
\newblock \emph{Io’s atmosphere}, pages 231--264.
\newblock Springer, 2007.
\newblock ISBN 978-3-540-48841-5.
\newblock \doi{10.1007/978-3-540-48841-5_10}.

\bibitem[Lellouch et~al.(2010)Lellouch, de~Bergh, Sicardy, Ferron, and K{\"a}ufl]{lellouch2010}
E.~Lellouch, C.~de~Bergh, B.~Sicardy, S.~Ferron, and H.-U. K{\"a}ufl.
\newblock Detection of co in triton's atmosphere and the nature of surface-atmosphere interactions.
\newblock \emph{Astronomy \& Astrophysics}, 512:\penalty0 L8, 2010.

\bibitem[Lellouch et~al.(2015)Lellouch, Ali-Dib, Jessup, Smette, K{\"a}ufl, and Marchis]{lellouch2015}
E.~Lellouch, M.~Ali-Dib, K.-L. Jessup, A.~Smette, H.-U. K{\"a}ufl, and F.~Marchis.
\newblock Detection and characterization of io’s atmosphere from high-resolution 4-$\mu$m spectroscopy.
\newblock \emph{Icarus}, 253:\penalty0 99--114, 2015.

\bibitem[Leonard et~al.(2022)Leonard, Howell, Mills, Senske, Patthoff, Hay, and Pappalardo]{Leonard2022}
E.~J. Leonard, S.~M. Howell, A.~Mills, D.~A. Senske, D.~A. Patthoff, H.~C. Hay, and R.~T. Pappalardo.
\newblock {Finding Order in Chaos: Quantitative Predictors of Chaos Terrain Morphology on Europa}.
\newblock \emph{Geophysical Research Letters}, 49\penalty0 (8), 4 2022.
\newblock ISSN 19448007.
\newblock \doi{10.1029/2021GL097309}.

\bibitem[Lesage et~al.(2021)Lesage, Schmidt, Andrieu, and Massol]{Lesage2021}
E.~Lesage, F.~Schmidt, F.~Andrieu, and H.~Massol.
\newblock {Constraints on effusive cryovolcanic eruptions on Europa using topography obtained from Galileo images}.
\newblock \emph{Icarus}, 361:\penalty0 114373, 6 2021.
\newblock ISSN 0019-1035.
\newblock \doi{10.1016/J.ICARUS.2021.114373}.

\bibitem[Lesage et~al.(2025)Lesage, Howell, Neveu, Miller, Naseem, Melwani~Daswani, Villette, and Vance]{Lesage2025}
E.~Lesage, S.~M. Howell, M.~Neveu, J.~W. Miller, M.~Naseem, M.~Melwani~Daswani, J.~Villette, and S.~D. Vance.
\newblock {Identifying signatures of past and present cryovolcanism on Europa}.
\newblock \emph{Nature Communications 2025 16:1}, 16\penalty0 (1):\penalty0 1--9, 2 2025.
\newblock ISSN 2041-1723.
\newblock \doi{10.1038/s41467-025-57070-8}.
\newblock URL \url{https://www.nature.com/articles/s41467-025-57070-8}.

\bibitem[Li et~al.(2017)Li, Danell, Pinnick, Grubisic, Van~Amerom, Arevalo~Jr, Getty, Brinckerhoff, Southard, Gonnsen, et~al.]{li2017mars}
X.~Li, R.~M. Danell, V.~T. Pinnick, A.~Grubisic, F.~Van~Amerom, R.~D. Arevalo~Jr, S.~A. Getty, W.~B. Brinckerhoff, A.~E. Southard, Z.~D. Gonnsen, et~al.
\newblock Mars organic molecule analyzer (moma) laser desorption/ionization source design and performance characterization.
\newblock \emph{International journal of mass spectrometry}, 422:\penalty0 177--187, 2017.

\bibitem[Ligier et~al.(2016)Ligier, Poulet, Carter, Brunetto, and Gourgeot]{Ligier2016}
N.~Ligier, F.~Poulet, J.~Carter, R.~Brunetto, and F.~Gourgeot.
\newblock {VLT/SINFONI OBSERVATIONS OF EUROPA: NEW INSIGHTS INTO THE SURFACE COMPOSITION}.
\newblock \emph{The Astronomical Journal}, 151\penalty0 (6):\penalty0 163, 5 2016.
\newblock ISSN 1538-3881.
\newblock \doi{10.3847/0004-6256/151/6/163}.
\newblock URL \url{https://iopscience.iop.org/article/10.3847/0004-6256/151/6/163 https://iopscience.iop.org/article/10.3847/0004-6256/151/6/163/meta}.

\bibitem[Ligterink et~al.(2020)Ligterink, Grimaudo, Moreno-Garc{\'\i}a, Lukmanov, Tulej, Leya, Lindner, Wurz, Cockell, Ehrenfreund, et~al.]{ligterink2020origin}
N.~F. Ligterink, V.~Grimaudo, P.~Moreno-Garc{\'\i}a, R.~Lukmanov, M.~Tulej, I.~Leya, R.~Lindner, P.~Wurz, C.~S. Cockell, P.~Ehrenfreund, et~al.
\newblock Origin: a novel and compact laser desorption--mass spectrometry system for sensitive in situ detection of amino acids on extraterrestrial surfaces.
\newblock \emph{Scientific reports}, 10\penalty0 (1):\penalty0 9641, 2020.

\bibitem[Ligterink et~al.(2022)Ligterink, Kipfer, Gruchola, Boeren, Keresztes~Schmidt, de~Koning, Tulej, Wurz, and Riedo]{ligterink2022origin}
N.~F. Ligterink, K.~A. Kipfer, S.~Gruchola, N.~J. Boeren, P.~Keresztes~Schmidt, C.~P. de~Koning, M.~Tulej, P.~Wurz, and A.~Riedo.
\newblock The origin space instrument for detecting biosignatures and habitability indicators on a venus life finder mission.
\newblock \emph{Aerospace}, 9\penalty0 (6):\penalty0 312, 2022.

\bibitem[Longo and Damer(2020)]{longo2020factoring}
A.~Longo and B.~Damer.
\newblock Factoring origin of life hypotheses into the search for life in the solar system and beyond.
\newblock \emph{Life}, 10\penalty0 (5):\penalty0 52, 2020.

\bibitem[Lovis et~al.(2024)Lovis, Blind, Chazelas, Shinde, Bugatti, Restori, Dinis, Genolet, Hughes, Sordet, Schnell, Rihs, Crausaz, Turbet, Billot, Fusco, Neichel, Sauvage, Diaz, Houelle, Blackman, Lanotte, K{\"{u}}hn, Hagelberg, Guyon, Martinez, Spang, Mordasini, Ehrenreich, Demory, and Bolmont]{Lovis2024}
C.~Lovis, N.~Blind, B.~Chazelas, M.~Shinde, M.~Bugatti, N.~Restori, I.~Dinis, L.~Genolet, I.~Hughes, M.~Sordet, R.~Schnell, S.~Rihs, A.~Crausaz, M.~Turbet, N.~Billot, T.~Fusco, B.~Neichel, J.-F. Sauvage, P.~S. Diaz, M.~Houelle, J.~Blackman, A.~Lanotte, J.~K{\"{u}}hn, J.~Hagelberg, O.~Guyon, P.~Martinez, A.~Spang, C.~Mordasini, D.~Ehrenreich, B.-O. Demory, and E.~Bolmont.
\newblock {RISTRETTO: reflected-light exoplanet spectroscopy at the diffraction limit of the VLT}.
\newblock In J.~J. Bryant, K.~Motohara, and J.~R.~D. Vernet, editors, \emph{Ground-based and Airborne Instrumentation for Astronomy X}, volume 13096, page 130961I. SPIE, 2024.
\newblock \doi{10.1117/12.3020142}.
\newblock URL \url{https://doi.org/10.1117/12.3020142}.

\bibitem[Lunine and Stevenson(1982)]{Lunine1982}
J.~I. Lunine and D.~J. Stevenson.
\newblock Formation of the galilean satellites in a gaseous nebula.
\newblock \emph{Icarus}, 52\penalty0 (1):\penalty0 14--39, 1982.

\bibitem[MacKenzie et~al.(2021)MacKenzie, Neveu, Davila, Lunine, Craft, Cable, Phillips-Lander, Hofgartner, Eigenbrode, Waite, et~al.]{mackenzie2021enceladus}
S.~M. MacKenzie, M.~Neveu, A.~F. Davila, J.~I. Lunine, K.~L. Craft, M.~L. Cable, C.~M. Phillips-Lander, J.~D. Hofgartner, J.~L. Eigenbrode, J.~H. Waite, et~al.
\newblock The enceladus orbilander mission concept: Balancing return and resources in the search for life.
\newblock \emph{The Planetary Science Journal}, 2\penalty0 (2):\penalty0 77, 2021.

\bibitem[Maeda et~al.(2024)Maeda, Ohtsuki, Suetsugu, Shibaike, Tanigawa, and Machida]{Maeda2024DeliveryPlanets}
N.~Maeda, K.~Ohtsuki, R.~Suetsugu, Y.~Shibaike, T.~Tanigawa, and M.~N. Machida.
\newblock {Delivery of Dust Particles from Protoplanetary Disks onto Circumplanetary Disks of Giant Planets}.
\newblock \emph{The Astrophysical Journal}, 968\penalty0 (2):\penalty0 62, 6 2024.
\newblock ISSN 0004-637X.
\newblock \doi{10.3847/1538-4357/AD4035}.
\newblock URL \url{https://iopscience.iop.org/article/10.3847/1538-4357/ad4035 https://iopscience.iop.org/article/10.3847/1538-4357/ad4035/meta}.

\bibitem[{Makarov} and {Efroimsky}(2023)]{2023A&A...672A..78M}
V.~V. {Makarov} and M.~{Efroimsky}.
\newblock {Pathways of survival for exomoons and inner exoplanets}.
\newblock \emph{\aap}, 672:\penalty0 A78, Apr. 2023.

\bibitem[Managadze et~al.(2016)Managadze, Engel, Getty, Wurz, Brinckerhoff, Shokolov, Sholin, Terent'ev, Chumikov, Skalkin, Blank, Prokhorov, Managadze, and Luchnikov]{Managadze2016}
G.~G. Managadze, M.~H. Engel, S.~Getty, P.~Wurz, W.~B. Brinckerhoff, A.~G. Shokolov, G.~V. Sholin, S.~A. Terent'ev, A.~E. Chumikov, A.~S. Skalkin, V.~D. Blank, V.~M. Prokhorov, N.~G. Managadze, and K.~A. Luchnikov.
\newblock Excess of l-alanine in amino acids synthesized in a plasma torch generated by a hypervelocity meteorite impact reproduced in the laboratory.
\newblock \emph{Planetary and Space Science}, 131:\penalty0 70--78, 2016.
\newblock ISSN 0032-0633.
\newblock \doi{https://doi.org/10.1016/j.pss.2016.07.005}.
\newblock URL \url{https://www.sciencedirect.com/science/article/pii/S0032063316300101}.

\bibitem[Mandt et~al.(2023)Mandt, Luspay-Kuti, Mousis, and Anderson]{Mandt2023}
K.~Mandt, A.~Luspay-Kuti, O.~Mousis, and S.~E. Anderson.
\newblock {Surface Volatile Composition as Evidence for Hydrothermal Processes Lasting Longer in Triton’s Interior than Pluto’s}.
\newblock \emph{The Astrophysical Journal}, 959\penalty0 (1):\penalty0 57, 12 2023.
\newblock ISSN 0004-637X.
\newblock \doi{10.3847/1538-4357/AD09B5}.
\newblock URL \url{https://iopscience.iop.org/article/10.3847/1538-4357/ad09b5 https://iopscience.iop.org/article/10.3847/1538-4357/ad09b5/meta}.

\bibitem[Manga and Wang(2007)]{Manga2007}
M.~Manga and C.~Y. Wang.
\newblock {Pressurized oceans and the eruption of liquid water on Europa and Enceladus}.
\newblock \emph{Geophysical Research Letters}, 34\penalty0 (7), 4 2007.
\newblock ISSN 00948276.
\newblock \doi{10.1029/2007GL029297}.

\bibitem[Marschall et~al.(2020)Marschall, Liao, Thomas, and Wu]{Marschall2020}
R.~Marschall, Y.~Liao, N.~Thomas, and J.~S. Wu.
\newblock {Limitations in the determination of surface emission distributions on comets through modelling of observational data - A case study based on Rosetta observations}.
\newblock \emph{Icarus}, 346:\penalty0 113742, 8 2020.
\newblock ISSN 0019-1035.
\newblock \doi{10.1016/J.ICARUS.2020.113742}.

\bibitem[Martin et~al.(2023)Martin, Whitten, Kattenhorn, Collins, Southworth, Wiser, and Prindle]{Martin2023}
E.~S. Martin, J.~L. Whitten, S.~A. Kattenhorn, G.~C. Collins, B.~S. Southworth, L.~S. Wiser, and S.~Prindle.
\newblock {Measurements of regolith thicknesses on Enceladus: Uncovering the record of plume activity}.
\newblock \emph{Icarus}, 392:\penalty0 115369, 3 2023.
\newblock ISSN 0019-1035.
\newblock \doi{10.1016/J.ICARUS.2022.115369}.

\bibitem[Martin et~al.(2008)Martin, Baross, Kelley, and Russell]{Martin2008}
W.~Martin, J.~Baross, D.~Kelley, and M.~J. Russell.
\newblock {Hydrothermal vents and the origin of life}.
\newblock \emph{Nature Reviews Microbiology 2008 6:11}, 6\penalty0 (11):\penalty0 805--814, 9 2008.
\newblock ISSN 1740-1534.
\newblock \doi{10.1038/nrmicro1991}.
\newblock URL \url{https://www.nature.com/articles/nrmicro1991}.

\bibitem[Martins et~al.(2013)Martins, Price, Goldman, Sephton, and Burchell]{Martins2013}
Z.~Martins, M.~C. Price, N.~Goldman, M.~A. Sephton, and M.~J. Burchell.
\newblock Shock synthesis of amino acids from impacting cometary and icy planet surface analogues.
\newblock \emph{Nature Geoscience}, 6\penalty0 (12):\penalty0 1045--1049, 2013.
\newblock \doi{10.1038/ngeo1930}.

\bibitem[Matson et~al.(1974)Matson, Johnson, and Fanale]{Matson1974}
D.~L. Matson, T.~V. Johnson, and F.~P. Fanale.
\newblock Sodium d-line emission from io: Sputtering and resonant scattering hypothesis.
\newblock \emph{Astrophysical Journal}, 192:\penalty0 L43--L46, 1974.

\bibitem[Mazarico et~al.(2023)Mazarico, Buccino, Castillo-Rogez, Dombard, Genova, Hussmann, Kiefer, Lunine, McKinnon, Nimmo, Park, Roberts, Srinivasan, Steinbr{\"{u}}gge, Tortora, and Withers]{Mazarico2023}
E.~Mazarico, D.~Buccino, J.~Castillo-Rogez, A.~J. Dombard, A.~Genova, H.~Hussmann, W.~S. Kiefer, J.~I. Lunine, W.~B. McKinnon, F.~Nimmo, R.~S. Park, J.~H. Roberts, D.~K. Srinivasan, G.~Steinbr{\"{u}}gge, P.~Tortora, and P.~Withers.
\newblock {The Europa Clipper Gravity and Radio Science Investigation}.
\newblock \emph{Space Science Reviews}, 219\penalty0 (4), 6 2023.
\newblock ISSN 15729672.
\newblock \doi{10.1007/S11214-023-00972-0}.

\bibitem[Mccord et~al.(1998)Mccord, Hansen, Fanale, Carlson, Matson, Johnson, Smythe, Crowley, Martin, Ocampo, Hibbitts, and Granahan]{Mccord1998}
T.~B. Mccord, G.~Hansen, F.~P. Fanale, R.~W. Carlson, D.~Matson, T.~V. Johnson, W.~Smythe, J.~K. Crowley, P.~D. Martin, A.~Ocampo, C.~A. Hibbitts, and J.~C. Granahan.
\newblock {SALTS ON EUROPA'S SURFACE FROM THE GALILEO NIMS INVESTIGATION}.
\newblock In \emph{Lunar and Planetary Science Conference XXIX}, page LSPC abstract N.1560, 1998.

\bibitem[McDonald et~al.(1994)McDonald, Thompson, Heinrich, Khare, and Sagan]{McDonald_1994}
G.~D. McDonald, W.~Thompson, M.~Heinrich, B.~N. Khare, and C.~Sagan.
\newblock Chemical investigation of titan and triton tholins.
\newblock \emph{Icarus}, 108\penalty0 (1):\penalty0 137--145, 1994.
\newblock ISSN 0019-1035.
\newblock \doi{https://doi.org/10.1006/icar.1994.1046}.
\newblock URL \url{https://www.sciencedirect.com/science/article/pii/S0019103584710463}.

\bibitem[McDoniel et~al.(2015)McDoniel, Goldstein, Varghese, and Trafton]{McDoniel2015}
W.~J. McDoniel, D.~B. Goldstein, P.~L. Varghese, and L.~M. Trafton.
\newblock {Three-dimensional simulation of gas and dust in Io’s Pele plume}.
\newblock \emph{Icarus}, 257:\penalty0 251--274, 9 2015.
\newblock ISSN 0019-1035.
\newblock \doi{10.1016/J.ICARUS.2015.03.019}.

\bibitem[McEwen et~al.(2000)McEwen, Belton, Breneman, Fagents, Geissler, Greeley, Head, Hoppa, Jaeger, Johnson, et~al.]{mcewen2000}
A.~McEwen, M.~Belton, H.~Breneman, S.~Fagents, P.~Geissler, R.~Greeley, J.~Head, G.~Hoppa, W.~Jaeger, T.~Johnson, et~al.
\newblock Galileo at io: Results from high-resolution imaging.
\newblock \emph{Science}, 288\penalty0 (5469):\penalty0 1193--1198, 2000.

\bibitem[McEwen(1986)]{McEwen1986}
A.~S. McEwen.
\newblock {Tidal reorientation and the fracturing of Jupiter's moon Europa}.
\newblock \emph{Nature}, 321\penalty0 (6065):\penalty0 49--51, 1986.
\newblock ISSN 00280836.
\newblock \doi{10.1038/321049a0}.

\bibitem[McEwen et~al.(1988)McEwen, Johnson, Matson, and Soderblom]{mcewen1988}
A.~S. McEwen, T.~V. Johnson, D.~L. Matson, and L.~A. Soderblom.
\newblock {The global distribution, abundance, and stability of SO2 on Io}.
\newblock \emph{Icarus}, 75\penalty0 (3):\penalty0 450--478, 9 1988.
\newblock ISSN 0019-1035.
\newblock \doi{10.1016/0019-1035(88)90157-1}.
\newblock URL \url{https://www.sciencedirect.com/science/article/abs/pii/0019103588901571?via%3Dihub}.

\bibitem[McEwen et~al.(2004)McEwen, Keszthelyi, Lopes, Schenk, and Spencer]{McEwen2004}
A.~S. McEwen, L.~P. Keszthelyi, R.~Lopes, P.~M. Schenk, and J.~R. Spencer.
\newblock {The lithosphere and surface of Io}.
\newblock In F.~Bagenal, T.~E. Dowling, and W.~B. McKinnon, editors, \emph{Jupiter. The Planet, Satellites and Magnetosphere}, volume~1, pages 307--328. 2004.

\bibitem[McGrath et~al.(2000)McGrath, Belton, Spencer, and Sartoretti]{mcgrath2000}
M.~A. McGrath, M.~J. Belton, J.~R. Spencer, and P.~Sartoretti.
\newblock Spatially resolved spectroscopy of io's pele plume and so2 atmosphere.
\newblock \emph{Icarus}, 146\penalty0 (2):\penalty0 476--493, 2000.

\bibitem[McKinnon(2013)]{McKinnon2013}
W.~B. McKinnon.
\newblock {The shape of Enceladus as explained by an irregular core: Implications for gravity, libration, and survival of its subsurface ocean}.
\newblock \emph{Journal of Geophysical Research: Planets}, 118\penalty0 (9):\penalty0 1775--1788, 9 2013.
\newblock ISSN 2169-9100.
\newblock \doi{10.1002/JGRE.20122}.
\newblock URL \url{https://onlinelibrary.wiley.com/doi/full/10.1002/jgre.20122 https://onlinelibrary.wiley.com/doi/abs/10.1002/jgre.20122 https://agupubs.onlinelibrary.wiley.com/doi/10.1002/jgre.20122}.

\bibitem[McKinnon and Kirk(2014)]{McKinnon_2014}
W.~B. McKinnon and R.~L. Kirk.
\newblock Chapter 40 - triton.
\newblock In T.~Spohn, D.~Breuer, and T.~V. Johnson, editors, \emph{Encyclopedia of the Solar System (Third Edition)}, pages 861--881. Elsevier, Boston, third edition edition, 2014.
\newblock ISBN 978-0-12-415845-0.
\newblock \doi{https://doi.org/10.1016/B978-0-12-415845-0.00040-2}.
\newblock URL \url{https://www.sciencedirect.com/science/article/pii/B9780124158450000402}.

\bibitem[McKinnon and Schenk(2025)]{Mckinnon2025}
W.~B. McKinnon and P.~M. Schenk.
\newblock {Is Mimas a Dyson Satellite? The Fate of Small Melting Moons}.
\newblock In \emph{56th Lunar and Planetary Science Conference}, volume 3090 of \emph{LPI Contributions}, page 2897, 3 2025.

\bibitem[McKinnon et~al.(1995)McKinnon, Lunine, and Banfield]{McKinnon_1995}
W.~B. McKinnon, J.~I. Lunine, and D.~Banfield.
\newblock Origin and evolution of triton.
\newblock In D.~P. Cruikshank, editor, \emph{Neptune and Triton}, pages 807--877. University of Arizona Press, 1995.

\bibitem[Meade and Jakosky(1991)]{Meade1991}
P.~E. Meade and B.~M. Jakosky.
\newblock {Thermally driven diffusion of SO2 within the surface of Io}.
\newblock \emph{Journal of Geophysical Research: Planets}, 96\penalty0 (E5):\penalty0 22729--22740, 12 1991.
\newblock ISSN 2156-2202.
\newblock \doi{10.1029/91JE02208}.
\newblock URL \url{/doi/pdf/10.1029/91JE02208 https://onlinelibrary.wiley.com/doi/abs/10.1029/91JE02208 https://agupubs.onlinelibrary.wiley.com/doi/10.1029/91JE02208}.

\bibitem[Meier and Loeffler(2020)]{meier2020sputtering}
R.~M. Meier and M.~J. Loeffler.
\newblock Sputtering of water ice by kev electrons at 60 k.
\newblock \emph{Surface Science}, 691:\penalty0 121509, 2020.

\bibitem[Melcher et~al.(1982)Melcher, LePoire, Cooper, and Tombrello]{Melcher1982}
C.~L. Melcher, D.~J. LePoire, B.~H. Cooper, and T.~A. Tombrello.
\newblock {Erosion of frozen sulfur dioxide by ion bombardment: Applications to Io}.
\newblock \emph{Geophysical Research Letters}, 9\penalty0 (10):\penalty0 1151--1154, 10 1982.
\newblock ISSN 1944-8007.
\newblock \doi{10.1029/GL009I010P01151}.
\newblock URL \url{/doi/pdf/10.1029/GL009i010p01151 https://onlinelibrary.wiley.com/doi/abs/10.1029/GL009i010p01151 https://agupubs.onlinelibrary.wiley.com/doi/10.1029/GL009i010p01151}.

\bibitem[Mendillo et~al.(2007)Mendillo, Laurent, Wilson, Baumgardner, Konrad, and Karl]{Mendillo2007}
M.~Mendillo, S.~Laurent, J.~Wilson, J.~Baumgardner, J.~Konrad, and W.~C. Karl.
\newblock {The sources of sodium escaping from Io revealed by spectral high definition imaging}.
\newblock \emph{Nature}, 448:\penalty0 330--332, 2007.
\newblock \doi{10.1038/nature06000}.

\bibitem[Meneghin et~al.(2022)Meneghin, Brucato, Fornaro, and Poggiali]{Meneghin2022}
A.~Meneghin, J.~R. Brucato, T.~Fornaro, and G.~Poggiali.
\newblock Life detection in martian returned samples: correlation between analytical techniques and biological signatures.
\newblock \emph{International Journal of Astrobiology}, 21\penalty0 (5):\penalty0 287–295, 2022.
\newblock \doi{10.1017/S1473550422000106}.

\bibitem[Meyer et~al.(2025)Meyer, Buffo, Nimmo, Wells, Boury, Fox-Powell, Tomlinson, Parkinson, and Vasil]{Meyer2025}
C.~R. Meyer, J.~J. Buffo, F.~Nimmo, A.~J. Wells, S.~Boury, M.~Fox-Powell, T.~C. Tomlinson, J.~R. Parkinson, and G.~M. Vasil.
\newblock {A Potential Mushy Source for the Geysers of Enceladus and Other Icy Satellites}.
\newblock \emph{Geophysical Research Letters}, 52\penalty0 (3):\penalty0 e2024GL111929, 2 2025.
\newblock ISSN 1944-8007.
\newblock \doi{10.1029/2024GL111929}.
\newblock URL \url{https://onlinelibrary.wiley.com/doi/full/10.1029/2024GL111929 https://onlinelibrary.wiley.com/doi/abs/10.1029/2024GL111929 https://agupubs.onlinelibrary.wiley.com/doi/10.1029/2024GL111929}.

\bibitem[Meyer and Wisdom(2008)]{meyer2008}
J.~Meyer and J.~Wisdom.
\newblock Tidal evolution of mimas, enceladus, and dione.
\newblock \emph{Icarus}, 193\penalty0 (1):\penalty0 213--223, 2008.

\bibitem[Meyer~zu Westram et~al.(2024)Meyer~zu Westram, Oza, and Galli]{MeyerZuWestram2024}
M.~Meyer~zu Westram, A.~V. Oza, and A.~Galli.
\newblock {Exomoon Phase Curves: Toroidal Exosphere Simulations of Exo-Ios Orbiting 8 Exoplanets in Alkali Spectroscopy}.
\newblock \emph{Journal of Geophysical Research: Planets}, 129\penalty0 (3):\penalty0 e2023JE007935, 3 2024.
\newblock ISSN 2169-9100.
\newblock \doi{10.1029/2023JE007935}.
\newblock URL \url{https://onlinelibrary.wiley.com/doi/full/10.1029/2023JE007935 https://onlinelibrary.wiley.com/doi/abs/10.1029/2023JE007935 https://agupubs.onlinelibrary.wiley.com/doi/10.1029/2023JE007935}.

\bibitem[Mieno et~al.(2025)Mieno, Nakamura, Sekiguchi, Hasegawa, Shibata, Kebukawa, and Kobayashi]{Mieno2025}
T.~Mieno, S.~Nakamura, S.~Sekiguchi, S.~Hasegawa, H.~Shibata, Y.~Kebukawa, and K.~Kobayashi.
\newblock Impact formation of amino acids in nitrogen gas using a light gas gun simulating the impact reactions of asteroids on planets/satellites with a gas atmosphere.
\newblock \emph{Discov. Life}, 55\penalty0 (29):\penalty0 1--18, 2025.
\newblock \doi{10.1007/s11084-025-09709-1}.

\bibitem[Milazzo et~al.(2005)Milazzo, Keszthelyi, Radebaugh, Davies, Turtle, Geissler, Klaasen, Rathbun, and McEwen]{milazzo2005}
M.~P. Milazzo, L.~P. Keszthelyi, J.~Radebaugh, A.~G. Davies, E.~P. Turtle, P.~Geissler, K.~P. Klaasen, J.~A. Rathbun, and A.~S. McEwen.
\newblock Volcanic activity at tvashtar catena, io.
\newblock \emph{Icarus}, 179\penalty0 (1):\penalty0 235--251, 2005.

\bibitem[Morabito et~al.(1979)Morabito, Synnott, Kupferman, and Collins]{morabito1979}
L.~Morabito, S.~Synnott, P.~Kupferman, and S.~A. Collins.
\newblock Discovery of currently active extraterrestrial volcanism.
\newblock \emph{Science}, 204\penalty0 (4396):\penalty0 972--972, 1979.

\bibitem[{Morbidelli} et~al.(2012){Morbidelli}, {Tsiganis}, {Batygin}, {Crida}, and {Gomes}]{Morby2012}
A.~{Morbidelli}, K.~{Tsiganis}, K.~{Batygin}, A.~{Crida}, and R.~{Gomes}.
\newblock {Explaining why the uranian satellites have equatorial prograde orbits despite the large planetary obliquity}.
\newblock \emph{\icarus}, 219\penalty0 (2):\penalty0 737--740, June 2012.
\newblock \doi{10.1016/j.icarus.2012.03.025}.

\bibitem[Moullet et~al.(2008)Moullet, Lellouch, Moreno, Gurwell, and Moore]{moullet2008}
A.~Moullet, E.~Lellouch, R.~Moreno, M.~A. Gurwell, and C.~Moore.
\newblock First disk-resolved millimeter observations of {Io}'s surface and {SO2} atmosphere.
\newblock \emph{Astronomy \& Astrophysics}, 482\penalty0 (1):\penalty0 279--292, 2008.

\bibitem[Moullet et~al.(2010)Moullet, Gurwell, Lellouch, and Moreno]{moullet2010}
A.~Moullet, M.~A. Gurwell, E.~Lellouch, and R.~Moreno.
\newblock Simultaneous mapping of so2, so, nacl in io’s atmosphere with the submillimeter array.
\newblock \emph{Icarus}, 208\penalty0 (1):\penalty0 353--365, 2010.

\bibitem[Moullet et~al.(2013)Moullet, Lellouch, Moreno, Gurwell, Black, and Butler]{moullet2013}
A.~Moullet, E.~Lellouch, R.~Moreno, M.~Gurwell, J.~H. Black, and B.~Butler.
\newblock Exploring io's atmospheric composition with apex: First measurement of 34so2 and tentative detection of kcl.
\newblock \emph{The Astrophysical Journal}, 776\penalty0 (1):\penalty0 32, 2013.

\bibitem[{Mousis} et~al.(2020){Mousis}, {Aguichine}, {Helled}, {Irwin}, and {Lunine}]{Mousis2020}
O.~{Mousis}, A.~{Aguichine}, R.~{Helled}, P.~G.~J. {Irwin}, and J.~I. {Lunine}.
\newblock {The role of ice lines in the formation of Uranus and Neptune}.
\newblock \emph{Philosophical Transactions of the Royal Society of London Series A}, 378\penalty0 (2187):\penalty0 20200107, Dec. 2020.
\newblock \doi{10.1098/rsta.2020.0107}.

\bibitem[Mousis et~al.(2022)Mousis, Bouquet, Langevin, André, Boithias, Durry, Faye, Hartogh, Helbert, Iess, Kempf, Masters, Postberg, Renard, Vernazza, Vorburger, Wurz, Atkinson, Barabash, Berthomier, Brucato, Cable, Carter, Cazaux, Coustenis, Danger, Dehant, Fornaro, Garnier, Gautier, Groussin, Hadid, Ize, Kolmasova, Lebreton, Maistre, Lellouch, Lunine, Mandt, Martins, Mimoun, Nenon, Caro, Rannou, Rauer, Schmitt-Kopplin, Schneeberger, Simons, Stephan, Hoolst, Vaverka, Wieser, and Wörner]{mousis2022moonraker}
O.~Mousis, A.~Bouquet, Y.~Langevin, N.~André, H.~Boithias, G.~Durry, F.~Faye, P.~Hartogh, J.~Helbert, L.~Iess, S.~Kempf, A.~Masters, F.~Postberg, J.-B. Renard, P.~Vernazza, A.~Vorburger, P.~Wurz, D.~H. Atkinson, S.~Barabash, M.~Berthomier, J.~Brucato, M.~Cable, J.~Carter, S.~Cazaux, A.~Coustenis, G.~Danger, V.~Dehant, T.~Fornaro, P.~Garnier, T.~Gautier, O.~Groussin, L.~Z. Hadid, J.-C. Ize, I.~Kolmasova, J.-P. Lebreton, S.~L. Maistre, E.~Lellouch, J.~I. Lunine, K.~E. Mandt, Z.~Martins, D.~Mimoun, Q.~Nenon, G.~M.~M. Caro, P.~Rannou, H.~Rauer, P.~Schmitt-Kopplin, A.~Schneeberger, M.~Simons, K.~Stephan, T.~V. Hoolst, J.~Vaverka, M.~Wieser, and L.~Wörner.
\newblock Moonraker: Enceladus multiple flyby mission.
\newblock \emph{The Planetary Science Journal}, 3\penalty0 (12):\penalty0 268, dec 2022.
\newblock \doi{10.3847/PSJ/ac9c03}.
\newblock URL \url{https://dx.doi.org/10.3847/PSJ/ac9c03}.

\bibitem[Mulder et~al.(2022)Mulder, Patty, Spadaccia, Pommerol, Demory, Keller, K{\"u}hn, Snik, and Stam]{mulder2022spectropolarimetry}
W.~Mulder, C.~L. Patty, S.~Spadaccia, A.~Pommerol, B.-O. Demory, C.~U. Keller, J.~G. K{\"u}hn, F.~Snik, and D.~M. Stam.
\newblock Spectropolarimetry of life: airborne measurements from a hot air balloon.
\newblock In \emph{Light in Nature IX}, volume 12214, pages 12--24. SPIE, 2022.

\bibitem[Mura et~al.(2020)Mura, Adriani, Tosi, Lopes, Sindoni, Filacchione, Williams, Davies, Plainaki, Bolton, et~al.]{Mura2020}
A.~Mura, A.~Adriani, F.~Tosi, R.~Lopes, G.~Sindoni, G.~Filacchione, D.~Williams, A.~Davies, C.~Plainaki, S.~Bolton, et~al.
\newblock Infrared observations of io from juno.
\newblock \emph{Icarus}, 341:\penalty0 113607, 2020.

\bibitem[{National Academies of Sciences, Engineering, and Medicine}(2022)]{OriginsWorldsandLife:ADecadalStrategy}
{National Academies of Sciences, Engineering, and Medicine}.
\newblock {Origins, Worlds, and Life: A Decadal Strategy for Planetary Science and Astrobiology 2023-2032}.
\newblock Technical report, Washington, DC, 2022.

\bibitem[Neveu and Rhoden(2019)]{Neveu2019}
M.~Neveu and A.~R. Rhoden.
\newblock {Evolution of Saturn’s mid-sized moons}.
\newblock \emph{Nature Astronomy}, 3\penalty0 (6):\penalty0 543--552, 6 2019.
\newblock ISSN 23973366.
\newblock \doi{10.1038/S41550-019-0726-Y}.

\bibitem[Newman et~al.(2008)Newman, Buratti, Brown, Jaumann, Bauer, and Momary]{Newman2008}
S.~F. Newman, B.~J. Buratti, R.~H. Brown, R.~Jaumann, J.~Bauer, and T.~Momary.
\newblock {Photometric and spectral analysis of the distribution of crystalline and amorphous ices on Enceladus as seen by Cassini}.
\newblock \emph{Icarus}, 193\penalty0 (2):\penalty0 397--406, 2 2008.
\newblock ISSN 0019-1035.
\newblock \doi{10.1016/J.ICARUS.2007.04.019}.

\bibitem[Nimmo and Manga(2009)]{Nimmo2004}
F.~Nimmo and M.~Manga.
\newblock {Geodynamics of Europa's Icy Shell}.
\newblock In R.~T. Pappalardo, W.~B. McKinnon, and K.~Khurana, editors, \emph{Europa}, pages 381--404. University of Arizona Press, 2009.

\bibitem[Nimmo and Pappalardo(2016)]{nimmo2016ocean}
F.~Nimmo and R.~T. Pappalardo.
\newblock Ocean worlds in the outer solar system.
\newblock \emph{Journal of Geophysical Research: Planets}, 121\penalty0 (8):\penalty0 1378--1399, 2016.

\bibitem[Nimmo et~al.(2007)Nimmo, Spencer, Pappalardo, and Mullen]{Nimmo2007}
F.~Nimmo, J.~R. Spencer, R.~T. Pappalardo, and M.~E. Mullen.
\newblock {Shear heating as the origin of the plumes and heat flux on Enceladus}.
\newblock \emph{Nature}, 447\penalty0 (7142):\penalty0 289--291, 5 2007.
\newblock ISSN 1476-4687.
\newblock \doi{10.1038/NATURE05783}.
\newblock URL \url{https://pubmed.ncbi.nlm.nih.gov/17507976/}.

\bibitem[Nimmo et~al.(2014)Nimmo, Porco, and Mitchell]{Nimmo2014}
F.~Nimmo, C.~Porco, and C.~Mitchell.
\newblock {TIDALLY MODULATED ERUPTIONS ON ENCELADUS: CASSINI ISS OBSERVATIONS AND MODELS}.
\newblock \emph{The Astronomical Journal}, 148\penalty0 (3):\penalty0 46, 7 2014.
\newblock ISSN 1538-3881.
\newblock \doi{10.1088/0004-6256/148/3/46}.
\newblock URL \url{https://iopscience.iop.org/article/10.1088/0004-6256/148/3/46 https://iopscience.iop.org/article/10.1088/0004-6256/148/3/46/meta}.

\bibitem[Nobis and Chevrier(2025)]{Nobis2025}
G.~Nobis and V.~Chevrier.
\newblock {Temperature, Albedo, and Emissivity of Triton and Proteus from Voyager 2 IRIS Data}.
\newblock \emph{The Planetary Science Journal}, 6\penalty0 (6):\penalty0 132, 6 2025.
\newblock ISSN 2632-3338.
\newblock \doi{10.3847/PSJ/ADD69C}.
\newblock URL \url{https://iopscience.iop.org/article/10.3847/PSJ/add69c https://iopscience.iop.org/article/10.3847/PSJ/add69c/meta}.

\bibitem[Obersnel et~al.(2025)Obersnel, Galli, Fausch, Ottersberg, and Wurz]{Obersnel2025}
L.~Obersnel, A.~Galli, R.~Fausch, R.~Ottersberg, and P.~Wurz.
\newblock Ozone production by electron irradiation of regolith ice.
\newblock \emph{Jou. Geophys. Res.}, page submitted, 2025.

\bibitem[{Ogilvie}(2013)]{2013MNRAS.429..613O}
G.~I. {Ogilvie}.
\newblock {Tides in rotating barotropic fluid bodies: the contribution of inertial waves and the role of internal structure}.
\newblock \emph{\mnras}, 429:\penalty0 613--632, Feb. 2013.

\bibitem[Olkin et~al.(1997)Olkin, Elliot, Hammel, Cooray, McDonald, Foust, Bosh, Buie, Millis, Wasserman, Dunham, Young, Howell, Hubbard, Hill, Marcialis, McDonald, Rank, Holbrook, and Reitsema]{Olkin1997}
C.~B. Olkin, J.~L. Elliot, H.~B. Hammel, A.~R. Cooray, S.~W. McDonald, J.~A. Foust, A.~S. Bosh, M.~W. Buie, R.~L. Millis, L.~H. Wasserman, E.~W. Dunham, L.~A. Young, R.~R. Howell, W.~B. Hubbard, R.~Hill, R.~L. Marcialis, J.~S. McDonald, D.~M. Rank, J.~C. Holbrook, and H.~J. Reitsema.
\newblock {The Thermal Structure of Triton's Atmosphere: Results from the 1993 and 1995 Occultations}.
\newblock \emph{Icarus}, 129\penalty0 (1):\penalty0 178--201, 9 1997.
\newblock ISSN 0019-1035.
\newblock \doi{10.1006/ICAR.1997.5757}.
\newblock URL \url{https://www.sciencedirect.com/science/article/pii/S0019103597957572?via%3Dihub}.

\bibitem[Orcutt et~al.(2011)Orcutt, Sylvan, Knab, and Edwards]{orcutt2011microbial}
B.~N. Orcutt, J.~B. Sylvan, N.~J. Knab, and K.~J. Edwards.
\newblock Microbial ecology of the dark ocean above, at, and below the seafloor.
\newblock \emph{Microbiology and molecular biology reviews}, 75\penalty0 (2):\penalty0 361--422, 2011.

\bibitem[Ottersberg(2024{\natexlab{a}})]{ottersberg2024irradiation}
R.~Ottersberg.
\newblock Vis-nir reflectance spectra collected before and after electron irradiation (2\,kev, 3.4\,$\times$\,10\textsuperscript{16}\,e\textsuperscript{–}\,cm\textsuperscript{–2}) of spherical salty ice particles (10–100\,\textmu{}m typical particle diameter) produced by freezing droplets of a solution of 5\,wt.\% nacl, 2024{\natexlab{a}}.

\bibitem[Ottersberg(2024{\natexlab{b}})]{ottersberg2024sublimation}
R.~Ottersberg.
\newblock Vis-nir reflectance spectra collected during low-temperature and near-vacuum sublimation of spherical salty ice particles (67\,\textmu{}m average diameter) produced by freezing droplets of a solution of 5\,wt.\% nacl, 2024{\natexlab{b}}.

\bibitem[Ottersberg et~al.(2025)Ottersberg, Pommerol, St{\"{o}}ckli, Obersnel, Galli, Murk, Wurz, and Thomas]{Ottersberg2025}
R.~Ottersberg, A.~Pommerol, L.~L. St{\"{o}}ckli, L.~Obersnel, A.~Galli, A.~Murk, P.~Wurz, and N.~Thomas.
\newblock {Evolution of granular salty ice analogs for Europa: Sublimation and irradiation}.
\newblock \emph{Icarus}, 439:\penalty0 116590, 10 2025.
\newblock ISSN 0019-1035.
\newblock \doi{10.1016/J.ICARUS.2025.116590}.
\newblock URL \url{https://linkinghub.elsevier.com/retrieve/pii/S001910352500137X}.

\bibitem[Oza and Johnson(2024)]{Oza2024CommonComets}
A.~V. Oza and R.~E. Johnson.
\newblock {Common origin of trapped volatiles in oxidized icy moons and comets}.
\newblock \emph{Icarus}, 411:\penalty0 115944, 3 2024.
\newblock ISSN 0019-1035.
\newblock \doi{10.1016/J.ICARUS.2024.115944}.

\bibitem[Oza et~al.(2019)Oza, Johnson, Lellouch, Schmidt, Schneider, Huang, Gamborino, Gebek, Wyttenbach, Demory, Mordasini, Saxena, Dubois, Moullet, and Thomas]{Oza2019}
A.~V. Oza, R.~E. Johnson, E.~Lellouch, C.~Schmidt, N.~Schneider, C.~Huang, D.~Gamborino, A.~Gebek, A.~Wyttenbach, B.-O. Demory, C.~Mordasini, P.~Saxena, D.~Dubois, A.~Moullet, and N.~Thomas.
\newblock {Sodium and Potassium Signatures of Volcanic Satellites Orbiting Close-in Gas Giant Exoplanets}.
\newblock \emph{The Astrophysical Journal}, 885\penalty0 (2):\penalty0 168, 11 2019.
\newblock ISSN 0004-637X.
\newblock \doi{10.3847/1538-4357/AB40CC}.
\newblock URL \url{https://iopscience.iop.org/article/10.3847/1538-4357/ab40cc https://iopscience.iop.org/article/10.3847/1538-4357/ab40cc/meta}.

\bibitem[Oza et~al.(2024)Oza, Seidel, Hoeijmakers, Unni, Kesseli, Schmidt, Thirupathi, Bello-Arufe, Gebek, Westram, Sousa, Lopes, Hu, de~Kleer, Fisher, Charnoz, Baker, Halverson, Schneider, Psaridi, Wyttenbach, Torres, Bhatnagar, and Johnson]{Oza2024RedshiftedTransit}
A.~V. Oza, J.~V. Seidel, H.~J. Hoeijmakers, A.~Unni, A.~Y. Kesseli, C.~A. Schmidt, S.~Thirupathi, A.~Bello-Arufe, A.~Gebek, M.~M.~z. Westram, S.~G. Sousa, R.~M.~C. Lopes, R.~Hu, K.~de~Kleer, C.~Fisher, S.~Charnoz, A.~D. Baker, S.~P. Halverson, N.~M. Schneider, A.~Psaridi, A.~Wyttenbach, S.~Torres, I.~Bhatnagar, and R.~E. Johnson.
\newblock {Redshifted Sodium Transient near Exoplanet Transit}.
\newblock \emph{The Astrophysical Journal Letters}, 973\penalty0 (2):\penalty0 L53, 9 2024.
\newblock ISSN 2041-8205.
\newblock \doi{10.3847/2041-8213/AD6B29}.
\newblock URL \url{https://iopscience.iop.org/article/10.3847/2041-8213/ad6b29 https://iopscience.iop.org/article/10.3847/2041-8213/ad6b29/meta}.

\bibitem[{Oza} et~al.(2026){Oza}, {Gebek}, {Meyer zu Westram}, {Tokadjian}, {Piro}, {Hu}, {Unni}, {Chari}, {Bello-Arufe}, {Schmidt}, {Louca}, {Miguel}, {Estrela}, {Yang}, {Damiano}, {Hasegawa}, {Welbanks}, {Powell}, {Garg}, {Gupta}, {Yung}, and {Lopes}]{Oza2026}
A.~V. {Oza}, A.~{Gebek}, M.~{Meyer zu Westram}, A.~{Tokadjian}, A.~L. {Piro}, R.~{Hu}, A.~{Unni}, R.~{Chari}, A.~{Bello-Arufe}, C.~A. {Schmidt}, A.~J. {Louca}, Y.~{Miguel}, R.~{Estrela}, J.~{Yang}, M.~{Damiano}, Y.~{Hasegawa}, L.~{Welbanks}, D.~{Powell}, R.~{Garg}, P.~{Gupta}, Y.~L. {Yung}, and R.~M.~C. {Lopes}.
\newblock {Volcanic satellites tidally venting Na, K, and SO$_{2}$ in optical and infrared light}.
\newblock \emph{\mnras}, 546\penalty0 (1):\penalty0 staf1526, Feb. 2026.
\newblock \doi{10.1093/mnras/staf1526}.

\bibitem[Pappalardo and Coon(1996)]{Pappalardo1996}
R.~Pappalardo and M.~D. Coon.
\newblock {A sea ice analog for the surface of Europa}.
\newblock \emph{Lunar and Planetary Science}, 27:\penalty0 997--998, 1996.

\bibitem[Pappalardo et~al.(2024)Pappalardo, Buratti, Korth, Senske, Blaney, Blankenship, Burch, Christensen, Kempf, Kivelson, Mazarico, Retherford, Turtle, Westlake, Paczkowski, Ray, Kampmeier, Craft, Howell, Klima, Leonard, Matiella Novak, Phillips, Daubar, Blacksberg, Brooks, Choukroun, Cochrane, Diniega, Elder, Ernst, Gudipati, Luspay-Kuti, Piqueux, Rymer, Roberts, Steinbr{\"{u}}gge, Cable, Scully, Castillo-Rogez, Hay, Persaud, Glein, McKinnon, Moore, Raymond, Schroeder, Vance, Wyrick, Zolotov, Hand, Nimmo, McGrath, Spencer, Lunine, Paty, Soderblom, Collins, Schmidt, Rathbun, Shock, Becker, Hayes, Prockter, Weiss, Hibbitts, Moussessian, Brockwell, Hsu, Jia, Gladstone, McEwen, Patterson, McNutt, Evans, Larson, Cangahuala, Havens, Buffington, Bradley, Campagnola, Hardman, Srinivasan, Short, Jedrey, St.~Vaughn, Clark, Vertesi, and Niebur]{Pappalardo2024}
R.~T. Pappalardo, B.~J. Buratti, H.~Korth, D.~A. Senske, D.~L. Blaney, D.~D. Blankenship, J.~L. Burch, P.~R. Christensen, S.~Kempf, M.~G. Kivelson, E.~Mazarico, K.~D. Retherford, E.~P. Turtle, J.~H. Westlake, B.~G. Paczkowski, T.~L. Ray, J.~Kampmeier, K.~L. Craft, S.~M. Howell, R.~L. Klima, E.~J. Leonard, A.~Matiella Novak, C.~B. Phillips, I.~J. Daubar, J.~Blacksberg, S.~M. Brooks, M.~N. Choukroun, C.~J. Cochrane, S.~Diniega, C.~M. Elder, C.~M. Ernst, M.~S. Gudipati, A.~Luspay-Kuti, S.~Piqueux, A.~M. Rymer, J.~H. Roberts, G.~Steinbr{\"{u}}gge, M.~L. Cable, J.~E.~C. Scully, J.~C. Castillo-Rogez, H.~C. F.~C. Hay, D.~M. Persaud, C.~R. Glein, W.~B. McKinnon, J.~M. Moore, C.~A. Raymond, D.~M. Schroeder, S.~D. Vance, D.~Y. Wyrick, M.~Y. Zolotov, K.~P. Hand, F.~Nimmo, M.~A. McGrath, J.~R. Spencer, J.~I. Lunine, C.~S. Paty, J.~M. Soderblom, G.~C. Collins, B.~E. Schmidt, J.~A. Rathbun, E.~L. Shock, T.~C. Becker, A.~G. Hayes, L.~M. Prockter, B.~P. Weiss, C.~A. Hibbitts, A.~Moussessian, T.~G. Brockwell, H.-W. Hsu,
  X.~Jia, G.~R. Gladstone, A.~S. McEwen, G.~W. Patterson, R.~L. McNutt, J.~P. Evans, T.~W. Larson, L.~A. Cangahuala, G.~G. Havens, B.~B. Buffington, B.~Bradley, S.~Campagnola, S.~H. Hardman, J.~M. Srinivasan, K.~L. Short, T.~C. Jedrey, J.~A. St.~Vaughn, K.~P. Clark, J.~Vertesi, and C.~Niebur.
\newblock {Science Overview of the Europa Clipper Mission}.
\newblock \emph{SSR}, 220\penalty0 (4):\penalty0 1--58, 5 2024.
\newblock ISSN 1572-9672.
\newblock \doi{10.1007/S11214-024-01070-5}.
\newblock URL \url{https://link.springer.com/article/10.1007/s11214-024-01070-5}.

\bibitem[Park et~al.(2025)Park, Jacobson, Gomez~Casajus, Nimmo, Ermakov, Keane, McKinnon, Stevenson, Akiba, Idini, Buccino, Magnanini, Parisi, Tortora, Zannoni, Mura, Durante, Iess, Connerney, Levin, and Bolton]{Park2025}
R.~S. Park, R.~A. Jacobson, L.~Gomez~Casajus, F.~Nimmo, A.~I. Ermakov, J.~T. Keane, W.~B. McKinnon, D.~J. Stevenson, R.~Akiba, B.~Idini, D.~R. Buccino, A.~Magnanini, M.~Parisi, P.~Tortora, M.~Zannoni, A.~Mura, D.~Durante, L.~Iess, J.~E. Connerney, S.~M. Levin, and S.~J. Bolton.
\newblock {Io’s tidal response precludes a shallow magma ocean}.
\newblock \emph{Nature}, 638:\penalty0 69--73, 2 2025.
\newblock ISSN 1476-4687.
\newblock \doi{10.1038/s41586-024-08442-5}.
\newblock URL \url{https://www.nature.com/articles/s41586-024-08442-5}.

\bibitem[Pasteur(1848)]{pasteur1848relations}
L.~Pasteur.
\newblock Sur les relations qui peuvent exister entre la forme crystalline, la composition chimique et le sens de la polarization rotatoire.
\newblock \emph{Annales Chimie Phys.}, 24:\penalty0 442--459, 1848.

\bibitem[Patty et~al.(2021)Patty, K{\"{u}}hn, Lambrev, Spadaccia, Jens~Hoeijmakers, Keller, Mulder, Pallichadath, Poch, Snik, Stam, Pommerol, and Demory]{Patty2021BiosignaturesLife}
C.~H. Patty, J.~G. K{\"{u}}hn, P.~H. Lambrev, S.~Spadaccia, H.~Jens~Hoeijmakers, C.~Keller, W.~Mulder, V.~Pallichadath, O.~Poch, F.~Snik, D.~M. Stam, A.~Pommerol, and B.~O. Demory.
\newblock {Biosignatures of the Earth: I. Airborne spectropolarimetric detection of photosynthetic life}.
\newblock \emph{Astronomy and Astrophysics}, 651, 7 2021.
\newblock ISSN 14320746.
\newblock \doi{10.1051/0004-6361/202140845}.

\bibitem[Patty et~al.(2017)Patty, Visser, Ariese, Buma, Sparks, van Spanning, R{\"o}ling, and Snik]{patty2017circular}
C.~L. Patty, L.~J. Visser, F.~Ariese, W.~J. Buma, W.~B. Sparks, R.~J. van Spanning, W.~F. R{\"o}ling, and F.~Snik.
\newblock Circular spectropolarimetric sensing of chiral photosystems in decaying leaves.
\newblock \emph{Journal of Quantitative Spectroscopy and Radiative Transfer}, 189:\penalty0 303--311, 2017.

\bibitem[Patty et~al.(2019)Patty, Ten~Kate, Buma, Van~Spanning, Steinbach, Ariese, and Snik]{patty2019circular}
C.~L. Patty, I.~L. Ten~Kate, W.~J. Buma, R.~J. Van~Spanning, G.~Steinbach, F.~Ariese, and F.~Snik.
\newblock Circular spectropolarimetric sensing of vegetation in the field: possibilities for the remote detection of extraterrestrial life.
\newblock \emph{Astrobiology}, 19\penalty0 (10):\penalty0 1221--1229, 2019.

\bibitem[Peale et~al.(1979)Peale, Cassen, and Reynolds]{peale1979}
S.~J. Peale, P.~Cassen, and R.~T. Reynolds.
\newblock Melting of io by tidal dissipation.
\newblock \emph{Science}, 203\penalty0 (4383):\penalty0 892--894, 1979.

\bibitem[Perez et~al.(2013)Perez, Cardenas, Martin, and Michel]{perez2013potential}
N.~Perez, R.~Cardenas, O.~Martin, and L.-M. Michel.
\newblock The potential for photosynthesis in hydrothermal vents: a new avenue for life in the universe?
\newblock \emph{Astrophysics and Space Science}, 346:\penalty0 327--331, 2013.

\bibitem[Peter et~al.(2024)Peter, Nordheim, and Hand]{peter2024detection}
J.~S. Peter, T.~A. Nordheim, and K.~P. Hand.
\newblock Detection of hcn and diverse redox chemistry in the plume of enceladus.
\newblock \emph{Nature Astronomy}, 8\penalty0 (2):\penalty0 164--173, 2024.

\bibitem[Petricca et~al.(2025{\natexlab{a}})Petricca, Castillo-Rogez, Genova, Melwani~Daswani, Styczinski, Cochrane, and Vance]{Petricca2025}
F.~Petricca, J.~C. Castillo-Rogez, A.~Genova, M.~Melwani~Daswani, M.~J. Styczinski, C.~J. Cochrane, and S.~D. Vance.
\newblock {Partial differentiation of Europa and implications for the origin of materials in the Jupiter system}.
\newblock \emph{Nature Astronomy}, 9\penalty0 (4):\penalty0 501--511, 1 2025{\natexlab{a}}.
\newblock ISSN 2397-3366.
\newblock \doi{10.1038/s41550-024-02469-4}.
\newblock URL \url{https://www.nature.com/articles/s41550-024-02469-4}.

\bibitem[Petricca et~al.(2025{\natexlab{b}})Petricca, Vance, Parisi, Buccino, Cascioli, Castillo-Rogez, Downey, Nimmo, Tobie, Journaux, Magnanini, Jones, Panning, Bagheri, Genova, and Lunine]{PetriccaTitans2025}
F.~Petricca, S.~D. Vance, M.~Parisi, D.~Buccino, G.~Cascioli, J.~Castillo-Rogez, B.~G. Downey, F.~Nimmo, G.~Tobie, B.~Journaux, A.~Magnanini, U.~Jones, M.~Panning, A.~Bagheri, A.~Genova, and J.~I. Lunine.
\newblock Titan's strong tidal dissipation precludes a subsurface ocean.
\newblock \emph{Nature}, 648\penalty0 (8094):\penalty0 556--561, Dec. 2025{\natexlab{b}}.
\newblock ISSN 1476-4687.
\newblock \doi{10.1038/s41586-025-09818-x}.
\newblock URL \url{https://doi.org/10.1038/s41586-025-09818-x}.

\bibitem[Pettine et~al.(2024)Pettine, Imbeah, Rathbun, Hayes, Lopes, Mura, Tosi, Zambon, and Bertolino]{Pettine2024}
M.~Pettine, S.~Imbeah, J.~Rathbun, A.~Hayes, R.~Lopes, A.~Mura, F.~Tosi, F.~Zambon, and S.~Bertolino.
\newblock Jiram observations of volcanic flux on io: Distribution and comparison to tidal heat flow models.
\newblock \emph{Geophysical Research Letters}, 51\penalty0 (17):\penalty0 e2023GL105782, 2024.

\bibitem[Phal et~al.(2021)Phal, Yeh, and Bhargava]{Phal2021ConcurrentImaging}
Y.~Phal, K.~Yeh, and R.~Bhargava.
\newblock {Concurrent vibrational circular dichroism measurements with infrared spectroscopic imaging}.
\newblock \emph{Analytical Chemistry}, 93\penalty0 (3):\penalty0 1294--1303, 1 2021.
\newblock ISSN 15206882.
\newblock \doi{10.1021/ACS.ANALCHEM.0C00323/SUPPL{\_}FILE/AC0C00323{\_}SI{\_}001.PDF}.
\newblock URL \url{https://pubs.acs.org/doi/abs/10.1021/acs.analchem.0c00323}.

\bibitem[Phal(2023)]{Phal2023QuantumCapabilities}
Y.~D. Phal.
\newblock \emph{{Quantum cascade laser-based mid-infrared spectroscopic imaging systems with polarization capabilities}}.
\newblock PhD thesis, 7 2023.
\newblock URL \url{https://hdl.handle.net/2142/121239}.

\bibitem[Piqueux and Christensen(2008)]{Piqueux_2008}
S.~Piqueux and P.~R. Christensen.
\newblock North and south subice gas flow and venting of the seasonal caps of mars: A major geomorphological agent.
\newblock \emph{Journal of Geophysical Research: Planets}, 113\penalty0 (E6):\penalty0 E06005, 2008.
\newblock \doi{10.1029/2007JE002969}.
\newblock URL \url{https://doi.org/10.1029/2007JE002969}.

\bibitem[Piqueux et~al.(2003)Piqueux, Kieffer, and Titus]{Piqueux_2003}
S.~Piqueux, H.~H. Kieffer, and T.~N. Titus.
\newblock Surface energy balance and CO$_2$ sublimation modeling of the south seasonal polar cap of mars.
\newblock \emph{Journal of Geophysical Research: Planets}, 108\penalty0 (E8):\penalty0 5084, 2003.
\newblock \doi{10.1029/2002JE002007}.
\newblock URL \url{https://doi.org/10.1029/2002JE002007}.

\bibitem[Pizzarello and Cronin(2000)]{pizzarello2000non}
S.~Pizzarello and J.~Cronin.
\newblock Non-racemic amino acids in the murray and murchison meteorites.
\newblock \emph{Geochimica et Cosmochimica Acta}, 64\penalty0 (2):\penalty0 329--338, 2000.

\bibitem[Pizzarello et~al.(2012)Pizzarello, Schrader, Monroe, and Lauretta]{pizzarello2012large}
S.~Pizzarello, D.~L. Schrader, A.~A. Monroe, and D.~S. Lauretta.
\newblock Large enantiomeric excesses in primitive meteorites and the diverse effects of water in cosmochemical evolution.
\newblock \emph{Proceedings of the National Academy of Sciences}, 109\penalty0 (30):\penalty0 11949--11954, 2012.

\bibitem[Plainaki et~al.(2018)Plainaki, Cassidy, Shematovich, Milillo, Wurz, Vorburger, Roth, Galli, Rubin, Bl{\"o}cker, et~al.]{plainaki2018}
C.~Plainaki, T.~A. Cassidy, V.~I. Shematovich, A.~Milillo, P.~Wurz, A.~Vorburger, L.~Roth, A.~Galli, M.~Rubin, A.~Bl{\"o}cker, et~al.
\newblock Towards a global unified model of europa’s tenuous atmosphere.
\newblock \emph{Space science reviews}, 214:\penalty0 1--71, 2018.

\bibitem[Poch et~al.(2018)Poch, Cerubini, Pommerol, Jost, and Thomas]{Poch2018}
O.~Poch, R.~Cerubini, A.~Pommerol, B.~Jost, and N.~Thomas.
\newblock {Polarimetry of Water Ice Particles Providing Insights on Grain Size and Degree of Sintering on Icy Planetary Surfaces}.
\newblock \emph{Journal of Geophysical Research: Planets}, 123\penalty0 (10):\penalty0 2564--2584, 10 2018.
\newblock ISSN 2169-9100.
\newblock \doi{10.1029/2018JE005753}.
\newblock URL \url{https://onlinelibrary.wiley.com/doi/full/10.1029/2018JE005753 https://onlinelibrary.wiley.com/doi/abs/10.1029/2018JE005753 https://agupubs.onlinelibrary.wiley.com/doi/10.1029/2018JE005753}.

\bibitem[{Pollack} et~al.(1978){Pollack}, {Witteborn}, {Erickson}, {Strecker}, {Baldwin}, and {Bunch}]{1978Icar...36..271P}
J.~B. {Pollack}, F.~C. {Witteborn}, E.~F. {Erickson}, D.~W. {Strecker}, B.~J. {Baldwin}, and T.~E. {Bunch}.
\newblock {Near-infrared spectra of the Galilean satellites: Observations and compositional implications}.
\newblock \emph{\icarus}, 36\penalty0 (3):\penalty0 271--303, Dec. 1978.
\newblock \doi{10.1016/0019-1035(78)90110-0}.

\bibitem[Pommerol et~al.(2019)Pommerol, Jost, Poch, Yoldi, Brouet, Gracia-Bern{\'{a}}, Cerubini, Galli, Wurz, Gundlach, Blum, Carrasco, Szopa, and Thomas]{Pommerol2019ExperimentingMaterial}
A.~Pommerol, B.~Jost, O.~Poch, Z.~Yoldi, Y.~Brouet, A.~Gracia-Bern{\'{a}}, R.~Cerubini, A.~Galli, P.~Wurz, B.~Gundlach, J.~Blum, N.~Carrasco, C.~Szopa, and N.~Thomas.
\newblock {Experimenting with Mixtures of Water Ice and Dust as Analogues for Icy Planetary Material}.
\newblock \emph{Space Science Reviews 2019 215:5}, 215\penalty0 (5):\penalty0 1--68, 6 2019.
\newblock ISSN 1572-9672.
\newblock \doi{10.1007/S11214-019-0603-0}.
\newblock URL \url{https://link.springer.com/article/10.1007/s11214-019-0603-0}.

\bibitem[Porco et~al.(2014)Porco, Dinino, and Nimmo]{Porco2014HOWRELATED}
C.~Porco, D.~Dinino, and F.~Nimmo.
\newblock {HOW THE GEYSERS, TIDAL STRESSES, AND THERMAL EMISSION ACROSS THE SOUTH POLAR TERRAIN OF ENCELADUS ARE RELATED}.
\newblock \emph{The Astronomical Journal}, 148\penalty0 (3):\penalty0 45, 7 2014.
\newblock ISSN 1538-3881.
\newblock \doi{10.1088/0004-6256/148/3/45}.
\newblock URL \url{https://iopscience.iop.org/article/10.1088/0004-6256/148/3/45 https://iopscience.iop.org/article/10.1088/0004-6256/148/3/45/meta}.

\bibitem[Porco et~al.(2006)Porco, Helfenstein, Thomas, Ingersoll, Wisdom, West, Neukum, Denk, Wagner, Roatsch, Kieffer, Turtle, McEwen, Johnson, Rathbun, Veverka, Wilson, Perry, Spitale, Brahic, Burns, DelGenio, Dones, Murray, and Squyres]{Porco2006}
C.~C. Porco, P.~Helfenstein, P.~C. Thomas, A.~P. Ingersoll, J.~Wisdom, R.~West, G.~Neukum, T.~Denk, R.~Wagner, T.~Roatsch, S.~Kieffer, E.~Turtle, A.~McEwen, T.~V. Johnson, J.~Rathbun, J.~Veverka, D.~Wilson, J.~Perry, J.~Spitale, A.~Brahic, J.~A. Burns, A.~D. DelGenio, L.~Dones, C.~D. Murray, and S.~Squyres.
\newblock {Cassini observes the active south pole of enceladus}.
\newblock \emph{Science}, 311\penalty0 (5766):\penalty0 1393--1401, 3 2006.
\newblock ISSN 00368075.
\newblock \doi{10.1126/science.1123013}.

\bibitem[Porco et~al.(2017)Porco, Dones, and Mitchell]{porco2017could}
C.~C. Porco, L.~Dones, and C.~Mitchell.
\newblock Could it be snowing microbes on enceladus? assessing conditions in its plume and implications for future missions.
\newblock \emph{Astrobiology}, 17\penalty0 (9):\penalty0 876--901, 2017.

\bibitem[Pospergelis(1969)]{pospergelis1969spectroscopic}
M.~Pospergelis.
\newblock Spectroscopic measurements of the four stokes parameters for light scattered by natural objects.
\newblock \emph{Soviet Astronomy}, 12:\penalty0 973, 1969.

\bibitem[{Postberg} et~al.(2006){Postberg}, {Frank}, {Kempf}, {Sascha}, {Srama}, {Ralf}, {Green}, F., {Hillier}, K., {McBride}, {Neil}, {Gr{\"{u}}n}, and {Eberhard}]{Postberg2006}
{Postberg}, {Frank}, {Kempf}, {Sascha}, {Srama}, {Ralf}, {Green}, S.~F., {Hillier}, J.~K., {McBride}, {Neil}, {Gr{\"{u}}n}, and {Eberhard}.
\newblock {Composition of jovian dust stream particles}.
\newblock \emph{Icarus}, 183\penalty0 (1):\penalty0 122--134, 7 2006.
\newblock ISSN 00191035.
\newblock \doi{10.1016/J.ICARUS.2006.02.001}.
\newblock URL \url{https://scixplorer.org/abs/2006Icar..183..122P/abstract}.

\bibitem[Postberg et~al.(2009)Postberg, Kempf, Schmidt, Brilliantov, Beinsen, Abel, Buck, and Srama]{Postberg2009}
F.~Postberg, S.~Kempf, J.~Schmidt, N.~Brilliantov, A.~Beinsen, B.~Abel, U.~Buck, and R.~Srama.
\newblock {Sodium salts in E-ring ice grains from an ocean below the surface of Enceladus}.
\newblock \emph{Nature}, 459\penalty0 (7250):\penalty0 1098--1101, 6 2009.
\newblock ISSN 00280836.
\newblock \doi{10.1038/NATURE08046;KWRD}.
\newblock URL \url{https://www.nature.com/articles/nature08046}.

\bibitem[Postberg et~al.(2011)Postberg, Schmidt, Hillier, Kempf, and Srama]{Postberg2011}
F.~Postberg, J.~Schmidt, J.~Hillier, S.~Kempf, and R.~Srama.
\newblock {A salt-water reservoir as the source of a compositionally stratified plume on Enceladus}.
\newblock \emph{Nature}, 474\penalty0 (7353):\penalty0 620--622, 6 2011.
\newblock ISSN 00280836.
\newblock \doi{10.1038/NATURE10175;SUBJMETA}.
\newblock URL \url{https://www.nature.com/articles/nature10175}.

\bibitem[Postberg et~al.(2018{\natexlab{a}})Postberg, Clark, Hansen, Coates, Dalle~Ore, Scipioni, Hedman, Waite, Postberg, Clark, Hansen, Coates, Dalle~Ore, Scipioni, Hedman, and Waite]{Postberg2018}
F.~Postberg, R.~N. Clark, C.~J. Hansen, A.~J. Coates, C.~M. Dalle~Ore, F.~Scipioni, M.~M. Hedman, J.~H. Waite, F.~Postberg, R.~N. Clark, C.~J. Hansen, A.~J. Coates, C.~M. Dalle~Ore, F.~Scipioni, M.~M. Hedman, and J.~H. Waite.
\newblock {Plume and Surface Composition of Enceladus}.
\newblock In P.~Schenk, R.~Clark, C.~Howett, A.~Verbiscer, and J.~Waite, editors, \emph{Enceladus and the Icy Moons of Saturn}, pages 129--162. The University of Arizona Press, 2018{\natexlab{a}}.
\newblock \doi{10.2458/AZU{\_}UAPRESS{\_}9780816537075-CH007}.
\newblock URL \url{https://ui.adsabs.harvard.edu/abs/2018eims.book..129P/abstract}.

\bibitem[Postberg et~al.(2018{\natexlab{b}})Postberg, Khawaja, Abel, Choblet, Glein, Gudipati, Henderson, Hsu, Kempf, Klenner, et~al.]{postberg2018macromolecular}
F.~Postberg, N.~Khawaja, B.~Abel, G.~Choblet, C.~R. Glein, M.~S. Gudipati, B.~L. Henderson, H.-W. Hsu, S.~Kempf, F.~Klenner, et~al.
\newblock Macromolecular organic compounds from the depths of enceladus.
\newblock \emph{Nature}, 558\penalty0 (7711):\penalty0 564--568, 2018{\natexlab{b}}.

\bibitem[Postberg et~al.(2023)Postberg, Sekine, Klenner, Glein, Zou, Abel, Furuya, Hillier, Khawaja, Kempf, Noelle, Saito, Schmidt, Shibuya, Srama, and Tan]{Postberg2023}
F.~Postberg, Y.~Sekine, F.~Klenner, C.~R. Glein, Z.~Zou, B.~Abel, K.~Furuya, J.~K. Hillier, N.~Khawaja, S.~Kempf, L.~Noelle, T.~Saito, J.~Schmidt, T.~Shibuya, R.~Srama, and S.~Tan.
\newblock {Detection of phosphates originating from Enceladus’s ocean}.
\newblock \emph{Nature}, 618\penalty0 (7965):\penalty0 489--493, 6 2023.
\newblock ISSN 14764687.
\newblock \doi{10.1038/S41586-023-05987-9;SUBJMETA}.
\newblock URL \url{https://www.nature.com/articles/s41586-023-05987-9}.

\bibitem[{Poston} et~al.(2017){Poston}, {Carlson}, and {Hand}]{2017JGRE..122.2644P}
M.~J. {Poston}, R.~W. {Carlson}, and K.~P. {Hand}.
\newblock {Spectral Behavior of Irradiated Sodium Chloride Crystals Under Europa-Like Conditions}.
\newblock \emph{Journal of Geophysical Research (Planets)}, 122\penalty0 (12):\penalty0 2644--2654, Dec. 2017.
\newblock \doi{10.1002/2017JE005429}.

\bibitem[Poston et~al.(2017)Poston, Carlson, and Hand]{Poston2017}
M.~J. Poston, R.~W. Carlson, and K.~P. Hand.
\newblock {Spectral Behavior of Irradiated Sodium Chloride Crystals Under Europa-Like Conditions}.
\newblock \emph{Journal of Geophysical Research: Planets}, 122\penalty0 (12):\penalty0 2644--2654, 12 2017.
\newblock ISSN 2169-9100.
\newblock \doi{10.1002/2017JE005429}.
\newblock URL \url{/doi/pdf/10.1002/2017JE005429 https://onlinelibrary.wiley.com/doi/abs/10.1002/2017JE005429 https://agupubs.onlinelibrary.wiley.com/doi/10.1002/2017JE005429}.

\bibitem[Prockter and Patterson(2009)]{ProckterPatterson2009}
L.~M. Prockter and G.~W. Patterson.
\newblock {Morphology and Evolution of Europa’s Ridges and Bands}.
\newblock \emph{in: Europa (Pappalardo et al.)}, pages 237--258, 2009.
\newblock \doi{10.2307/j.ctt1xp3wdw.16}.

\bibitem[{Quarles} et~al.(2020){Quarles}, {Li}, and {Rosario-Franco}]{2020ApJ...902L..20Q}
B.~{Quarles}, G.~{Li}, and M.~{Rosario-Franco}.
\newblock {Application of Orbital Stability and Tidal Migration Constraints for Exomoon Candidates}.
\newblock \emph{\apjl}, 902\penalty0 (1):\penalty0 L20, Oct. 2020.

\bibitem[Quick and Hedman(2020)]{Quick2020CharacterizingEuropa}
L.~C. Quick and M.~M. Hedman.
\newblock {Characterizing deposits emplaced by cryovolcanic plumes on Europa}.
\newblock \emph{Icarus}, 343, 6 2020.
\newblock ISSN 10902643.
\newblock \doi{10.1016/J.ICARUS.2020.113667}.

\bibitem[Quick et~al.(2013)Quick, Barnouin, Prockter, and Patterson]{Quick2013}
L.~C. Quick, O.~S. Barnouin, L.~M. Prockter, and G.~W. Patterson.
\newblock {Constraints on the detection of cryovolcanic plumes on Europa}.
\newblock \emph{Planetary and Space Science}, 86:\penalty0 1--9, 9 2013.
\newblock ISSN 0032-0633.
\newblock \doi{10.1016/J.PSS.2013.06.028}.

\bibitem[Quirico et~al.(1999)Quirico, Douté, Schmitt, {de Bergh}, Cruikshank, Owen, Geballe, and Roush]{Quirico_1999}
E.~Quirico, S.~Douté, B.~Schmitt, C.~{de Bergh}, D.~P. Cruikshank, T.~C. Owen, T.~R. Geballe, and T.~L. Roush.
\newblock Composition, physical state, and distribution of ices at the surface of triton.
\newblock \emph{Icarus}, 139\penalty0 (2):\penalty0 159--178, 1999.
\newblock ISSN 0019-1035.
\newblock \doi{https://doi.org/10.1006/icar.1999.6111}.
\newblock URL \url{https://www.sciencedirect.com/science/article/pii/S0019103599961110}.

\bibitem[Radebaugh et~al.(2004)Radebaugh, McEwen, Milazzo, Keszthelyi, Davies, Turtle, and Dawson]{radebaugh2004}
J.~Radebaugh, A.~S. McEwen, M.~P. Milazzo, L.~P. Keszthelyi, A.~G. Davies, E.~P. Turtle, and D.~D. Dawson.
\newblock Observations and temperatures of io's pele patera from cassini and galileo spacecraft images.
\newblock \emph{Icarus}, 169\penalty0 (1):\penalty0 65--79, 2004.

\bibitem[Ramirez and Kaltenegger(2017)]{ramirez2017volcanic}
R.~M. Ramirez and L.~Kaltenegger.
\newblock A volcanic hydrogen habitable zone.
\newblock \emph{The Astrophysical Journal Letters}, 837\penalty0 (1):\penalty0 L4, 2017.

\bibitem[{Rathbun} et~al.(2010){Rathbun}, {Rodriguez}, and {Spencer}]{2010Icar..210..763R}
J.~A. {Rathbun}, N.~J. {Rodriguez}, and J.~R. {Spencer}.
\newblock {Galileo PPR observations of Europa: Hotspot detection limits and surface thermal properties}.
\newblock \emph{\icarus}, 210\penalty0 (2):\penalty0 763--769, Dec. 2010.
\newblock \doi{10.1016/j.icarus.2010.07.017}.

\bibitem[Rathbun et~al.(2010)Rathbun, Rodriguez, and Spencer]{Rathbun2010}
J.~A. Rathbun, N.~J. Rodriguez, and J.~R. Spencer.
\newblock Galileo ppr observations of europa: Hotspot detection limits and surface thermal properties.
\newblock \emph{Icarus}, 210\penalty0 (2):\penalty0 763--769, 2010.

\bibitem[Ray et~al.(2024)Ray, Ar{\'e}valo~Jr, Southard, Willhite, Bardyn, Ni, Danell, Grubisic, Gundersen, Llano, et~al.]{ray2024characterization}
S.~Ray, R.~Ar{\'e}valo~Jr, A.~Southard, L.~Willhite, A.~Bardyn, Z.~Ni, R.~Danell, A.~Grubisic, C.~Gundersen, J.~Llano, et~al.
\newblock Characterization of regolith and trace economic resources (crater): An orbitrap-based laser desorption mass spectrometry instrument for in situ exploration of the moon.
\newblock \emph{Rapid Communications in Mass Spectrometry}, 38\penalty0 (6):\penalty0 e9657, 2024.

\bibitem[Redwing et~al.(2022)Redwing, de~Pater, Luszcz-Cook, de~Kleer, Moullet, and Rojo]{Redwing2022}
E.~Redwing, I.~de~Pater, S.~Luszcz-Cook, K.~de~Kleer, A.~Moullet, and P.~M. Rojo.
\newblock {NaCl and KCl in Io’s Atmosphere}.
\newblock \emph{The Planetary Science Journal}, 3\penalty0 (10):\penalty0 238, 10 2022.
\newblock ISSN 2632-3338.
\newblock \doi{10.3847/PSJ/AC9784}.
\newblock URL \url{https://iopscience.iop.org/article/10.3847/PSJ/ac9784 https://iopscience.iop.org/article/10.3847/PSJ/ac9784/meta}.

\bibitem[Reinhardt et~al.(2020)Reinhardt, Chau, Stadel, and Helled]{Reinhardt2020}
C.~Reinhardt, A.~Chau, J.~Stadel, and R.~Helled.
\newblock {Bifurcation in the history of Uranus and Neptune: the role of giant impacts}.
\newblock \emph{Monthly Notices of the Royal Astronomical Society}, 492\penalty0 (4):\penalty0 5336--5353, 3 2020.
\newblock ISSN 0035-8711.
\newblock \doi{10.1093/MNRAS/STZ3271}.
\newblock URL \url{https://dx.doi.org/10.1093/mnras/stz3271}.

\bibitem[Remick and Helmann(2023)]{remick2023elements}
K.~A. Remick and J.~D. Helmann.
\newblock The elements of life: A biocentric tour of the periodic table.
\newblock In \emph{Advances in Microbial Physiology}, volume~82, pages 1--127. Elsevier, 2023.

\bibitem[Ren et~al.(2018)Ren, Guo, Cheng, Wang, Sun, Zhang, Dong, and Li]{Ren2018}
Z.~Ren, M.~Guo, Y.~Cheng, Y.~Wang, W.~Sun, H.~Zhang, M.~Dong, and G.~Li.
\newblock A review of the development and application of space miniature mass spectrometers.
\newblock \emph{Vacuum}, 155:\penalty0 108--117, 2018.
\newblock ISSN 0042-207X.
\newblock \doi{https://doi.org/10.1016/j.vacuum.2018.05.048}.
\newblock URL \url{https://www.sciencedirect.com/science/article/pii/S0042207X18302458}.

\bibitem[Retherford et~al.(2007)Retherford, Spencer, Stern, Saur, Strobel, Steffl, Gladstone, Weaver, Cheng, Parker, et~al.]{retherford2007}
K.~Retherford, J.~Spencer, S.~Stern, J.~Saur, D.~Strobel, A.~Steffl, G.~Gladstone, H.~Weaver, A.~Cheng, J.~W. Parker, et~al.
\newblock Io's atmospheric response to eclipse: Uv aurorae observations.
\newblock \emph{science}, 318\penalty0 (5848):\penalty0 237--240, 2007.

\bibitem[{Revol} et~al.(2024){Revol}, {Bolmont}, {Sastre}, {Tobie}, {Libert}, {Kervazo}, and {Blanco-Cuaresma}]{2024A&A...691L...3R}
A.~{Revol}, {\'E}.~{Bolmont}, M.~{Sastre}, G.~{Tobie}, A.-S. {Libert}, M.~{Kervazo}, and S.~{Blanco-Cuaresma}.
\newblock {Drifts of the substellar points of the TRAPPIST-1 planets}.
\newblock \emph{\aap}, 691:\penalty0 L3, Nov. 2024.

\bibitem[Rhoden and Hurford(2013)]{Rhoden2013}
A.~R. Rhoden and T.~A. Hurford.
\newblock {Lineament azimuths on Europa: Implications for obliquity and non-synchronous rotation}.
\newblock \emph{Icarus}, 226\penalty0 (1):\penalty0 841--859, 9 2013.
\newblock ISSN 0019-1035.
\newblock \doi{10.1016/J.ICARUS.2013.06.029}.

\bibitem[Rhoden et~al.(2015)Rhoden, Hurford, Roth, and Retherford]{Rhoden2015}
A.~R. Rhoden, T.~A. Hurford, L.~Roth, and K.~Retherford.
\newblock {Linking Europa’s plume activity to tides, tectonics, and liquid water}.
\newblock \emph{Icarus}, 253:\penalty0 169--178, 6 2015.
\newblock ISSN 0019-1035.
\newblock \doi{10.1016/J.ICARUS.2015.02.023}.

\bibitem[Rhoden et~al.(2020)Rhoden, Hurford, Spitale, Henning, Huff, Bland, and Sajous]{Rhoden2020}
A.~R. Rhoden, T.~A. Hurford, J.~Spitale, W.~Henning, E.~M. Huff, M.~T. Bland, and S.~Sajous.
\newblock {The formation of Enceladus' Tiger Stripe Fractures from eccentricity tides}.
\newblock \emph{Earth and Planetary Science Letters}, 544:\penalty0 116389, 8 2020.
\newblock ISSN 0012-821X.
\newblock \doi{10.1016/J.EPSL.2020.116389}.

\bibitem[Rhoden et~al.(2024)Rhoden, Walker, Rudolph, Bland, and Manga]{Rhoden2024}
A.~R. Rhoden, M.~E. Walker, M.~L. Rudolph, M.~T. Bland, and M.~Manga.
\newblock {The evolution of a young ocean within Mimas}.
\newblock \emph{Earth and Planetary Science Letters}, 635:\penalty0 118689, 6 2024.
\newblock ISSN 0012-821X.
\newblock \doi{10.1016/J.EPSL.2024.118689}.

\bibitem[Riedo et~al.(2013)Riedo, Neuland, Meyer, Tulej, and Wurz]{Riedo2013}
A.~Riedo, M.~Neuland, S.~Meyer, M.~Tulej, and P.~Wurz.
\newblock Coupling of lms with a fs-laser ablation ion source: elemental and isotope composition measurements.
\newblock \emph{J. Anal. At. Spectrom.}, 28:\penalty0 1256--1269, 2013.
\newblock \doi{10.1039/C3JA50117E}.
\newblock URL \url{http://dx.doi.org/10.1039/C3JA50117E}.

\bibitem[Riedo et~al.(2025{\natexlab{a}})Riedo, Boeren, Keresztes~Schmidt, Tulej, and Wurz]{Riedo2025}
A.~Riedo, N.~Boeren, P.~Keresztes~Schmidt, M.~Tulej, and P.~Wurz.
\newblock Life detection beyond earth: Laser-based mass spectrometry for organics detection on solar system objects.
\newblock \emph{CHIMIA}, 79:\penalty0 70--76, 2025{\natexlab{a}}.
\newblock \doi{10.2533/chimia.2025.70}.

\bibitem[Riedo et~al.(2025{\natexlab{b}})Riedo, Gruchola, N.J., Keresztes~Schmidt, Knecht, Sellam, Tulej, and Wurz]{Riedo2025SSR}
A.~Riedo, S.~Gruchola, B.~N.J., P.~Keresztes~Schmidt, L.~Knecht, Y.~Sellam, M.~Tulej, and P.~Wurz.
\newblock Laser-based mass spectrometry for the detection of signatures of life within our solar system.
\newblock \emph{Sp. Sci. Rev.}, this compilation:\penalty0 submitted, 2025{\natexlab{b}}.

\bibitem[Rohner et~al.(2003)Rohner, Whitby, and Wurz]{Rohner2003}
U.~Rohner, J.~A. Whitby, and P.~Wurz.
\newblock A miniature laser ablation time-of-flight mass spectrometer for in situ planetary exploration.
\newblock \emph{Measurement Science and Technology}, 14\penalty0 (12):\penalty0 2159, oct 2003.
\newblock \doi{10.1088/0957-0233/14/12/017}.
\newblock URL \url{https://dx.doi.org/10.1088/0957-0233/14/12/017}.

\bibitem[{Ronnet} and {Johansen}(2020)]{Ronnet20}
T.~{Ronnet} and A.~{Johansen}.
\newblock {Formation of moon systems around giant planets. Capture and ablation of planetesimals as foundation for a pebble accretion scenario}.
\newblock \emph{\aap}, 633:\penalty0 A93, Jan. 2020.
\newblock \doi{10.1051/0004-6361/201936804}.

\bibitem[Roth et~al.(2014{\natexlab{a}})Roth, Retherford, Saur, Strobel, Feldman, McGrath, and Nimmo]{Roth2014b}
L.~Roth, K.~D. Retherford, J.~Saur, D.~F. Strobel, P.~D. Feldman, M.~A. McGrath, and F.~Nimmo.
\newblock {Orbital apocenter is not a sufficient condition for HST/STIS detection of Europa's water vapor aurora}.
\newblock \emph{Proceedings of the National Academy of Sciences of the United States of America}, 111\penalty0 (48):\penalty0 E5123 -- E5132, 11 2014{\natexlab{a}}.
\newblock ISSN 10916490.
\newblock \doi{10.1073/pnas.1416671111}.

\bibitem[Roth et~al.(2014{\natexlab{b}})Roth, Saur, Retherford, Feldman, and Strobel]{roth2014a}
L.~Roth, J.~Saur, K.~D. Retherford, P.~D. Feldman, and D.~F. Strobel.
\newblock A phenomenological model of io’s uv aurora based on hst/stis observations.
\newblock \emph{Icarus}, 228:\penalty0 386--406, 2014{\natexlab{b}}.

\bibitem[Roth et~al.(2014{\natexlab{c}})Roth, Saur, Retherford, Strobel, Feldman, McGrath, and Nimmo]{Roth2014}
L.~Roth, J.~Saur, K.~D. Retherford, D.~F. Strobel, P.~D. Feldman, M.~A. McGrath, and F.~Nimmo.
\newblock {Transient water vapor at Europa's south pole}.
\newblock \emph{Science}, 343\penalty0 (6167):\penalty0 171--174, 2014{\natexlab{c}}.
\newblock ISSN 10959203.
\newblock \doi{10.1126/science.1247051}.

\bibitem[Roth et~al.(2016)Roth, Saur, Retherford, Strobel, Feldman, McGrath, Spencer, Bl{\"o}cker, and Ivchenko]{Roth2016}
L.~Roth, J.~Saur, K.~D. Retherford, D.~F. Strobel, P.~D. Feldman, M.~A. McGrath, J.~R. Spencer, A.~Bl{\"o}cker, and N.~Ivchenko.
\newblock Europa's far ultraviolet oxygen aurora from a comprehensive set of hst observations.
\newblock \emph{Journal of Geophysical Research: Space Physics}, 121\penalty0 (3):\penalty0 2143--2170, 2016.

\bibitem[Roth et~al.(2017)Roth, Retherford, Ivchenko, Schlatter, Strobel, Becker, and Grava]{Roth2017}
L.~Roth, K.~D. Retherford, N.~Ivchenko, N.~Schlatter, D.~F. Strobel, T.~M. Becker, and C.~Grava.
\newblock Detection of a hydrogen corona in hst ly$\alpha$ images of europa in transit of jupiter.
\newblock \emph{The Astronomical Journal}, 153\penalty0 (2):\penalty0 67, 2017.

\bibitem[Roth et~al.(2020)Roth, Boissier, Moullet, S{\`{a}}nchez-Monge, de~Kleer, Yoneda, Hikida, Kita, Tsuchiya, Bl{\"{o}}cker, Gladstone, Grodent, Ivchenko, Lellouch, Retherford, Saur, Schilke, Strobel, and Thorwirth]{roth2020}
L.~Roth, J.~Boissier, A.~Moullet, A.~S{\`{a}}nchez-Monge, K.~de~Kleer, M.~Yoneda, R.~Hikida, H.~Kita, F.~Tsuchiya, A.~Bl{\"{o}}cker, G.~R. Gladstone, D.~Grodent, N.~Ivchenko, E.~Lellouch, K.~D. Retherford, J.~Saur, P.~Schilke, D.~Strobel, and S.~Thorwirth.
\newblock {An attempt to detect transient changes in Io’s SO2 and NaCl atmosphere}.
\newblock \emph{Icarus}, 350:\penalty0 113925, 11 2020.
\newblock ISSN 0019-1035.
\newblock \doi{10.1016/J.ICARUS.2020.113925}.

\bibitem[Rudolph et~al.(2022)Rudolph, Manga, Walker, and Rhoden]{Rudolph2022}
M.~L. Rudolph, M.~Manga, M.~Walker, and A.~R. Rhoden.
\newblock {Cooling Crusts Create Concomitant Cryovolcanic Cracks}.
\newblock \emph{Geophysical Research Letters}, 49\penalty0 (5):\penalty0 e2021GL094421, 3 2022.
\newblock ISSN 1944-8007.
\newblock \doi{10.1029/2021GL094421}.
\newblock URL \url{https://onlinelibrary.wiley.com/doi/full/10.1029/2021GL094421 https://onlinelibrary.wiley.com/doi/abs/10.1029/2021GL094421 https://agupubs.onlinelibrary.wiley.com/doi/10.1029/2021GL094421}.

\bibitem[Rudolph et~al.(2025)Rudolph, Rhoden, Manga, and Walker]{Rudolph2025}
M.~L. Rudolph, A.~R. Rhoden, M.~Manga, and M.~E. Walker.
\newblock {Ocean Underpressure and Eruptions of Liquid Water on Icy Ocean Worlds}.
\newblock In \emph{56th Lunar and Planetary Science Conference}, volume 3090 of \emph{LPI Contributions}, page 1799, 3 2025.

\bibitem[{Sagan}(1990)]{Sagan1990}
C.~{Sagan}.
\newblock {Triton's streaks as windblown dust}.
\newblock \emph{Nature}, 346:\penalty0 546--548, 1990.
\newblock \doi{10.1038/346546a0}.

\bibitem[Salmon and Canup(2017)]{Salmon2017}
J.~Salmon and R.~M. Canup.
\newblock Accretion of saturn's inner mid-sized moons from a massive primordial ice ring.
\newblock \emph{The Astrophysical Journal}, 836\penalty0 (1):\penalty0 109, 2017.

\bibitem[{Sasaki} et~al.(2010){Sasaki}, {Stewart}, and {Ida}]{Sasaki10}
T.~{Sasaki}, G.~R. {Stewart}, and S.~{Ida}.
\newblock {Origin of the Different Architectures of the Jovian and Saturnian Satellite Systems}.
\newblock \emph{\apj}, 714\penalty0 (2):\penalty0 1052--1064, May 2010.
\newblock \doi{10.1088/0004-637X/714/2/1052}.

\bibitem[Saur et~al.(2004)Saur, Neubauer, Connerney, Zarka, and Kivelson]{saur2004}
J.~Saur, F.~M. Neubauer, J.~Connerney, P.~Zarka, and M.~G. Kivelson.
\newblock Plasma interaction of io with its plasma torus.
\newblock \emph{Jupiter: The planet, satellites and magnetosphere}, 1:\penalty0 537--560, 2004.

\bibitem[Schenk and Zahnle(2007)]{Schenk_Zahnle_2007}
P.~M. Schenk and K.~Zahnle.
\newblock On the negligible surface age of triton.
\newblock \emph{Icarus}, 192\penalty0 (1):\penalty0 135--149, 2007.
\newblock ISSN 0019-1035.
\newblock \doi{https://doi.org/10.1016/j.icarus.2007.07.004}.
\newblock URL \url{https://www.sciencedirect.com/science/article/pii/S0019103507003004}.

\bibitem[Schlarmann et~al.(2024)Schlarmann, Vorburger, Carberry~Mogan, and Wurz]{Schlarmann2024}
L.~Schlarmann, A.~Vorburger, S.~R. Carberry~Mogan, and P.~Wurz.
\newblock {Investigating influences of collisions on the icy Galilean moon atmospheres}, 9 2024.
\newblock URL \url{https://meetingorganizer.copernicus.org/EPSC2024/EPSC2024-509.html}.

\bibitem[Schmidt(2022)]{Schmidt2022Doppler-ShiftedExomoons}
C.~A. Schmidt.
\newblock {Doppler-Shifted Alkali D Absorption as Indirect Evidence for Exomoons}.
\newblock \emph{FrASS}, 9:\penalty0 801873, 3 2022.
\newblock ISSN 2296987X.
\newblock \doi{10.3389/fspas.2022.801873}.

\bibitem[Schubert et~al.(2004)Schubert, Anderson, Spohn, and McKinnon]{Schubert2004}
G.~Schubert, J.~Anderson, T.~Spohn, and W.~McKinnon.
\newblock {Interior composition, structure and dynamics of the Galilean satellites}.
\newblock In F.~Bagenal, T.~E. Dowling, and W.~B. McKinnon, editors, \emph{Jupiter. The Planet, Satellites and Magnetosphere}, volume~1, pages 281--306. 2004.

\bibitem[Schubert et~al.(2009)Schubert, Sohl, and Hussmann]{Schubert2009}
G.~Schubert, F.~Sohl, and H.~Hussmann.
\newblock {Part III : Interior , Icy Shell , and Ocean}.
\newblock \emph{Europa}, 2009.

\bibitem[Schulson(2006)]{Schulson2006}
E.~M. Schulson.
\newblock {The fracture of water ice Ih: A short overview}.
\newblock \emph{Meteoritics {\&} Planetary Science}, 41\penalty0 (10):\penalty0 1497--1508, 10 2006.
\newblock ISSN 1945-5100.
\newblock \doi{10.1111/J.1945-5100.2006.TB00432.X}.
\newblock URL \url{/doi/pdf/10.1111/j.1945-5100.2006.tb00432.x https://onlinelibrary.wiley.com/doi/abs/10.1111/j.1945-5100.2006.tb00432.x https://onlinelibrary.wiley.com/doi/10.1111/j.1945-5100.2006.tb00432.x}.

\bibitem[Schwander et~al.(2022)Schwander, Ligterink, Kipfer, Lukmanov, Grimaudo, Tulej, de~Koning, Keresztes~Schmidt, Gruchola, Boeren, et~al.]{schwander2022correlation}
L.~Schwander, N.~F. Ligterink, K.~A. Kipfer, R.~A. Lukmanov, V.~Grimaudo, M.~Tulej, C.~P. de~Koning, P.~Keresztes~Schmidt, S.~Gruchola, N.~J. Boeren, et~al.
\newblock Correlation network analysis for amino acid identification in soil samples with the origin space-prototype instrument.
\newblock \emph{Frontiers in Astronomy and Space Sciences}, 9:\penalty0 909193, 2022.

\bibitem[Schwieterman et~al.(2018)Schwieterman, Kiang, Parenteau, Harman, DasSarma, Fisher, Arney, Hartnett, Reinhard, Olson, et~al.]{schwieterman2018exoplanet}
E.~W. Schwieterman, N.~Y. Kiang, M.~N. Parenteau, C.~E. Harman, S.~DasSarma, T.~M. Fisher, G.~N. Arney, H.~E. Hartnett, C.~T. Reinhard, S.~L. Olson, et~al.
\newblock Exoplanet biosignatures: a review of remotely detectable signs of life.
\newblock \emph{Astrobiology}, 18\penalty0 (6):\penalty0 663--708, 2018.

\bibitem[Scipioni et~al.(2017)Scipioni, Schenk, Tosi, D'Aversa, Clark, Combe, and Ore]{Scipioni2017}
F.~Scipioni, P.~Schenk, F.~Tosi, E.~D'Aversa, R.~Clark, J.~P. Combe, and C.~M. Ore.
\newblock {Deciphering sub-micron ice particles on Enceladus surface}.
\newblock \emph{Icarus}, 290:\penalty0 183--200, 7 2017.
\newblock ISSN 0019-1035.
\newblock \doi{10.1016/J.ICARUS.2017.02.012}.

\bibitem[Segatz et~al.(1988)Segatz, Spohn, Ross, and Schubert]{segatz1988}
M.~Segatz, T.~Spohn, M.~Ross, and G.~Schubert.
\newblock Tidal dissipation, surface heat flow, and figure of viscoelastic models of io.
\newblock \emph{Icarus}, 75\penalty0 (2):\penalty0 187--206, 1988.

\bibitem[Shibaike(2025)]{Shibaike2025Callisto}
Y.~Shibaike.
\newblock {Partial Differentiation of Callisto as Possible Evidence for Pebble Accretion}.
\newblock \emph{The Astrophysical Journal Letters}, 988\penalty0 (1):\penalty0 L32, 7 2025.
\newblock ISSN 2041-8205.
\newblock \doi{10.3847/2041-8213/ADEB6D}.
\newblock URL \url{https://iopscience.iop.org/article/10.3847/2041-8213/adeb6d https://iopscience.iop.org/article/10.3847/2041-8213/adeb6d/meta}.

\bibitem[Shibaike and Mordasini(2024)]{Shibaike2024ConstraintsEvolution}
Y.~Shibaike and C.~Mordasini.
\newblock {Constraints on PDS 70 b and c from the dust continuum emission of the circumplanetary discs considering in situ dust evolution}.
\newblock \emph{Astronomy {\&} Astrophysics}, 687:\penalty0 A166, 7 2024.
\newblock ISSN 0004-6361.
\newblock \doi{10.1051/0004-6361/202449522}.
\newblock URL \url{https://www.aanda.org/articles/aa/full_html/2024/07/aa49522-24/aa49522-24.html https://www.aanda.org/articles/aa/abs/2024/07/aa49522-24/aa49522-24.html}.

\bibitem[Shibaike and Mori(2023)]{Shibaike2023}
Y.~Shibaike and S.~Mori.
\newblock Effective dust growth in laminar circumplanetary discs with magnetic wind-driven accretion.
\newblock \emph{Monthly Notices of the Royal Astronomical Society}, 518\penalty0 (4):\penalty0 5444--5456, 2023.

\bibitem[Shibaike et~al.(2017)Shibaike, Okuzumi, Sasaki, and Ida]{Shibaike2017}
Y.~Shibaike, S.~Okuzumi, T.~Sasaki, and S.~Ida.
\newblock Satellitesimal formation via collisional dust growth in steady circumplanetary disks.
\newblock \emph{The Astrophysical Journal}, 846\penalty0 (1):\penalty0 81, 2017.

\bibitem[Shibaike et~al.(2019)Shibaike, Ormel, Ida, Okuzumi, and Sasaki]{Shibaike2019}
Y.~Shibaike, C.~W. Ormel, S.~Ida, S.~Okuzumi, and T.~Sasaki.
\newblock {The Galilean Satellites Formed Slowly from Pebbles}.
\newblock \emph{The Astrophysical Journal}, 885\penalty0 (1):\penalty0 79, 11 2019.
\newblock ISSN 0004-637X.
\newblock \doi{10.3847/1538-4357/AB46A7}.
\newblock URL \url{https://iopscience.iop.org/article/10.3847/1538-4357/ab46a7 https://iopscience.iop.org/article/10.3847/1538-4357/ab46a7/meta}.

\bibitem[Shibaike et~al.(2025)Shibaike, Hashimoto, Dong, Mordasini, Fukagawa, and Muto]{Shibaike2025}
Y.~Shibaike, J.~Hashimoto, R.~Dong, C.~Mordasini, M.~Fukagawa, and T.~Muto.
\newblock {Predictions of Dust Continuum Emission from a Potential Circumplanetary Disk: A Case Study of the Planet Candidate AB Aurigae b}.
\newblock \emph{The Astrophysical Journal}, 979\penalty0 (1):\penalty0 24, 1 2025.
\newblock ISSN 0004-637X.
\newblock \doi{10.3847/1538-4357/AD9B21}.
\newblock URL \url{https://iopscience.iop.org/article/10.3847/1538-4357/ad9b21 https://iopscience.iop.org/article/10.3847/1538-4357/ad9b21/meta}.

\bibitem[Singer et~al.(2021)Singer, McKinnon, and Schenk]{Singer2021}
K.~N. Singer, W.~B. McKinnon, and P.~M. Schenk.
\newblock {Pits, uplifts and small chaos features on Europa: Morphologic and morphometric evidence for intrusive upwelling and lower limits to ice shell thickness}.
\newblock \emph{Icarus}, 364:\penalty0 114465, 8 2021.
\newblock ISSN 0019-1035.
\newblock \doi{10.1016/J.ICARUS.2021.114465}.

\bibitem[Slattery et~al.(1992)Slattery, Benz, and Cameron]{Wayne1992}
W.~L. Slattery, W.~Benz, and A.~G. Cameron.
\newblock {Giant impacts on a primitive Uranus}.
\newblock \emph{Icarus}, 99\penalty0 (1):\penalty0 167--174, 9 1992.
\newblock ISSN 0019-1035.
\newblock \doi{10.1016/0019-1035(92)90180-F}.
\newblock URL \url{https://www.sciencedirect.com/science/article/abs/pii/001910359290180F?via%3Dihub}.

\bibitem[Smith et~al.(1979)Smith, Soderblom, Johnson, Ingersoll, Collins, Shoemaker, Hunt, Masursky, Carr, Davies, et~al.]{smith1979}
B.~A. Smith, L.~A. Soderblom, T.~V. Johnson, A.~P. Ingersoll, S.~A. Collins, E.~M. Shoemaker, G.~Hunt, H.~Masursky, M.~H. Carr, M.~E. Davies, et~al.
\newblock The jupiter system through the eyes of voyager 1.
\newblock \emph{Science}, 204\penalty0 (4396):\penalty0 951--972, 1979.

\bibitem[Smith et~al.(1989)Smith, Soderblom, Banfield, Barnet, Basilevksy, Beebe, Bollinger, Boyce, Brahic, Briggs, Brown, Chyba, Collins, Colvin, Cook, Crisp, Croft, Cruikshank, Cuzzi, Danielson, Davies, De~Jong, Dones, Godfrey, Goguen, Grenier, Haemmerle, Hammel, Hansen, Helfenstein, Howell, Hunt, Ingersoll, Johnson, Kargel, Kirk, Kuehn, Limaye, Masursky, McEwen, Morrison, Owen, Owen, Pollack, Porco, Rages, Rogers, Rudy, Sagan, Schwartz, Shoemaker, Showalter, Sicardy, Simonelli, Spencer, Sromovsky, Stoker, Strom, Suomi, Synott, Terrile, Thomas, Thompson, Verbiscer, and Veverka]{Smith1989}
B.~A. Smith, L.~A. Soderblom, D.~Banfield, C.~Barnet, A.~T. Basilevksy, R.~F. Beebe, K.~Bollinger, J.~M. Boyce, A.~Brahic, G.~A. Briggs, R.~H. Brown, C.~Chyba, S.~A. Collins, T.~Colvin, A.~F. Cook, D.~Crisp, S.~K. Croft, D.~Cruikshank, J.~N. Cuzzi, G.~E. Danielson, M.~E. Davies, E.~De~Jong, L.~Dones, D.~Godfrey, J.~Goguen, I.~Grenier, V.~R. Haemmerle, H.~Hammel, C.~J. Hansen, C.~P. Helfenstein, C.~Howell, G.~E. Hunt, A.~P. Ingersoll, T.~V. Johnson, J.~Kargel, R.~Kirk, D.~I. Kuehn, S.~Limaye, H.~Masursky, A.~McEwen, D.~Morrison, T.~Owen, W.~Owen, J.~B. Pollack, C.~C. Porco, K.~Rages, P.~Rogers, D.~Rudy, C.~Sagan, J.~Schwartz, E.~M. Shoemaker, M.~Showalter, B.~Sicardy, D.~Simonelli, J.~Spencer, L.~A. Sromovsky, C.~Stoker, R.~G. Strom, V.~E. Suomi, S.~P. Synott, R.~J. Terrile, P.~Thomas, W.~R. Thompson, A.~Verbiscer, and J.~Veverka.
\newblock {Voyager 2 at Neptune: Imaging Science Results}.
\newblock \emph{Science}, 246\penalty0 (4936):\penalty0 1422, 12 1989.
\newblock ISSN 0036-8075.
\newblock \doi{10.1126/SCIENCE.246.4936.1422}.
\newblock URL \url{https://ui.adsabs.harvard.edu/abs/1989Sci...246.1422S/abstract}.

\bibitem[Sou{\v{c}}ek et~al.(2024)Sou{\v{c}}ek, B{\v{e}}hounkov{\'a}, Lanzend{\"o}rfer, Tobie, and Choblet]{Soucek2024}
O.~Sou{\v{c}}ek, M.~B{\v{e}}hounkov{\'a}, M.~Lanzend{\"o}rfer, G.~Tobie, and G.~Choblet.
\newblock Variations in plume activity reveal the dynamics of water-filled faults on enceladus.
\newblock \emph{Nature Communications}, 15\penalty0 (1):\penalty0 7405, 2024.

\bibitem[Southworth et~al.(2015)Southworth, Kempf, and Schmidt]{Southworth2015ModelingPlumes}
B.~S. Southworth, S.~Kempf, and J.~Schmidt.
\newblock {Modeling Europa's dust plumes}.
\newblock \emph{Geophysical Research Letters}, 42\penalty0 (24):\penalty0 541--10, 12 2015.
\newblock ISSN 1944-8007.
\newblock \doi{10.1002/2015GL066502}.
\newblock URL \url{https://onlinelibrary.wiley.com/doi/full/10.1002/2015GL066502 https://onlinelibrary.wiley.com/doi/abs/10.1002/2015GL066502 https://agupubs.onlinelibrary.wiley.com/doi/10.1002/2015GL066502}.

\bibitem[Southworth et~al.(2019)Southworth, Kempf, and Spitale]{Southworth2019}
B.~S. Southworth, S.~Kempf, and J.~Spitale.
\newblock {Surface deposition of the Enceladus plume and the zenith angle of emissions}.
\newblock \emph{Icarus}, 319:\penalty0 33--42, 2 2019.
\newblock ISSN 0019-1035.
\newblock \doi{10.1016/J.ICARUS.2018.08.024}.

\bibitem[Sparks et~al.(2005)Sparks, Hough, and Bergeron]{sparks2005search}
W.~B. Sparks, J.~H. Hough, and L.~E. Bergeron.
\newblock A search for chiral signatures on mars.
\newblock \emph{Astrobiology}, 5\penalty0 (6):\penalty0 737--748, 2005.

\bibitem[Sparks et~al.(2016)Sparks, Hand, McGrath, Bergeron, Cracraft, and Deustua]{Sparks2016}
W.~B. Sparks, K.~P. Hand, M.~A. McGrath, E.~Bergeron, M.~Cracraft, and S.~E. Deustua.
\newblock {Probing for Evidence of Plumes on Europa With HST/STIS}.
\newblock \emph{The Astrophysical Journal}, 829\penalty0 (2):\penalty0 121, 2016.
\newblock ISSN 0004-637X.
\newblock \doi{10.3847/0004-637x/829/2/121}.
\newblock URL \url{http://dx.doi.org/10.3847/0004-637X/829/2/121}.

\bibitem[Sparks et~al.(2017)Sparks, Schmidt, McGrath, Hand, Spencer, Cracraft, and Deustua]{Sparks2017}
W.~B. Sparks, B.~E. Schmidt, M.~A. McGrath, K.~P. Hand, J.~R. Spencer, M.~Cracraft, and S.~E. Deustua.
\newblock {Active Cryovolcanism on Europa?}
\newblock \emph{The Astrophysical Journal}, 839\penalty0 (2):\penalty0 L18, 4 2017.
\newblock ISSN 2041-8213.
\newblock \doi{10.3847/2041-8213/AA67F8}.

\bibitem[Spencer(1987)]{Spencer1987ThermalSatellites}
J.~R. Spencer.
\newblock {Thermal segregation of water ice on the Galilean satellites}.
\newblock \emph{Icarus}, 69\penalty0 (2):\penalty0 297--313, 1987.

\bibitem[Spencer and Calvin(2002)]{spencer2002condensed}
J.~R. Spencer and W.~M. Calvin.
\newblock Condensed o2 on europa and callisto.
\newblock \emph{The Astronomical Journal}, 124\penalty0 (6):\penalty0 3400, 2002.

\bibitem[Spencer and Nimmo(2013)]{spencer2013}
J.~R. Spencer and F.~Nimmo.
\newblock Enceladus: An active ice world in the saturn system.
\newblock \emph{Annu. Rev. Earth Planet. Sci.}, 41:\penalty0 693–--717, 2013.
\newblock \doi{10.1146/annurev-earth-050212-124025}.

\bibitem[Spencer et~al.(1997)Spencer, Sartoretti, Ballester, McEwen, Clarke, and McGrath]{Spencer1997}
J.~R. Spencer, P.~Sartoretti, G.~E. Ballester, A.~S. McEwen, J.~T. Clarke, and M.~A. McGrath.
\newblock The pele plume (io): Observations with the hubble space telescope.
\newblock \emph{Geophysical Research Letters}, 24\penalty0 (20):\penalty0 2471--2474, 1997.

\bibitem[Spencer et~al.(2000)Spencer, Jessup, McGrath, Ballester, and Yelle]{spencer2000}
J.~R. Spencer, K.~L. Jessup, M.~A. McGrath, G.~E. Ballester, and R.~Yelle.
\newblock Discovery of gaseous s2 in io's pele plume.
\newblock \emph{Science}, 288\penalty0 (5469):\penalty0 1208--1210, 2000.

\bibitem[Spencer et~al.(2005)Spencer, Lellouch, Richter, L{\'o}pez-Valverde, Jessup, Greathouse, and Flaud]{spencer2005}
J.~R. Spencer, E.~Lellouch, M.~J. Richter, M.~A. L{\'o}pez-Valverde, K.~L. Jessup, T.~K. Greathouse, and J.-M. Flaud.
\newblock Mid-infrared detection of large longitudinal asymmetries in io's so2 atmosphere.
\newblock \emph{Icarus}, 176\penalty0 (2):\penalty0 283--304, 2005.

\bibitem[Spencer et~al.(2006)Spencer, Pearl, Segura, Flasar, Mamoutkine, Romani, Buratti, Hendrix, Spilker, and Lopes]{Spencer2006}
J.~R. Spencer, J.~C. Pearl, M.~Segura, F.~M. Flasar, A.~Mamoutkine, P.~Romani, B.~J. Buratti, A.~R. Hendrix, L.~J. Spilker, and R.~M. Lopes.
\newblock {Cassini Encounters Enceladus: Background and the Discovery of a South Polar Hot Spot}.
\newblock \emph{Science}, 311\penalty0 (5766):\penalty0 1401--1405, 3 2006.
\newblock ISSN 00368075.
\newblock \doi{10.1126/SCIENCE.1121661}.
\newblock URL \url{/doi/pdf/10.1126/science.1121661}.

\bibitem[{Spencer} et~al.(2006){Spencer}, {Pearl}, {Segura}, {Flasar}, {Mamoutkine}, {Romani}, {Buratti}, {Hendrix}, {Spilker}, and {Lopes}]{2006Sci...311.1401S}
J.~R. {Spencer}, J.~C. {Pearl}, M.~{Segura}, F.~M. {Flasar}, A.~{Mamoutkine}, P.~{Romani}, B.~J. {Buratti}, A.~R. {Hendrix}, L.~J. {Spilker}, and R.~M.~C. {Lopes}.
\newblock {Cassini Encounters Enceladus: Background and the Discovery of a South Polar Hot Spot}.
\newblock \emph{Science}, 311\penalty0 (5766):\penalty0 1401--1405, Mar. 2006.
\newblock \doi{10.1126/science.1121661}.

\bibitem[Spencer et~al.(2018)Spencer, Nimmo, Ingersoll, Hurford, Kite, Rhoden, Schmidt, and Howett]{Spencer2018}
J.~R. Spencer, F.~Nimmo, A.~P. Ingersoll, T.~A. Hurford, E.~S. Kite, A.~R. Rhoden, J.~Schmidt, and C.~J.~A. Howett.
\newblock {Plume Origins and Plumbing: From Ocean to Surface}.
\newblock In P.~M. Schenk, R.~N. Clark, C.~J.~A. Howett, A.~J. Verbiscer, and J.~H. Waite, editors, \emph{Enceladus and the Icy Moons of Saturn}, pages 163--174. The University of Arizona Press, 2018.
\newblock \doi{10.2458/azu{\_}uapress{\_}9780816537075-ch008}.

\bibitem[Spiers and Schmidt(2023)]{Spiers2023}
E.~M. Spiers and B.~E. Schmidt.
\newblock {Variable Salinity and Hydrogen Production in Europa's Ocean}.
\newblock \emph{Journal of Geophysical Research: Planets}, 128\penalty0 (11):\penalty0 e2023JE008028, 11 2023.
\newblock ISSN 2169-9100.
\newblock \doi{10.1029/2023JE008028}.
\newblock URL \url{/doi/pdf/10.1029/2023JE008028 https://onlinelibrary.wiley.com/doi/abs/10.1029/2023JE008028 https://agupubs.onlinelibrary.wiley.com/doi/10.1029/2023JE008028}.

\bibitem[Spitale and Porco(2007)]{Spitale2007}
J.~N. Spitale and C.~C. Porco.
\newblock {Association of the jets of Enceladus with the warmest regions on its south-polar fractures}.
\newblock \emph{Nature}, 449\penalty0 (7163):\penalty0 695--697, 10 2007.
\newblock ISSN 1476-4687.
\newblock \doi{10.1038/NATURE06217}.
\newblock URL \url{https://pubmed.ncbi.nlm.nih.gov/17928854/}.

\bibitem[Spitale et~al.(2015)Spitale, Hurford, Rhoden, Berkson, and Platts]{Spitale2015}
J.~N. Spitale, T.~A. Hurford, A.~R. Rhoden, E.~E. Berkson, and S.~S. Platts.
\newblock {Curtain eruptions from Enceladus’ south-polar terrain}.
\newblock \emph{Nature 2015 521:7550}, 521\penalty0 (7550):\penalty0 57--60, 5 2015.
\newblock ISSN 1476-4687.
\newblock \doi{10.1038/nature14368}.
\newblock URL \url{https://www.nature.com/articles/nature14368}.

\bibitem[Spitale et~al.(2025)Spitale, Tigges, Berne, Rhoden, Hurford, and Webster]{Spitale2025}
J.~N. Spitale, M.~D. Tigges, A.~Berne, A.~Rhoden, T.~A. Hurford, and K.~D. Webster.
\newblock {Curtain-based Maps of Eruptive Activity in Enceladus’s South-polar Terrain at 15 Cassini Epochs}.
\newblock \emph{The Planetary Science Journal}, 6\penalty0 (3):\penalty0 67, 3 2025.
\newblock ISSN 2632-3338.
\newblock \doi{10.3847/PSJ/ADB7D7}.
\newblock URL \url{https://iopscience.iop.org/article/10.3847/PSJ/adb7d7 https://iopscience.iop.org/article/10.3847/PSJ/adb7d7/meta}.

\bibitem[Srama et~al.(2004)Srama, Ahrens, Altobelli, Auer, Bradley, Burton, Dikarev, Economou, Fechtig, G{\"{o}}rlich, Grande, Graps, Gr{\"{u}}n, Havnes, Helfert, Horanyi, Igenbergs, Jessberger, Johnson, Kempf, Krivov, Kr{\"{u}}ger, Mocker-Ahlreep, Moragas-Klostermeyer, Lamy, Landgraf, Linkert, Linkert, Lura, Mcdonnell, M{\"{o}}hlmann, Morfill, M{\"{u}}ller, Roy, Sch{\"{a}}fer, Schlotzhauer, Schwehm, Spahn, St{\"{u}}big, Svestka, Tschernjawski, Tuzzolino, W{\"{a}}sch, and Zook]{Srama2004}
R.~Srama, T.~J. Ahrens, N.~Altobelli, S.~Auer, J.~G. Bradley, M.~Burton, V.~V. Dikarev, T.~Economou, H.~Fechtig, M.~G{\"{o}}rlich, M.~Grande, A.~Graps, E.~Gr{\"{u}}n, O.~Havnes, S.~Helfert, M.~Horanyi, E.~Igenbergs, E.~K. Jessberger, T.~V. Johnson, S.~Kempf, A.~V. Krivov, H.~Kr{\"{u}}ger, A.~Mocker-Ahlreep, G.~Moragas-Klostermeyer, P.~Lamy, M.~Landgraf, D.~Linkert, G.~Linkert, F.~Lura, J.~A. Mcdonnell, D.~M{\"{o}}hlmann, G.~E. Morfill, M.~M{\"{u}}ller, M.~Roy, G.~Sch{\"{a}}fer, G.~Schlotzhauer, G.~H. Schwehm, F.~Spahn, M.~St{\"{u}}big, J.~Svestka, V.~Tschernjawski, A.~J. Tuzzolino, R.~W{\"{a}}sch, and H.~A. Zook.
\newblock {The Cassini cosmic dust analyzer}.
\newblock \emph{Space Science Reviews}, 114\penalty0 (1-4):\penalty0 465--518, 12 2004.
\newblock ISSN 00386308.
\newblock \doi{10.1007/S11214-004-1435-Z}.

\bibitem[Steinbr{\"{u}}gge and Patterson(2025)]{Steinbrugge2025ShallowRidges}
G.~Steinbr{\"{u}}gge and G.~Patterson.
\newblock {Shallow Subsurface Water at the Base of Europa's Double Ridges}.
\newblock \emph{Journal of Geophysical Research: Planets}, 130\penalty0 (2):\penalty0 e2024JE008673, 2 2025.
\newblock ISSN 2169-9100.
\newblock \doi{10.1029/2024JE008673}.
\newblock URL \url{/doi/pdf/10.1029/2024JE008673 https://onlinelibrary.wiley.com/doi/abs/10.1029/2024JE008673 https://agupubs.onlinelibrary.wiley.com/doi/10.1029/2024JE008673}.

\bibitem[Steinke et~al.(2020)Steinke, Hu, H{\"o}ning, Van~der Wal, and Vermeersen]{steinke2020}
T.~Steinke, H.~Hu, D.~H{\"o}ning, W.~Van~der Wal, and B.~Vermeersen.
\newblock Tidally induced lateral variations of io's interior.
\newblock \emph{Icarus}, 335:\penalty0 113299, 2020.

\bibitem[Stephan et~al.(2021)Stephan, Roatsch, Tosi, Matz, Kersten, Wagner, Palumbo, Poulet, Hussmann, Barabash, Bruzzone, Dougherty, Gladstone, Gurvits, Hartogh, Iess, Wahlund, Wurz, Witasse, Grasset, Altobelli, Carter, d'Aversa, Corte, Filacchione, Galli, Galuzzi, Gwinner, Hauber, Jaumann, Langevin, Lucchetti, Migliorini, Piccioni, Solomonidou, Stark, Tobie, Vallat, van Hoolst, , and the JUICE SWT~team]{Stephan2017}
K.~Stephan, T.~Roatsch, F.~Tosi, K.-D. Matz, E.~Kersten, R.~Wagner, P.~Palumbo, F.~Poulet, H.~Hussmann, S.~Barabash, L.~Bruzzone, M.~Dougherty, R.~Gladstone, L.~Gurvits, P.~Hartogh, L.~Iess, J.-E. Wahlund, P.~Wurz, O.~Witasse, O.~Grasset, N.~Altobelli, J.~Carter, E.~d'Aversa, V.~D. Corte, G.~Filacchione, A.~Galli, V.~Galuzzi, K.~Gwinner, E.~Hauber, R.~Jaumann, Y.~Langevin, A.~Lucchetti, A.~Migliorini, G.~Piccioni, A.~Solomonidou, A.~Stark, G.~Tobie, C.~Vallat, T.~van Hoolst, , and the JUICE SWT~team.
\newblock Regions of interest on ganymede's and callisto's surface as potential targets for esa's juice mission.
\newblock \emph{Planet. Sp. Sci.}, 208:\penalty0 105324, 2021.
\newblock \doi{10.1016/j.pss.2021.105324}.

\bibitem[Strobel and Wolven(2001)]{strobel2001}
D.~F. Strobel and B.~C. Wolven.
\newblock The atmosphere of {Io}: Abundances and sources of sulfur dioxide and atomic hydrogen.
\newblock \emph{Astrophysics and Space Science}, 277:\penalty0 271--287, 2001.

\bibitem[Strobel and Zhu(2017)]{Strobel2017}
D.~F. Strobel and X.~Zhu.
\newblock Comparative planetary nitrogen atmospheres: Density and thermal structures of pluto and triton.
\newblock \emph{Icarus}, 291:\penalty0 55--64, 2017.
\newblock ISSN 0019-1035.
\newblock \doi{https://doi.org/10.1016/j.icarus.2017.03.013}.
\newblock URL \url{https://www.sciencedirect.com/science/article/pii/S0019103516306583}.

\bibitem[Strobel et~al.(1994)Strobel, Zhu, and Summers]{Strobel1994}
D.~F. Strobel, X.~Zhu, and M.~E. Summers.
\newblock {On the Vertical Thermal Structure of Io's Atmosphere}.
\newblock \emph{Icarus, Volume 111, Issue 1, p. 18-30.}, 111\penalty0 (1):\penalty0 18, 9 1994.
\newblock ISSN 0019-1035.
\newblock \doi{10.1006/ICAR.1994.1130}.
\newblock URL \url{https://ui.adsabs.harvard.edu/abs/1994Icar..111...18S/abstract}.

\bibitem[Sucerquia and Cuello(2025)]{Sucerquia2025}
M.~Sucerquia and N.~Cuello.
\newblock {Extreme exomoons in WASP-49 Ab: Dynamics and detectability}.
\newblock \emph{Astronomy {\&} Astrophysics}, 2025.
\newblock ISSN 0004-6361.
\newblock \doi{10.1051/0004-6361/202452968}.
\newblock URL \url{https://www.aanda.org/articles/aa/abs/forth/aa52968-24/aa52968-24.html}.

\bibitem[Swedlund et~al.(1972)Swedlund, Kemp, and Wolstencroft]{swedlund1972circular}
J.~B. Swedlund, J.~C. Kemp, and R.~D. Wolstencroft.
\newblock Circular polarization of saturn.
\newblock \emph{Astrophysical Journal, Vol. 178, pp. 257-266 (1972)}, 178:\penalty0 257--266, 1972.

\bibitem[Szalay et~al.(2024)Szalay, Allegrini, Ebert, Bagenal, Bolton, Fatemi, McComas, Pontoni, Saur, Smith, Strobel, Vance, Vorburger, and Wilson]{Szalay2024}
J.~R. Szalay, F.~Allegrini, R.~W. Ebert, F.~Bagenal, S.~J. Bolton, S.~Fatemi, D.~J. McComas, A.~Pontoni, J.~Saur, H.~T. Smith, D.~F. Strobel, S.~D. Vance, A.~Vorburger, and R.~J. Wilson.
\newblock {Oxygen production from dissociation of Europa’s water-ice surface}.
\newblock \emph{Nature Astronomy 2024 8:5}, 8\penalty0 (5):\penalty0 567--576, 3 2024.
\newblock ISSN 2397-3366.
\newblock \doi{10.1038/s41550-024-02206-x}.
\newblock URL \url{https://www.nature.com/articles/s41550-024-02206-x}.

\bibitem[{Szul{\'a}gyi}(2017)]{Szulagyi17gap}
J.~{Szul{\'a}gyi}.
\newblock {Effects of the Planetary Temperature on the Circumplanetary Disk and on the Gap}.
\newblock \emph{\apj}, 842\penalty0 (2):\penalty0 103, June 2017.
\newblock \doi{10.3847/1538-4357/aa7515}.

\bibitem[{Szul{\'a}gyi} et~al.(2014){Szul{\'a}gyi}, {Morbidelli}, {Crida}, and {Masset}]{Szulagyi14}
J.~{Szul{\'a}gyi}, A.~{Morbidelli}, A.~{Crida}, and F.~{Masset}.
\newblock {Accretion of Jupiter-mass Planets in the Limit of Vanishing Viscosity}.
\newblock \emph{\apj}, 782\penalty0 (2):\penalty0 65, Feb. 2014.
\newblock \doi{10.1088/0004-637X/782/2/65}.

\bibitem[{Szul{\'a}gyi} et~al.(2018){Szul{\'a}gyi}, {Cilibrasi}, and {Mayer}]{Szulagyi18}
J.~{Szul{\'a}gyi}, M.~{Cilibrasi}, and L.~{Mayer}.
\newblock {In Situ Formation of Icy Moons of Uranus and Neptune}.
\newblock \emph{\apjl}, 868\penalty0 (1):\penalty0 L13, Nov. 2018.
\newblock \doi{10.3847/2041-8213/aaeed6}.

\bibitem[Szul{\'a}gyi et~al.(2022)Szul{\'a}gyi, Binkert, and Surville]{Szulagyi2022}
J.~Szul{\'a}gyi, F.~Binkert, and C.~Surville.
\newblock Meridional circulation of dust and gas in the circumstellar disk: Delivery of solids onto the circumplanetary region.
\newblock \emph{The Astrophysical Journal}, 924\penalty0 (1):\penalty0 1, 2022.

\bibitem[Tacconi et~al.(2021)Tacconi, Arridge, Buonanno, Cruise, Grasset, Amina~Helmi, et~al.]{tacconi2021voyage}
L.~Tacconi, C.~Arridge, A.~Buonanno, M.~Cruise, O.~Grasset, A.~Amina~Helmi, et~al.
\newblock Voyage 2050. final recommendations from the voyage 2050 senior committee.
\newblock \emph{Tech. rep., European Space Agency.}, 2021.

\bibitem[Tan et~al.(2022)Tan, Sekine, and Kuzuhara]{Tan_2022}
S.~Tan, Y.~Sekine, and M.~Kuzuhara.
\newblock Spatially resolved observations of europa’s surface with subaru/ircs at 1.0–1.8 $\mu$m: Upper limits to the abundances of hydrated cl-bearing salts.
\newblock \emph{The Planetary Science Journal}, 3\penalty0 (3):\penalty0 70, mar 2022.
\newblock \doi{10.3847/PSJ/ac596c}.
\newblock URL \url{https://dx.doi.org/10.3847/PSJ/ac596c}.

\bibitem[{Tanigawa} et~al.(2012){Tanigawa}, {Ohtsuki}, and {Machida}]{Tanigawa12}
T.~{Tanigawa}, K.~{Ohtsuki}, and M.~N. {Machida}.
\newblock {Distribution of Accreting Gas and Angular Momentum onto Circumplanetary Disks}.
\newblock \emph{\apj}, 747\penalty0 (1):\penalty0 47, Mar. 2012.
\newblock \doi{10.1088/0004-637X/747/1/47}.

\bibitem[Tate et~al.(2023)Tate, Rathbun, Hayes, Spencer, and Pettine]{Tate2023}
C.~D. Tate, J.~A. Rathbun, A.~G. Hayes, J.~R. Spencer, and M.~Pettine.
\newblock Discovery of seven volcanic outbursts on io from an infrared telescope facility observation campaign, 2016--2022.
\newblock \emph{The Planetary Science Journal}, 4\penalty0 (10):\penalty0 189, 2023.

\bibitem[Teachey and Kipping(2018)]{Teachey2018}
A.~Teachey and D.~M. Kipping.
\newblock {Evidence for a large exomoon orbiting Kepler-1625b}.
\newblock \emph{Science Advances}, 4\penalty0 (10), 10 2018.
\newblock ISSN 23752548.
\newblock \doi{10.1126/SCIADV.AAV1784/SUPPL{\_}FILE/AAV1784{\_}SM.PDF}.
\newblock URL \url{https://www.science.org/doi/10.1126/sciadv.aav1784}.

\bibitem[Teolis et~al.(2017)Teolis, Plainaki, Cassidy, and Raut]{teolis2017water}
B.~Teolis, C.~Plainaki, T.~Cassidy, and U.~Raut.
\newblock Water ice radiolytic o2, h2, and h2o2 yields for any projectile species, energy, or temperature: a model for icy astrophysical bodies.
\newblock \emph{Journal of Geophysical Research: Planets}, 122\penalty0 (10):\penalty0 1996--2012, 2017.

\bibitem[Thaller(2000)]{PDS3_official}
T.~F. Thaller.
\newblock {GALILEO ORBITAL OPERATIONS SOLID STATE IMAGING RAW EDR V1.0 [Dataset]}, 2000.
\newblock URL \url{https://pds.nasa.gov/ds-view/pds/viewDataset.jsp?dsid=GO-J%2FJSA-SSI-2-REDR-V1.0}.

\bibitem[Thelen et~al.(2024)Thelen, de~Kleer, Camarca, Akins, Gurwell, Butler, and de~Pater]{Thelen2024}
A.~E. Thelen, K.~de~Kleer, M.~Camarca, A.~Akins, M.~Gurwell, B.~Butler, and I.~de~Pater.
\newblock {Subsurface Thermophysical Properties of Europa’s Leading and Trailing Hemispheres as Revealed by ALMA}.
\newblock \emph{The Planetary Science Journal}, 5\penalty0 (2):\penalty0 56, 2 2024.
\newblock ISSN 2632-3338.
\newblock \doi{10.3847/PSJ/AD251C}.
\newblock URL \url{https://iopscience.iop.org/article/10.3847/PSJ/ad251c https://iopscience.iop.org/article/10.3847/PSJ/ad251c/meta}.

\bibitem[Thomas et~al.(2004)Thomas, Bagenal, Hill, and Wilson]{thomas2004}
N.~Thomas, F.~Bagenal, T.~Hill, and J.~Wilson.
\newblock The io neutral clouds and plasma torus.
\newblock \emph{Jupiter. The planet, satellites and magnetosphere}, 1:\penalty0 561--591, 2004.

\bibitem[Tinner et~al.(2024)Tinner, Galli, B{\"{a}}r, Pommerol, Rubin, Vorburger, and Wurz]{Tinner2024Electron-InducedOxygen}
C.~Tinner, A.~Galli, F.~B{\"{a}}r, A.~Pommerol, M.~Rubin, A.~Vorburger, and P.~Wurz.
\newblock {Electron-Induced Radiolysis of Water Ice and the Buildup of Oxygen}.
\newblock \emph{Journal of Geophysical Research: Planets}, 129\penalty0 (12):\penalty0 e2024JE008393, 12 2024.
\newblock ISSN 2169-9100.
\newblock \doi{10.1029/2024JE008393}.
\newblock URL \url{https://onlinelibrary.wiley.com/doi/full/10.1029/2024JE008393 https://onlinelibrary.wiley.com/doi/abs/10.1029/2024JE008393 https://agupubs.onlinelibrary.wiley.com/doi/10.1029/2024JE008393}.

\bibitem[Tobie et~al.(2005)Tobie, Mocquet, and Sotin]{tobie2005}
G.~Tobie, A.~Mocquet, and C.~Sotin.
\newblock Tidal dissipation within large icy satellites: Applications to europa and titan.
\newblock \emph{Icarus}, 177\penalty0 (2):\penalty0 534--549, 2005.

\bibitem[Tobie et~al.(2025)Tobie, Auclair-Desrotour, B{\v{e}}hounkov{\'{a}}, Kervazo, Sou{\v{c}}ek, and Kalousov{\'{a}}]{Tobie2025}
G.~Tobie, P.~Auclair-Desrotour, M.~B{\v{e}}hounkov{\'{a}}, M.~Kervazo, O.~Sou{\v{c}}ek, and K.~Kalousov{\'{a}}.
\newblock {Tidal Deformation and Dissipation Processes in Icy Worlds}.
\newblock \emph{Space Science Reviews 2025 221:1}, 221\penalty0 (1):\penalty0 1--57, 1 2025.
\newblock ISSN 1572-9672.
\newblock \doi{10.1007/S11214-025-01136-Y}.
\newblock URL \url{https://link.springer.com/article/10.1007/s11214-025-01136-y}.

\bibitem[Tosi et~al.(2020)Tosi, Mura, Lopes, Filacchione, Ciarniello, Zambon, Adriani, Bolton, Brooks, Noschese, Sordini, Turrini, Altieri, Cicchetti, Grassi, Hansen, Migliorini, Moriconi, Piccioni, Plainaki, and Sindoni]{Tosi2020}
F.~Tosi, A.~Mura, R.~M. Lopes, G.~Filacchione, M.~Ciarniello, F.~Zambon, A.~Adriani, S.~J. Bolton, S.~M. Brooks, R.~Noschese, R.~Sordini, D.~Turrini, F.~Altieri, A.~Cicchetti, D.~Grassi, C.~J. Hansen, A.~Migliorini, M.~L. Moriconi, G.~Piccioni, C.~Plainaki, and G.~Sindoni.
\newblock {Mapping Io's Surface Composition With Juno/JIRAM}.
\newblock \emph{Journal of Geophysical Research: Planets}, 125\penalty0 (11):\penalty0 e2020JE006522, 11 2020.
\newblock ISSN 2169-9100.
\newblock \doi{10.1029/2020JE006522}.
\newblock URL \url{/doi/pdf/10.1029/2020JE006522 https://onlinelibrary.wiley.com/doi/abs/10.1029/2020JE006522 https://agupubs.onlinelibrary.wiley.com/doi/10.1029/2020JE006522}.

\bibitem[Tosi et~al.(2024)Tosi, Roatsch, Galli, Hauber, Lucchetti, Molyneux, Stephan, Achilleos, Bovolo, Carter, Cavali{\'{e}}, Cim{\`{o}}, D’Aversa, Gwinner, Hartogh, Huybrighs, Langevin, Lellouch, Migliorini, Palumbo, Piccioni, Plaut, Postberg, Poulet, Retherford, Rezac, Roth, Solomonidou, Tobie, Tortora, Tubiana, Wagner, Wirstr{\"{o}}m, Wurz, Zambon, Zannoni, Barabash, Bruzzone, Dougherty, Gladstone, Gurvits, Hussmann, Iess, Wahlund, Witasse, Vallat, and Lorente]{Tosi2024}
F.~Tosi, T.~Roatsch, A.~Galli, E.~Hauber, A.~Lucchetti, P.~Molyneux, K.~Stephan, N.~Achilleos, F.~Bovolo, J.~Carter, T.~Cavali{\'{e}}, G.~Cim{\`{o}}, E.~D’Aversa, K.~Gwinner, P.~Hartogh, H.~Huybrighs, Y.~Langevin, E.~Lellouch, A.~Migliorini, P.~Palumbo, G.~Piccioni, J.~J. Plaut, F.~Postberg, F.~Poulet, K.~Retherford, L.~Rezac, L.~Roth, A.~Solomonidou, G.~Tobie, P.~Tortora, C.~Tubiana, R.~Wagner, E.~Wirstr{\"{o}}m, P.~Wurz, F.~Zambon, M.~Zannoni, S.~Barabash, L.~Bruzzone, M.~Dougherty, R.~Gladstone, L.~I. Gurvits, H.~Hussmann, L.~Iess, J.~E. Wahlund, O.~Witasse, C.~Vallat, and R.~Lorente.
\newblock {Characterization of the Surfaces and Near-Surface Atmospheres of Ganymede, Europa and Callisto by JUICE}.
\newblock \emph{Space Science Reviews}, 220\penalty0 (5), 8 2024.
\newblock ISSN 15729672.
\newblock \doi{10.1007/S11214-024-01089-8}.

\bibitem[Trafton et~al.(1996)Trafton, Caldwell, Barnet, and Cunningham]{trafton1996}
L.~Trafton, J.~Caldwell, C.~Barnet, and C.~Cunningham.
\newblock The gaseous sulfur dioxide abundance over io's leading and trailing hemispheres: Hst spectra of io's c 1b 2--x 1a 1 band of so 2 near 2100 angstrom.
\newblock \emph{Astrophysical Journal v. 456, p. 384}, 456:\penalty0 384, 1996.

\bibitem[{Trani} et~al.(2020){Trani}, {Hamers}, {Geller}, and {Spera}]{2020MNRAS.499.4195T}
A.~A. {Trani}, A.~S. {Hamers}, A.~{Geller}, and M.~{Spera}.
\newblock {The ominous fate of exomoons around hot Jupiters in the high-eccentricity migration scenario}.
\newblock \emph{\mnras}, 499\penalty0 (3):\penalty0 4195--4205, Oct. 2020.

\bibitem[Trumbo and Brown(2023)]{Trumbo2023}
S.~K. Trumbo and M.~E. Brown.
\newblock {The distribution of CO2 on Europa indicates an internal source of carbon}.
\newblock \emph{Science}, 381:\penalty0 1308--1311, 9 2023.
\newblock ISSN 10959203.
\newblock \doi{10.1126/science.adg4155}.

\bibitem[Trumbo et~al.(2019)Trumbo, Brown, and Hand]{Trumbo2019}
S.~K. Trumbo, M.~E. Brown, and K.~P. Hand.
\newblock {Sodium chloride on the surface of Europa}.
\newblock \emph{Science Advances}, 5\penalty0 (6), 6 2019.
\newblock ISSN 23752548.
\newblock \doi{10.1126/SCIADV.AAW7123}.

\bibitem[{Trumbo} et~al.(2022){Trumbo}, {Becker}, {Brown}, {Denman}, {Molyneux}, {Hendrix}, {Retherford}, {Roth}, and {Alday}]{2022PSJ.....3...27T}
S.~K. {Trumbo}, T.~M. {Becker}, M.~E. {Brown}, W.~T.~P. {Denman}, P.~{Molyneux}, A.~{Hendrix}, K.~D. {Retherford}, L.~{Roth}, and J.~{Alday}.
\newblock {A New UV Spectral Feature on Europa: Confirmation of NaCl in Leading-hemisphere Chaos Terrain}.
\newblock \emph{The Planetary Science Journal}, 3\penalty0 (2):\penalty0 27, Feb. 2022.
\newblock \doi{10.3847/PSJ/ac4580}.

\bibitem[Tryka et~al.(1993)Tryka, Brown, and Anicich]{Tryka_1993}
K.~A. Tryka, R.~H. Brown, and V.~G. Anicich.
\newblock Spectroscopic determination of the phase composition and temperature of nitrogen ice on triton.
\newblock \emph{Science}, 261\penalty0 (5122):\penalty0 751--754, 1993.
\newblock \doi{10.1126/science.261.5122.751}.

\bibitem[Tsang et~al.(2012)Tsang, Spencer, Lellouch, L{\'o}pez-Valverde, Richter, and Greathouse]{tsang2012}
C.~C. Tsang, J.~R. Spencer, E.~Lellouch, M.~A. L{\'o}pez-Valverde, M.~J. Richter, and T.~K. Greathouse.
\newblock Io’s atmosphere: Constraints on sublimation support from density variations on seasonal timescales using nasa irtf/texes observations from 2001 to 2010.
\newblock \emph{Icarus}, 217\penalty0 (1):\penalty0 277--296, 2012.

\bibitem[Tsang et~al.(2016)Tsang, Spencer, Lellouch, Lopez-Valverde, and Richter]{tsang2016}
C.~C. Tsang, J.~R. Spencer, E.~Lellouch, M.~A. Lopez-Valverde, and M.~J. Richter.
\newblock The collapse of io's primary atmosphere in jupiter eclipse.
\newblock \emph{Journal of Geophysical Research: Planets}, 121\penalty0 (8):\penalty0 1400--1410, 2016.

\bibitem[Tulej et~al.(2014)Tulej, Riedo, Neuland, Meyer, Wurz, Thomas, Grimaudo, Moreno-García, Broekmann, Neubeck, and Ivarsson]{Tulej2014}
M.~Tulej, A.~Riedo, M.~B. Neuland, S.~Meyer, P.~Wurz, N.~Thomas, V.~Grimaudo, P.~Moreno-García, P.~Broekmann, A.~Neubeck, and M.~Ivarsson.
\newblock Camam: A miniature laser ablation ionisation mass spectrometer and microscope-camera system for in situ investigation of the composition and morphology of extraterrestrial materials.
\newblock \emph{Geostandards and Geoanalytical Research}, 38\penalty0 (4):\penalty0 441--466, 2014.
\newblock \doi{https://doi.org/10.1111/j.1751-908X.2014.00302.x}.
\newblock URL \url{https://onlinelibrary.wiley.com/doi/abs/10.1111/j.1751-908X.2014.00302.x}.

\bibitem[Tyler et~al.(2015)Tyler, Henning, and Hamilton]{tyler2015}
R.~H. Tyler, W.~G. Henning, and C.~W. Hamilton.
\newblock Tidal heating in a magma ocean within jupiter’s moon io.
\newblock \emph{The Astrophysical Journal Supplement Series}, 218\penalty0 (2):\penalty0 22, 2015.

\bibitem[Ulibarri et~al.(2023)Ulibarri, Munsat, Voss, Fontanese, Horanyi, Kempf, and Sternovsky]{Ulibarri2023}
Z.~Ulibarri, T.~Munsat, M.~Voss, J.~Fontanese, M.~Horanyi, S.~Kempf, and Z.~Sternovsky.
\newblock Detection of the amino acid histidine and its breakup products in hypervelocity impact ice spectra.
\newblock \emph{Icarus}, 391:\penalty0 115319, 2023.
\newblock \doi{10.1016/j.icarus.2022.115319}.

\bibitem[{Unni} et~al.(2025){Unni}, {Oza}, {Hoijmakers}, {Seidel}, {Sivarani}, {Schmidt}, {Kesseli}, {de Kleer}, {Baker}, {Gebek}, {Westram}, {Fisher}, {Sallum}, {Bestha}, and {Bello Arufe}]{Unni2025}
A.~{Unni}, A.~V. {Oza}, H.~J. {Hoijmakers}, J.~V. {Seidel}, T.~{Sivarani}, C.~A. {Schmidt}, A.~Y. {Kesseli}, K.~{de Kleer}, A.~D. {Baker}, A.~{Gebek}, M.~M.~z. {Westram}, C.~{Fisher}, S.~{Sallum}, M.~{Bestha}, and A.~{Bello Arufe}.
\newblock {Doppler Shifted Transient Sodium Detection by KECK/HIRES}.
\newblock \emph{arXiv e-prints}, art. arXiv:2504.03974, Apr. 2025.
\newblock \doi{10.48550/arXiv.2504.03974}.

\bibitem[Van~Hoolst et~al.(2024)Van~Hoolst, Tobie, Vallat, Altobelli, Bruzzone, Cao, Dirkx, Genova, Hussmann, Iess, Kimura, Khurana, Lucchetti, Mitri, Moore, Saur, Stark, Vorburger, Wieczorek, Aboudan, Bergman, Bovolo, Breuer, Cappuccio, Carrer, Cecconi, Choblet, De~Marchi, Fayolle, Fienga, Futaana, Hauber, Kofman, Kumamoto, Lainey, Molyneux, Mousis, Plaut, Puccio, Retherford, Roth, Seignovert, Steinbr{\"{u}}gge, Thakur, Tortora, Tosi, Zannoni, Barabash, Dougherty, Gladstone, Gurvits, Hartogh, Palumbo, Poulet, Wahlund, Grasset, and Witasse]{VanHoolst2024GeophysicalExplorer}
T.~Van~Hoolst, G.~Tobie, C.~Vallat, N.~Altobelli, L.~Bruzzone, H.~Cao, D.~Dirkx, A.~Genova, H.~Hussmann, L.~Iess, J.~Kimura, K.~Khurana, A.~Lucchetti, G.~Mitri, W.~Moore, J.~Saur, A.~Stark, A.~Vorburger, M.~Wieczorek, A.~Aboudan, J.~Bergman, F.~Bovolo, D.~Breuer, P.~Cappuccio, L.~Carrer, B.~Cecconi, G.~Choblet, F.~De~Marchi, M.~Fayolle, A.~Fienga, Y.~Futaana, E.~Hauber, W.~Kofman, A.~Kumamoto, V.~Lainey, P.~Molyneux, O.~Mousis, J.~Plaut, W.~Puccio, K.~Retherford, L.~Roth, B.~Seignovert, G.~Steinbr{\"{u}}gge, S.~Thakur, P.~Tortora, F.~Tosi, M.~Zannoni, S.~Barabash, M.~Dougherty, R.~Gladstone, L.~I. Gurvits, P.~Hartogh, P.~Palumbo, F.~Poulet, J.~E. Wahlund, O.~Grasset, and O.~Witasse.
\newblock {Geophysical Characterization of the Interiors of Ganymede, Callisto and Europa by ESA’s JUpiter ICy moons Explorer}.
\newblock \emph{Space Science Reviews}, 220\penalty0 (5), 8 2024.
\newblock ISSN 15729672.
\newblock \doi{10.1007/S11214-024-01085-Y}.

\bibitem[Veeder et~al.(1994)Veeder, Matson, Johnson, Blaney, and Goguen]{veeder1994}
G.~J. Veeder, D.~L. Matson, T.~V. Johnson, D.~L. Blaney, and J.~D. Goguen.
\newblock Io's heat flow from infrared radiometry: 1983--1993.
\newblock \emph{Journal of Geophysical Research: Planets}, 99\penalty0 (E8):\penalty0 17095--17162, 1994.

\bibitem[Velluz et~al.(1965)Velluz, Le~Grand, and Grosjean]{velluz1965optical}
L.~Velluz, M.~Le~Grand, and M.~Grosjean.
\newblock Optical circular dichroism: principles, measurements, and applications.
\newblock \emph{Verlag Chemie}, 1965.

\bibitem[Villanueva et~al.(2023{\natexlab{a}})Villanueva, Hammel, Milam, Faggi, Kofman, Roth, Hand, Paganini, Stansberry, Spencer, Protopapa, Strazzulla, Cruz-Mermy, Glein, Cartwright, and Liuzzi]{Villanueva2023}
G.~L. Villanueva, H.~B. Hammel, S.~N. Milam, S.~Faggi, V.~Kofman, L.~Roth, K.~P. Hand, L.~Paganini, J.~Stansberry, J.~Spencer, S.~Protopapa, G.~Strazzulla, G.~Cruz-Mermy, C.~R. Glein, R.~Cartwright, and G.~Liuzzi.
\newblock {Endogenous CO2 ice mixture on the surface of Europa and no detection of plume activity}.
\newblock \emph{Science}, 381\penalty0 (6664):\penalty0 1305--1308, 9 2023{\natexlab{a}}.
\newblock ISSN 10959203.
\newblock \doi{10.1126/science.adg4270}.
\newblock URL \url{https://www.science.org/doi/10.1126/science.adg4270}.

\bibitem[Villanueva et~al.(2023{\natexlab{b}})Villanueva, Hammel, Milam, Kofman, Faggi, Glein, Cartwright, Roth, Hand, Paganini, Spencer, Stansberry, Holler, Rowe-Gurney, Protopapa, Strazzulla, Liuzzi, Cruz-Mermy, El~Moutamid, Hedman, and Denny]{Villanueva2023Enc}
G.~L. Villanueva, H.~B. Hammel, S.~N. Milam, V.~Kofman, S.~Faggi, C.~R. Glein, R.~Cartwright, L.~Roth, K.~P. Hand, L.~Paganini, J.~Spencer, J.~Stansberry, B.~Holler, N.~Rowe-Gurney, S.~Protopapa, G.~Strazzulla, G.~Liuzzi, G.~Cruz-Mermy, M.~El~Moutamid, M.~Hedman, and K.~Denny.
\newblock {JWST molecular mapping and characterization of Enceladus’ water plume feeding its torus}.
\newblock \emph{Nature Astronomy 2023 7:9}, 7\penalty0 (9):\penalty0 1056--1062, 6 2023{\natexlab{b}}.
\newblock ISSN 2397-3366.
\newblock \doi{10.1038/s41550-023-02009-6}.
\newblock URL \url{https://www.nature.com/articles/s41550-023-02009-6}.

\bibitem[Vorburger and Wurz(2018)]{Vorburger2018}
A.~Vorburger and P.~Wurz.
\newblock {Europa's ice-related atmosphere: The sputter contribution}.
\newblock \emph{Icarus}, 311:\penalty0 135--145, 2018.
\newblock ISSN 10902643.
\newblock \doi{10.1016/j.icarus.2018.03.022}.
\newblock URL \url{https://doi.org/10.1016/j.icarus.2018.03.022}.

\bibitem[Vorburger et~al.(2020)Vorburger, Wurz, and Waite]{Vorburger2020}
A.~Vorburger, P.~Wurz, and H.~Waite.
\newblock Chemical and isotopic composition measurements on atmospheric probes exploring uranus and neptune.
\newblock \emph{Space Science Reviews}, 216\penalty0 (57):\penalty0 1--31, 2020.
\newblock ISSN 1572-9672.
\newblock \doi{10.1007/s11214-020-00684-9}.
\newblock URL \url{https://doi.org/10.1007/s11214-020-00684-9}.

\bibitem[Waite et~al.(2004)Waite, Lewis, Kasprzak, Anicich, Block, Cravens, Fletcher, Ip, Luhmann, McNutt, et~al.]{waite2004}
J.~H. Waite, W.~Lewis, W.~Kasprzak, V.~Anicich, B.~Block, T.~E. Cravens, G.~Fletcher, W.-H. Ip, J.~G. Luhmann, R.~McNutt, et~al.
\newblock The cassini ion and neutral mass spectrometer (inms) investigation.
\newblock \emph{The Cassini-Huygens Mission: Orbiter In Situ Investigations Volume 2}, pages 113--231, 2004.

\bibitem[Waite et~al.(2017)Waite, Glein, Perryman, Teolis, Magee, Miller, Grimes, Perry, Miller, Bouquet, et~al.]{waite2017cassini}
J.~H. Waite, C.~R. Glein, R.~S. Perryman, B.~D. Teolis, B.~A. Magee, G.~Miller, J.~Grimes, M.~E. Perry, K.~E. Miller, A.~Bouquet, et~al.
\newblock Cassini finds molecular hydrogen in the enceladus plume: evidence for hydrothermal processes.
\newblock \emph{Science}, 356\penalty0 (6334):\penalty0 155--159, 2017.

\bibitem[Wald(1957)]{wald1957origin}
G.~Wald.
\newblock The origin of optical activity.
\newblock \emph{Annals of the New York Academy of Sciences}, 69\penalty0 (2):\penalty0 352--368, 1957.

\bibitem[Walker et~al.(2012)Walker, Moore, Goldstein, Varghese, and Trafton]{Walker2012}
A.~C. Walker, C.~H. Moore, D.~B. Goldstein, P.~L. Varghese, and L.~M. Trafton.
\newblock {A parametric study of Io’s thermophysical surface properties and subsequent numerical atmospheric simulations based on the best fit parameters}.
\newblock \emph{Icarus}, 220\penalty0 (1):\penalty0 225--253, 7 2012.
\newblock ISSN 0019-1035.
\newblock \doi{10.1016/J.ICARUS.2012.05.001}.

\bibitem[Walker et~al.(2021)Walker, Bassis, and Schmidt]{CWalker2021}
C.~C. Walker, J.~N. Bassis, and B.~E. Schmidt.
\newblock {Propagation of Vertical Fractures through Planetary Ice Shells: The Role of Basal Fractures at the Ice–Ocean Interface and Proximal Cracks}.
\newblock \emph{The Planetary Science Journal}, 2\penalty0 (4):\penalty0 135, 7 2021.
\newblock ISSN 2632-3338.
\newblock \doi{10.3847/PSJ/AC01EE}.
\newblock URL \url{https://iopscience.iop.org/article/10.3847/PSJ/ac01ee https://iopscience.iop.org/article/10.3847/PSJ/ac01ee/meta}.

\bibitem[Willhite et~al.(2021)Willhite, Ni, Arevalo, Bardyn, Gundersen, Minasola, Southard, Briois, Thirkell, Colin, et~al.]{willhite2021corals}
L.~Willhite, Z.~Ni, R.~Arevalo, A.~Bardyn, C.~Gundersen, N.~Minasola, A.~Southard, C.~Briois, L.~Thirkell, F.~Colin, et~al.
\newblock Corals: a laser desorption/ablation orbitrap mass spectrometer for in situ exploration of europa.
\newblock In \emph{2021 IEEE Aerospace Conference (50100)}, pages 1--13. IEEE, 2021.

\bibitem[Wolstencroft(1974)]{Wolstencroft1974}
R.~Wolstencroft.
\newblock \emph{Planets, stars and nebulae: studied with photopolarimetry}, volume~23.
\newblock University of Arizona Press, 1974.

\bibitem[Wong and Smyth(2000)]{Wong2000}
M.~C. Wong and W.~H. Smyth.
\newblock Model calculations for io's atmosphere at eastern and western elongations.
\newblock \emph{Icarus}, 146\penalty0 (1):\penalty0 60--74, 2000.
\newblock ISSN 0019-1035.
\newblock \doi{https://doi.org/10.1006/icar.2000.6362}.
\newblock URL \url{https://www.sciencedirect.com/science/article/pii/S0019103500963620}.

\bibitem[Woo et~al.(2022)Woo, Reinhardt, Cilibrasi, Chau, Helled, and Stadel]{Woo2022}
J.~M.~Y. Woo, C.~Reinhardt, M.~Cilibrasi, A.~Chau, R.~Helled, and J.~Stadel.
\newblock {Did Uranus' regular moons form via a rocky giant impactor?}
\newblock \emph{Icarus}, 375:\penalty0 114842, 3 2022.
\newblock ISSN 0019-1035.
\newblock \doi{10.1016/J.ICARUS.2021.114842}.

\bibitem[Wurz et~al.(2021)Wurz, Tulej, Riedo, Grimaudo, Lukmanov, and Thomas]{Wurz2021}
P.~Wurz, M.~Tulej, A.~Riedo, V.~Grimaudo, R.~Lukmanov, and N.~Thomas.
\newblock Investigation of the surface composition by laser ablation/ionization mass spectrometry.
\newblock In \emph{2021 IEEE Aerospace Conference (50100)}, pages 1--15, 2021.
\newblock \doi{10.1109/AERO50100.2021.9438486}.

\bibitem[Wurz et~al.(2022)Wurz, Tulej, Lukmanov, Grimaudo, Gruchola, Kipfer, de~Koning, Boeren, Schwander, Schmidt, Ligterink, and Riedo]{Wurz2022}
P.~Wurz, M.~Tulej, R.~Lukmanov, V.~Grimaudo, S.~Gruchola, K.~Kipfer, C.~de~Koning, N.~Boeren, L.~Schwander, P.~K. Schmidt, N.~F.~W. Ligterink, and A.~Riedo.
\newblock Identifying biosignatures on planetary surfaces with laser-based mass spectrometry.
\newblock In \emph{2022 IEEE Aerospace Conference (AERO)}, pages 01--16, 2022.
\newblock \doi{10.1109/AERO53065.2022.9843803}.

\bibitem[Wyttenbach et~al.(2017)Wyttenbach, Lovis, Ehrenreich, Bourrier, Pino, Allart, Astudillo-Defru, Cegla, Heng, Lavie, Melo, Murgas, Santerne, S{\'{e}}gransan, Udry, and Pepe]{Wyttenbach2017HotWASP-49b}
A.~Wyttenbach, C.~Lovis, D.~Ehrenreich, V.~Bourrier, L.~Pino, R.~Allart, N.~Astudillo-Defru, H.~M. Cegla, K.~Heng, B.~Lavie, C.~Melo, F.~Murgas, A.~Santerne, D.~S{\'{e}}gransan, S.~Udry, and F.~Pepe.
\newblock {Hot Exoplanet Atmospheres Resolved with Transit Spectroscopy (HEARTS): I. Detection of hot neutral sodium at high altitudes on WASP-49b}.
\newblock \emph{A{\&}A}, 602:\penalty0 A36, 6 2017.
\newblock ISSN 14320746.
\newblock \doi{10.1051/0004-6361/201630063}.

\bibitem[Xu et~al.(2025)Xu, Liu, Zhang, Lau, Cleaves, Huang, Glein, and Hao]{Xu2025}
W.~Xu, C.~Liu, A.~Zhang, M.~Lau, H.~J. Cleaves, F.~Huang, C.~R. Glein, and J.~Hao.
\newblock {Enough Sulfur and Iron for Potential Life Make Enceladus’s Ocean Fully Habitable}.
\newblock \emph{The Astrophysical Journal Letters}, 980\penalty0 (1):\penalty0 L10, 2 2025.
\newblock ISSN 2041-8205.
\newblock \doi{10.3847/2041-8213/ADAD65}.
\newblock URL \url{https://iopscience.iop.org/article/10.3847/2041-8213/adad65}.

\bibitem[Young et~al.(2004)Young, Berthelier, Blanc, Burch, Coates, Goldstein, Grande, Hill, Johnson, Kelha, Mccomas, Sittler, Svenes, Szeg{\"{o}}, Tanskanen, Ahola, Anderson, Bakshi, Baragiola, Barraclough, Black, Bolton, Booker, Bowman, Casey, Crary, Delapp, Dirks, Eaker, Funsten, Furman, Gosling, Hannula, Holmlund, Huomo, Illiano, Jensen, Johnson, Linder, Luntama, Maurice, Mccabe, Mursula, Narheim, Nordholt, Preece, Rudzki, Ruitberg, Smith, Szalai, Thomsen, Viherkanto, Vilppola, Vollmer, Wahl, W{\"{u}}est, Ylikorpi, and Zinsmeyer]{Young2004}
D.~T. Young, J.~J. Berthelier, M.~Blanc, J.~L. Burch, A.~J. Coates, R.~Goldstein, M.~Grande, T.~W. Hill, R.~E. Johnson, V.~Kelha, D.~J. Mccomas, E.~C. Sittler, K.~R. Svenes, K.~Szeg{\"{o}}, P.~Tanskanen, K.~Ahola, D.~Anderson, S.~Bakshi, R.~A. Baragiola, B.~L. Barraclough, R.~K. Black, S.~Bolton, T.~Booker, R.~Bowman, P.~Casey, F.~J. Crary, D.~Delapp, G.~Dirks, N.~Eaker, H.~Funsten, J.~D. Furman, J.~T. Gosling, H.~Hannula, C.~Holmlund, H.~Huomo, J.~M. Illiano, P.~Jensen, M.~A. Johnson, D.~R. Linder, T.~Luntama, S.~Maurice, K.~P. Mccabe, K.~Mursula, B.~T. Narheim, J.~E. Nordholt, A.~Preece, J.~Rudzki, A.~Ruitberg, K.~Smith, S.~Szalai, M.~F. Thomsen, K.~Viherkanto, J.~Vilppola, T.~Vollmer, T.~E. Wahl, M.~W{\"{u}}est, T.~Ylikorpi, and C.~Zinsmeyer.
\newblock {Cassini plasma spectrometer investigation}.
\newblock \emph{Space Science Reviews}, 114\penalty0 (1-4):\penalty0 1--112, 12 2004.
\newblock ISSN 00386308.
\newblock \doi{10.1007/S11214-004-1406-4}.

\bibitem[Zambon et~al.(2023)Zambon, Mura, Lopes, Rathbun, Tosi, Sordini, Noschese, Ciarniello, Cicchetti, Adriani, et~al.]{Zambon2023}
F.~Zambon, A.~Mura, R.~Lopes, J.~Rathbun, F.~Tosi, R.~Sordini, R.~Noschese, M.~Ciarniello, A.~Cicchetti, A.~Adriani, et~al.
\newblock Io hot spot distribution detected by juno/jiram.
\newblock \emph{Geophysical Research Letters}, 50\penalty0 (1):\penalty0 e2022GL100597, 2023.

\bibitem[Zhang and Hamilton(2007)]{Zhang2007}
K.~Zhang and D.~P. Hamilton.
\newblock {Orbital resonances in the inner neptunian system: I. The 2:1 Proteus–Larissa mean-motion resonance}.
\newblock \emph{Icarus}, 188\penalty0 (2):\penalty0 386--399, 6 2007.
\newblock ISSN 0019-1035.
\newblock \doi{10.1016/J.ICARUS.2006.12.002}.
\newblock URL \url{https://www.sciencedirect.com/science/article/pii/S0019103506004428}.

\bibitem[Zhang and Hamilton(2008)]{Zhang2008}
K.~Zhang and D.~P. Hamilton.
\newblock {Orbital resonances in the inner neptunian system: II. Resonant history of Proteus, Larissa, Galatea, and Despina}.
\newblock \emph{Icarus}, 193\penalty0 (1):\penalty0 267--282, 1 2008.
\newblock ISSN 0019-1035.
\newblock \doi{10.1016/J.ICARUS.2007.08.024}.
\newblock URL \url{https://www.sciencedirect.com/science/article/pii/S0019103507003776?via%3Dihub}.

\bibitem[Zhang and Nimmo(2009)]{Zhang2009}
K.~Zhang and F.~Nimmo.
\newblock {Recent orbital evolution and the internal structures of Enceladus and Dione}.
\newblock \emph{Icarus}, 204\penalty0 (2):\penalty0 597--609, 12 2009.
\newblock ISSN 0019-1035.
\newblock \doi{10.1016/J.ICARUS.2009.07.007}.
\newblock URL \url{https://www.sciencedirect.com/science/article/pii/S0019103509002887?casa_token=1utsnFYSoKsAAAAA:u2DE6gCzj_VjY3N8qaNPrBe8Is1kKcsrLuwPTte7jI0W7z6qS572ypVB2oEKpI59-e1d61LcMzgo}.

\bibitem[Zolotov(2017)]{Zolotov2017Ceres}
M.~Y. Zolotov.
\newblock {Aqueous origins of bright salt deposits on Ceres}.
\newblock \emph{Icarus}, 296:\penalty0 289--304, 11 2017.
\newblock ISSN 0019-1035.
\newblock \doi{10.1016/J.ICARUS.2017.06.018}.

\bibitem[Zsom et~al.(2013)Zsom, Seager, De~Wit, and Stamenkovi{\'c}]{zsom2013toward}
A.~Zsom, S.~Seager, J.~De~Wit, and V.~Stamenkovi{\'c}.
\newblock Toward the minimum inner edge distance of the habitable zone.
\newblock \emph{The Astrophysical Journal}, 778\penalty0 (2):\penalty0 109, 2013.

\end{thebibliography}

\begin{table*}[hbtp]
\centering
\caption{Observations of different water-related species in the atmospheres of Io, Europa and Enceladus.}
\resizebox{\textwidth}{!}{%
\begin{tabular}{cccccc}
\hline
\hline
Species  & Column density               & Instrument     & Observation               & Notes                           & Reference \\
         & [cm$^{-2}$]                  &                &                           & (e.g., hemisphere )             & \\
\hline
\multicolumn{6}{c}{Europa} \\
\hline
H$_2$O & $\sim1.5\cdot10^{16}$        & HST/STIS        & UV (Ly-$\alpha$)       & south polar region           & \cite{Roth2014}$^{pl}$ \\        
       & $(1.2 \pm 0.5)\cdot10^{14}$  & Keck/HIRES      & optical                & sub-Jovian           		& \cite{deKleer2023TheCallisto} \\ 
H$_2$  & $(1.7{-}1.9) \cdot 10^{13}$  & Juno/JADE       & --                     & pickup ion observations      & \cite{Szalay2024} \\             
O$_2$  & $(1.5\pm 0.5)\cdot10^{15}$   & HST/GHRS        & far-UV                 & trailing                     & \cite{Hall1995} \\               
       & $(2.4{-}14)\cdot10^{14}$     & HST/GHRS        & far-UV                 & leading \& trailing          & \cite{hall1998} \\               
       & $(7.4{-}12.4)\cdot10^{14}$   & Cassini/UVIS    & far-UV                 & --                           & \cite{Hansen2005} \\             
       & $\sim5\cdot10^{15}$          & HST/STIS        & UV (Ly-$\alpha$)       & south polar region           & \cite{Roth2014}$^{pl}$ \\
       & $(3{-}6)\cdot10^{14}$        & HST/ACS         & far-UV                 & \textit{leading \& trailing} & \cite{Roth2016} \\               
       & $(4.1 \pm 0.1)\cdot10^{14}$  & Keck/HIRES      & optical                & sub-Jovian                   & \cite{deKleer2023TheCallisto} \\ 
O      & ${<}2\cdot10^{14}$           & HST/GHRS        & far-UV                 & trailing                     & \cite{Hall1995} \\               
       & ${<}(1.6{-}3.4)\cdot10^{13}$ & HST/GHRS        & far-UV                 & leading \& trailing          & \cite{hall1998} \\               
       & $(1.7{-}3.1)\cdot10^{13}$    & Cassini/UVIS    & far-UV                 & --                           & \cite{Hansen2005} \\             
       & ${<}1\cdot10^{13}$           & Keck/HIRES      & optical                & sub-Jovian                   & \cite{deKleer2023TheCallisto} \\ 
H      & $(2.3{-}3.5)\cdot10^{11}$    & HST/STIS        & UV (Ly-$\alpha$)       & \textit{leading \& trailing} & \cite{Roth2017} \\
\hline
\multicolumn{6}{c}{Io} \\
\hline
SO$_2$ & $6 \cdot 10^{17}$                				   & IRAM          & mm      &  --                & \cite{lellouch1992} \\
       & $(0.6{-}1.0)\cdot 10^{16}$/${<}3 \cdot 10^{17}$ 	& HST/FOS       & UV      & trailing           & \cite{ballester1994} \\
       & ${<}4 \cdot 10^{16}$             					& HST/FOS       & UV      &  --                & \cite{clarke1994} \\
       & $(5.0{/}7.0)\cdot 10^{15}$/${<}9.3 \cdot 10^{16}$ & HST/GHRS      & UV      & leading/trailing   & \cite{trafton1996} \\
       & $3.7 \cdot 10^{17}$              					& HST/WFPC      & near-UV &  --               & \cite{Spencer1997}$^{pl}$ \\
       & $(0.70{-}3.25)\cdot 10^{16}$     					& HST/FOS       & near-UV &  --                & \cite{mcgrath2000} \\
       & $(1{-}4)\cdot 10^{16}$           					& HST/STIS      & UV (Ly-$\alpha$)      &   --                   & \cite{feldman2000} \\
       & $(7{\pm}3)\cdot 10^{16}$         					& HST/STIS/WFPC & UV      & Pele               & \cite{spencer2000}$^{pl}$ \\
       & $(1{-}2)\cdot 10^{16}$           					& HST/STIS      & UV (Ly-$\alpha$)      &    --                  & \cite{strobel2001} \\
       & $(0.5{-}1.0)\cdot 10^{16}$       					& HST/STIS      & UV      &   --                   & \cite{feaga2002} \\
       & $(1.4{-}2.4)\cdot 10^{17}$       					& HST/STIS      & UV      & Prometheus         & \cite{jessup2004}$^{pl}$ \\
       & $(0.15{-}1.50)\cdot 10^{17}$     					& IRTF/TEXES    & mid-IR  &   --                   & \cite{spencer2005} \\
       & $(1{-}5)\cdot 10^{16}$           					& HST/STIS      & UV      & Pele               & \cite{jessup2007}$^{pl}$ \\
       & ${<}1.5 \cdot 10^{17}$           					& IRAM/PdBI     & mm      &   --                   & \cite{moullet2008} \\
       & $(0.018{-}5)\cdot 10^{16}$       					& HST/STIS      & UV (Ly-$\alpha$)      &                    & \cite{feaga2009} \\
       & $(2.3{-}4.6)\cdot 10^{16}$       					& SMA           & mm      & leading            & \cite{moullet2010} \\
       & $(0.7{-}1.1)\cdot 10^{16}$       					& SMA           & mm      & trailing           & \cite{moullet2010} \\
       & $(0.61{/}1.51)\cdot 10^{17}$     					& IRTF/TEXES    & mid-IR  & aphel./peri. (anti-Jov.) & \cite{tsang2012} \\
       & $(5.5{\pm}0.7)\cdot 10^{15}$     					& APEX          & sub mm  &   --                   & \cite{moullet2013} \\
       & $(0.3{-}1.5)\cdot 10^{17}$       					& VLT/CRIRES    & near-IR &    --                  & \cite{lellouch2015} \\
       & $(1.5{\pm}0.3)\cdot 10^{16}$     					& ALMA          & sub mm  &    --                  & \cite{depater2020} \\
       & $(0.75{-}1.19)\cdot 10^{16}$     					& IRAM/NOEMA    & mm      &    --                  & \cite{roth2020} \\
       & ${<}2 \cdot 10^{17}$             					& HST/STIS      & UV (Ly-$\alpha$)      &    --                  & \cite{giono2021} \\
       & $(1.030{\pm}0.032)\cdot 10^{16}$ 					& ALMA          & sub mm  & leading            & \cite{dekleer2024} \\
       & $(3.53{\pm}0.21)\cdot 10^{15}$   					& ALMA          & sub mm  & trailing           & \cite{dekleer2024} \\
       & $(0.9{/}2.1)\cdot 10^{17}$       					& IRTF/TEXES    & mid-IR  & aphel./peri. (anti-Jov.) & \cite{Giles2024} \\
       & $(1{/}4)\cdot 10^{16}$           					& IRTF/TEXES    & mid-IR  & aphel./peri.  (sub-Jov.) & \cite{Giles2024} \\
SO     & $(2.0{-}6.0)\cdot 10^{14}$       				   & IRAM          & mm      &   --                   & \cite{lellouch1996} \\
       & $(0.5{-}2.5)\cdot 10^{15}$       					& HST/FOS       & near-UV &   --                   & \cite{mcgrath2000} \\
       & $(1.0{-}7.0)\cdot 10^{14}$       					& SMA           & mm      &    --                  & \cite{moullet2010} \\
       & $(3.8{-}4.5)\cdot 10^{14}$       					& APEX          & sub mm  &    --                  & \cite{moullet2013} \\
       & $1 \cdot 10^{15}$                					& ALMA          & sub mm  &    --                  & \cite{depater2020} \\
S$_2$  & $(1.0{\pm}0.2)\cdot 10^{16}$     				   & HST/STIS/WFPC & UV      & Pele               & \cite{spencer2000}$^{pl}$ \\
       & ${<}7.5 \cdot 10^{14}$           					& HST/STIS      & UV      & Prometheus         & \cite{jessup2004}$^{pl}$ \\
       & $(1.0{-}4.0)\cdot 10^{15}$       					& HST/STIS      & UV      & Pele               & \cite{jessup2007}$^{pl}$ \\
S      & $(0.5{-}1.0)\cdot 10^{14}$       				   & HST/FOS       & near-UV &    --                  & \cite{mcgrath2000} \\
       & $(0.36{-}1.70)\cdot 10^{13}$     					& HST/STIS      & UV      &     --                 & \cite{feaga2002} \\
NaCl   & $(0.8{-}20)\cdot 10^{13}$        				   & IRAM          & mm      &     --                 & \cite{lellouch2003} \\
       & $(1.2{-}1.4)\cdot 10^{13}$       					& IRAM/NOEMA    & mm      &    --                  & \cite{roth2020} \\
       & $(5.1{\pm}2.0)\cdot 10^{13}$     					& ALMA          & sub mm  & leading            & \cite{dekleer2024} \\
       & $(3.3{\pm}1.8)\cdot 10^{13}$     					& ALMA          & sub mm  & trailing           & \cite{dekleer2024} \\
KCl    & $(3.0{\pm}1.0)\cdot 10^{12}$     				   & APEX          & sub mm     &     --                 & \cite{moullet2013} \\
       & $(9.9{\pm}3.9)\cdot 10^{12}$     					& ALMA          & sub mm  & leading            & \cite{dekleer2024} \\
       & $(3.5{\pm}2.0)\cdot 10^{12}$     					& ALMA          & sub mm  & trailing           & \cite{dekleer2024} \\
\hline
\multicolumn{6}{c}{Enceladus} \\
\hline
H$_2$O & $(0.9{\pm}0.23) \cdot 10^{16}$  & Cassini/UVIS & extreme UV                   & --                           & \cite{Hansen2011}$^{pl}$ \\
       & $4.0 \cdot 10^{13}$        & Herschel     & sub-mm                       & torus                        & \cite{hartogh2011} \\
       & $4.5 \cdot 10^{13}$        & JWST/NIRSpec & near-IR                      & torus                        & \cite{Villanueva2023} \\
       & $3.6 \cdot 10^{15}$        & Cassini/VIMS & --                           & --                           & \cite{Denny2024ConstrainingSpectrometer}$^{pl}$ \\
N$_2$  & ${<}5 \cdot 10^{13}$         & Cassini/UVIS & extreme UV                   & --                           & \cite{Hansen2011}$^{pl}$\\
\hline
\multicolumn{6}{c}{Notes. $^{pl}$plume observation}\\
\end{tabular}
}
\label{tab:data}
\end{table*}

\end{document}